\documentclass[12pt]{iopart}
\usepackage{graphicx}
 
\def\beq{\begin{equation}} 
\def\eeq{\end{equation}} 
\def\beqn{\begin{eqnarray}} 
\def\eeqn{\end{eqnarray}} 
\def\as{\alpha_S} 
 
\def\sighat{\hat{\sigma}}

\def\qq{q \bar{q}}

\def\c2w{\cos^2{\theta_W}} 
\def\lapproxeq{\lower .7ex\hbox{$\;\stackrel{\textstyle <}{\sim}\;$}} 
\def\gapproxeq{\lower .7ex\hbox{$\;\stackrel{\textstyle >}{\sim}\;$}} 
 
\def\la{\langle}
\def\ra{\rangle}

\def\msb{{\overline{\rm MS}}}

\begin{document}

\review{Hard Interactions of Quarks and Gluons: a Primer for LHC Physics}
\author{J. M. Campbell\\
Department of Physics and Astronomy\\
University of Glasgow\\
Glasgow G12 8QQ\\
United Kingdom\\}

\author{J. W. Huston\\
Department of Physics and Astronomy\\
Michigan State University\\
East Lansing, MI 48824\\
USA\\
}

\author{W. J. Stirling\\
Institute for Particle Physics Phenomenology\\
University of Durham\\
Durham DH1 3LE\\
United Kingdom\\}

\begin{abstract}	
In this review article, we will develop the perturbative framework for the calculation of hard scattering processes. We
will undertake to provide both  a reasonably rigorous development of the formalism of  hard scattering of quarks and gluons
as well as an intuitive understanding of the physics behind the scattering. We will emphasize the role of logarithmic
corrections as well as power counting in $\as$ in order to understand the behaviour of  hard scattering processes.  We
will include ``rules of thumb'' as well as ``official recommendations'', and where possible will seek to dispel some myths.
We will also discuss  the impact  of soft processes on the measurements  of hard scattering processes. Experiences that
have been gained at the Fermilab Tevatron will be recounted and, where appropriate, extrapolated to the LHC.
\end{abstract}

\submitto{\RPP}

\maketitle
\tableofcontents
			      

\section{Introduction}
\label{sec:intro}

Scattering processes at high energy hadron colliders can be classified as either hard or soft. Quantum Chromodynamics (QCD)
is the underlying theory for all such processes, but the approach and level of understanding is very different for the two
cases. For hard processes, e.g. Higgs boson or high $p_T$ jet production, the rates and event properties can be predicted with
good precision using perturbation theory. For soft processes, e.g. the total cross section, the underlying event etc., the
rates and properties are dominated by non-perturbative QCD effects, which are less well understood.  For many hard
processes, soft interactions are occurring along with the hard interactions and their effects must be understood for
comparisons to be made to perturbative predictions. An understanding of the rates and characteristics of predictions for
hard processes, both signals and backgrounds, using perturbative QCD (pQCD) is crucial for both the Tevatron and LHC. 

In this review article, we will develop the perturbative framework for the calculation of hard scattering processes. We
will undertake to provide both  a reasonably rigorous development of the formalism of  hard scattering of quarks and gluons
as well as an intuitive understanding of the physics behind the scattering. We will emphasize the role of logarithmic
corrections as well as power counting in $\as$ in order to understand the behaviour of  hard scattering processes.  We
will include ``rules of thumb'' as well as ``official recommendations'', and where possible will seek to dispel some myths.
We will also discuss  the impact  of soft processes on the measurements  of hard scattering processes. Given the limitations
of space, we will concentrate on a few processes, mostly inclusive jet, $W/Z$ production, and $W/Z+$jets, but the lessons
should be useful for other processes at the Tevatron and LHC as well.
As a bonus feature, this paper is accompanied by a ``benchmark 
website''~\footnote{www.pa.msu.edu/\~{}huston/Les\_Houches\_2005/Les\_Houches\_SM.html}, where updates and more
detailed discussions than are possible in this limited space will be available. We will refer to this website on several
occasions in the course of this review article. 

In Section~\ref{sec:formalism}, we introduce the hard scattering formalism and the QCD factorization theorem. In
Section~\ref{sec:partonxsecs}, we apply this formalism to some basic processes at leading order, next-to-leading order and
next-to-next-to-leading order. Section~\ref{sec:pdfs} provides a detailed discussion of parton  distribution functions (pdfs) and
global pdf fits and in Section~\ref{sec:tevatron} we compare the predictions of pQCD to measurements at the Tevatron. Lastly, in
Section~\ref{sec:lhc}, we  provide some benchmarks and predictions for  measurements to be performed at the LHC. 

\section{Hard scattering formalism and the QCD factorization theorem} 
\label{sec:formalism}
\subsection{Introduction}
\label{sec:formintro}

In this section we will discuss in more detail how the QCD factorization theorem can be used to calculate a wide
variety of hard scattering cross sections in hadron-hadron collisions. For simplicity we will restrict our
attention to leading-order processes and calculations; the extension of the formalism to more complicated
processes and to include higher-order perturbative contributions will be discussed in
Sections~\ref{sec:partonxsecs} and~\ref{sec:lhc}.

\begin{figure}[htbp]
\begin{center}    
\includegraphics[width=6cm]{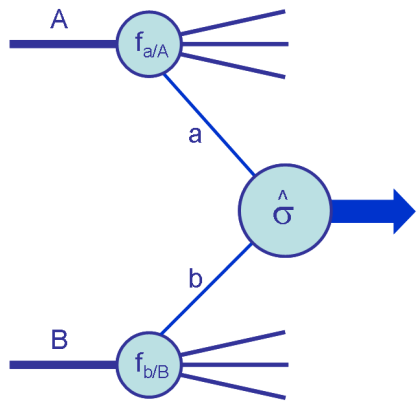}
\end{center}    
\vspace*{-0.5cm}
\caption{Diagrammatic structure of a generic hard scattering process.}
\label{basicstruct}
\end{figure}

We begin with a brief review of the factorization theorem. It was first pointed out by Drell and Yan
\cite{DRELL}  more than 30 years ago that parton model ideas developed for deep inelastic scattering could be
extended to certain processes in hadron-hadron collisions. The paradigm process was the production of a massive
lepton pair by quark-antiquark annihilation --- the Drell--Yan process --- and it was postulated that the
hadronic cross section $\sigma(AB \to \mu^+\mu^- + X)$ could be obtained by weighting the subprocess cross
section $\sighat$ for $\qq\rightarrow\mu^+\mu^-$ with the parton distribution functions (pdfs) $f_{q/A}(x)$ extracted
from deep inelastic scattering:
\begin{equation}
\sigma_{AB} = \int dx_a dx_b\; f_{a/A}(x_a) f_{b/B}(x_b)\;
\sighat_{ab\to X}\; ,
\label{siglo}
\end{equation}
where for the Drell--Yan process, $X=l^+l^-$ and $ab = q\bar{q},\;
\bar{q}q$.
The domain of  validity is  the asymptotic ``scaling'' limit
(the analogue of the Bjorken scaling limit in deep inelastic scattering)
$M_X \equiv M_{l^+l^-}^2, s \to
\infty$, $\tau = M_{l^+l^-}^2/s$ fixed.    The good agreement  between
theoretical predictions and the measured cross sections provided
confirmation of the parton model formalism, and allowed for the first
time a rigorous, quantitative treatment of certain hadronic cross sections.
Studies were extended to other ``hard scattering'' processes, for
example the production of hadrons and photons with large transverse momentum,
with equally successful results. Problems, however,
appeared to arise when perturbative corrections
from real and virtual gluon emission were calculated. Large
logarithms from gluons emitted collinear with the incoming quarks
appeared to spoil the convergence of the perturbative expansion.
It was subsequently  realized that these logarithms were the same as those
that arise in deep inelastic scattering structure function
calculations,
and could therefore be absorbed, via the DGLAP
equations, in the definition of the
parton distributions, giving rise to logarithmic violations of
scaling. The key point was that {\it all} logarithms appearing in the Drell--Yan
corrections could be factored into renormalized parton distributions
in this way, and {\it factorization theorems} which showed that this
was a general feature of hard scattering processes were
derived~\cite{FACT}. Taking into account the leading logarithm corrections, \eref{siglo} simply becomes:
\begin{equation}
\sigma_{AB} = \int dx_a dx_b\; f_{a/A}(x_a,Q^2) f_{b/B}(x_b,Q^2)\;
\sighat_{ab\to X}\; .
\label{sigll}
\end{equation}
corresponding to the structure depicted in Figure~\ref{basicstruct}. The $Q^2$ that appears in the parton
distribution functions (pdfs) is a large momentum scale that characterizes the hard scattering, e.g.
$M_{l^+l^-}^2$, $p_T^2$, ... . Changes to the $Q^2$  scale of ${\cal O}(1)$, e.g. $Q^2 = 2 M_{l^+l^-}^2, \;
M_{l^+l^-}^2/2 $ are equivalent in this leading logarithm approximation.

The final step in the theoretical development was the recognition that the {\it finite}
corrections left behind after the logarithms had been factored were
not universal and had to be calculated separately for each process, giving rise
to perturbative ${\cal O}(\as^n)$ corrections to the leading logarithm
cross section of \eref{sigll}. Schematically
\begin{equation}
\fl
\sigma_{AB} = \int dx_a dx_b\>  f_{a/A}(x_a,\mu_F^2) \>
f_{b/B}(x_b,\mu_F^2)\;
\times \;  [\> \sighat_0\> +\> \as(\mu_R^2)\> \sighat_1\> +\> ...\> ]_{ab\to X} \ .
\label{signlo}
\end{equation}
Here $\mu_F$ is the {\it factorization scale}, which can be thought of as the scale that separates the long- and
short-distance physics, and $\mu_R$ is the {\it renormalization scale} for the QCD running coupling. Formally,
the cross section calculated to all orders in perturbation theory is invariant under changes in these
parameters, the $\mu_F^2$ and $\mu_R^2$ dependence of the coefficients,  e.g. $\sighat_1$, exactly compensating
the explicit scale dependence of the parton distributions and the coupling constant. This compensation becomes
more exact as more terms are included in the perturbation series, as will be discussed in more detail in
Section~\ref{sec:scaledep}. In the absence of a complete set of higher order corrections, it is necessary to make a
specific choice for the two scales in order to make cross section predictions. Different choices will yield
different (numerical) results, a reflection of the uncertainty in the prediction due to unknown higher order
corrections, see Section~3. To avoid unnaturally large logarithms reappearing in the perturbation series it is
sensible to choose $\mu_F$ and $\mu_R$ values of the order of the typical momentum scales of the hard scattering
process, and $\mu_F = \mu_R$ is also often assumed. For the Drell-Yan process, for example, the standard choice
is $\mu_F = \mu_R = M$, the mass of the lepton pair.

The recipe for using the above leading-order formalism to calculate a cross section for a given (inclusive)
final state $X + $anything is very simple: (i) identify the leading-order partonic process that contributes to
$X$, (ii) calculate the corresponding $\sighat_0$, (iii) combine with an appropriate combination (or
combinations) of pdfs for the initial-state partons $a$ and $b$, (iv) make a specific choice for the scales
$\mu_F$ and $\mu_R$, and (v) perform a numerical integration over the variables $x_a$, $x_b$ and any other
phase-space variables associated with the final state $X$. Some simple examples are 
\bigskip
\begin{center}
\begin{tabular}{|l|l|}
\hline
$Z$ production & $q \bar q \to Z$ \\
top quark production & $ q \bar q \to t \bar t, \ g g \to t \bar t$ \\
large $E_T$ jet production &  $ gg \to gg, \ qg \to qg, \ qq \to qq$ etc. \\
\hline
\end{tabular} 
\end{center}
%
\bigskip
\noindent where appropriate scale choices are $M_Z$, $m_t$, $E_T$ respectively. Expressions for the
corresponding subprocess cross  sections $\sighat_0$ are widely available in the literature, see for example~\cite{Ellis:1991qj}.

The parton distributions used in these hard-scattering calculations are solutions of the DGLAP
equations~\cite{DGLAP}~\footnote{The DGLAP equations effectively sum leading powers of $[\as \log \mu^2]^n$
generated by multiple gluon emission in a region of phase space where the gluons are strongly ordered in
transverse momentum. These are the dominant contributions when $\log(\mu)\gg \log(1/x)$.} 
\beqn 
\fl
{\partial q_i(x,\mu^2) \over \partial \log \mu^2} = {\as\over 2\pi} 
\int_x^1 \frac{dz}{z} \Big\{
    P_{q_i q_j}(z,\as) q_j(\frac{x}{z},\mu^2)
     +  P_{q_i g}(z,\as)\; g(\frac{x}{z},\mu^2)
    \Big\}    \nonumber \\
\fl
 {\partial g(x,\mu^2) \over \partial \log \mu^2} =  {\as\over 2\pi} \int_x^1
\frac{dz}{z} \Big\{
    P_{g q_j}(z,\as) q_j(\frac{x}{z},\mu^2)  +
    P_{g g}(z,\as) g(\frac{x}{z},\mu^2)
    \Big\}
\label{eq:DGLAP}
\eeqn
where the splitting functions have perturbative expansions: 
\beq
 P_{ab}(x,\as) = P_{ab}^{(0)}(x) + {\as\over 2\pi} P_{ab}^{(1)}(x) + ...
\label{eq:splitting}
\eeq
Expressions for the leading order (LO) and next-to-leading order (NLO) splitting functions can be found in~\cite{Ellis:1991qj}.
The DGLAP equations
determine the $Q^2$ dependence of the pdfs. The $x$ dependence, on the other hand, has to be obtained from
fitting deep inelastic and other hard-scattering data. This will be discussed in more
detail in Section~\ref{sec:pdfs}. Note that for consistency, the order of the expansion of the splitting
functions should be the same as that of the
subprocess cross section, see~\eref{signlo}. Thus, for example, a full NLO calculation will include both
the $\sighat_1$ term in~\eref{signlo} and the $P_{ab}^{(1)}$ terms in the determination of the pdfs via
\eref{eq:DGLAP} and~\eref{eq:splitting}.

\begin{figure}[t]                                                               
\begin{center}
\includegraphics[width=12cm]{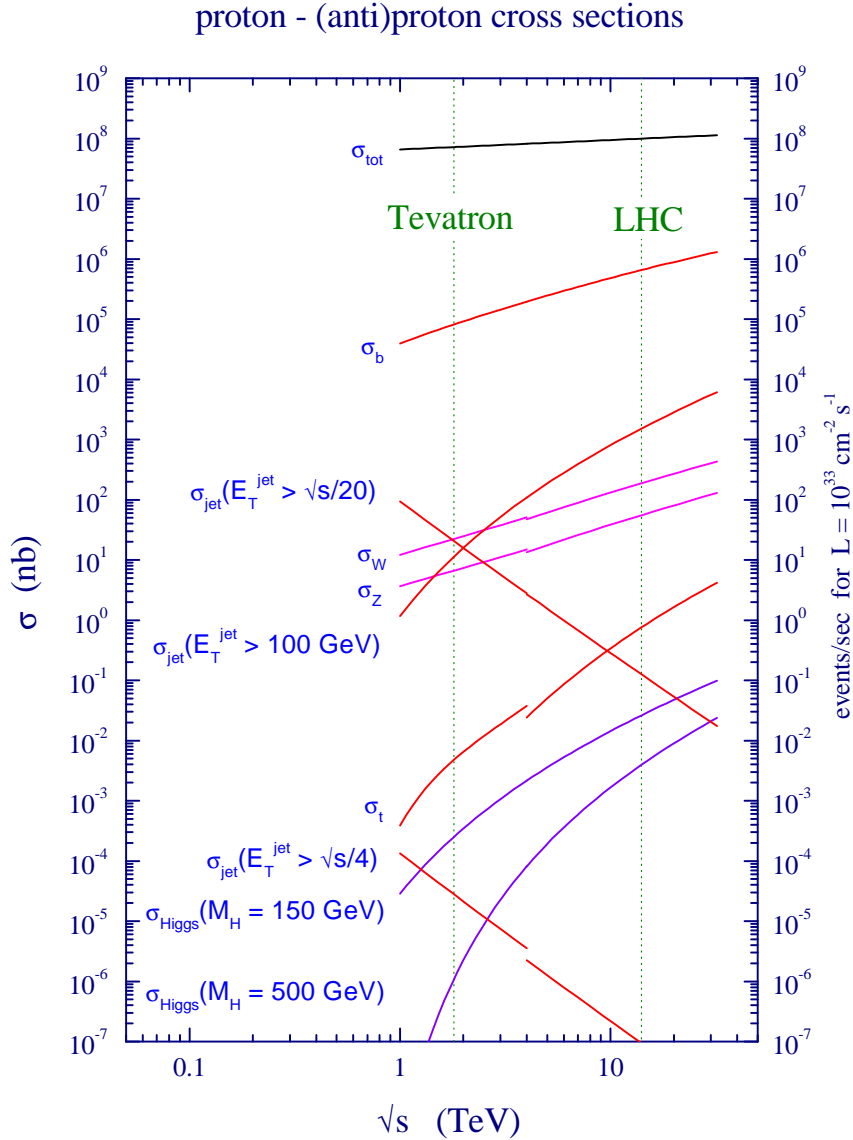}
\end{center}
\vspace*{-1cm}
\caption{Standard Model cross sections at the Tevatron
and LHC colliders.\label{fig:lhcall}}                    
\end{figure}

Figure~\ref{fig:lhcall} shows the predictions for
some important Standard Model cross sections at  $p \bar p$ and  $pp$  colliders, calculated
using the above formalism (at next-to-leading order in perturbation theory, i.e. including also the 
$\sighat_1$ term in \eref{signlo}). 

We have already mentioned that the Drell--Yan process is the paradigm
hadron--collider hard scattering process, and so we will discuss this in
some detail in what follows. Many of the remarks apply also
to other processes, in particular those shown in Figure~\ref{fig:lhcall},
although of course the higher--order corrections and the initial--state
parton combinations are process dependent.

\subsection{The Drell--Yan process}
\label{sec:DY}

The Drell--Yan process is the production of a lepton pair ($e^+e^-$ or
$ \mu^+ \mu^-$ in practice) of large invariant mass $M$ in hadron-hadron collisions by the
mechanism of quark--antiquark annihilation~\cite{DRELL}.
In the basic  Drell--Yan mechanism, a quark and antiquark
annihilate to produce  a virtual photon, $q \bar q \to \gamma^* \to 
l^+l^-$.  At high-energy colliders, such as the Tevatron and LHC, there is
of course sufficient
centre--of--mass energy for the production of on--shell $W$ and $Z$ bosons
as well.
The cross section for quark-antiquark annihilation to a lepton
pair via an intermediate massive photon is easily obtained from the
fundamental QED $e^+e^-\rightarrow \mu^+\mu^-$ cross section, with the addition
of the appropriate colour and charge factors.
\beq
\hat\sigma(q\bar{q} \rightarrow e^+e^-) 
=\frac{4 \pi \alpha^2}{3 \hat{s}}
\frac{1}{N} Q_q^2  ,
\label{dy:1}
\eeq
where $Q_q$ is the quark charge: $Q_u = +2/3$, $Q_d = -1/3$ {\it etc}.
The overall colour factor of $1/N= 1/3$ is due to the fact that only when the 
colour of the quark matches with the colour of the antiquark can 
annihilation into a colour--singlet final state take place.

In general, the incoming quark and antiquark will have a spectrum of
centre--of--mass energies $\sqrt{\hat{s}}$, and so it is more appropriate to consider
the differential mass distribution:
\beq 
\frac{d \hat{\sigma} }{d M^2}  = \frac{\hat{\sigma}_0}{N} Q_q^2 \delta (\hat{s}
-M^2), \;\;\hat{\sigma}_0 = \frac{4 \pi \alpha^2}{3 M^2} ,
\label{dy:2}
\eeq
where $M$ is the mass of the lepton pair.
In the centre--of--mass frame of the two hadrons, the components of momenta of
the incoming partons may be written as
\beqn 
p_1^\mu &=& \frac{\sqrt{s}}{2} ( x_1,0,0,x_1) \nonumber \\
p_2^\mu &=& \frac{\sqrt{s}}{2} ( x_2,0,0,-x_2)\; . 
\label{dy:3}
\eeqn

\begin{figure}[t]                                                               
\begin{center}     
\includegraphics[width=8cm]{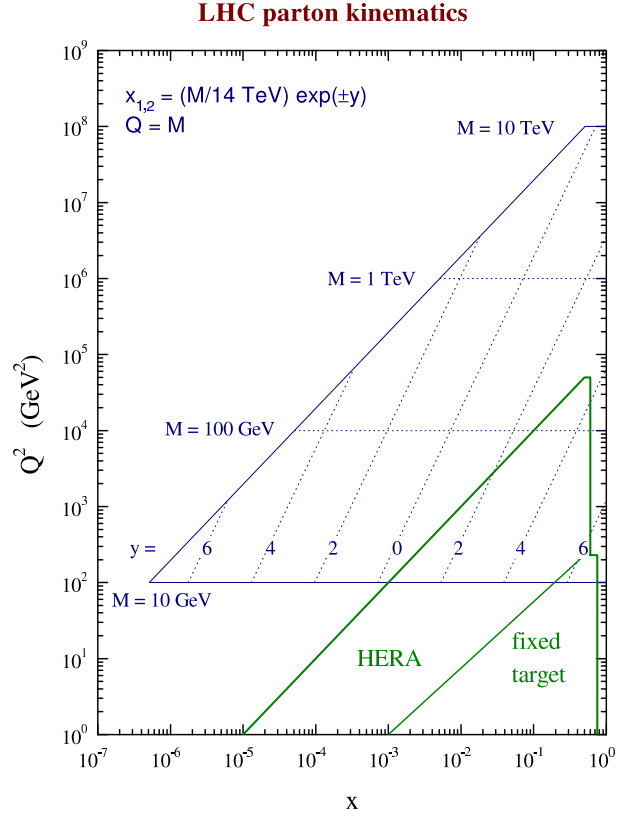}     
\end{center}    
\caption{Graphical representation of the relationship between parton $(x,Q^2)$ variables and the kinematic
variables corresponding to a final state of mass $M$ produced with rapidity $y$
at the LHC collider with $\sqrt{s} = 14$~TeV.\label{fig:lhcxq}}                    
\end{figure}

The square of the parton centre--of--mass energy $\hat{s}$ is related to the corresponding  hadronic quantity
by  $\hat{s}= x_1 x_2 s$. Folding in the pdfs for the initial state quarks and antiquarks in the colliding beams
gives  the hadronic cross section:
\beqn
 \frac{d \sigma}{d M^2} &=& \frac{\hat{\sigma}_0}{N} \int_0^1 {dx_1} {dx_2} 
 \delta(x_1 x_2 s -M^2)\nonumber \\
& &\times \quad \Big[ \sum_k\;  Q_k^2\; \big( q_k(x_1,M^2) \bar{q}_k(x_2,M^2)
                   +\big[ 1 \leftrightarrow 2 \big]\big)  \Big]\; .
\label{dy:4}
\eeqn
From \eref{dy:3}, the rapidity of the produced lepton pair 
is found to be $y= \textstyle{1/2 } \log(x_1/x_2)$, and hence  
\beq
x_1 =\frac{M}{\sqrt{s}}\;  {\rm e}^y\ ,\qquad x_2 =\frac{M}{\sqrt{s}}\;  {\rm e}^{-y}.
\label{dy:6}
\eeq
The double--differential cross section is therefore
\beq
\frac{d \sigma}{d M^2 dy } = \frac{\hat{\sigma}_0}{N s } 
\Big[ \sum_k\;  Q_k^2 \big( q_k(x_1,M^2) \bar{q}_k(x_2,M^2)
                   +\big[ 1 \leftrightarrow 2 \big]\big)  \Big].
\label{dy:7}
\eeq
with $x_1$ and $x_2$ given by \eref{dy:6}. Thus different values of $M$ and $y$ probe different values of the
parton $x$ of the colliding beams. The formulae relating $x_1$ and $x_2$ to $M$ and $y$ of course also apply to
the production of {\it any} final state with this mass and rapidity. Assuming the factorization scale ($Q$) is
equal to $M$, the mass of the final state, the relationship between the parton $(x,Q^2)$ values and the
kinematic variables $M$ and $y$ is illustrated pictorially in Figure~\ref{fig:lhcxq}, for the LHC collision energy
$\sqrt{s} = 14$~TeV.
For a given rapidity $y$ there are two (dashed) lines, corresponding to the values of $x_1$
and $x_2$. For $y=0$, $x_1 = x_2 = M/\sqrt{s}$. 

In analogy with the Drell--Yan cross section derived above,
the subprocess cross sections for (on--shell) $W$ and $Z$ production are readily
calculated to be
\beqn
\hat{\sigma}^{q\bar{q}'\rightarrow W}&=& 
\frac{\pi}{3} \sqrt{2} G_F M_W^2 \vert V_{q q^\prime} \vert^2
\delta (\hat{s} -M_W^2) , \nonumber \\
\hat{\sigma}^{q\bar{q}\rightarrow Z}&=& \frac{\pi }{3}
\sqrt{2} G_F M_Z^2 (v_q^2+a_q^2) 
\delta (\hat{s} -M_Z^2),
\label{eq:WZlo}
\eeqn
where $V_{q q^\prime}$ is the appropriate  Cabibbo--Kobayashi--Maskawa matrix element, and $v_q$ ($a_q$) is the
vector (axial vector) coupling of the $Z$ to the quarks. These formulae are valid in the {\it narrow width
production} in which  the decay width of the gauge boson is neglected. The resulting cross sections can then be
multiplied by the branching ratio for any particular hadronic or leptonic final state of interest.

\begin{figure}[t]
\begin{center}    
\includegraphics[width=5cm,angle=90]{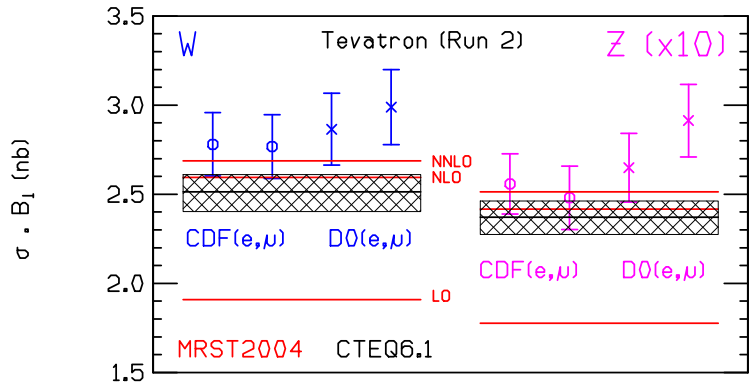}    
\end{center}    
\caption{Predictions for the $W$ and $Z$ total cross sections at the Tevatron and LHC, 
using MRST2004~\protect\cite{Martin:2004ir} and CTEQ6.1 pdfs~\cite{Stump:2003yu},
compared with recent data from CDF and D0.
The MRST predictions are shown at LO, NLO and NNLO.
The CTEQ6.1 NLO predictions and the accompanying pdf error bands are also shown.}
\label{fig:wz}
\end{figure}

High-precision measurements of $W$ and $Z$ production cross sections from the Fermilab Tevatron $p \bar p$
collider are available and allow the above formalism to be tested quantitatively. Thus Figure~\ref{fig:wz} shows
the cross sections for $ W^\pm$ and $Z^0$ production and decay
into various leptonic final states from the CDF~\cite{Acosta:2004uq}
and D0~\cite{D0WZ} collaborations at the Tevatron. The theoretical predictions are
calculated at LO (i.e. using~\eref{eq:WZlo}),
NLO and NNLO (next-to-next-to-leading order) in perturbation theory using the  $\msb$ scheme MRST parton distributions
of~\cite{Martin:2004ir},
with renormalization and factorization scales $\mu_F=\mu_R = M_W,M_Z$. The net effect of
the NLO and NNLO corrections, which will be discussed in more detail in Sections~\ref{sec:nlo} and~\ref{sec:nnlo},
is to increase the lowest-order cross section by about  $25\%$ and $5\%$ respectively.

Perhaps the most important point to note from Figure~\ref{fig:wz} is that, aside from unknown (and presumably
small) $O(\as^3)$ corrections, there is virtually no theoretical uncertainty associated with the predictions --
the parton distributions are being probed in a range of $x \sim M_W/\sqrt{s}$ where they are constrained from
deep inelastic scattering, see Figure~\ref{fig:lhcxq}, and the scale dependence is weak~\cite{Martin:2004ir}.  This
overall agreement with experiment, therefore, provides a powerful test of the whole theoretical edifice that
goes into the calculation. Figure~\ref{fig:wz} also illustrates the importance of higher-order perturbative
corrections when making detailed comparisons of data and theory. 

\subsection{Heavy quark production}
\label{sec:hqrk}

\begin{figure}[t]
\begin{center}    
\includegraphics[width=9cm]{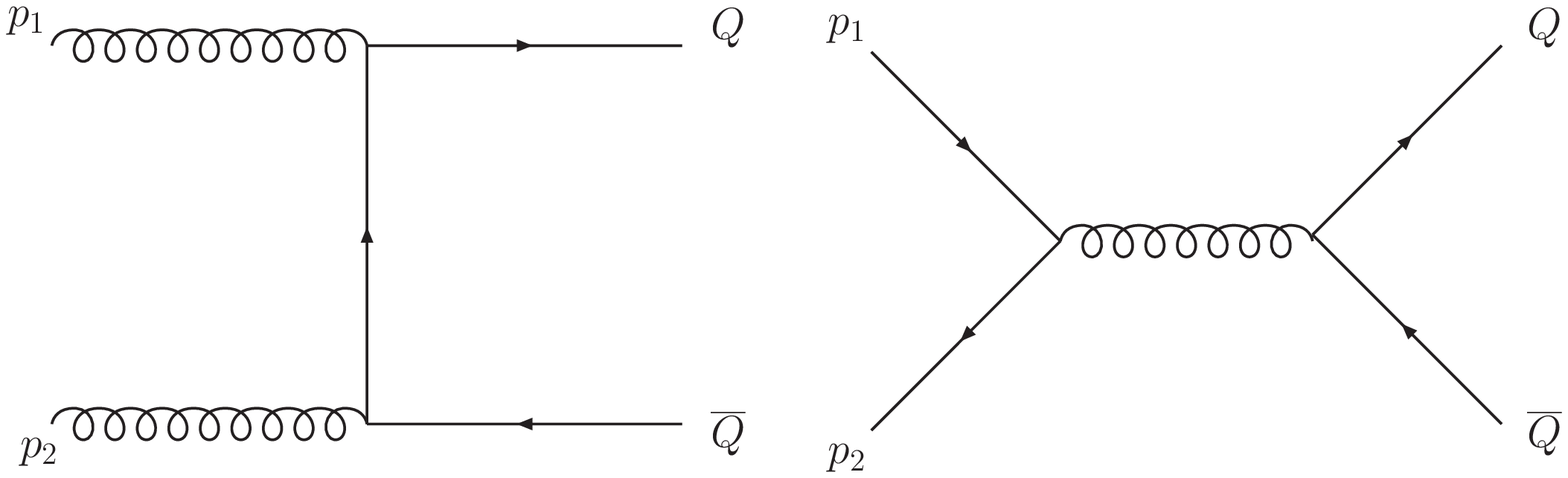}    
\end{center}    
\caption{Representative Feynman diagrams for the production of a pair of heavy quarks at hadron colliders,
via $gg$ (left) and $q{\bar q}$ (right) initial states.}
\label{fig:hqdiags}
\end{figure}

The production of heavy quarks at hadron colliders proceeds via Feynman diagrams such as the ones shown in
Figure~\ref{fig:hqdiags}.
Therefore, unlike the Drell-Yan process that we have just discussed, in this case the cross section is
sensitive to the gluon content of the incoming hadrons as well as the valence and sea quark distributions.
The pdfs are probed at values of $x_1$ and $x_2$ given by (c.f. equation~\eref{dy:6}),
\beq
x_1 = \frac{m_T}{\sqrt{s}} \left( e^{y_Q} + e^{y_{\bar Q}} \right) \qquad \mbox{and} \qquad
x_2 = \frac{m_T}{\sqrt{s}} \left( e^{-y_Q} + e^{-y_{\bar Q}} \right) \; ,
\label{eq:hqx1x2}
\eeq
where $m_T$ is the transverse mass given by $m_T = \sqrt{m_Q^2 + p_T^2}$, $p_T$ is the transverse momentum of
the quarks and $y_Q$, $y_{\overline Q}$ are the quark and antiquark rapidities. Although more complicated than in the
Drell-Yan case, these relations may be simply derived using the same frame and notation as in \eref{dy:3} and
writing, for instance, the 4-momentum of the outgoing heavy quark as,
\beq
\label{eq:hqmom}
p_Q^\mu = (m_T \cosh y_Q, \vec{p_T}, m_T \sinh y_Q ) \; ,
\eeq
where ${\vec p_T}$ is the 2-component transverse momentum.
From examining \eref{eq:hqx1x2} it is clear that the dependence on the quark and gluon pdfs
can vary considerably at different colliders ($\sqrt{s}$) and when producing different flavours of heavy
quark (for instance, $m_c \approx 1.5$~GeV compared to $m_t \approx 175$~GeV).

In this frame the heavy quark propagator that appears in the left-hand diagram
of Figure~\ref{fig:hqdiags} can easily be evaluated. It is given by,
\beq
(p_Q - p_1)^2 - m_Q^2 = -2 p_Q \cdot p_1 = - \sqrt{s} \; x_1 m_T (\cosh y_Q - \sinh y_Q) \; ,
\eeq
which, when inserting the expression for $x_1$ in~\eref{eq:hqx1x2} reduces to the simple relation,
\beq
(p_Q - p_1)^2 - m_Q^2 = -m_T^2 \left( 1 + e^{(y_Q-y_{\bar Q})} \right) \; .
\eeq 
Thus the propagator always remains off-shell, since $m_T^2 \geq m_Q^2$. This is in fact true for all the propagators
that appear in the diagrams for heavy quark production. The addition of the mass scale $m_Q$ sets a lower bound for the
propagators -- which would not occur if we considered the production of massless (or light) quarks, where the
appropriate cut-off would be the scale $\Lambda_{\rm QCD}$. Since the calculation would then enter the non-perturbative domain,
such processes cannot be calculated in the same way as for heavy quarks;
instead one must introduce a separate hard scale to render the calculation
perturbative, as we shall discuss at more length in Section~\ref{sec:localcs}. In contrast, as long as the quark
is sufficiently heavy, $m_Q \gg \Lambda_{\rm QCD}$ (as is certainly the case for top and bottom quarks), the mass sets a scale at which
perturbation theory is expected to hold.

Although we shall not concentrate on the many aspects of heavy quark processes in this article, we will
examine the success of perturbation theory for the case of top production at the Tevatron in Section~\ref{sec:tT_tev}.

\subsection{Higgs boson production}
\label{sec:higgs}

\begin{figure}[t]
\begin{center}    
\includegraphics[width=9cm]{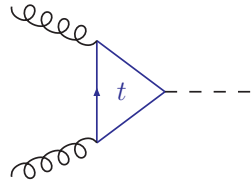}    
\end{center}    
\caption{The one-loop diagram representing Higgs production via gluon fusion at hadron colliders. The dominant contribution
is from a top quark circulating in the loop, as illustrated.}
\label{fig:Hgg}
\end{figure}

The search for the elusive Higgs boson has been the focus of much analysis at both the Tevatron and the LHC. As such, many
different channels have been proposed in which to observe events containing Higgs bosons, including the production of a Higgs
boson in association with a $W$ or a $Z$ as well as Higgs production with a pair of heavy quarks. However, the largest rate
for a putative Higgs boson at both the Tevatron and the LHC results from the gluon fusion process depicted in
Figure~\ref{fig:Hgg}.
Since the Higgs boson is responsible for giving mass to the particles in the Standard Model, it couples to fermions with a
strength proportional to the fermion mass. Therefore, although any quark may circulate in the loop, the largest contribution by
far results from the top quark. Since the LO diagram already contains a loop, the production of a Higgs boson in
this way is considerably harder to calculate than the tree level processes mentioned thus far -- particularly when one starts to
consider higher orders in perturbation theory or the radiation of additional hard jets.

For this reason it is convenient to formulate
the diagram in Figure~\ref{fig:Hgg} as an effective coupling of the Higgs boson to two gluons in the limit that the top
quark is infinitely massive. Although formally one would expect that this approximation is valid only when all other scales in
the problem are much smaller than $m_t$, in fact one finds that only $m_H < m_t$ (and $p_T({\rm jet}) < m_t$, when additional
jets are present) is necessary for an accurate approximation~\cite{DelDuca:2001fn}. Using this approach the Higgs boson
cross section via gluon fusion has been calculated to NNLO~\cite{Harlander:2002wh,Anastasiou:2002yz}, as we shall discuss
further in Section~\ref{sec:nnlo}.

\begin{figure}[t]
\begin{center}    
\includegraphics[width=7cm]{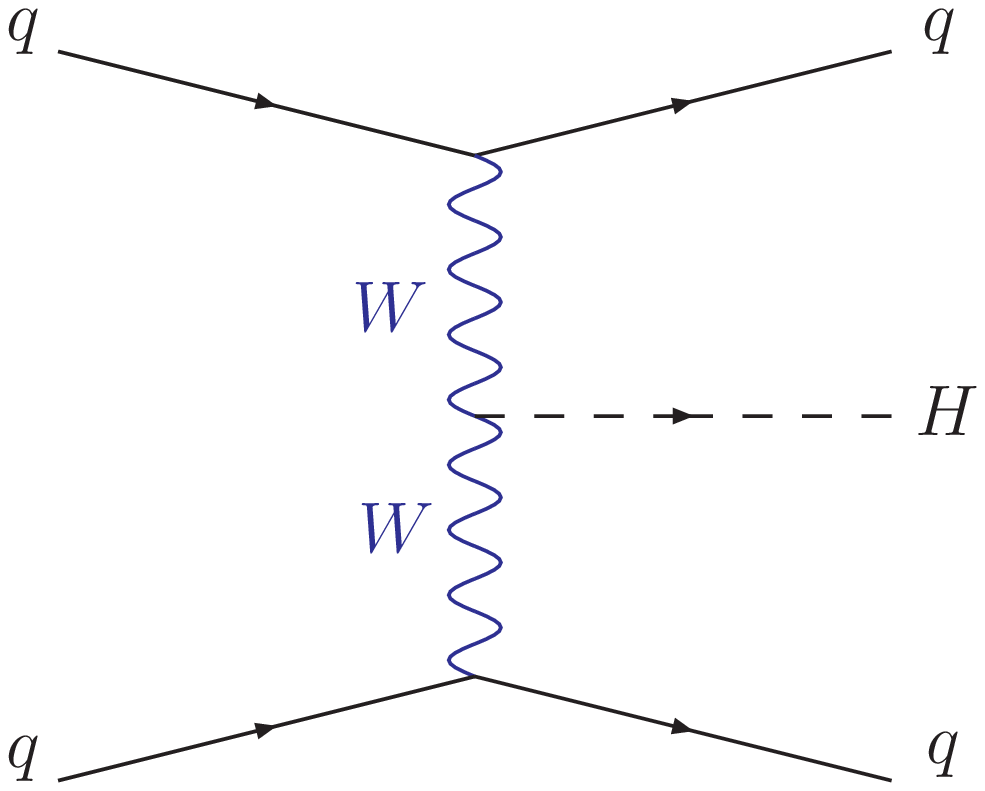}    
\includegraphics[width=7cm]{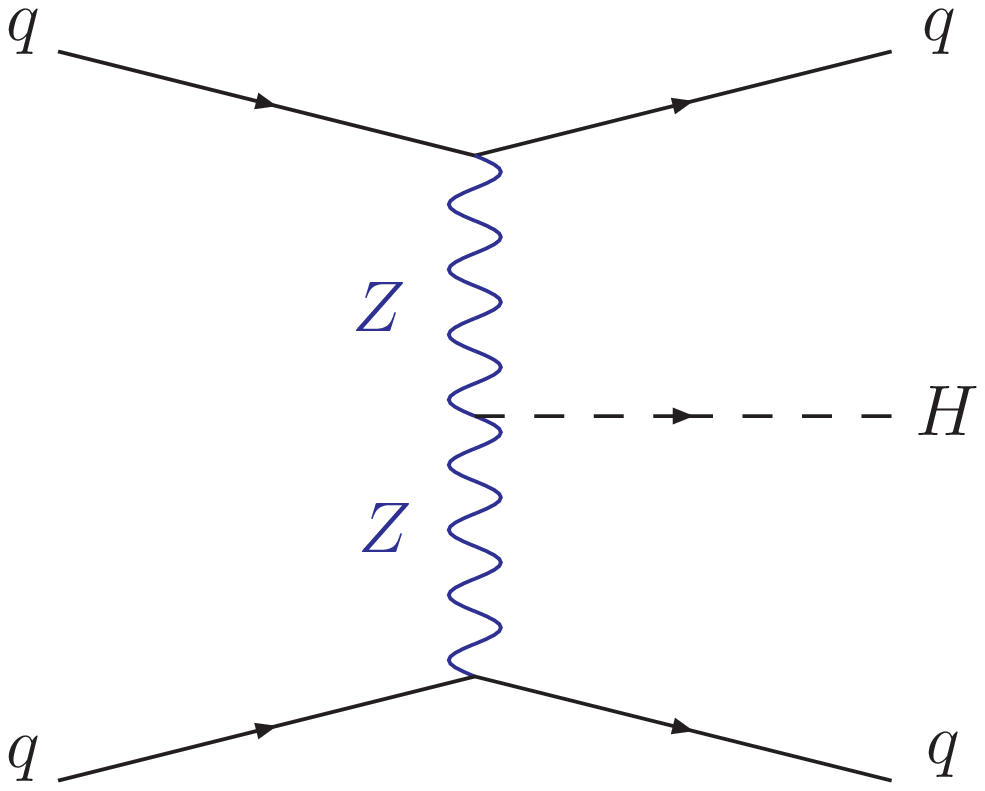}    
\end{center}    
\caption{Diagrams representing the production of a Higgs boson via the weak boson fusion mechanism.}
\label{fig:wbf}
\end{figure}

The second-largest Higgs boson cross section at the LHC is provided by the weak-boson fusion (WBF) mechanism, which proceeds
via the exchange of $W$ or $Z$ bosons from incoming quarks, as shown in Figure~\ref{fig:wbf}.
Although this process is an electroweak one and therefore proceeds at a slower rate (about an order of magnitude lower than
gluon fusion) it has a
very clear experimental signature. The incoming quarks only receive a very small kick in their direction when radiating the
$W$ or $Z$ bosons, so they can in principle be detected as jets very forward and backward at large absolute rapidities.
At the same time, since no
coloured particles are exchanged between the quark lines, very little hadronic radiation is expected in the central region of
the detector. Therefore the type of event that is expected from this mechanism is often characterized by a ``rapidity gap''
in the hadronic calorimeters of the experiment. As well as forming part of the search strategy for the Higgs boson, this channel
opens up the possibility of measuring the nature of the Higgs coupling to vector bosons~\cite{Zeppenfeld:2000td}.

Although the scope of this review does not allow a lengthy discussion of the many facets of Higgs physics, including all its
production mechanisms, decay modes, search strategies and properties, we will touch on a few important aspects of Higgs boson
phenomenology, particularly in Section~\ref{sec:higgs}. For a recent and more complete review of Higgs physics we refer the
reader to~\cite{Djouadi:2005gi}.

\subsection{$W$ and $Z$ transverse momentum distributions}
\label{sec:wzpt}

Like Drell-Yan lepton pairs, most $W$ and $Z$ bosons (here collectively  denoted by $V$) are produced with
relatively little transverse momentum, i.e. $p_T \ll M_V$. In the leading-order model discussed
in Section~\ref{sec:DY}, in which
the colliding partons are assumed to be exactly collinear with the colliding beam particles, the gauge bosons
are produced with zero transverse momentum. This approach does not take account of the intrinsic
(non-perturbative) transverse motion of the quarks and gluons inside the colliding hadrons, nor of the
possibility of generating large transverse momentum by recoil against additional energetic partons produced in
the hard scattering.

At very small $p_T$, the intrinsic transverse motion of the quarks and gluons inside the colliding hadrons, $k_T
\sim \Lambda_{\rm QCD}$,  cannot be neglected. Indeed the measured $p_T$ distribution of Drell--Yan lepton pairs
produced in fixed-target $pN$ collisions is well parametrized by assuming a Gaussian distribution for the
intrinsic transverse momentum with  $\langle k_T\rangle \sim 700$~MeV, see for example~\cite{Ellis:1991qj}. 
However the data on the $p_T$ distribution also show clear evidence of a hard, power-law tail, and it is natural
to attribute this to the (perturbative) emission of one or more hard partons, i.e. $q \bar q \to V  g$, $q g \to
V q$ etc. The Feynman diagrams for these processes  are identical to those for large $p_T$ direct photon
production, and the corresponding annihilation and Compton matrix elements are, for $W$ production,
\beqn
\label{eq:sec2w1j}
 {\sum}\vert{\cal M}^{q\bar q'\rightarrow Wg}\vert^2  
&=& \pi\as \sqrt{2} G_F M^2_W \vert V_{qq'} \vert^2\ 
\frac{8}{9}\ {{\hat t}^2+{\hat u}^2+2 M_W^2 {\hat s} \over {\hat t} {\hat u} } \; , \nonumber \\
{\sum}\vert {\cal M}^{gq\rightarrow Wq'}\vert^2 
&=&  \pi\as \sqrt{2} G_F M^2_W \vert V_{qq'} \vert^2\  
\frac{1}{3}\ 
{{\hat s}^2+{\hat u}^2+2 {\hat t} M_W^2 \over - {\hat s} {\hat u} } \; ,
\eeqn
with similar results for the $Z$ boson and for Drell--Yan lepton pairs. The sum is over colours and spins in the
final and initial states, with appropriate averaging factors for the latter. The  transverse momentum
distribution $d\sigma/dp_T^2$  is then obtained by convoluting these matrix elements with parton distributions
in the usual way. In principle, one can combine the hard (perturbative) and intrinsic (non-perturbative)
contributions, for example using a convolution integral in transverse momentum space, to give a theoretical
prediction valid for all $p_T$. A more refined prediction would then include next-to-leading-order ($O(\as^2)$)
perturbative corrections, for example from processes like $q \bar q \to V gg$, to the high $p_T$ tail. Some
fraction of the $O(\as)$ and $O(\as^2)$ contributions could be expected to correspond to distinct $V+1$~jet and
$V+2$~jet final states respectively.

However, a major problem in carrying out the above procedure is that the $2\to 2$ matrix elements are singular
when the final-state partons become soft and/or are emitted collinear with the initial-state partons. These
singularities are related to the poles at ${\hat t}=0$ and ${\hat u}=0$ in the above matrix elements. In addition, processes
like $q \bar q \to V gg$ are singular when the two final-state gluons become collinear. This means in practice
that the lowest-order perturbative contribution to the $p_T$ distribution is singular as $p_T \to 0$, and that
higher-order contributions from processes like $q \bar q \to V gg$   are singular for any $p_T$.

The fact that the predictions of perturbative QCD {\it are} in fact finite for physical processes is due to a
number of deep and powerful theorems (applicable to any quantum field theory) that guarantee that for suitably
defined cross sections and distributions, the singularities arising from real and virtual parton emissions at
intermediate stages of the calculation cancel when all contributions are included. We have already seen an
example of this in the discussion above. The $O(\as)$ contribution to the total $W$ cross section from the process
$q \bar q \to W g$ is singular when $p_T(W) = 0$, but this singularity is exactly cancelled by a $O(\as)$
contribution from a virtual gluon loop  correction  to $q \bar q \to W$. The net result is the finite NLO
contribution to the cross section displayed in Figure~\ref{fig:wz}.  The details of how and under what
circumstances these cancellations take place will be discussed in the following section.








\section{Partonic cross sections}
\label{sec:partonxsecs}

\subsection{Introduction}

At the heart of the prediction of any hadron collider observable lies the
calculation of the relevant hard scattering process. In this section we will
outline the perturbative approaches that are employed to calculate these processes
and describe some of their features and limitations. In addition, we will describe
how the partonic calculations can be used to make predictions for an exclusive
hadronic final state.

\subsection{Lowest order calculations}
\label{sec:localcs}

The simplest predictions can be obtained by calculating the lowest order in the perturbative
expansion of the observable, as discussed in the previous section.  This is performed by
calculating the squared matrix element represented by tree-level Feynman diagrams and
integrating this over the appropriate phase space.  For the simplest cases and for certain
observables only, the phase space integration can be performed analytically. For example, in
Section~\ref{sec:formalism}, we calculated the lowest order cross section for Drell-Yan production. However, to
obtain fully differential predictions in general, the integration must be carried out
numerically. For most calculations, it is necessary to impose restrictions on the phase space
in order that divergences in the matrix elements are avoided.  This can best be understood by
consideration of one of the simplest such cases, $W+1$~jet production at a hadron collider.

\subsubsection{$W+1$~jet production}
\label{sec:w1j}

\begin{figure}[t]
\begin{center}
\includegraphics[width=12cm]{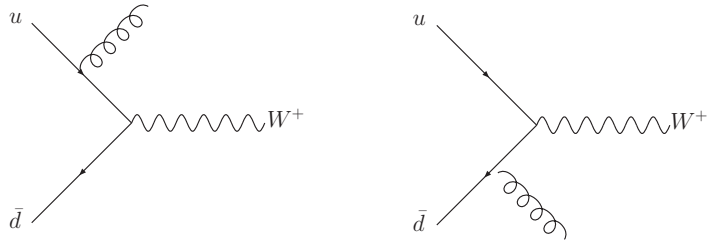}
\end{center}
\caption{Lowest order diagrams for the production of a $W$ and one jet
at hadron colliders.}
\label{fig:dy_real}
\end{figure}

In Figure~\ref{fig:dy_real}, we have extended the LO diagrams for Drell-Yan production
(for the specific initial state $u\bar{d}$) by
adding a final state gluon to each of the initial state quark legs. This is one  of the
subprocesses responsible for $W+1$~jet production, with the other crossed process being $gq \rightarrow Wq$.
After application of the Feynman rules, the squared
matrix elements obtained from the sum of the diagrams take the form:
\begin{equation}
|{\cal M}^{u{\bar d} \to W+g}|^2 \sim \left(
\frac{{\hat t}^2 + {\hat u}^2 + 2 Q^2 \, {\hat s}}{{\hat t}{\hat u}} \right),
\label{eq:wreal}
\end{equation}
where $Q^2$ is the virtuality of the $W$ boson,
${\hat s}=s_{u{\bar d}}$, ${\hat t}=s_{ug}$, and ${\hat u}=s_{{\bar d}g}$
, c.f.~\eref{eq:sec2w1j} of Section~\ref{sec:formalism}.
This expression diverges in the limit where the gluon is unresolved --
either it is collinear to one of the quarks (${\hat t} \to 0$ or
${\hat u} \to 0$), or it is soft (both invariants vanish, so $E_g \to 0$).
Let us consider the impact of these divergences on the calculation of this
cross section. In order to turn the matrix elements into a cross section,
one must convolute with pdfs and perform the integration over the appropriate
phase space,
\begin{equation}
\fl
\sigma = \int dx_1 dx_2 f_u(x_1,Q^2) f_{\bar d}(x_2,Q^2) \,
 \frac{|{\cal M}|^2}{32\pi^2{\hat s}}
 \frac{d^3p_W}{E_W} \frac{d^3p_g}{E_g} \delta(p_u+p_{\bar d}-p_g-p_W),
\label{eq:wrealps}
\end{equation}
where $x_1$, $x_2$ are the momentum fractions of the $u$ and ${\bar d}$ quarks.
These momentum fractions are of course related to the centre-of-mass energy squared of
the collider $s$ by the relation, ${\hat s}=x_1 x_2 s$.

After suitable manipulations, this can be transformed into
a cross section that is differential in $Q^2$ and the transverse momentum ($p_T$)
and rapidity ($y$) of the $W$ boson~\cite{Kajantie:1978qv},
\begin{equation}
\frac{d\sigma}{dQ^2 dy dp_T^2} \sim \frac{1}{s} \int dy_g  \,
 f_u(x_1,Q^2) f_{\bar d}(x_2,Q^2) \frac{|{\cal M}|^2}{{\hat s}}
\end{equation} 
The remaining integral to be done is over the rapidity of the gluon, $y_g$. Note that
the $p_T$ of the gluon is related to the invariants of the process by
$p_T^2={\hat t}{\hat u}/{\hat s}$. Thus the leading divergence represented by the third
term of~\eref{eq:wreal}, where ${\hat t}$ and ${\hat u}$ both approach zero and the gluon
is soft, can be written as $1/p_T^2$. Furthermore, in this limit ${\hat s} \to Q^2$, so that 
the behaviour of the cross section becomes,
\begin{equation}
\fl
\frac{d\sigma}{dQ^2 dy dp_T^2} \sim \frac{2}{s} \frac{1}{p_T^2} \int dy_g \,
 f_u(x_1,Q^2) f_{\bar d}(x_2,Q^2) + (\mbox{sub-leading in }p_T^2) \; .
\end{equation} 
As the $p_T$ of the $W$ boson becomes small, the limits on the $y_g$ integration are given
by $\pm \log(\sqrt{s}/p_T)$. Under
the assumption that the rest of the integrand is approximately constant, the integral
can be simply performed. This yields,
\begin{equation}
\frac{d\sigma}{dQ^2 dy dp_T^2} \sim \frac{\log(s/p_T^2)}{p_T^2},
\label{eq:dylog}
\end{equation}
so that the differential cross section contains a logarithmic dependence on $p_T$. If no cut
is applied on the gluon $p_T$ then the integral over $p_T$ diverges -- only after applying a
minimum value of the $p_T$ do we obtain a finite result. Once we apply a cutoff at
$p_T=p_{T,{\rm min}}$ and then perform the integration, we find a result proportional to
$\log^2(s/p_{T,{\rm min}}^2)$. This is typical of a fixed order expansion -- it is not merely an
expansion in $\as$, but in $\as \log(\ldots)$, where the argument of the logarithm
depends upon the process and cuts under consideration. As we shall discuss later, these
logarithms may be systematically accounted for in various all-orders treatments.

\begin{figure}[t]
\begin{center}
\includegraphics[width=7cm]{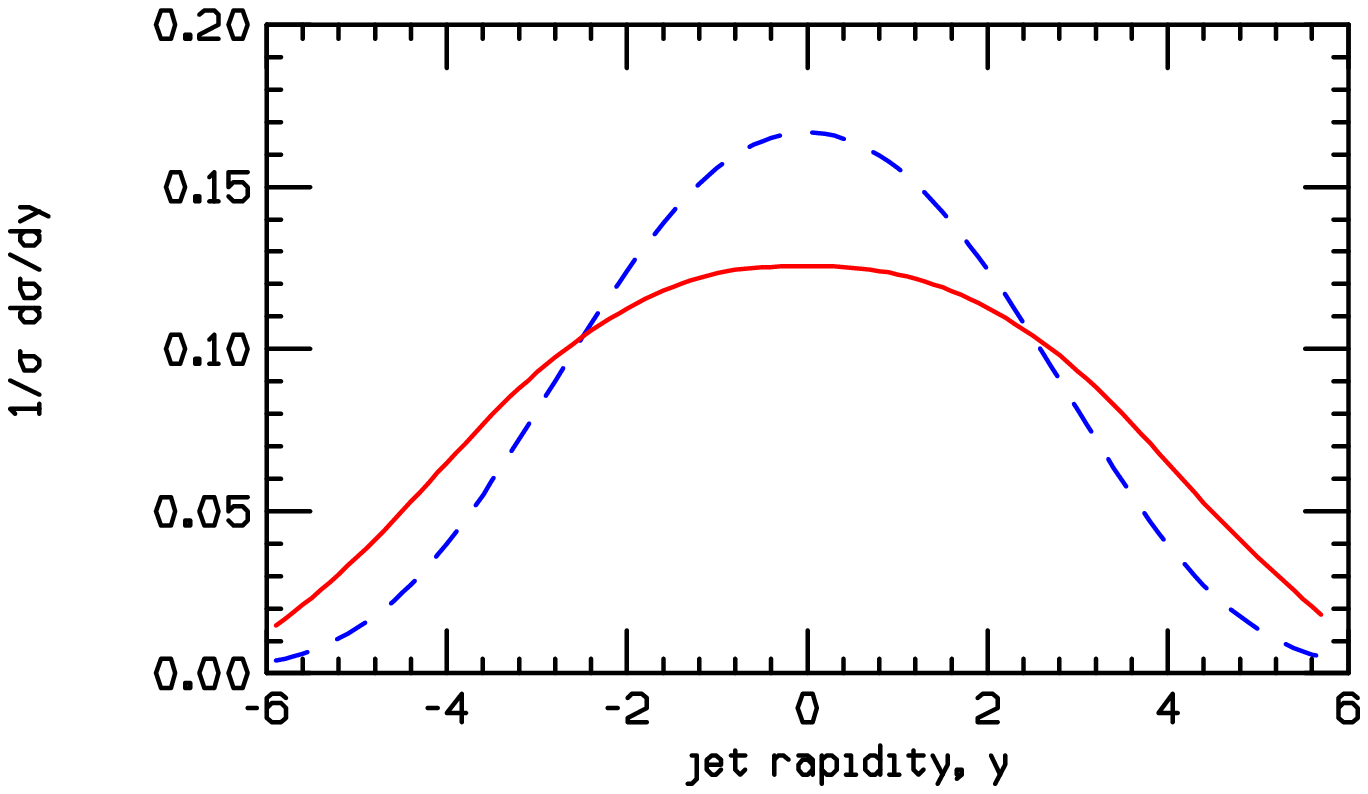}
\includegraphics[width=7cm]{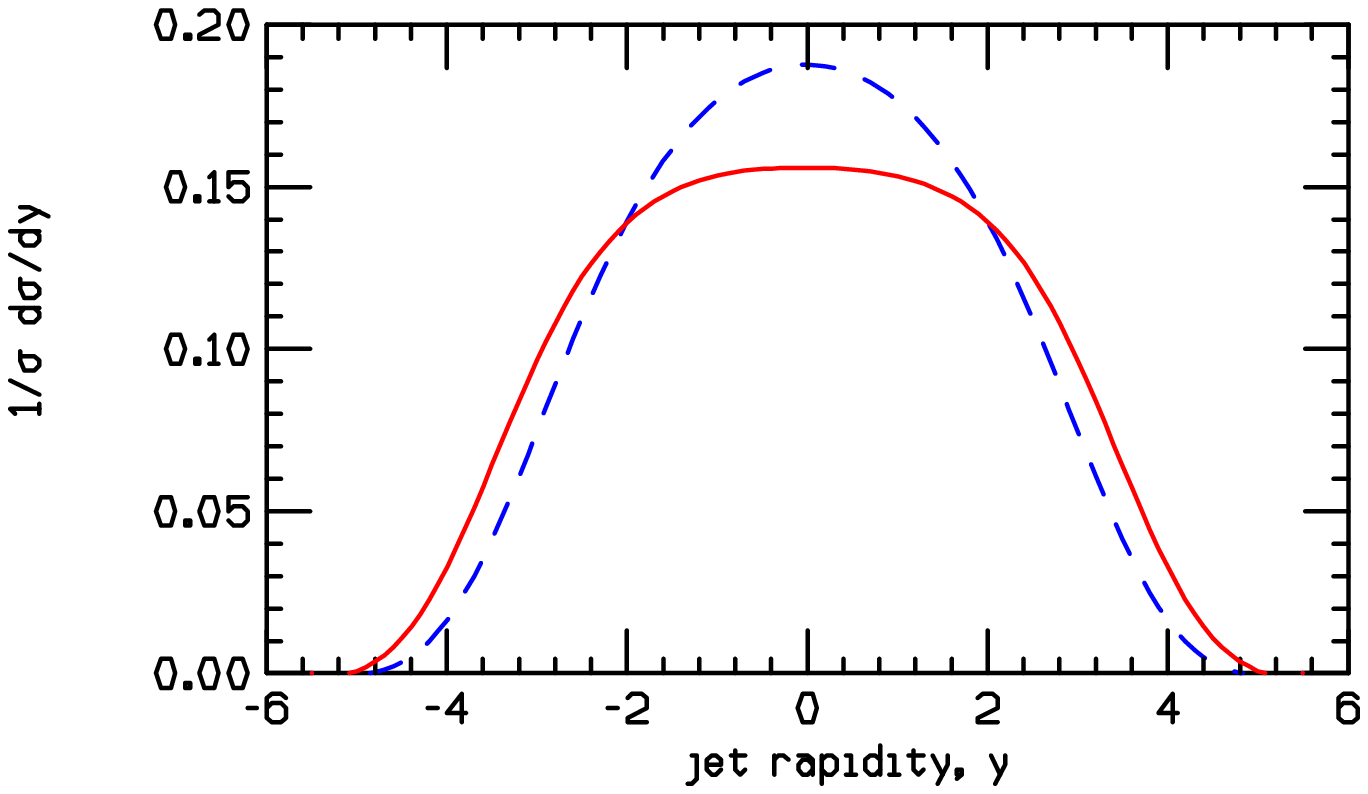}
\end{center}
\caption{The rapidity distribution of the final state parton found in a lowest order calculation
of the $W+1$~jet cross section at the LHC. The parton is required to have
a $p_T$ larger than $2$~GeV (left) or $50$~GeV(right). Contributions from $q{\bar q}$ annihilation (solid red
line) and the $qg$ process (dashed blue line) are shown separately.}
\label{fig:raps}
\end{figure}

\begin{figure}[t]
\begin{center}
\includegraphics[width=12cm]{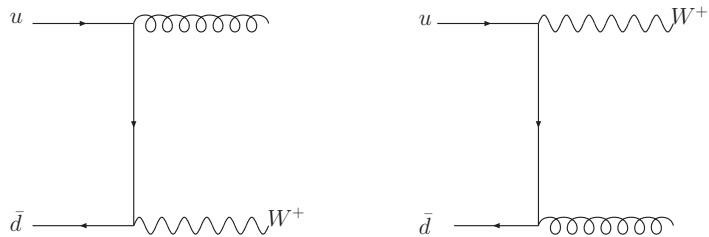}
\end{center}
\caption{An alternative way of drawing the diagrams of Figure~\ref{fig:dy_real}.}
\label{fig:dy_real2to2}
\end{figure}

In Figure~\ref{fig:raps} we show the rapidity distribution of the jet, calculated using
this lowest order process.
In the calculation, a sum over all species of quarks has been performed and the contribution from
the quark-gluon process included. The rapidity distribution is shown for two different choices of
minimum jet transverse momentum, which is the cut-off used to regulate the collinear divergences
discussed above. For very small values of $p_T$, we can view the radiated gluon as being emitted
from the quark line at an early time, typically termed ``initial-state radiation''. From the
left-hand plot, this radiation is indeed produced quite often at large rapidities, although it is
also emitted centrally with a large probability. The canonical ``wisdom'' is that initial-state
radiation is primarily found in the forward region. There is indeed a collinear pole in the matrix
element so that a fixed energy gluon tends to be emitted close to the original parton direction.
However, we are interested not in fixed energy but rather in fixed transverse momentum. When using
a higher $p_T$ cut-off the gluon is emitted less often at large rapidities and is more central, as
shown by the plot on the right-hand side. In this case, one can instead think of the diagrams as a
$2 \to 2$ scattering as depicted in Figure~\ref{fig:dy_real2to2}.
Of course, the manner in which such Feynman diagrams are drawn is purely a matter of convention. 
The diagrams are exactly the same as in Figure~\ref{fig:dy_real}, but re-drawing them in this way
is suggestive of the different kinematic region that is probed with a gluon at high $p_T$.

There is also a collinear pole involved for the emission of gluons from final state partons. Thus, the gluons will
be emitted preferentially near the direction of the emitting parton. In fact, it is just such emissions that give
rise to the finite size of the jet arising from a single final state parton originating from the hard scatter. Much
of the jet structure is determined by the hardest gluon emission; thus NLO theory, in which a jet consists of at
most 2 partons, provides a good description of the jet shape observed in data~\cite{Ellis:1992qq}. 

\subsubsection{$W+2$~jet production}
\label{sec:w2j}

By adding a further parton, one can simulate the production of a $W+2$~jet final state. Many
different partonic processes contribute in general, so for the sake of illustration
we just consider the production of a $W$ boson in association with two gluons.

\begin{figure}[t]
\begin{center}
\includegraphics[width=12cm]{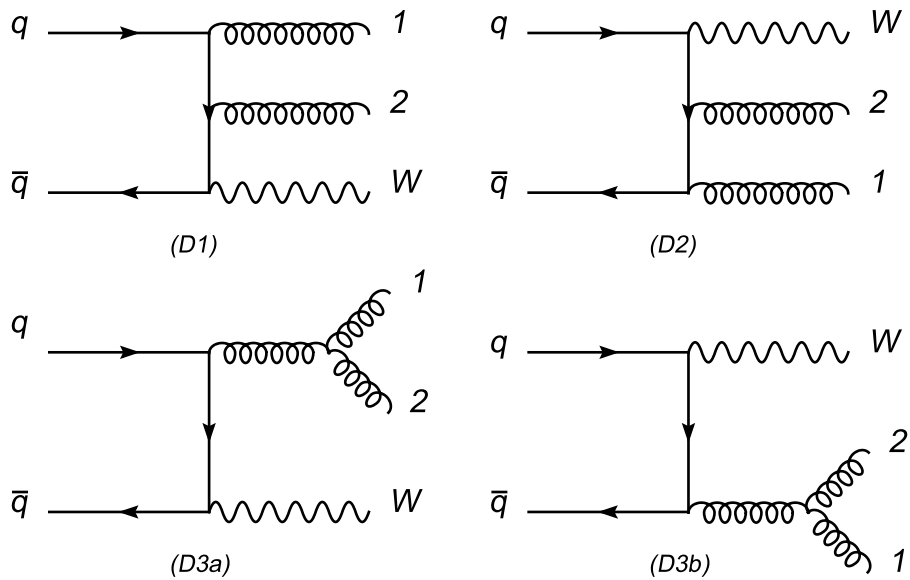}
\end{center}
\caption{The 4 diagrams that contribute to the matrix elements for the production
of $W+2$~gluons when gluon $1$ is soft. \label{fig:w2jsoft}}
\end{figure}

First, we shall study the singularity structure of the matrix elements in more detail.
In the limit that one of the gluons, $p_1$, is soft the singularities in the
matrix elements occur in 4 diagrams only. These diagrams, in which gluon $p_1$
is radiated from an external line, are depicted in Figure~\ref{fig:w2jsoft}.
The remaining diagrams, in which gluon $p_1$ is attached to an internal line,
do not give rise to singularities because the adjacent propagator does not
vanish in this limit.

This is the first of our examples in which the matrix elements contain
non-trivial colour structure. Denoting the colour labels of gluons $p_1$ and $p_2$
by $t^A$ and $t^B$ respectively, diagram (1) is proportional to $t^Bt^A$, whilst
(2) is proportional to $t^At^B$. The final two diagrams, (3a) and (3b) are each
proportional to $f^{ABC}t^C$, which can of course be written as $(t^At^B-t^Bt^A)$.
Using this identity, the amplitude (in the limit that $p_1$ is soft) can be
written in a form in which the dependence on the colour matrices is factored out,
\begin{equation}
{\cal M}^{q{\bar q} \to Wgg} = t^At^B(D_2+D_3) + t^Bt^A(D_1-D_3)
\label{eq:colstruc}
\end{equation}
so that the kinematic structures obtained from the Feynman rules are collected
in the functions $D_1$, $D_2$ (for diagrams (1) and (2)) and $D_3$ (the sum of
diagrams (3a) and (3b)). The combinations of these that appear in~\eref{eq:colstruc}
are often referred to as colour-ordered amplitudes.

\begin{figure}[t]
\begin{center}
\includegraphics[width=12cm]{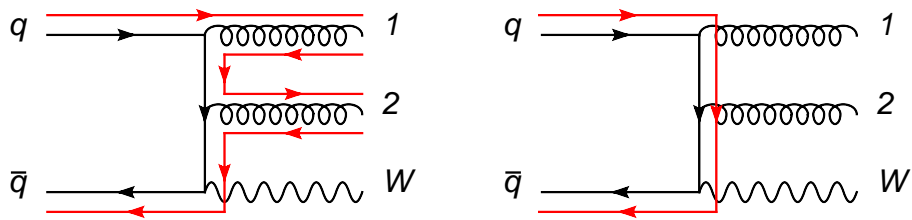}
\end{center}
\caption{Two examples of colour flow in a $W+2$~jet event, shown in red. In the left hand diagram, a
leading colour flow is shown. The right-hand diagram depicts the sub-leading colour flow
resulting from interference. \label{fig:colflow}}
\end{figure}

With the colour factors stripped out, it is
straightforward to square the amplitude in~\eref{eq:colstruc} using the
identities ${\rm tr}(t^At^Bt^Bt^A)=N C_F^2$ and ${\rm tr}(t^At^Bt^At^B)=-C_F/2$,
\begin{eqnarray}
\fl
|{\cal M}^{q{\bar q} \to Wgg}|^2 &= NC_F^2 \left[ |D_2+D_3|^2 + |D_1-D_3|^2 \right]
 - C_F \; {\rm Re} \left[ (D_2+D_3)(D_1-D_3)^\star \right] \nonumber \\
 &= \frac{C_F N^2}{2} \left[ |D_2+D_3|^2 + |D_1-D_3|^2
     - \frac{1}{N^2} |D_1+D_2|^2 \right].
\label{eq:colstrucsq}
\end{eqnarray}
Moreover, these colour-ordered amplitudes possess special factorization properties in the
limit that gluon $p_1$ is soft. They can be written as the
product of an eikonal term and the matrix elements containing only one gluon,
\begin{eqnarray}
D_2+D_3 &\longrightarrow& \epsilon_\mu
 \left( \frac{q^\mu}{p_1.q}-\frac{p_2^\mu}{p_1.p_2} \right) {\cal M}_{q{\bar q} \to Wg}
 \nonumber \\
D_1-D_3 &\longrightarrow& \epsilon_\mu
 \left( \frac{p_2^\mu}{p_1.p_2}-\frac{{\bar q}^\mu}{p_1.{\bar q}} \right) {\cal M}_{q{\bar q} \to Wg}
\label{eq:eikonal}
\end{eqnarray}
where $\epsilon_\mu$ is the polarization vector for gluon $p_1$.
The squares of these eikonal terms are easily computed using
the replacement $\epsilon_\mu \epsilon_\nu^\star \to -g_{\mu\nu}$ to sum over the gluon
polarizations. This yields terms of the form,
\begin{equation}
\frac{a.b}{p_1.a \, p_1.b} \equiv [a~b],
\label{eq:eikonalsq}
\end{equation}
so that the final result is,
\begin{equation}
|{\cal M}^{q{\bar q} \to Wgg}|^2 \stackrel{{\rm soft}}{~\longrightarrow~}
   \frac{C_F N^2}{2} \left[ [q~p_2] + [p_2~{\bar q}]
     - \frac{1}{N^2} [q~{\bar q}] \right] {\cal M}^{q{\bar q} \to Wg}.
\end{equation}
Inspecting this equation, one can see that the leading term (in the number of colours)
contains singularities along two lines of colour flow -- one connecting the gluon $p_2$ to the
quark, the other connecting it to the antiquark. On the other hand, the sub-leading term has
singularities along the line connecting the quark and antiquark. It is these lines of colour
flow that indicate the preferred directions for the emission of additional gluons. In the
sub-leading term the colour flow does not relate the gluon colour to the parent quarks at all.
The matrix elements are in fact the same as those for the emission of two photons from a quark
line (apart from overall coupling factors) with no unique assignment to either diagram 1 or
diagram 2, unlike the leading term. For this reason only the information about the
leading colour flow is used by parton shower Monte Carlos such as HERWIG~\cite{Corcella:2000bw}
and PYTHIA~\cite{Sjostrand:2006za}. These lines of colour flow generalize in an obvious manner to
higher multiplicity final states. As as example, the lines of colour flow in a $W+2$~jet event are
shown in Figure~\ref{fig:colflow}.

Since all the partons are massless, it is trivial to re-write the eikonal factor of~\eref{eq:eikonalsq}
in terms of the energy of the radiated gluon, $E$ and the
angle it makes with the hard partons, $\theta_a$, $\theta_b$. It can then be
combined with the phase space for the emitted gluon to yield a contribution such as,
\begin{equation}
[a~b] \, dPS_{\rm gluon} =
 \frac{1}{E^2} \frac{1}{1-\cos\theta_a}
 \, E dE \; d\cos\theta_a \; .
\end{equation}
In this form, it is clear that the cross section diverges as either
$\cos\theta_a \to 1$ (the gluon is emitted collinear to parton $a$) or
$E \to 0$ (for any angle of radiation). Moreover, each divergence is logarithmic
and regulating the divergence, by providing a fixed cutoff (either in angle or energy),
will produce a single logarithm from
collinear configurations and another from soft ones -- just as we found when considering the
specific case of $W+1$~jet production in the previous subsection.

This argument can be applied at successively higher orders of perturbation theory.
Each gluon that is added yields an additional power of $\as$ and, via the eikonal
factorization outlined above, can produce an additional two logarithms. This means
that we can write the $W+1$~jet cross section schematically as a sum of contributions,
\begin{eqnarray}
\label{eq:w1jlogs}
d\sigma&=&\sigma_0(W+1~{\rm jet}) \left[
 1+\as(c_{12}L^2+c_{11}L+c_{10}) \right. \nonumber \\
&& \left. +\as^2(c_{24}L^4+c_{23}L^3+c_{22}L^2+c_{21}L+c_{20}) + \ldots \right]
\end{eqnarray}
where $L$ represents the logarithm controlling the divergence, either soft or collinear.  The size of
$L$ depends upon the criteria used to define the jets -- the minimum transverse energy of a jet and
the jet cone size. The coefficients $c_{ij}$ in front of the logarithms depend upon colour factors.
Note that the addition of each gluon results not just in an additional factor of 
$\as$, but  in a factor of $\as$ times logarithms. For many important kinematic
configurations, the logs can be large, leading to an enhanced probability for additional gluon
emissions to occur. For inclusive quantities, where the same cuts are applied to every jet, the logs
tends to be small, and counting powers of $\as$ becomes a valid estimator for the rate of
production of additional jets. 

\setcounter{footnote}{0}

Noticing that the factor $(\as L)$ appears throughout~\eref{eq:w1jlogs}, it is useful to
re-write the expansion in brackets as,
\begin{eqnarray}
\Bigl[ \ldots \Bigr] & =
 1+\as L^2c_{12}+(\as L^2)^2c_{24} 
 +\as Lc_{11}(1+\as L^2 c_{23}/c_{11} + \ldots) + \ldots \nonumber \\
 & = \exp \left[ c_{12} \as L^2 + c_{11} \as L \right] \; ,
\label{eq:w1resum}
\end{eqnarray}
where the infinite series have been resummed into an exponential form\footnote{
Unfortunately, systematically collecting the terms in this way is far from trivial and
only possible when considering certain observables and under specific choices of jet
definition (such as when using the $k_T$-clustering algorithm).}.
The first term in the exponent is commonly referred to as the leading logarithmic term,
with the second being required in order to reproduce next-to-leading logarithms.
This reorganization of the perturbative expansion is especially useful when the
product $\as L$ is large, for instance when the logarithm is a ratio of two
physical scales that are very different such as $\log(m_H/m_b)$. This exponential
form is the basis of all orders predictions and can be interpreted in terms of
Sudakov probabilities, both subjects that we will return to in later discussions.

It is instructive to recast the discussion of the total $W$ cross section in these terms,
where the calculation is decomposed into components that each contain a given number of jets:
\begin{equation}
\label{eq:wxsecdecomp}
\sigma_W = \sigma_{W+0j}+\sigma_{W+1j}+\sigma_{W+2j}+\sigma_{W+3j}+\ldots
\end{equation}
Now, as in~\eref{eq:w1jlogs}, we can further write out each contribution
as an expansion in powers of $\as$ and logarithms,
\begin{eqnarray}
\label{eq:wxseclogs}
\sigma_{W+0j} & = & a_0+\as(a_{12}L^2+a_{11}L+a_{10}) \nonumber \\
 &&+\as^2(a_{24}L^4+a_{23}L^3+a_{22}L^2+a_{21}L+a_{20}) + \ldots \nonumber \\
\sigma_{W+1j} & = &  \as(b_{12}L^2+b_{11}L+b_{10}) \nonumber \\
 &&+\as^2(b_{24}L^4+b_{23}L^3+b_{22}L^2+b_{21}L+b_{20}) + \ldots \nonumber \\
\sigma_{W+2j} & = & \ldots .
\end{eqnarray}
As the jet definitions change, the size of the logarithms shuffle the contributions from one jet cross section to
another, whilst keeping the sum over all jet contributions the same. For example, as the jet cone size is decreased the
logarithm $L$ increases. As a result,
the average jet multiplicity goes up and terms in~\eref{eq:wxsecdecomp} that represent
relatively higher multiplicities will become more important.

\begin{figure}[t]
\begin{center}
\includegraphics[width=6cm]{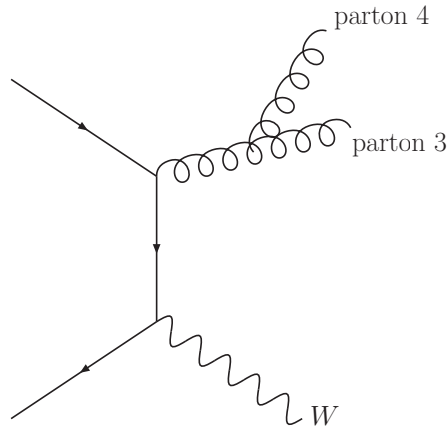}
\end{center}
\caption{A final state configuration containing a $W$ and $2$ partons. After the jet
definition has been applied, either zero, one or two jets may be reconstructed. \label{fig:jetexample}}
\end{figure}

This is illustrated in Figure~\ref{fig:jetexample}.
Such a configuration may be reconstructed as an event containing up to two jets, depending upon the
jet definition and the momenta of the partons. The matrix elements for this process contain terms
proportional to $\as \log(p_{T,3}/p_{T,4})$ and $\as\log(1/\Delta R_{34})$ which is the
reason that minimum values for the transverse energy and separation must be imposed. We shall see
later that this is not the case in a full next-to-leading order calculation where these soft and
collinear divergences are cancelled.

Finally, we note that although the decomposition in~\eref{eq:wxsecdecomp}
introduces quantities which are dependent upon the jet definition, we can recover
results that are independent of these parameters by simply summing up the
terms in the expansion that enter at the same order of perturbation theory, i.e.
the $a_{ij}$ and $b_{ij}$ in equation~\ref{eq:wxseclogs} are not independent.
As we will discuss shortly, in Section~\ref{sec:nlo}, at a given order of
perturbation theory the sum of the logarithms vanishes and we just recover
the perturbative expansion of the total cross section,
\begin{eqnarray*}
\sigma_W^{LO} & = & a_0 \nonumber \\
\sigma_W^{NLO} & = & \as \left( a_{10}+b_{10} \right) \; .
\end{eqnarray*}

\subsubsection{Leading order tools}
\label{sec:LOtools}

Once suitable cuts have been applied, as we have discussed extensively above, leading order cross
sections can be calculated using a number of computer programs.

There is a wide range of programs available, most notably
ALPGEN~\cite{Mangano:2001xp,Mangano:2002ea},
the COMPHEP package~\cite{Pukhov:1999gg,Boos:2004kh} and
MADGRAPH~\cite{Stelzer:1994ta,Maltoni:2002qb}.
All of these programs implement the calculation of the diagrams
numerically and provide a suitable phase space over which they can
be integrated. ALPGEN uses an approach which is not based on a traditional
Feynman diagram evaluation~\cite{Caravaglios:1998yr}, whereas the other two
programs rely on more conventional methods such as the helicity amplitudes
evaluation of HELAS~\cite{Murayama:1992gi} in MADGRAPH.

Although in principle these programs can be used to calculate any
tree-level prediction, in practice the complexity of the process that may
be studied is limited by the number of particles that is produced in the
final state. This is largely due to the factorial growth in the number of
Feynman diagrams that must be calculated. Even in approaches which do
not rely directly on the Feynman diagrams, the growth is still as a power
of the number of particles. For processes which involve a
large number of quarks and gluons, as is the case when attempting
to describe a multi-jet final state at a hadron collider such as the Tevatron
or the LHC, an additional concern is the calculation of colour matrices
which appear as coefficients in the amplitudes~\cite{Maltoni:2002mq}.

In many cases, such as in the calculation of amplitudes representing multiple
gluon scattering, the final result is remarkably simple. Motivated by such
results, the last couple of years has seen remarkable progress in the
development of new approaches to QCD tree-level calculations. Some of the
structure behind the amplitudes can be understood by transforming to ``twistor
space''~\cite{Witten:2003nn}, in which amplitudes are represented by
intersecting lines. This idea can be taken further with the introduction of
``MHV'' rules~\cite{Cachazo:2004kj}, which use the simplest
(maximally helicity-violating, or MHV) amplitudes as the building blocks of
more complicated ones. Although these rules at first only applied to
amplitudes containing gluons, they were soon extended to cases of more
general interest at hadron colliders~\cite{Georgiou:2004wu,Wu:2004jx,
Georgiou:2004by,Dixon:2004za,Badger:2004ty,Bern:2004ba}.
Even more recently, further simplification of amplitudes has been obtained
by using ``on-shell recursion relations''~\cite{Britto:2004ap,Britto:2005fq}.
As well as providing very compact expressions, this approach has the advantage
of being both easily proven and readily extendible to processes involving
fermions and vector bosons.

\subsection{Next-to-leading order calculations}
\label{sec:nlo}

Although lowest order calculations can in general describe broad features of
a particular process and provide the first estimate of its cross section, in
many cases this approximation is insufficient. The inherent uncertainty in
a lowest order calculation derives from its dependence on the unphysical
renormalization and factorization scales, which is often large. In addition,
some processes may contain large logarithms that need to be resummed or extra
partonic processes may contribute only when going beyond the first
approximation. Thus, in order to compare with predictions that have smaller
theoretical uncertainties, next-to-leading order calculations are imperative
for experimental analyses in Run II of the Tevatron and at the LHC.

\subsubsection{Virtual and real radiation}
\label{sec:virtreal}

\begin{figure}[t]
\begin{center}
\includegraphics[width=15cm]{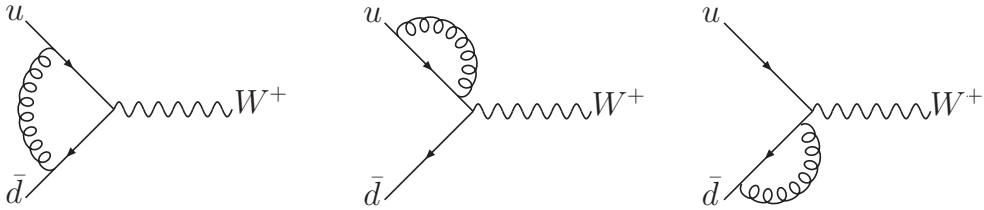}
\end{center}
\vspace*{-0.5cm}
\caption{Virtual diagrams included in the next-to-leading order corrections
to Drell-Yan production of a $W$ at hadron colliders.}
\label{fig:dy_virt}
\end{figure}

A next-to-leading order QCD calculation requires the consideration of all diagrams
that contribute an additional strong coupling factor, $\as$.
These diagrams are obtained from the lowest order ones by adding additional quarks
and gluons and they can be divided into two categories, virtual (or
loop) contributions and the real radiation component. We shall illustrate this
by considering the next-to-leading order corrections to Drell-Yan
production at a hadron collider. The virtual diagrams for this process
are shown in Figure~\ref{fig:dy_virt} whilst the real diagrams are exactly
the ones that enter the $W+1$ jet calculation (in Figure~\ref{fig:dy_real}). 

Let us first consider the virtual contributions. In order to evaluate the
diagrams in Figure~\ref{fig:dy_virt}, it is necessary to introduce an additional
loop momentum $\ell$ which circulates around the loop in each diagram and
is unconstrained. To complete the evaluation of these diagrams, one must
therefore integrate over the momentum $\ell$. However, the
resulting contribution is not finite but contains infrared divergences -- in the
same way that the diagrams of Figure~\ref{fig:dy_real} contain infrared (soft and collinear)
singularities. By isolating the singularities appropriately, one can see that the divergences
that appear in each contribution are equal, but of opposite sign. The fact
that the sum is finite is a demonstration of the theorems of Bloch
and Nordsieck~\cite{Bloch:1937pw} and Kinoshita, Lee and
Nauenberg~\cite{Kinoshita:1962ur,Lee:1964is},
which guarantee that this is the case at all orders in perturbation theory and
for any number of final state particles.

The real contribution consists of the diagrams in Figure~\ref{fig:dy_real},
together with a quark-gluon scattering piece that can be obtained from these
diagrams by interchanging the gluon in the final state with a quark (or
antiquark) in the initial state. As discussed in Section~\ref{sec:w1j}, the
quark-antiquark matrix elements contain a singularity as the gluon transverse
momentum vanishes.

In our NLO calculation we want to carefully regulate and then isolate these
singularities in order to extend the treatment down to zero transverse momentum.
The most common method to regulate the singularities is dimensional
regularization. In this approach the number of dimensions is continued
to $D=4-2\epsilon$, where $\epsilon < 0$, so that in intermediate stages the singularities appear
as single and double poles in $\epsilon$. After they have cancelled, the
limit $D \to 4$ can be safely taken. Within this scheme, the cancellation
of divergences between real and virtual terms can be seen schematically by
consideration of a toy calculation~\cite{Kunszt:1992tn},
\begin{equation}
{\cal I} = \lim_{\epsilon \to 0} \left(
\int_0^1 \frac{dx}{x} x^{-\epsilon} {\cal M}(x)
 + \frac{1}{\epsilon} {\cal M}(0) \right).
\end{equation}
Here, ${\cal M}(x)$ represents the real radiation matrix elements which are
integrated over the extra phase space of the gluon emission,
which contains a regulating factor $x^{-\epsilon}$. $x$ represents
a kinematic invariant which vanishes as the gluon becomes unresolved. The second
term is representative of the virtual contribution, which contains an
explicit pole, $1/\epsilon$, multiplying the lowest order matrix elements,
${\cal M}(0)$.

Two main techniques have been developed for isolating the singularities, which are commonly
referred to as the subtraction
method~\cite{Ellis:1980wv,Frixione:1995ms,Catani:1996vz,Nagy:1996bz}
and phase-space slicing~\cite{Fabricius:1981sx,Giele:1991vf}. For the sake
of illustration, we shall consider only the subtraction method. In this
approach, one explicitly adds and subtracts a divergent term such that
the new real radiation integral is manifestly finite. In the toy integral this
corresponds to,
\begin{eqnarray}
{\cal I} & = & \lim_{\epsilon \to 0} \left(
\int_0^1 \frac{dx}{x} x^{-\epsilon} \left[ {\cal M}(x) - {\cal M}(0) \right]
 +{\cal M}(0) \int_0^1 \frac{dx}{x} x^{-\epsilon}
 + \frac{1}{\epsilon} {\cal M}(0) \right) \nonumber \\
 & = & \int_0^1 \frac{dx}{x} \left[ {\cal M}(x) - {\cal M}(0) \right].
\end{eqnarray}
This idea can be generalized in order to render finite the real radiation contribution to any process, with a
separate counter-term for each singular region of phase space. Processes with a complicated phase space, such as
$W+2$~jet production, can end up with a large number of counterterms. NLO calculations are often set up to generate
cross sections by histogramming ``events'' generated with the relevant matrix  elements. Such events can not be
directly interfaced to parton shower programs, which we will discuss later in Section~\ref{sec:allorders},
as the presence of virtual corrections means that many of the events
will have (often large) negative weights. Only the total sum of events over all relevant subprocesses will lead to
a physically meaningful cross section. 

The inclusion of real radiation diagrams in a NLO calculation extends
the range of predictions that may be described by a lowest order calculation. For
instance, in the example above the $W$ boson is produced with zero transverse
momentum at lowest order and only acquires a finite $p_T$ at NLO. Even then, the
$W$ transverse momentum is exactly balanced by that of a single parton. In a real
event, the $W$ $p_T$ is typically balanced by the sum of several jet transverse momenta. 
In a fixed order calculation, these contributions would be included by moving to
even higher orders so that, for instance, configurations where the $W$ transverse
momentum is balanced by two jets enter at NNLO. Although this feature is clear
for the $p_T$ distribution of the $W$, the same argument applies for
other distributions and for more complex processes.

\subsubsection{Scale dependence}
\label{sec:scaledep}

\begin{figure}[t]
\begin{center}
\includegraphics[width=8cm]{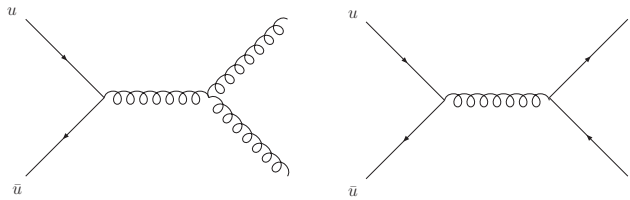}
\end{center}
\vspace*{-0.5cm}
\caption{The leading order diagrams representing inclusive jet production from a
quark antiquark initial state.}
\label{fig:jets}
\end{figure}

One of the benefits of performing a calculation to higher order in perturbation
theory is the reduction of the dependence of related predictions on the
unphysical renormalization ($\mu_R$) and factorization scales ($\mu_F$).
This can be demonstrated by considering inclusive jet production from a
quark antiquark initial state~\cite{Stump:2003yu}, which is represented by the lowest order
diagrams shown in Figure~\ref{fig:jets}.
This is a simplification of the full calculation, but is the dominant contribution when the typical
jet transverse momentum is large.

For this process, we can write the lowest order prediction for the single jet
inclusive distribution as,
\begin{equation}
\frac{d\sigma}{dE_T}=\as^2(\mu_R) \, \sigma_0 \, 
 \otimes f_q(\mu_F) \otimes f_{\bar q}(\mu_F),
\end{equation}
where $\sigma_0$ represents the lowest order partonic cross section calculated
from the diagrams of Figure~\ref{fig:jets} and $f_i(\mu_F)$ is the parton
distribution function for a parton $i$. Similarly, after including the
next-to-leading order corrections, the prediction can be written as,
\begin{eqnarray}
\fl
\frac{d\sigma}{dE_T} = \Biggl[ \as^2(\mu_R) \sigma_0
 + \as^3(\mu_R) \Bigl(
   \sigma_1 + 2b_0 \log(\mu_R/E_T) \sigma_0
   -2 P_{qq} \log(\mu_F/E_T) \sigma_0 \Bigr) \Biggr] 
   \nonumber \\
 \otimes f_q(\mu_F) \otimes f_{\bar q}(\mu_F).
\label{eq:nlomudep}
\end{eqnarray}
In this expression the logarithms that explicitly involve the renormalization
and factorization scales have been exposed. The remainder of the
${\cal O}(\as^3)$ corrections lie in the function $\sigma_1$.

From this expression, the sensitivity of the distribution to the renormalization
scale is easily calculated using,
\begin{equation}
\mu_R \frac{\partial \as(\mu_R)}{\partial \mu_R} 
 = -b_0 \as^2(\mu_R) - b_1 \as^3(\mu_R) + {\cal O}(\as^4),
\end{equation}
where the two leading coefficients in the beta-function, $b_0$ and $b_1$, are given
by $b_0=(33-2n_f)/6\pi$, $b_1=(102-38n_f/3)/8\pi^2$.
The contributions from the first and third
terms in~\eref{eq:nlomudep} cancel and the result vanishes, up to
${\cal O}(\as^4)$ .

In a similar fashion, the factorization scale dependence can be calculated
using the non-singlet DGLAP equation,
\begin{equation}
\mu_F \frac{\partial f_i(\mu_F)}{\partial \mu_F} =
 \as(\mu_F) P_{qq} \otimes f_i(\mu_F).
\end{equation}
This time, the partial derivative of each parton distribution function, 
multiplied by the first term in~\eref{eq:nlomudep}, cancels with 
the final term. Thus, once again, the only remaining terms are of
order $\as^4$.

This is a generic feature of a next-to-leading order calculation. An observable
that is predicted to order $\as^n$ is independent of the choice of either
renormalization or factorization scale, up to the next higher order in $\as$.

\begin{figure}[t]
\begin{center}
\includegraphics[width=12cm]{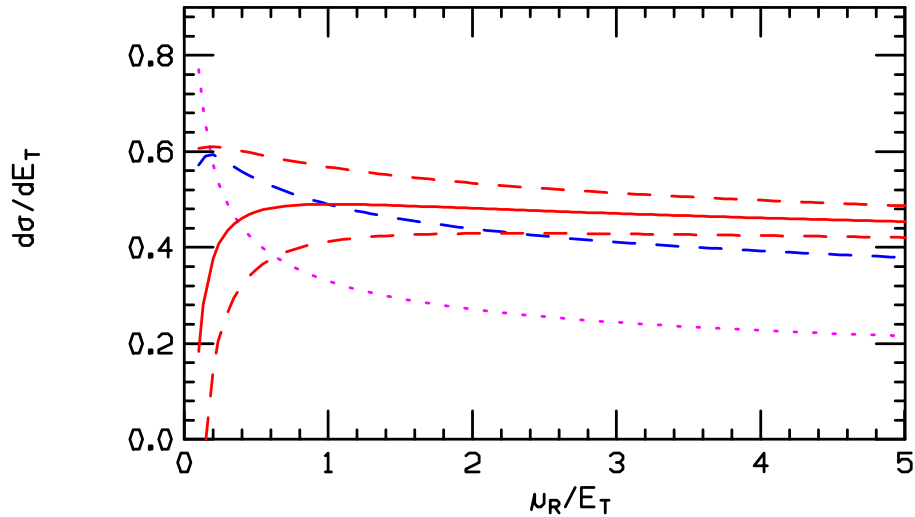}
\end{center}
\caption{The single jet inclusive distribution at $E_T=100$~GeV, appropriate
for Run~I of the Tevatron. Theoretical predictions are shown at LO (dotted
magenta), NLO (dashed blue) and NNLO (red). Since the full NNLO calculation is
not complete, three plausible possibilities are shown.\label{fig:mudepplot}}
\end{figure}

This discussion can be made more concrete by inserting numerical results into the
formulae indicated above. For simplicity, we will consider only the renormalization
scale dependence, with the factorization scale held fixed at $\mu_F=E_T$.
In this case it is simple to extend~\eref{eq:nlomudep} one higher order in
$\as$~\cite{Glover:2002gz}, 
\begin{eqnarray}
\fl
\frac{d\sigma}{dE_T} = \Biggl[ \as^2(\mu_R) \, \sigma_0 \,
 + \as^3(\mu_R) \Bigl(
   \sigma_1 + 2b_0 L\, \sigma_0
   \Bigr) 
   \nonumber \\
 + \as^4(\mu_R) \Bigl(
   \sigma_2 + 3b_0 L \, \sigma_1 + (3b_0^2 L^2 + 2b_1L) \, \sigma_0 
   \Bigr)
   \Biggr] \otimes f_q(\mu_F) \otimes f_{\bar q}(\mu_F),
\end{eqnarray}
where the logarithm is abbreviated as $L \equiv \log(\mu_R/E_T)$. For a realistic
example at the Tevatron Run I, $\sigma_0=24.4$ and $\sigma_1=101.5$. With these values
the LO and NLO scale dependence can be calculated; the result is shown
in Figure~\ref{fig:mudepplot}, adapted from~\cite{Glover:2002gz}.
At the moment the value of $\sigma_2$ is unknown (see Section~\ref{sec:nnlo}). However,
a range of predictions based on plausible values that it could take are also shown
in the figure, $\sigma_2=0$ (solid) and $\sigma_2 = \pm \sigma_1^2/\sigma_0$ (dashed).
It is clear that the renormalization scale dependence is reduced
when going from LO and NLO and will become smaller still at NNLO.

Although Figure~\ref{fig:mudepplot} is representative of the situation found at NLO, the exact details depend upon
the kinematics of the process under study and on choices such as the running of $\as$ and the pdfs used. Of
particular interest are the positions on the NLO curve which correspond to often-used scale choices. Due to the
structure of~\eref{eq:nlomudep} there will normally be a peak in the NLO curve, around which the scale
dependence is minimized. The scale at which this peak occurs is often favoured as a choice. For example, for
inclusive jet production at the Tevatron, a scale of $E_T^{jet}/2$ is usually chosen. This is near the peak of the
NLO cross section for many kinematic regions. It is also usually near the scale at which the LO and NLO curves
cross, i.e. when the NLO corrections do not change the LO cross section.
Finally, a rather different motivation comes from the consideration of a ``physical'' scale for the process. For
instance, in the case of $W$ production, one might think that a natural scale is the $W$ mass. Clearly, these three
typical methods for choosing the scale at which cross sections should be calculated do not in general agree. If
they do, one may view it as a sign that the perturbative expansion is well-behaved. If they do not agree then the
range of predictions provided by the different choices can be ascribed to the ``theoretical error'' on the
calculation.

\subsubsection{The NLO $K$-factor}
\label{sec:kfac}

The $K$-factor for a given process is a useful shorthand which encapsulates the strength of the
NLO corrections to the lowest order cross section. It is calculated by simply taking the ratio of
the NLO to the LO cross section. In principle, the $K$-factor may be very different for various
kinematic regions of the same process. In practice, the $K$-factor often varies slowly and may be
approximated as one number. 

However, when referring to a given $K$-factor one must take care to consider
the cross section predictions that entered its calculation. For instance,
the ratio can depend quite strongly on the pdfs that were used in both the LO
and NLO evaluations. It is by now standard practice to use a NLO pdf (for instance,
the CTEQ6M set) in evaluating the NLO cross section and a LO pdf (such as CTEQ6L)
in the lowest order calculation. Sometimes this is not the case, instead the same
pdf set may be used for both predictions. Of course, if one wants to estimate
the NLO effects on a lowest order cross section, one should take care to match
the appropriate $K$-factor.

A further complication is caused by the fact that the $K$-factor can depend quite strongly on the region of
phase space that is being studied. The $K$-factor which is appropriate for the total cross section of a
given process may be quite different from the one when stringent analysis cuts are applied. For processes in
which basic cuts must be applied in order to obtain a finite cross section, the $K$-factor again depends
upon the values of those cuts. Lastly, of course, as can be seen from Figure~\ref{fig:mudepplot} the
$K$-factor depends very strongly upon the renormalization and factorization scales at which it is evaluated.
A $K$-factor can be less than, equal to, or greater than 1, depending on all of the factors described
above. 

As examples, in Table~\ref{tab:kfac} we show the $K$-factors that have been obtained for a few
interesting processes at the Tevatron and the LHC.
\Table{\label{tab:kfac}$K$-factors for various processes at the Tevatron
and the LHC calculated using a selection of input parameters. In all cases, the CTEQ6M
pdf set is used at NLO. ${\cal K}$ uses the CTEQ6L1 set at leading order, whilst
${\cal K}^\prime$ uses the same set, CTEQ6M, as at NLO.
Jets satisfy the requirements $p_T>15$~GeV and $|\eta|<2.5$ ($5.0$) at the
Tevatron (LHC). In the $W+2$~jet process the jets are separated by $\Delta R>0.52$, whilst the
weak boson fusion (WBF) calculations are performed for a Higgs boson of mass $120$~GeV. Both renormalization
and factorization scales are equal to the scale indicated.}
\br & \centre{2}{Typical scales} & 
 \centre{3}{Tevatron $K$-factor} & \centre{3}{LHC $K$-factor} \\ \ns
& \crule{2} & \crule{3} & \crule{3} \\
Process & $\mu_0$ & $\mu_1$ &
 ${\cal K}(\mu_0)$ & ${\cal K}(\mu_1)$ & ${\cal K}^\prime(\mu_0)$ &
 ${\cal K}(\mu_0)$ & ${\cal K}(\mu_1)$ & ${\cal K}^\prime(\mu_0)$ \\
\mr
$W$                & $m_W$ & $2m_W$		    & 1.33 & 1.31 & 1.21 & 1.15 & 1.05 & 1.15 \\
$W+1$~jet          & $m_W$ & $\la p_T^{\rm jet}\ra$ & 1.42 & 1.20 & 1.43 & 1.21 & 1.32 & 1.42 \\
$W+2$~jets & $m_W$ & $\la p_T^{\rm jet}\ra$	    & 1.16 & 0.91 & 1.29 & 0.89 & 0.88 & 1.10 \\
$t{\bar t}$        & $m_t$ & $2m_t$		    & 1.08 & 1.31 & 1.24 & 1.40 & 1.59 & 1.48 \\
$b{\bar b}$        & $m_b$ & $2m_b$		    & 1.20 & 1.21 & 2.10 & 0.98 & 0.84 & 2.51 \\
Higgs via WBF      & $m_H$ & $\la p_T^{\rm jet}\ra$ & 1.07 & 0.97 & 1.07 & 1.23 & 1.34 & 1.09 \\
\br
\endTable
In each case the value of the $K$-factor is compared at two often-used scale choices, where the
scale indicated is used for both renormalization and factorization scales. For
comparison, we also note the $K$-factor that is obtained when using the same (CTEQ6M) pdf set at
leading order and at NLO. In general, the difference when using CTEQ6L1 and CTEQ6M at leading
order is not great. However, for the case of bottom production, the combination of the large
difference in $\as(m_b^2)$ and the gluon distribution at small $x$, result in very different
$K$-factors. The values ${\cal K}^\prime$ may, for instance, be useful in performing a NLO
normalization of parton shower predictions, as we shall discuss in later sections. 

\begin{figure}[t]
\begin{center}
\includegraphics[width=12cm]{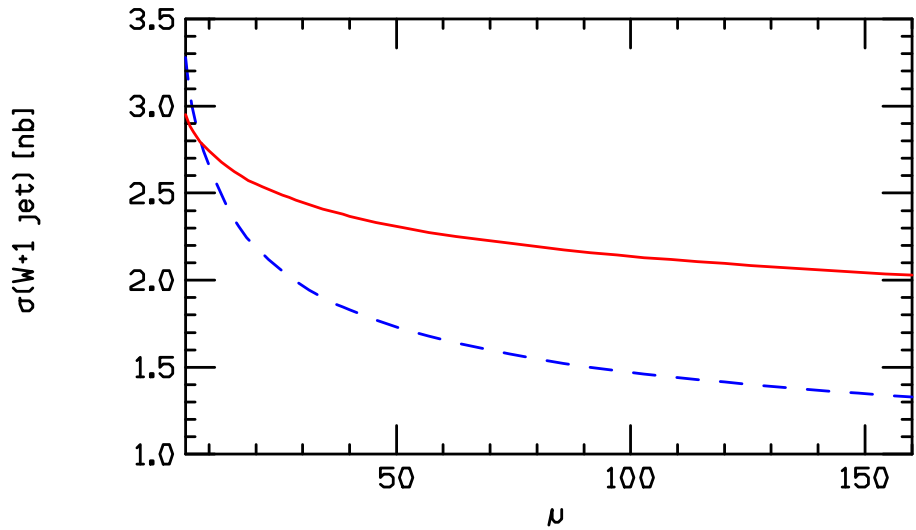}
\end{center}
\caption{The LO (dashed) and NLO (solid) scale variation of the $W+1$~jet cross section at the
Tevatron, using the same inputs as in Table~\ref{tab:kfac}.\label{fig:w1jmudep}}
\end{figure}

Such $K$-factors can be used as estimators for the NLO corrections for the listed processes in
situations where only the leading order cross sections are available (for instance, when using a
parton shower prediction).  Note that, for the case of $W+$~jet production, we have two relevant
scales for the hard scattering process: $m_W$ and the minimum allowed $p_T^{\rm jet}$. If this
threshold is quite low, as is the case in most studies at the Tevatron, these scales are quite
different. Thus, there can be a fairly large variation in the size of the predicted cross section
even at NLO, as illustrated in Figure~\ref{fig:w1jmudep}. In the leading order calculation, the
cross section varies by about a factor of $2.5$ over the range of scales shown. Although this
variation is reduced considerably at NLO, the cross section still increases by about $40\%$ when
moving from the highest scale shown to the lowest.

\subsection{Next-to-next-to-leading order}
\label{sec:nnlo}

With all the advantages of NLO, it is only natural to consider going deeper into
the perturbative expansion. In the same sense that one only gains a reliable prediction
of an observable at NLO, the first meaningful estimate of the theoretical error
comes at NNLO. Further reduction of scale uncertainties is expected and, as we shall see,
in cases where NLO corrections are large, it is a chance to check the convergence of the
perturbative expansion.

\begin{figure}[t]
\begin{center}
\includegraphics[width=8cm,angle=-90]{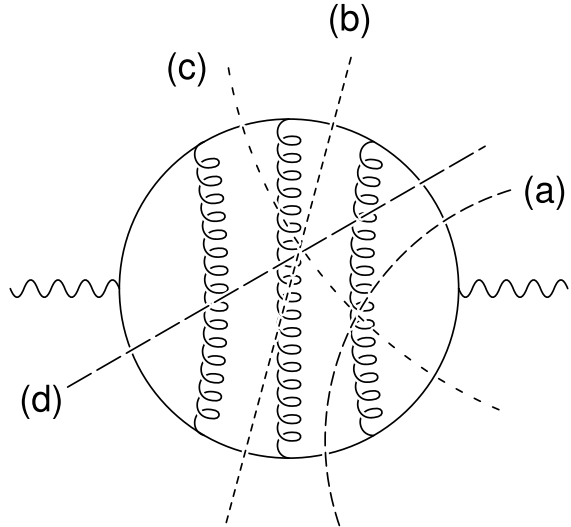}
\end{center}
\caption{A three-loop diagram which, when cut in all possible ways, shows
the partonic contributions that must be calculated to perform a NNLO prediction
of the $e^+ e^- \to 3$~jets rate. A description of the contribution represented
by each of the cuts (a)-(d) can be found in the text.}
\label{fig:nnlocuts}
\end{figure}

With these sorts of justifications in mind, a recent goal of theoretical effort
has been the calculation of the $3$~jet rate in $e^+e^-$ annihilation to NNLO.
Together with data from LEP and the SLC, this could be used to reduce the error on
the measurement of $\as(M_Z^2)$ to a couple of percent. However, the
ingredients of a NNLO calculation are both more numerous and more complicated than
those entering at NLO. The different contributions can best be understood by
considering all possible cuts of a relevant ${\cal O}(\as^3)$ three-loop
diagram, as shown in Figure~\ref{fig:nnlocuts}.

The first contribution, represented by cut (a) in Figure~\ref{fig:nnlocuts},
corresponds to 2-loop 3-parton diagrams. As the result of much innovative work
over the last few years, this contribution is now known~(see, for example,
references [4]--[6] of~\cite{Gehrmann-DeRidder:2005cm}). The contribution
labelled by (b) corresponds to the square of the 1-loop 3-parton matrix elements,
the same ones which appear (interfered with tree-level) in the NLO
calculation~\cite{Ellis:1980wv,Fabricius:1981sx}. The third contribution (c)
also contains 1-loop matrix elements, this time with 4 partons in the final
state, one of which is unresolved. As in a NLO calculation, when one parton is
unresolved this contribution diverges and a method must be developed to extract
all singularities. Both these matrix
elements~\cite{Glover:1996eh,Bern:1996ka,Campbell:1997tv,Bern:1997sc} and such
methods~(for instance,~\cite{Badger:2004uk} and references therein) have been
known for some time. The final contribution (d) involves only tree-level
5-parton matrix elements, but has so far proven the stumbling block to a complete
NNLO $3$-jet calculation. This piece contains two unresolved partons and, just
as before, this gives rise singularities that must be subtracted. However,
at present no general procedure for doing this exists and instead calculations
can only be performed on a case-by-case basis. Quite recently a method has
been developed for $e^+ e^- \to$~jets calculations which has been used to
calculate the doubly-unresolved sub-leading in $N_c$ contribution to the
$3$-jet rate~\cite{Gehrmann-DeRidder:2005cm}. Such progress bodes well for the
completion both of this calculation and the closely related $2$-jet rate at
hadron colliders~\footnote{
A consistent NNLO calculation at a hadron collider also requires parton densities evolved at the same order, which
is now possible thanks to the calculation of the QCD 3-loop splitting functions~\cite{Moch:2004pa,Vogt:2004mw}. The
differences between NLO and NNLO parton densities are reasonably small though, throughout most of the $x$ range.}.  

\begin{figure}[t]
\begin{center}
\includegraphics[width=8cm,angle=-90]{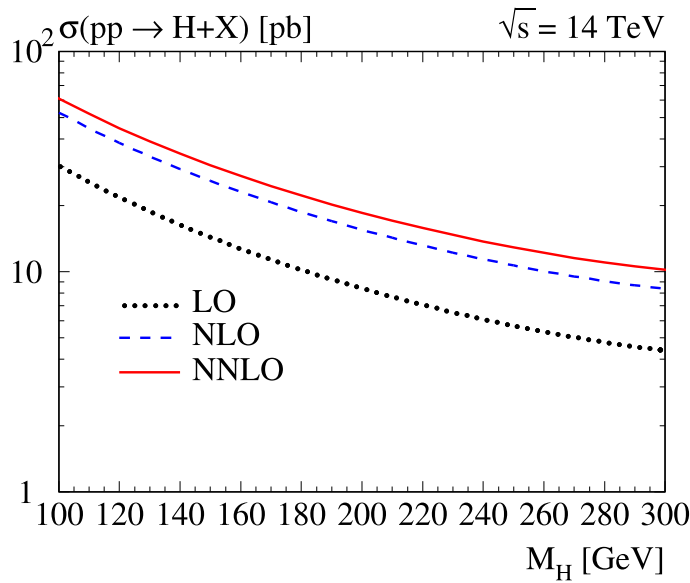}
\end{center}
\caption{The inclusive Higgs boson cross section as a function of the
Higgs boson mass.}
\label{fig:hxsec}
\end{figure}

The calculation that we have described represents the current frontier of NNLO
predictions. For slightly simpler $2 \to 1$ and $2 \to 2$ processes, NNLO results
are already available. The total inclusive cross section for the Drell-Yan
process, production of a lepton pair by a $W$ or $Z$ in a hadronic collision, has
long been known to NNLO accuracy~\cite{Hamberg:1990np}. In recent years the
inclusive Higgs boson cross section, which is also a one-scale problem in the
limit of large $m_t$, has also been computed at
NNLO~\cite{Harlander:2002wh,Anastasiou:2002yz}. For both these processes, the NLO
corrections had already been observed to be large and the inclusion of the NNLO
terms only provided a small further increase, thus stabilizing the perturbative
expansion of these cross sections. This is illustrated in Figure~\ref{fig:hxsec},
taken from~\cite{Harlander:2002wh},
which shows the inclusive Higgs boson cross section at the LHC at each order of
perturbation theory.

The above calculations have now been extended to include rapidity cuts on the
leptons in the Drell-Yan process, in order to be more applicable for studies at
the LHC~\cite{Anastasiou:2003yy}. These calculations extend the method used
in~\cite{Anastasiou:2002yz}, which uses an ingenious trick to bypass the problems
associated with doubly-unresolved radiation that we have described above. In this
approach, the phase space integrals are related to 2-loop integrals that are known
and whose calculation can be automated. In this way, NNLO predictions can be
provided for simple quantities such as rapidities. Further developments now allow
for the introduction of generic cuts, paving the way for more detailed
experimental analyses~\cite{Anastasiou:2004xq}.

\subsection{All orders approaches}
\label{sec:allorders}

Rather than systematically calculating to higher and higher orders in the perturbative
expansion of a given observable, a number of different ``all-orders'' approaches are
also commonly used to describe the phenomena observed at high-energy colliders. These
alternative descriptions are typically most useful under a different set of conditions
than a fixed order approach. The merging of such a description with fixed-order
calculations, in order to offer the best of both worlds, is of course highly desirable.

Resummation is one such approach, in which the dominant contributions from each order in perturbation theory are
singled out and ``resummed'' by the use of an evolution equation. Near the boundaries of phase space, fixed order
predictions break down due to large logarithmic corrections, as we have seen above. A straightforward example is
provided by the production of a vector boson at hadron colliders. In this case, two large logarithms can be
generated. One is associated with the production of the vector boson close to threshold (${\hat s}=Q^2$) and takes
the form $\as^n \log^{2n-1}(1-z)/(1-z)$, where $z=Q^2/{\hat s}-1$. The other logarithm, as illustrated
earlier, is associated with the recoil of the vector boson at very small transverse momenta $p_T$, so that
logarithms appear as $\as^n \log^{2n-1}(Q^2/p_T^2)$, c.f.~\eref{eq:dylog}. Various methods for
performing these resummations are available~\cite{Sterman:1986aj, Catani:1989ne, Catani:1990rp,
 Parisi:1979se, Altarelli:1984pt, Collins:1981uk, Collins:1981va, Collins:1984kg}, with some techniques
 including both effects at the
same time~\cite{Sterman:2004yk,Kulesza:2003wn,Kulesza:2002vk}. As we shall see later, the inclusion of such
effects is crucial in order to describe data at
the Tevatron and to estimate genuine non-perturbative effects. The ResBos program~\footnote{
http://hep.pa.msu.edu/people/cao/ResBos-A.html}~\cite{Balazs:1997xd}
is a publicly available program that provides NLO resummed predictions for processes such as
$W,Z,\gamma\gamma$ and Higgs boson production at hadron-hadron colliders. 
 Resummation is of course not restricted to the study
of these processes alone, with much progress recently in the resummation of event shape variables at hadron
colliders~(for a recent review, see~\cite{Banfi:2005mt}). 

\begin{figure}[t]
\begin{center}
\includegraphics[width=8cm,angle=90]{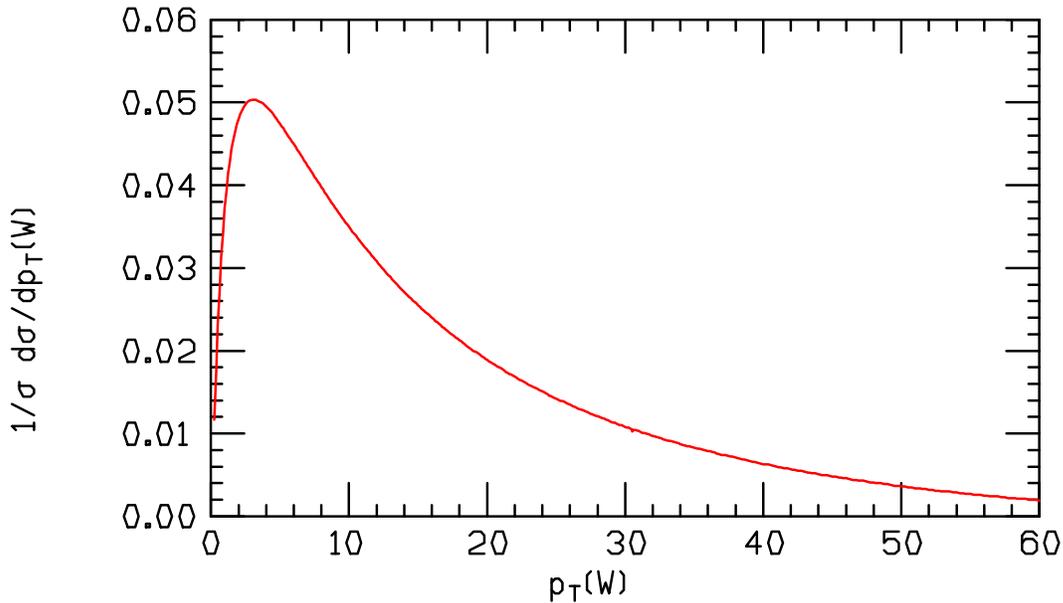}
\end{center}
\caption{The resummed (leading log) $W$ boson transverse momentum distribution.}
\label{fig:pt}
\end{figure}

The expression for the $W$ boson transverse momentum in which the leading logarithms have
been resummed to all orders is given by (c.f. \eref{eq:dylog} and \eref{eq:w1resum}), 
\begin{equation}
\frac{d\sigma}{dp_T^2} = \sigma \frac{d}{dp_T^2}
 \exp \left( -\frac{\as C_F}{2\pi} \log^2M_W^2/p_T^2 \right) \; .
\label{eq:wpt}
\end{equation}
This describes the basic  shape for the transverse distribution for $W$ production, which is shown in
Figure~\ref{fig:pt}.
Note that in this approximation the $p_T^2$ distribution vanishes as $p_T \to 0$, a feature
which is not seen experimentally. However this can be explained by the fact that the only configuration
included as $p_T \to 0$ is the one in which all emitted gluons are soft. In reality (and in a more
complete resummed prediction), multiple gluon emissions with a vector sum equal to $p_T$ contribute
and fill in the dip at $p_T=0$.

A different, but related, approach is provided by parton showers. The numerical implementation of a
parton shower, for instance in the programs PYTHIA,  HERWIG (HERWIG++~ \cite{Gieseke:2006ga}) and
SHERPA~\cite{Gleisberg:2003xi}, is a common tool used in many current
physics analyses.  By the use of the parton showering process, a few partons produced in a hard
interaction at  a high energy scale can be related  to partons at  an energy scale close to
$\Lambda_{\rm QCD}$.  At this lower energy scale, a universal non-perturbative model can then be used to
provide  the transition from partons to the hadrons that are observed experimentally. This is
possible  because the parton showering allows for the evolution, using the DGLAP formalism, of the
parton fragmentation function. The solution of this DGLAP evolution equation can be rewritten with
the help of the Sudakov form factor, which indicates the probability of evolving from a higher
scale to a lower scale
without the emission of a gluon greater than a given value. For the case of parton showers from the
initial state, the evolution proceeds backwards from the hard scale of the process to the cutoff
scale, with the Sudakov form factors being weighted by the parton distribution functions at the
relevant scales.

In the parton showering process, successive values of an evolution variable $t$, a momentum fraction
$z$ and an azimuthal angle $\phi$ are generated, along with the flavours of the partons emitted during
the showering. The evolution variable $t$ can be the virtuality of the parent parton (as in PYTHIA
versions 6.2 and earlier and in SHERPA), $E^2(1-\cos\theta)$, where $E$ is the energy of the parent parton and
$\theta$ is the opening angle between the two partons
(as in HERWIG)\footnote{An extension of this angular variable that allows for showering from heavy
objects has been implemented in HERWIG++.}, or the square of the relative
transverse momentum of the two partons in the splitting (as in PYTHIA~6.3). The HERWIG evolution
variable has angular ordering built in, angular ordering is implicit in the PYTHIA
6.3~\cite{Sjostrand:2004ef} evolution variable, and angular ordering has to be imposed after the fact
for the PYTHIA~6.2 evolution variable.  Angular ordering represents an attempt to simulate more
precisely those higher order contributions that are enhanced due to soft gluon emission (colour coherence).
Fixed order calculations explicitly account for colour coherence, while parton shower Monte Carlos
that include colour flow information model it only approximately.  

Note that with parton showering, we in principle introduce two new scales, one for initial state parton
showering and one for the shower in the final state. In the PYTHIA Monte Carlo, the scale used is most often
related to the maximum virtuality in the hard scattering, although a larger ad hoc scale, such as the total
centre-of-mass energy, can also be chosen by the user. The HERWIG showering scale is determined by the specific colour
flow in the hard process and is related to the invariant mass of the colour connected partons.

We can write an expression for the Sudakov form factor of an initial state parton in the form shown in
\eref{eq:sudakov}, where $t$ is the hard scale, $t_0$ is the cutoff scale and $P(z)$ is the
splitting function for the branching under consideration. 
\begin{equation}
\Delta(t) \equiv \exp \left[ - \int_{t_0}^t \frac{dt^\prime}{t^\prime}
 \int \frac{dz}{z} \frac{\as}{2\pi} \; P(z) \;
 \frac{f(x/z,t)}{f(x,t)} \right]
\label{eq:sudakov}
\end{equation}
The Sudakov form factor has a similar form for the final state but without the pdf weighting.  The introduction of
the Sudakov form factor resums all the effects of soft and collinear gluon emission, which leads to well-defined
predictions even in these regions. However, this ability comes at a price. Although the soft and collinear regions
are logarithmically enhanced and thus the dominant effect, this method does not attempt to correctly include the
non-singular contributions that are due to large energy, wide angle gluon emission.
We shall return to this discussion later.

\subsubsection{Sudakov form factors}
\label{sec:sudakov}

As discussed in the previous section, the Sudakov form factor gives the probability for a parton to evolve from a
harder scale to a softer scale without emitting a parton harder than some resolution scale, either in the initial
state or in the final state. Sudakov form factors form the basis for both  parton showering and resummation.
Typically, the details of the form factors are buried inside the interior of such programs. It is useful, however,
to generate plots  of  the initial state Sudakov form factors for the kinematic conditions encountered at both the
Tevatron  and LHC. Such plots  indicate  the likelihood for the non-radiation of gluons from the initial state
partons, and thus conversely for  the radiation of at least one such gluon. Thus, they can also serve as a handy
reference for the probability of jets from initial state radiation. A Sudakov form factor will depend on: (1) the
parton type (quark or gluon), (2)  the momentum fraction $x$ of the  initial state parton, (3) the hard and cutoff
scales for the process and (4) the resolution scale for the emission. Several examples are discussed below. These
plots were generated with the HERWIG++ parton shower
formalism~\cite{Gieseke:2004tc}~\footnote{We thank Stefan Gieseke for providing us with the relevant information.}. 

\begin{figure}[t]
\begin{center}
\includegraphics[width=10cm,angle=0]{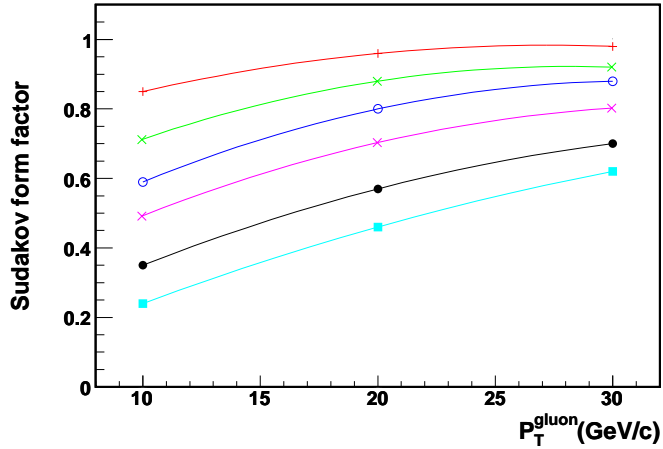}
\end{center}
\vspace*{-0.5cm}
\caption{The Sudakov form factors for initial state gluons at a hard scale of $100$~GeV as a
function of the transverse momentum of the emitted gluon. The form factors are for (top to bottom)
parton $x$ values of $0.3$, $0.1$, $0.03$, $0.01$, $0.001$ and $0.0001$.}
\label{fig:gluon_100}
\end{figure}

In Figure~\ref{fig:gluon_100} are plotted the Sudakov form factors for the splitting $g \rightarrow gg$, at a hard
scale of $100$~GeV, and for several different values of the parton $x$ value. The form factors are plotted versus
the resolution scale for the emitted gluon, which can be thought of roughly as the transverse momentum of the
emitted gluon. The probability for no emission decreases as the transverse momentum of the emitted gluon decreases
and as the parton $x$ decreases. The former is  fairly obvious; the latter  may not be so. The smaller the value
of  the initial parton momentum fraction, the larger is  the possible phase space for gluon emission. 

\begin{figure}[t]
\begin{center}
\includegraphics[width=10cm,angle=0]{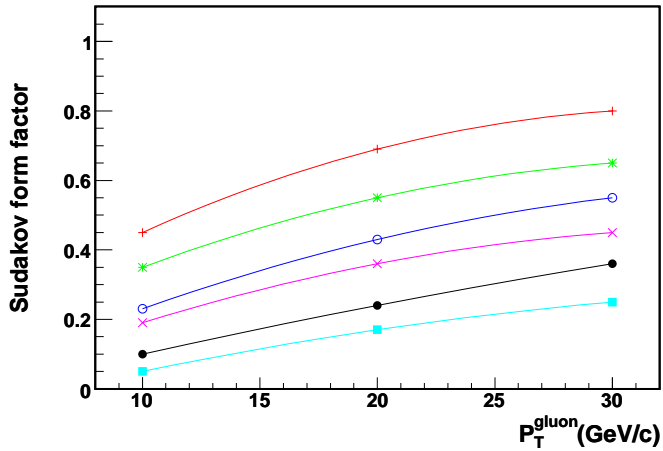}
\end{center}
\vspace*{-0.5cm}
\caption{The Sudakov form factors for initial state gluons at a hard scale of $500$~GeV as a
function of the transverse momentum of the emitted gluon. The form factors are for (top to bottom)
parton $x$ values of $0.3$, $0.1$, $0.03$, $0.01$, $0.001$ and $0.0001$.}
\label{fig:gluon_500}
\end{figure}

For example, the probability for a gluon with an $x$ value of $0.03$ to evolve from $100$~GeV down to $10$~GeV
without emitting a gluon of $10$~GeV or greater can be read off the plot as being $60\%$; thus the probability for at
least one such emission is $40\%$. This is another example where the probability of emission of a hard gluon is
enhanced by a logarithm (in this case the ratio of the hard scale to the resolution scale) compared to the naive
expectation of a factor of $\as$. The probability of emission of such a gluon from an initial state gluon on
the opposite side of the collision would of course be the same. 

\begin{figure}[t]
\begin{center}
\includegraphics[width=10cm,angle=0]{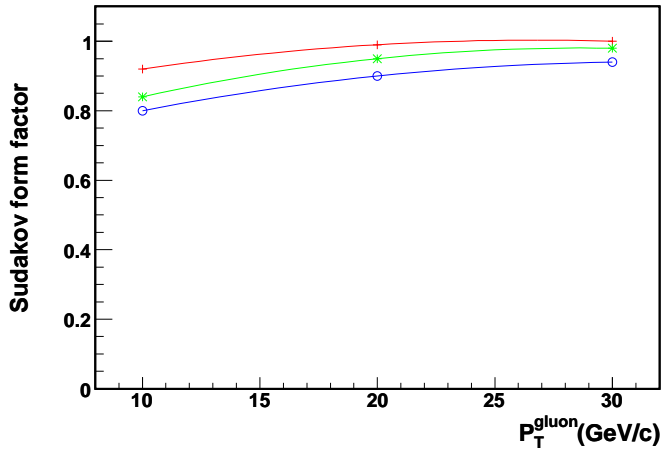}
\end{center}
\vspace*{-0.5cm}
\caption{The Sudakov form factors for initial state quarks at a hard scale of $100$~GeV as a function of the
transverse momentum of the emitted gluon. The form factors are for (top to bottom) parton $x$ values of
$0.3$, $0.1$ and $0.03$.
\label{fig:quark_100}}
\end{figure}

\begin{figure}[t]
\begin{center}
\includegraphics[width=10cm,angle=0]{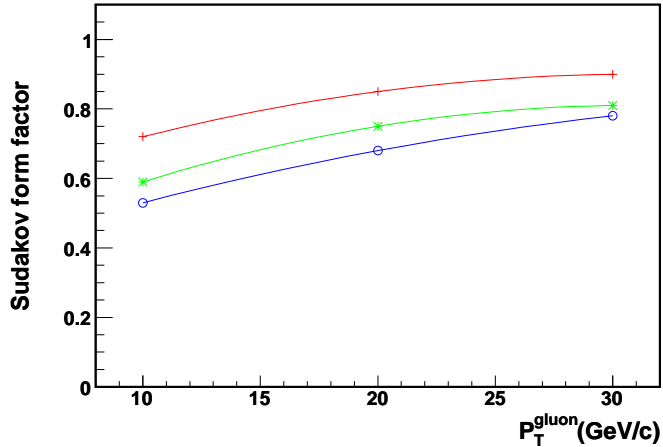}
\end{center}
\vspace*{-0.5cm}
\caption{The Sudakov form factors for initial state quarks at a hard scale of $500$~GeV as a function of the
transverse momentum of the emitted gluon. The form factors are for (top to bottom) parton $x$ values of
$0.3$, $0.1$ and $0.03$.
\label{fig:quark_500}}
\end{figure}

In Figure~\ref{fig:gluon_500}, the same Sudakov form factors are plotted but now using a hard
scale of $500$~GeV. The increased probability of a hard gluon emission can be observed. In
Figures~\ref{fig:quark_100} and~\ref{fig:quark_500}, the Sudakov form factors are plotted for the
hard scales of $100$~GeV and $500$~GeV as before, but now for  the splitting $q \rightarrow qg$.
The probability of no emission is larger, due to the smaller colour factor of the initial state
quark compared to the gluon.  Note that the form factor curves for $x$ values of less than $0.03$
have not been plotted as they would essentially lie on top of the $x=0.03$ curve. It is not the
smaller colour factor that causes the difference with the gluon but rather the splitting function.
The splitting function for $g \rightarrow gg$ has singularities both as $z \rightarrow 0$ and as
$z \rightarrow 1$, while the $q \rightarrow qg$ has only the $z \rightarrow 1$ singularity.  Thus,
for the $q \rightarrow qg$ splitting, there  is not much to gain from decreasing $x$ on a
logarithmic scale, as there is no singularity at $z=0$ in  the splitting function. 


\subsection{Partons and jet algorithms}
\label{sec:jetalgs}

In the detectors of experiments at the Tevatron and the LHC, collimated beams of particles are observed. In order
to categorize these events, the hadrons are collected into jets using a jet algorithm. To make a comparison with a
theoretical calculation of the types we have been discussing, it is necessary to also apply a jet algorithm at the
parton level. Ideally, one would like an algorithm which yields similar results at the experimental (hadron) and
theoretical (parton) levels. The goal is to characterize the short-distance physics event-by-event, in terms of the
jets formed by the algorithm. 

There are two essential stages for any jet algorithm. First, the objects belonging to a  cluster are identified. 
Second, the kinematic variables defining the jet are calculated  from the objects defining the cluster. The two
stages are independent.  For the latter  stage, using the jet algorithms developed for Run 2 at the Tevatron, the
jet kinematic properties are defined (using a 4-vector  recombination scheme) in terms of: $p_{jet}$, $p_T^{jet}$,
$y^{jet}$ and $\phi^{jet}$. 


At the experimental or simulated data level, jet algorithms cluster together objects  such as particles or energies
measured in calorimeter cells. At the theoretical level,   partons  are clustered. The goal of a jet algorithm is
to produce similar results no matter  the level it is applied. For a $2\rightarrow2$ LO calculation, a jet consists
simply of 1  parton and no jet algorithm is necessary. As more partons are added to a calculation, the  complexity
of a jet grows and approaches the complexity found either in parton shower  Monte Carlos or in data. For all
situations in which a jet can consist of more than 1 parton,   a completely specified jet algorithm is needed. The
clustering algorithms rely on the  association of these objects based on transverse momentum
(the $k_T$ algorithm)~\cite{Blazey:2000qt} or angles (the cone algorithm), relative to a jet axis. 

For NLO calculations, as for example $W + 2$~jets, a jet can consist of either 1 or 2  partons. Cone jet 
algorithms as currently used by the Tevatron experiments require the use of seeds (initial directions for  jet
cones)as the  starting points for jet searches.  For a partonic level final state, the seeds are the partons 
themselves. The Run 2 cone algorithm (midpoint)  places additional seeds  between stable cones having a separation
of less than twice the size of the  clustering cones; the use of these additional seeds removes problems with
infrared instabilities  in theoretical calculations. Without a midpoint seed, a jet could be formed/not formed 
depending on the presence of a soft gluon between two hard partons; this leads to an  undesirable logarithmic
dependence of the cross section on the energy of this soft gluon. 

With a cone algorithm, two partons are nominally included in the same  jet if they are within $R_{cone}$ of the
$p_T$-weighted jet centroid, and so within a maximum $\Delta R$ of $1.4$ of each other if a cone 
radius of $0.7$ is used. However, it was noted that with the experimental jet algorithms used 
at the Tevatron, that two jets would not be merged into a single jet if they were separated 
by a distance greater than $1.3$ times the cone radius. Thus, a phenomenological parameter 
$R_{sep}=1.3$ was added to the theoretical prediction; two partons would not be merged 
into a single jet if they were separated by more than $R_{sep} \times R_{cone}$ from each other. 
So, in a parton level calculation having at most $3$ partons in the final state, two partons
are merged into the same jet if they are within 
$R_{cone}$ of the $p_T$-weighted jet centroid and within  $R_{sep} \times R_{cone}$ of  each other; 
otherwise the two partons are termed separate jets.  Thus, for $W+2$~jet production at 
NLO, the final state can consist of either $2$ or $3$ partons. The $2$ parton final state will 
always be reconstructed as $2$ jets; the $3$ parton final state may be reconstructed as either 
$2$ or $3$ jets depending on whether the $2$ lowest $p_T$ partons satisfy the clustering 
criteria described above. Note that for some partonic level programs such as JETRAD~\cite{Giele:1994gf}
, 
the clustering is not performed prior to the evaluation of the matrix element. Thus, 
the individual transverse momenta of the jets are also not known at this time; only 
the transverse momentum of the highest $p_T$ parton, which by momentum conservation 
must remain unclustered, is known. For this reason, a renormalization/factorization 
scale of $p_T^{max}$ (the $p_T$ of this parton) is used in calculating the cross 
section for each jet. 

\begin{figure}[t]
\begin{center}
\includegraphics[width=12cm]{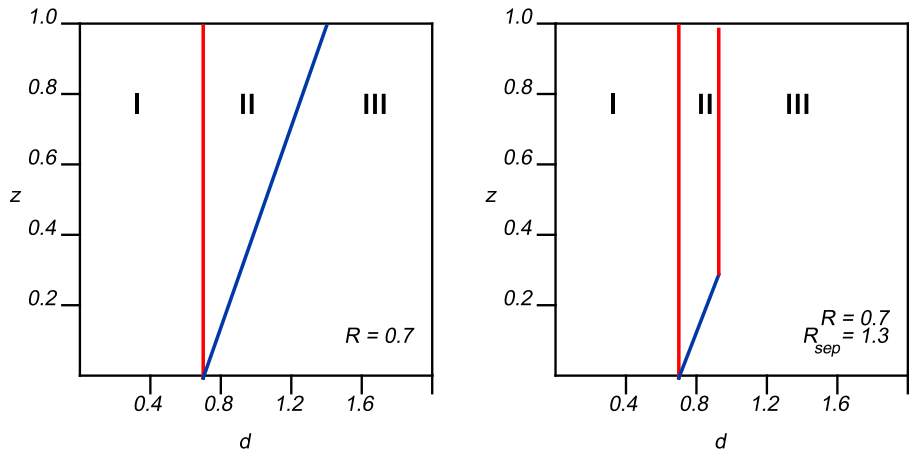}
\end{center}
\vspace*{-0.5cm}
\caption{
The parameter space $(d,Z)$ for which two partons  will be merged into a single jet.} 
\label{fig:dvsz}
\end{figure}

 A schematic diagram indicating the regions in which two partons will be included 
in the same jet is shown in Figure~\ref{fig:dvsz}~\cite{Ellis:2001aa}.  All partons within 
$R_{cone}$ of each other will always be clustered in the same jet. This corresponds to 
the region labeled I. An ideal cone algorithm acting on data would cluster together only 
the underlying parton level configurations corresponding to region II and not configurations 
in region III. However, as will be seen in Section~\ref{sec:tevatron}, the stochastic character of the parton 
showering and hadronization process makes such a clean division difficult in either real data 
or with a parton shower Monte Carlo. Because the matrix element for the emission of an 
additional real parton has both a collinear and soft pole, configurations in regions 
II and III with two partons having  $\Delta R$ near $0.7$ and $z$ near $0$ will be most  
heavily populated. The fractional contribution to the inclusive jet cross section of the merged two
parton configurations in Region II is proportional to $\as(p_T^2)$ and thus should decrease with
increasing jet transverse momentum.


%

The $k_T$ algorithm~\cite{Blazey:2000qt}  is conceptually simpler at all levels.  Two partons
(or particles, or energies) in
calorimeter towers are combined  if their relative transverse momentum is less than a given
measure. At the  parton level, each parton is considered as a proto-jet. The quantities 
$k^2_{T,i}=P^2_{T,i}$ and $k_{T,(i,j)}=\min(P^2_{T,i},P^2_{T,j})^.\Delta R^2_{i,j}/D^2$  are
computed for each parton and each pair of partons respectively. $P_{T,i}$ is the  transverse
momentum of the $i^{th}$ parton, $\Delta R_{i,j}$ is the distance (in $y$, $\phi$ space)  between
each pair of partons, and $D$ is a parameter that controls the size of the jet.  If the smallest
of the above quantities is a $k_{T,i}$ then that parton becomes a jet;  if the smallest quantity
is a $k_{T,(i,j)}$, then the two partons are combined into a single  proto-jet by summing their
four-vector components. In a NLO inclusive jet  calculation, the  two lowest $p_T$ partons may be
combined into a single jet, and thus the final state can consist of either $2$ or $3$ jets, as
was also true for the case of the cone algorithm. The $k_T$ algorithm merges two partons into a single jet
if $\Delta R_{ij}$ < $D$. In Figure~\ref{fig:dvsz}, if D is set to $R_{cone}$, the $k_T$ algorithm would
merge partons for which $d$ < $R$, or in the language used above for the cone algorithm, $R_{sep}=1$.
Since the cone algorithm merges the partons not only for $d$ < $R$, but also in the triangular region
above the line $d=R(1+z)$, a cone jet cross section at the parton level, for cone radius R, will necessarily
be larger than a $k_T$ jet cross section with D=R. If we use an $R_{sep}$ value of 1.3 rather than 2 for the
cone algorithm, the difference in relative cross sections between the two algorithms will be reduced. As we
will see in Section~\ref{sec:tevatron}, this relative behaviour of the two algorithms does not hold at the
hadron level. 

The assumption we are making above is that the jets of hadrons measured in a collider experiment can
be represented by the 1 or 2 partons that comprise a jet at the NLO theoretical level. That is, the 1
or 2 partons present in a NLO jet effectively represent the many partons produced by a parton shower.
For example, an equivalent description of the jet shape is provided by the two types of calculation. 
This approximation has been borne out in practice with one remaining correction being necessary.
Partons whose trajectories lie inside the jet cone eventually produce hadrons, some of which may land
outside the cone due to the fragmentation process. The fragmentation correction takes a particularly
simple form. For a cone of radius $0.7$, each jet loses approximately $1$~GeV due to fragmentation,
basically independent of the jet transverse energy. The na\"ive assumption might be that the energy
outside the cone due to fragmentation would rise with the jet energy; however, the jet becomes
narrower at higher $E_T$, leading to a roughly constant amount of energy in the outermost portions of
the jet. As will be described in Section~\ref{sec:tevatron}, corrections  also need to be applied to the data or to
the theory to take into account the underlying event energy. 

A quantity related to the jet shape is the jet mass. To first order, the jet mass, like the jet shape, is
determined by the hardest gluon emission. This emission is governed by the Sudakov form factor, and thus
the jet mass distribution should fall roughly as $1/m^2$, with modification at the two endpoints. There will be
a Sudakov suppression for low jet masses (corresponding to little or no gluon emission), and at high jet masses
the jet algorithm will tend to break the jet up into two separate jets. The average mass for this falling
distribution will be approximately $10$--$15$\% of the transverse momentum of the jet. 

\subsection{Merging parton showers and fixed order}
\label{sec:merge}

As we have discussed previously in Section~\ref{sec:allorders}, parton showers provide an excellent
description in regions which are dominated by soft and collinear gluon emission.
On the other hand, matrix element calculations provide a good description of processes 
where the partons are energetic and widely separated and, in addition, include the effects 
of interference between amplitudes with the same external partons. But, on the other hand, 
the matrix element calculations do not take into account the interference effects in soft 
and collinear gluon emissions which cannot be resolved, and  which lead to a Sudakov 
suppression of such emissions.  

Clearly, a description of a hard interaction which combines the two types of calculations would be
preferable. For this combination to take place, there first needs to be  a universal formalism that
allows the matrix element calculation to ``talk'' to the parton shower Monte Carlo. Such a universal
formalism was crafted during the Les Houches Workshop on Collider Physics in 2001 and the resulting
``Les Houches Accord'' is in common use~\cite{Boos:2001cv}.  The accord specifies an interface between
the matrix element and the parton shower program which provides information on the parton
$4$-vectors, the mother-daughter relationships, and the spin/helicities and colour flow. It also points
to intermediate particles whose mass should be preserved in the parton showering. All of the details
are invisible to the casual user and are intended for the matrix element/parton shower authors.

Some care must be taken however,  as a straight addition of the two techniques would lead to
double-counting in kinematic regions where the two calculations overlap. There have been many examples
where matrix element information has been used to correct the first or the hardest emission in a
parton shower. There are also more general techniques that allow  matrix element calculations and
parton showers  to each be used in kinematic regions where they provide the best description of the
event properties and that avoid double-counting. One such technique is  termed
CKKW~\cite{Catani:2001cc}.

\begin{figure}
\begin{center}
\includegraphics[width=10cm]{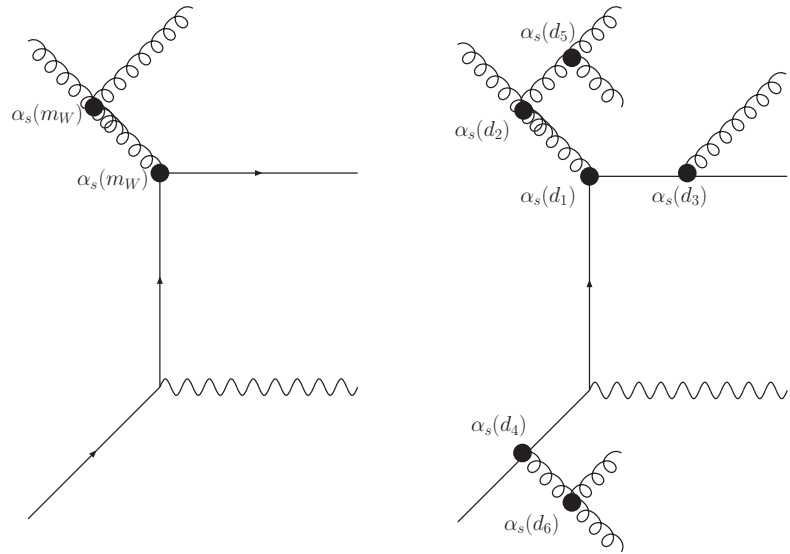}
\end{center}
\caption{
In the NLO formalism, the same scale, proportional to the hardness of the process,  is used for each QCD vertex. For the case of the
$W$+~2 jet diagram shown above to the left, a scale related to the mass of the $W$ boson, or to the average transverse momentum of the
produced jets, is typically used. The figure to the right shows the results of a simulation using the CKKW formalism. Branchings occur
at the vertices with resolution parameters $d_i$, where $d_1>d_2>>d_{ini}>d_3>d_4>d_5>d_6$ Branchings at the vertices 1-2 are produced
with matrix element information while the branchings at vertices 3-6 are produced by the parton shower. 
} 
\label{fig:w_jets_ckkw}
\end{figure}

With the CKKW technique,  the matrix element  description is used to describe parton branchings at 
large angle and/or energy, while the parton shower description is used for the smaller angle, lower 
energy emissions. The phase space for parton emission is thus divided into two regions, matrix element
dominated and parton shower dominated, using a resolution parameter $d_{ini}$. The argument of
$\as$ at all of the vertices is chosen to be equal to the resolution parameter $d_i$ at which the
branching has taken place and Sudakov form factors are inserted on all of the quark and gluon lines to
represent the lack of any emissions with a  scale larger than $d_{ini}$ between vertices. The $d_i$
represent a virtuality or energy scale. Parton showering is used to produce additional emissions at
scales less than $d_{ini}$. For a typical matching scale, approximately 10\% of the n-jet cross section is produced by parton showering from the n-1 parton matrix element; the rest arises mostly from the n-parton matrix element. A schematic representation of the CKKW scheme is shown in
Figure~\ref{fig:w_jets_ckkw} for the case of $W +$~jets production at a hadron-hadron collider. A
description of a $W + 2$~jet event in the NLO formalism is also shown for comparison.

The CKKW procedure provides a matching between the matrix  element and parton shower that should be
correct to the next-to-leading-logarithm (NLL) level. There are, however, a number of choices that
must be made in the matching procedure that do not formally affect the logarithmic behaviour but do
affect the numerical predictions, on the order of $20$--$30\%$. The CKKW procedure gives the right amount
of radiation but tends to put some of it in the wrong place with the wrong colour flow. Variations that
result from  these choices must be considered as part of the systematic error inherent in the CKKW
process. This will be discussed further in Section~\ref{sec:tevatron}.

For the CKKW formalism to work, matrix element information must in principle be available for any
number $n$ of partons in the final state. Practically speaking, having information available for $n$ up
to $4$ is sufficient for  the description of most events at the Tevatron or LHC. The CKKW formalism is
implemented in the parton shower Monte Carlo SHERPA~\cite{Gleisberg:2004hm} and has also been used for
event generation at the Tevatron and LHC using the Mrenna-Richardson formalism~\cite{Mrenna:2003if}.
An approximate version of CKKW matching (the ``MLM approach''~\footnote{...which one of the authors takes
credit for naming.}) is available in ALPGEN 2.0~\cite{Mangano:2002ea}.

\subsection{Merging NLO calculations and parton showers}
\label{sec:mergenlo}

A combination of NLO calculations with parton shower Monte Carlos leads to the best of both worlds. The NLO aspect
leads to a correct prediction for the rate of the process and also improves the description of the first hard
parton emission. The parton shower aspect provides a sensible description of multiple/soft collinear emissions with
a final state consisting of hadrons, which can then be input to a detector simulation. In a parton shower
interface, a  specific subtraction  scheme must be implemented to  preserve the NLO cross section. As each parton
shower Monte Carlo may produce a different real radiation component,  the subtraction  scheme must necessarily
depend on the Monte Carlo program to which the matrix element program is matched.  The presence of  interference
effects with NLO calculations requires that a  relatively small fraction ($\sim 10\%$) of events have negative
weights (of value $-1$). 

Several groups have worked on the subject to consistently 
combine partonic NLO calculations with parton showers. 
\begin{itemize}
\item  Collins, Zu~\cite{Chen:2001nf,Collins:2004vq}
\item  Frixione, Nason, Webber (MC@NLO) 
\cite{Frixione:2002ik,Frixione:2003ei,Frixione:2005gz} 
\item  Kurihara, Fujimoto, Ishikawa, Kato, Kawabata,
Munehisa, Tanaka~\cite{Kurihara:2002ne}
\item  Kr\"amer, Soper~\cite{Kramer:2003jk,Soper:2003ya,Kramer:2005hw}
\item  Nagy, Soper~\cite{Nagy:2005aa,Nagy:2006kb}
\end{itemize}

MC@NLO is the only publicly available program that combines NLO calculations with parton showering and
hadronization. The HERWIG Monte Carlo is used for  the latter. The use of a different Monte Carlo, such as PYTHIA,
would require a different subtraction scheme for  the NLO matrix elements. The processes included to date are:
($W,Z,\gamma^*,H,b\overline{b}, t\overline{t},HW,HZ,WW,WZ,ZZ$). Recently, single top hadroproduction has been added
to MC@NLO~\cite{Frixione:2005vw}. This is the first implementation of a process that has both initial- and
final-state singularities. This allows a more general category of additional processes to be added in the future. 
Work is proceeding on the addition of inclusive jet production and the production of a Higgs boson via $WW$ fusion.
Adding spin correlations to a process increases the level of difficulty but is important for processes
such as single top production. 

If, in addition, the CKKW formalism could be used for the description of hard parton emissions, the utility and
accuracy of a NLO Monte Carlo could be greatly increased. The merger of these two techniques should be possible in
Monte Carlos available by the time of the LHC turn-on. 

\section{Parton distribution functions}
\label{sec:pdfs}

\subsection{Introduction}
\label{sec:pdfintro}

As mentioned in Section~\ref{sec:formalism}, the calculation of the production cross sections at hadron colliders
for both interesting physics processes and their backgrounds relies upon a knowledge of the
distribution of the momentum fraction $x$ of the partons
(quarks and gluons) in a proton in the relevant  kinematic range. These parton
distribution functions (pdfs) can not be
calculated perturbatively but rather  are  determined by global fits to data from deep inelastic
scattering (DIS), Drell-Yan
(DY), and jet production at current energy ranges. Two major groups, CTEQ~\cite{Stump:2003yu}
and MRST~\cite{Martin:2004ir},
provide semi-regular updates to the parton distributions when new data and/or theoretical developments
become available. In
addition, there are also pdfs available from Alekhin~\cite{Alekhin:2005gq} and from the two HERA
experiments~\cite{Adloff:2000qk, Adloff:2003uh,Chekanov:2002pv, Chekanov:2005nn}.  The newest pdfs, in most cases, provide the
most accurate description of the world's data, and should be utilized in preference to older pdf sets. 

\subsection{Processes involved in global analysis fits}
\label{sec:global}

Measurements of deep-inelastic scattering (DIS) structure functions ($F_2,F_3$) in lepton-hadron scattering
and of lepton pair
production cross sections in hadron-hadron collisions provide the main source of information on quark distributions
$f_{q/p}(x,Q^2)$
inside hadrons. At leading order, the gluon distribution function 
$f_{g/p}(x,Q^2)$ enters directly
in hadron-hadron
scattering processes with jet final states.  Modern global parton distribution fits are carried out
to NLO and in some cases to NNLO, which allows $\as(Q^2)$, 
$f_{q/p}(x,Q^2)$
and 
$f_{g/p}(x,Q^2)$
to all mix and  contribute in the  theoretical formulae for all processes.
Nevertheless, the broad picture described above still holds to some degree in  global pdf  analyses.

The data from DIS, DY and jet processes utilized in pdf fits cover a wide range in $x$ and $Q^2$. HERA data
(H1~\cite{H1}+ZEUS~\cite{ZEUS}) are predominantly at low $x$, while the fixed target DIS~\cite{CCFR2,BCDMSp,BCDMSd,NMC,CCFR3} and
DY~\cite{E605,E866} data are at higher $x$. Collider jet data~\cite{CDFjet,D0jet} cover a broad range in $x$ and $Q^2$ by themselves and
are particularly important in the determination of the high $x$ gluon distribution. There is considerable  overlap, however, among the
datasets with the degree of overlap increasing with time as the statistics of the HERA experiments increase. Parton distributions
determined at a given $x$ and $Q^2$ `feed-down' or evolve to lower $x$ values at higher $Q^2$ values.  DGLAP-based NLO (and NNLO) pQCD
should  provide an accurate description of the data (and of the evolution of the parton distributions) over the entire kinematic  range
present in current global fits. At very low $x$ and $Q$, DGLAP evolution is believed to be no longer 
applicable and a BFKL~\cite{Fadin:1975cb,Kuraev:1976ge,Kuraev:1977fs,Balitsky:1978ic} description 
must be used.  No clear evidence of BFKL physics is seen in the current range of data;  thus all global analyses use
conventional DGLAP evolution of pdfs.

Many processes have been calculated to NLO and there is the possibility of including data from these processes in global fits.
For example, the rapidity distributions for $W^+$,  $W^-$ and $Z$ production at the Tevatron and  LHC should prove to be very
useful in constraining $u$ and $d$ valence and sea quarks. 

There is a remarkable consistency between the data in the pdf fits and the perturbative QCD theory fit to them. Both the CTEQ and
MRST groups use over 2000 data points in  their global pdf analyses  and the $\chi^2$/DOF for the fit of theory to data is on the
order of unity. For most of the data points, the statistical errors are smaller than the systematic errors, so a proper treatment
of the systematic errors and their bin-to-bin correlations is important. 

The accuracy of the extrapolation to higher $Q^2$ depends on the accuracy of the original measurement, any uncertainty on
$\as(Q^2)$ and the accuracy of the evolution code.  Most global pdf analyses are carried out at NLO; recently, the DGLAP
evolution kernels have been calculated at NNLO~\cite{Moch:2004sf}, allowing a full NNLO evolution to be carried out, and NNLO
pdfs calculated in this manner are available~\cite{Alekhin:2005gq,Martin:2002dr}. However, not all processes in the global
fits, and specifically inclusive jet production, are available at NNLO. Thus, any current NNLO global pdf analyses are still
approximate for this reason, but in practice the approximation should work well.  Current programs in use by CTEQ and MRST should
be able to carry out the evolution using NLO DGLAP to an accuracy of a few percent over the hadron collider kinematic range,
except perhaps at very large $x$ and very small $x$. (See the discussion in Section~\ref{sec:lhc} regarding the validity of NLO analysis at
the LHC.) The kinematics appropriate for the production of a state of mass  $M$ and rapidity $y$ at the LHC was shown in 
Figure~\ref{fig:lhcxq} in Section~\ref{sec:formalism}. For example,
to produce a state of mass $100$~GeV and rapidity $2$ requires partons of $x$
values $0.05$ and $0.001$ at a $Q^2$ value of $1 \times 10^4$~GeV$^2$. Also shown in the figure is a view of the kinematic
coverage of the fixed target and HERA experiments used in the global pdf fits. 

\subsection{Parameterizations and schemes}
\label{sec:schemes}

A global pdf analysis  carried out at next-to-leading order needs to be performed in a specific renormalization and
factorization   scheme. The evolution kernels are calculated in a specific scheme and  to maintain consistency, any hard scattering cross
section calculations used for the input processes or  utilizing the resulting pdfs need to have been implemented in that same 
renormalization scheme. Almost universally, the $\overline{MS}$ scheme is used;  pdfs are also available in the DIS scheme, a
fixed flavour scheme (see, for example,  GRV~\cite{Gluck:1998xa}) and several schemes that differ in their specific treatment of the charm
quark mass.

It is also possible to use only leading-order matrix element calculations in the global fits which results in leading-order
parton  distribution functions, which have been made available by both the CTEQ and MRST groups. For many hard matrix elements
for processes used in the global analysis, there exist $K$ factors significantly different from unity. Thus, one expects
there to be noticeable differences between the LO and NLO parton distributions. 

All global analyses use a generic form for the parameterization of both the quark and gluon distributions at some reference value
$Q_0$:
\begin{equation}
\ F(x,{Q_0})={A_0}x^{A_1}(1-x)^{A_2}P(x;A_3,...) 
\end{equation}
The reference value $Q_0$ is usually chosen in the range of $1$--$2$ GeV. The parameter $A_1$ is associated with small-$x$ Regge
behaviour while $A_2$ is associated with large-$x$ valence counting rules.

The first two factors, in general, are not  sufficient to describe either quark or gluon distributions. The term
$P(x;A_3,...)$ is a suitably chosen smooth  function, depending on one or more parameters, that adds more flexibility to the
pdf parameterization. In general, both the number of free parameters and the functional form can have an influence on the
global fit. The pdfs made available to the world from the global analysis groups can either be in a form where the $x$ and
$Q^2$ dependence is parameterized, or the pdfs for a given $x$ and $Q^2$ range can be interpolated from a grid that is
provided, or the grid can be generated given the starting parameters for the pdfs (see the discussion on
LHAPDF in Section~\ref{sec:LHAPDF}). All techniques should provide an accuracy on the output pdf distributions
on the order of a few percent. 

\begin{figure}[t]
\begin{center}
\includegraphics[width=8cm]{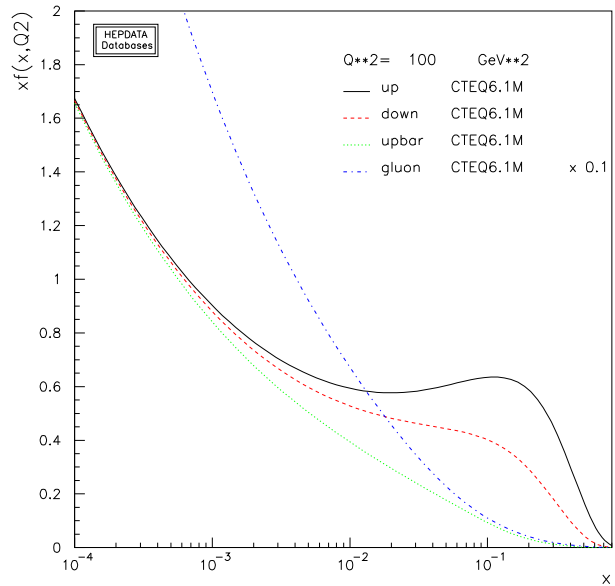}
\end{center}
\caption{
The CTEQ6.1 parton distribution functions evaluated at a $Q$ of $10$~GeV. 
} 
\label{fig:Allpdf}
\end{figure}

The parton distributions from the CTEQ6.1 pdfs release are  plotted in Figure~\ref{fig:Allpdf} at a $Q$ value of $10$~GeV. The
gluon  distribution is dominant at $x$ values of less than $0.01$ with the valence quark distributions dominant at higher $x$. One
of the major influences of the HERA data has been to steepen the gluon distribution at low $x$. 

\subsection{Uncertainties on pdfs}
\label{sec:uncertainties}

In addition to having the best estimates for the values of the pdfs in a given kinematic range, it is also important to
understand the allowed range of variation of  the pdfs, i.e. their uncertainties. A conventional method of estimating parton
distribution uncertainties has been to compare different published parton  distributions.  This is unreliable since most
published sets of parton distributions (for example from CTEQ and MRST) adopt  similar assumptions and the differences between
the sets do not fully explore the uncertainties that actually exist.  

The sum of the quark distributions 
($\Sigma f_{q/p}(x,Q^2)+ f_{g/p}(x,Q^2))$ 
is, in general,
well-determined over a wide range of $x$ and $Q^2$.  As stated above, the quark distributions are predominantly determined by the
DIS and DY data sets which have large statistics, and systematic errors in the few percent range ($\pm3\%$ for 
$10^{-4}<x<0.75$).  Thus the sum of the quark distributions is basically known to a similar accuracy. The individual  quark
flavours, though,  may have a greater uncertainty than the sum. This can be important, for example, in predicting distributions
that depend on specific quark flavours, like the $W$ asymmetry distribution~\cite{ Abe:1998rv} and the $W$ and $Z$ rapidity
distributions.

The largest uncertainty of any parton distribution, however, is that on the gluon distribution. The gluon distribution can be
determined indirectly at low $x$ by measuring the scaling violations in the quark distributions, but a direct measurement is
necessary at moderate to high $x$. The best direct information on the gluon distribution at moderate to high $x$ comes from jet
production at the Tevatron. 

There has been a great deal of recent activity on the subject of pdf uncertainties. Two techniques in particular, the Lagrange
Multiplier and Hessian techniques, have been used by CTEQ and MRST to estimate pdf uncertainties~\cite{Stump:2001gu,
Pumplin:2001ct, Martin:2002aw}. The Lagrange Multiplier technique is useful for probing the pdf uncertainty of a given process,
such as the $W$ cross section, while the Hessian technique provides a more general framework for estimating the pdf uncertainty for
any cross section. In addition, the Hessian technique results in tools more accessible to the general user. 

\begin{figure}[t]
\begin{center}
\includegraphics[width=12cm]{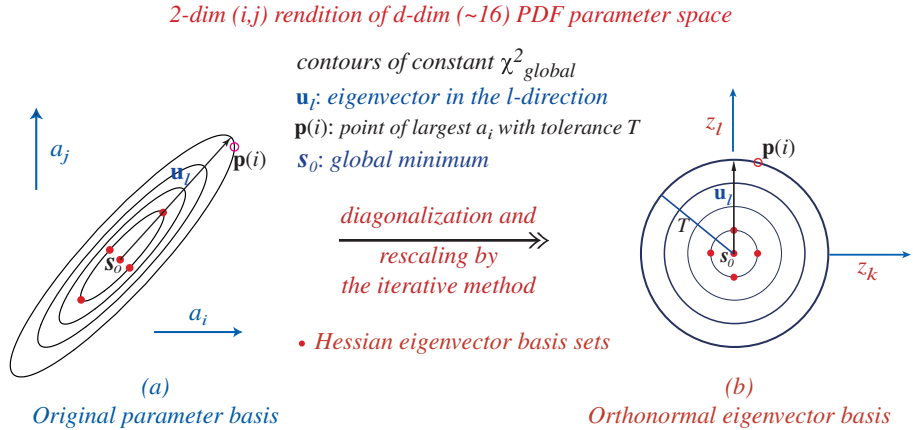}
\end{center}
\caption{
A schematic representation of the transformation from the pdf parameter basis to the orthonormal eigenvector basis. 
} 
\label{fig:Hesse}
\end{figure}

In the Hessian method a large matrix ($20\times20$ for CTEQ, $15\times15$ for MRST), with dimension equal to the number of free
parameters in the fit, has to be diagonalized. The result is 20 (15) orthonormal eigenvector directions for CTEQ (MRST) which
provide the basis for the determination of the pdf error for any cross section. This process is shown schematically in
Figure~\ref{fig:Hesse}. The eigenvectors are now admixtures of the 20 pdf parameters left free in the global fit. There is a
broad range for the eigenvalues, over a factor of one million. The  eigenvalues are distributed roughly linearly as $\log
\epsilon_i$, where $\epsilon_i$ is the eigenvalue  for the $i$-th direction. The larger eigenvalues correspond to directions which
are well-determined; for example, eigenvectors 1 and 2 are sensitive primarily to the valence quark distributions at moderate
$x$, a region where they are well-constrained. The theoretical uncertainty on the determination of the $W$ mass at both the Tevatron and the LHC
depends primarily on these 2 eigenvector directions, as $W$ production at the Tevatron proceeds primarily through collisions of
valence quarks. The most significant eigenvector directions for determination of the $W$ mass at the LHC correspond to larger
eigenvector numbers, which are primarily determined by sea quark distributions. In most cases, the eigenvector can not be
directly tied to the behaviour of a particular pdf in a specific kinematic region. A longer discussion of the meaning of the
eigenvectors for the CTEQ6.1 pdf analysis can be found at the benchmark website. 

There are two things that can happen when new pdfs (eigenvector directions) are added: a new direction in parameter space can be
opened to which some cross sections will be sensitive to (such as eigenvector 15 in the CTEQ6.1 error pdf set which is sensitive
to the high $x$ gluon behaviour and thus influences the high $p_T$ jet cross section at the Tevatron). In this case, a smaller
parameter space is an underestimate of the true pdf error since it did not sample a direction important for some physics. In the
second case, adding new eigenvectors does not appreciably open new parameter space and the new parameters should not contribute
much pdf error to most physics processes (although the error may be redistributed somewhat among the new and old eigenvectors).

\begin{figure}[t]
\begin{center}
\includegraphics[width=12cm]{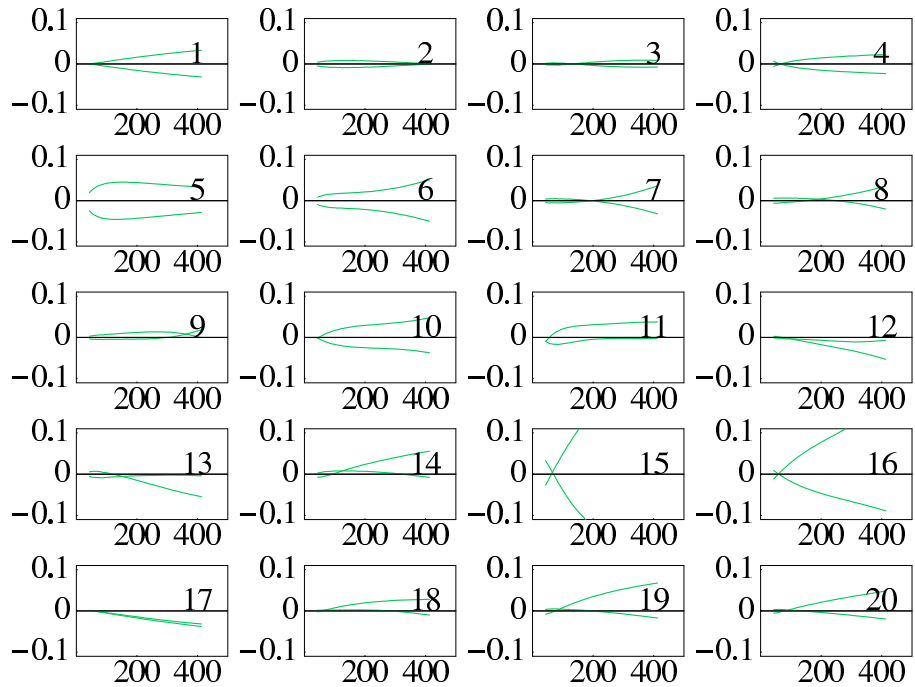}
\end{center}
\caption{
The pdf errors for the CDF inclusive jet cross section in Run 1 for the $20$ different eigenvector directions.
The vertical axes show the fractional deviation from the central prediction and the horizontal axes the jet
transverse momentum in GeV. 
\label{fig:jet_errors}
}
\end{figure}

Each error pdf results from an excursion along the ``$+$'' and ``$-$'' directions for each eigenvector. The excursions are symmetric
for the larger eigenvalues, but may be asymmetric for the more poorly determined directions. There are 40 pdfs for the CTEQ6.1
error set and 30 for the MRST error set.  In Figure~\ref{fig:jet_errors}, the pdf errors are shown in the ``$+$'' and ``$-$''
directions for the 20 CTEQ eigenvector directions for predictions for inclusive jet production at the Tevatron. The excursions
are symmetric for the first 10 eigenvectors but can be asymmetric for the last 10. 

Consider a variable $X$; its value using the central pdf for an error set (say CTEQ6.1M) is given by $X_0$. $X_i^+$
is the value of that variable using the pdf corresponding to the ``$+$'' direction for eigenvector $i$ and $X_i^-$ the
value for the variable using the pdf corresponding to the ``$-$'' direction. 
In order to calculate the pdf error for an observable, a {\it Master Equation} should be used: 
\begin{eqnarray}  
\Delta X^{+}_{max}&=&\sqrt{\sum_{i=1}^N[max(X^{+}_{i}-X_{0},X^{-}_{i}-X_{0},0)]^{2}} \nonumber \\
\Delta X^{-}_{max}&=&\sqrt{\sum_{i=1}^N[max(X_{0}-X^{+}_{i},X_{0}-X^{-}_{i},0)]^{2}}
\end{eqnarray}  

$\Delta X^{+}$ adds in quadrature the pdf error contributions that lead to an increase in the observable $X$ and
$\Delta X^{-}$ the pdf error contributions that lead to a decrease. The addition in quadrature is justified by the
eigenvectors forming an orthonormal basis. The sum is over all $N$ eigenvector directions, or $20$ in the case of
CTEQ6.1. Ordinarily, $X^{+}_{i}-X_{0}$ will be positive and $X^{-}_{i}-X_{0}$ will be negative, and thus it is
trivial as to which term is to be included in each quadratic sum. For the higher number eigenvectors, however, we
have seen that the ``$+$'' and ``$-$'' contributions may be in the same direction (see for example eigenvector 17 in
Figure~\ref{fig:jet_errors}). In this case, only the most positive term will be included in the calculation of
$\Delta X^{+}$ and the most negative in the calculation of $\Delta X^{-}$.
Thus, there may be less than $N$ terms for either
the ``$+$'' or ``$-$'' directions. There are other versions of the {\it Master Equation} in current use but the
version listed above is the ``official'' recommendation of the authors. 
Either $X_0$ and $X_i^{\pm}$ can be calculated separately in a matrix element/Monte Carlo program (requiring the
program to be run $2N+1$ times) or $X_0$ can be calculated with the program and at the same time the ratio of the pdf
luminosities (the product of the two pdfs at the $x$ values used in the generation of the event) for eigenvector $i$
($\pm$) to that of the central fit can be calculated and stored. This results in an effective sample with $2N+1$
weights, but identical kinematics, requiring a substantially reduced amount of time to generate.  

Perhaps the most controversial aspect of pdf uncertainties is the determination of the $\Delta\chi^2$ excursion from the central
fit that is representative of a reasonable error. CTEQ chooses a $\Delta\chi^2$ value of $100$ (corresponding to a 90\% CL limit)
while MRST uses a value of $50$.  Thus, in general, the pdf uncertainties for any cross section will be larger for the CTEQ set
than for the MRST set. The difference in the criterion indicates the difficulty of treating the error analysis for such a large
disparate sample of data in a statistically rigorous and unique manner. But in any case, both groups are in agreement that a
$\chi^2$ excursion of 1 (for a $1\sigma$ error) is too low of a value in a global pdf fit. The global fits use data sets
arising from a number of different processes and different experiments; there is a non-negligible tension between some of the
different data sets. pdf fits performed within a single experiment, or with a single data type may be able to set tighter
tolerances. The uncertainties for all predictions should be linearly dependent on the tolerance parameter used; thus, it should
be reasonable to scale the uncertainty for an observable from the $90\%$ CL limit provided by the CTEQ/MRST error pdfs to a
one-sigma error by dividing by a factor of $1.6$. Such a scaling will be a better approximation for observables more dependent on
the low number eigenvectors, where the $\chi^2$ function is closer to a quadratic form. Note that such a scaling may result in
an  underestimate of the $\it true$ pdf uncertainty, as the central results for CTEQ and MRST often differ by an amount similar
to this one-sigma error. (See the discussion in Section~\ref{sec:DY} regarding the $W$ cross section predictions at the Tevatron.)

\begin{figure}[t]
\begin{center}
\includegraphics[width=7cm]{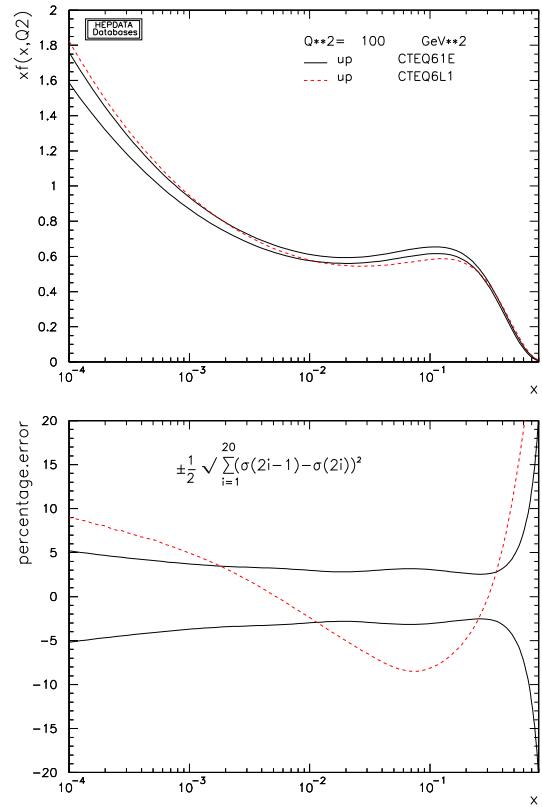}
\end{center}
\caption{
The pdf uncertainty for the up quark distribution from the CTEQ6.1 pdf set and a comparison to the CTEQ6L1 up quark distribution. 
\label{fig:plotpaw_up}
}
\end{figure}

\begin{figure}[t]
\begin{center}
\includegraphics[width=7cm]{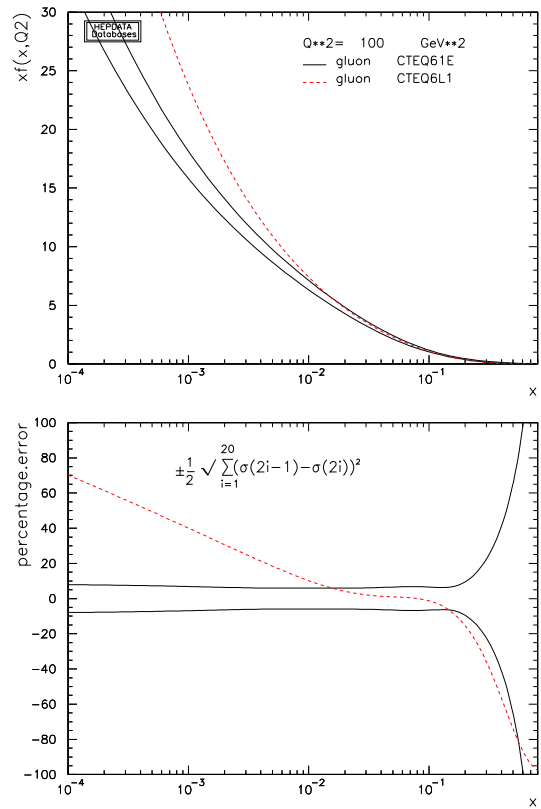}
\end{center}
\caption{
The pdf uncertainty for the gluon distribution from the CTEQ6.1 pdf set and a comparison to the CTEQ6L1 gluon distribution. 
\label{fig:plotpaw_gluon}
}
\end{figure}

\begin{figure}[t]
\begin{center}
\includegraphics[width=7cm]{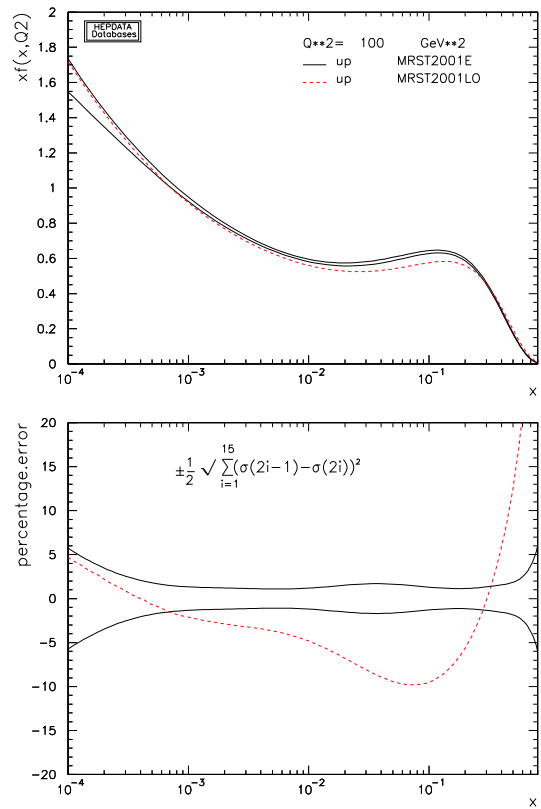}
\end{center}
\caption{
The pdf uncertainty for the up quark distribution from the MRST2001 pdf set and a comparison to the MRST2001LO up quark distribution. 
\label{fig:plotpaw_mrst_up}
}
\end{figure}

\begin{figure}[t]
\begin{center}
\includegraphics[width=7cm]{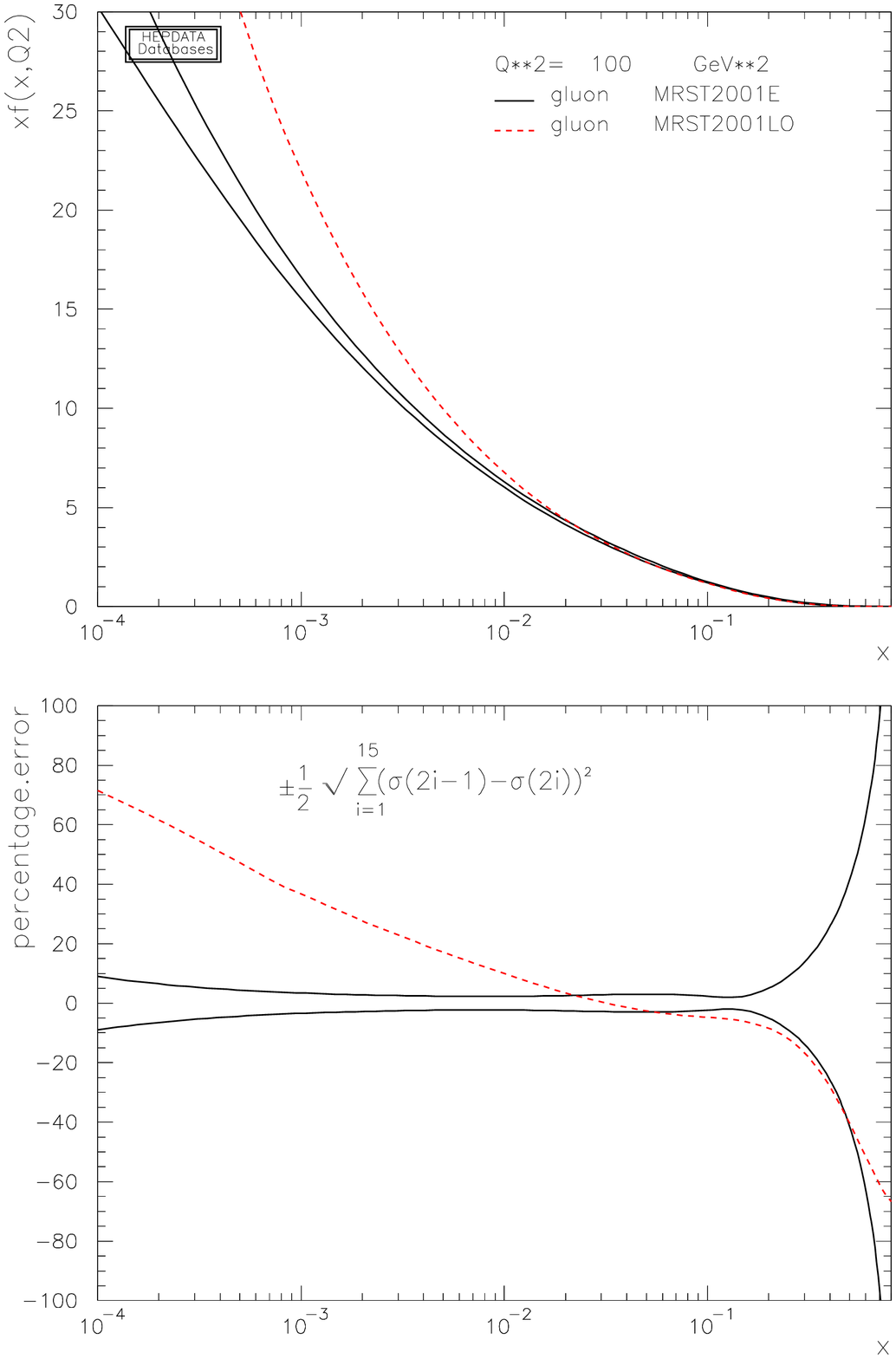}
\end{center}
\caption{
The pdf uncertainty for the gluon distribution from the MRST2001 pdf set and a comparison to the MRST2001LO gluon distribution. 
\label{fig:plotpaw_mrst_gluon}
}
\end{figure}

The CTEQ and MRST uncertainties for the up quark and gluon distributions are shown in
Figures~\ref{fig:plotpaw_up},~\ref{fig:plotpaw_gluon},~\ref{fig:plotpaw_mrst_up} and~\ref{fig:plotpaw_mrst_gluon}. The pdf
luminosity uncertainties for various pdf combinations for some kinematic situations appropriate to the LHC will be discussed in
Section~\ref{sec:lhc}. 

\begin{figure}[t]
\begin{center}
\includegraphics[width=8cm,height=10cm]{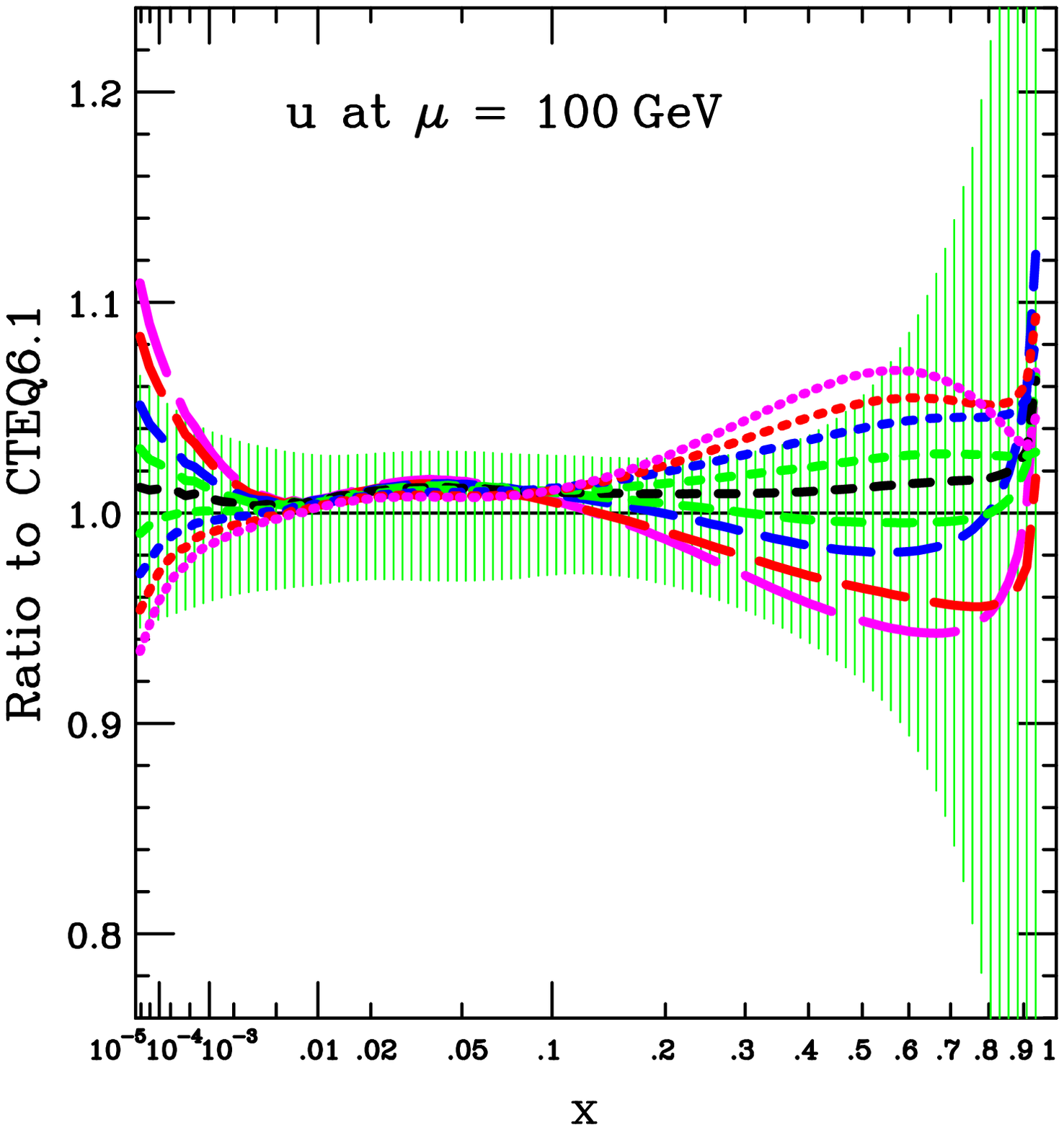}
\end{center}
\caption{
Variation of the up quark distribution for different values of $\as$. $\as$ varies from
0.110 (short-dash) to 0.124 (long-dash) in increments of 0.002.  
\label{fig:up_alphas}
}
\end{figure}

\begin{figure}[t]
\begin{center}
\includegraphics[width=8cm,height=10cm]{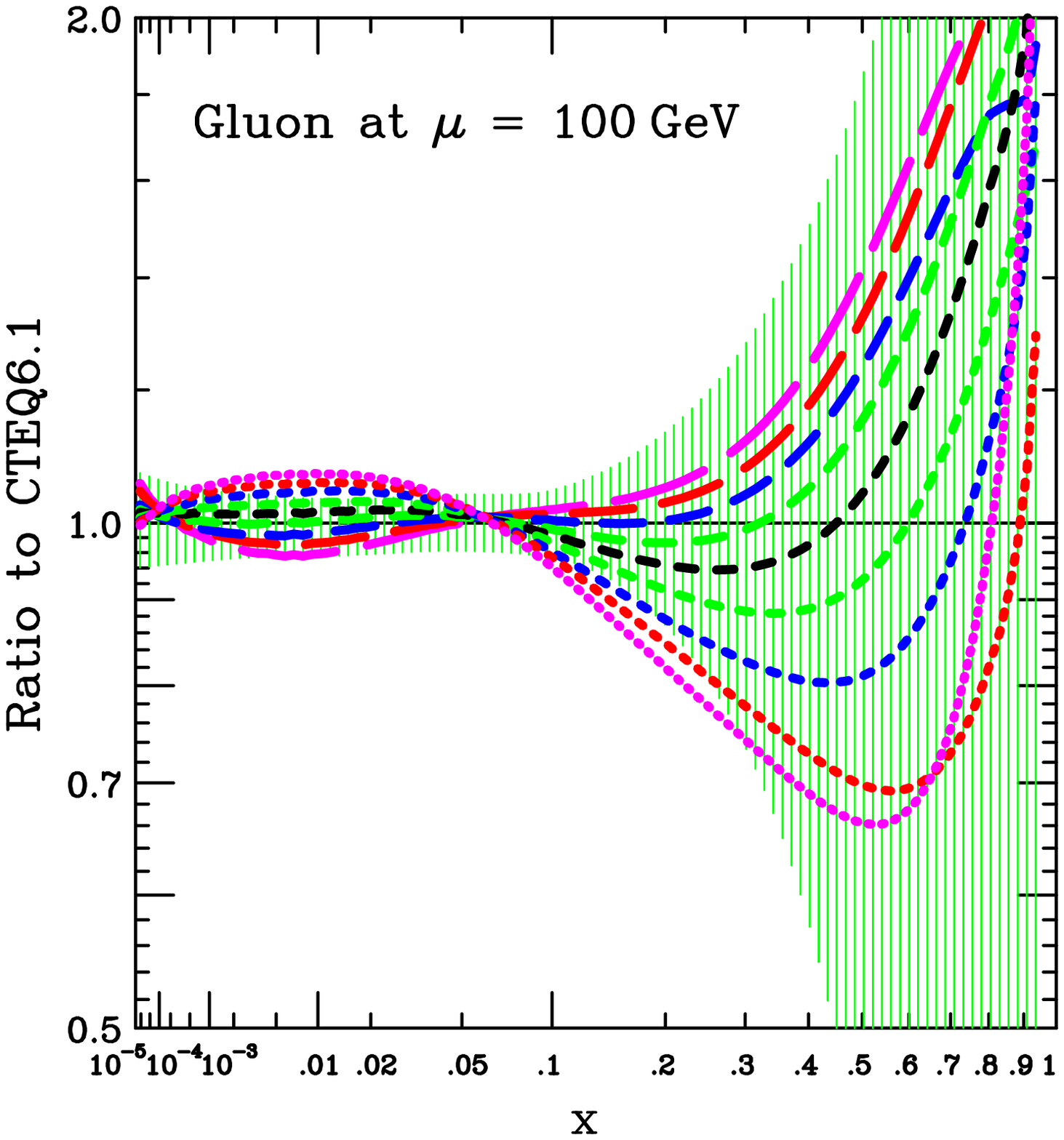}
\end{center}
\caption{
Variation of the gluon distribution for different values of $\as$.
$\as$ varies from 0.110 (short-dash) to 0.124 (long-dash) in increments of 0.002.
\label{fig:gluon_alphas}
}
\end{figure}

Both the  CTEQ and MRST groups typically use a fixed value of $\as(M_Z^2)$ (equal to the PDG world
average~\cite{Eidelman:2004wy} in their global fits. An additional uncertainty in the determination of pdfs results from the
uncertainty in the value of $\as(M_Z^2)$ used in the global fits. Figure~\ref{fig:up_alphas}  and
Figure~\ref{fig:gluon_alphas}  show the results of varying the value of $\as$ on the gluon and up quark
distributions~\cite{Pumplin:2005rh}. As expected, the gluon distribution has a greater sensitivity on $\as$ due  to the
coupling mentioned earlier. 

\subsection{NLO and LO pdfs}
\label{sec:nlolopdfs}

In the past, the global pdf fitting groups have produced sets of pdfs in which leading order rather than  next-to-leading order
matrix elements, along with the 1-loop $\as$ rather than the 2-loop $\as$, have been used to fit  the input datasets.
The resultant leading order pdfs have most often  been used in conjunction  with leading  order matrix element programs or
parton shower Monte Carlos. However, the leading order pdfs of a given  set will tend to differ  from the central pdfs 
in the NLO fit, and in fact will most often lie outside the pdf error band.  Such is the case for the up quark distribution shown
in Figures~\ref{fig:plotpaw_up} and~\ref{fig:plotpaw_mrst_up} and the gluon distribution shown in
Figures~\ref{fig:plotpaw_gluon} and~\ref{fig:plotpaw_mrst_gluon},  where the LO pdfs are plotted along with the NLO pdf error
bands. 

\setcounter{footnote}{0}

The global pdf fits are dominated by the high statistics, low systematic error deep inelastic scattering data and the differences
between the LO and NLO pdfs are determined most often by the differences between the LO and NLO matrix elements for deep
inelastic scattering. As the NLO corrections for most processes of interest at the LHC are reasonably small, the use of NLO
pdfs in conjunction with LO matrix elements will most often give a closer approximation of the full NLO result (although the
result remains formally LO). 
In many cases in which a relatively large $K$-factor results from a calculation of collider processes, the primary cause
is the difference between LO and NLO pdfs, rather than the differences between LO and NLO matrix elements. 

In addition, it is often useful to examine variations in acceptances in Monte Carlos using the
families of NLO error pdfs; thus, it is important to also compare to the predictions using the central (NLO) pdf.  It is our
recommendation, then,  that NLO pdfs be used for predictions at the LHC, even with LO matrix element programs and parton shower Monte
Carlos. There are two consequences: the pdfs must be positive-definite in the kinematic regions of interest as they will be
used to develop the initial state showering history and (2) underlying event tunes must be available using  the NLO pdfs. An
underlying event model that uses multiple parton interactions depends strongly on the slope of the low $x$  gluon  distribution.
The NLO gluon distribution  tends to have a much shallower slope than does the LO gluon and thus a different set of parameters
will  be needed for the tune~\footnote{The low $x$ behaviour for the 40 CTEQ6.1 error pdfs are similar
enough to that of the central CTEQ6.1 pdf that the same underlying event tune will work for all.}.
A NLO tune  is currently available for PYTHIA using the (NLO) CTEQ6.1 pdfs that is equivalent to
Tune A (see Section~\ref{sec:UE}), which uses CTEQ5L. An equivalent tune for HERWIG (with Jimmy)  is also available~\cite{Group:2006rt}.
At the end of the day, though,  the accuracy of the LO prediction is still only LO~\footnote{
It has been pointed out in~\cite{Collins:2002ey}
that $\overline{MS}$ pdfs are appropriate for calculations of inclusive cross sections but are not suitable for the exclusive
predictions inherent with Monte Carlo event generators. It is in principle necessary to define pdfs specific to the parton showering
algorithm for each Monte Carlo. However, the error introduced by using $\overline{MS}$ pdfs with leading order parton shower
Monte Carlos should be of a higher logarithmic order than the accuracy we are seeking.} 

\begin{figure}[t]
\begin{center}
\includegraphics[width=7cm]{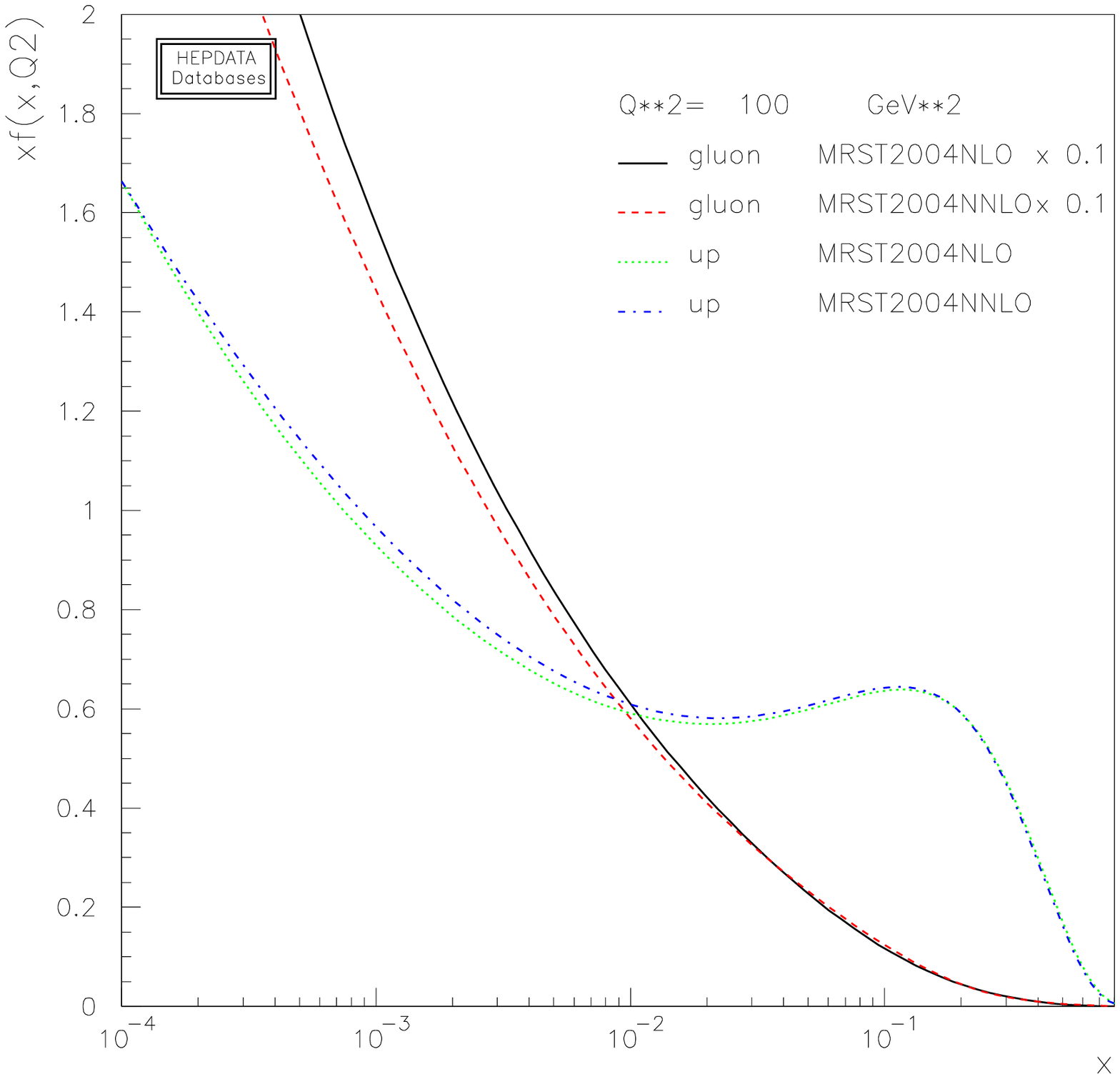}
\end{center}
\caption{
A comparison of the NLO and NNLO gluon and up quark distributions for the MRST2004 sets of pdfs. 
\label{fig:plotpaw_nlo_nnlo}
}
\end{figure}

The transition from NLO to NNLO results in much smaller changes to the pdfs as can be observed in
Figure~\ref{fig:plotpaw_nlo_nnlo}.

\subsection{Pdf uncertainties and Sudakov form factors}
\label{sec:pdfsud}

As discussed in the above section, it is often useful to use the error pdf sets with parton shower Monte Carlos.
The caveat still remains that a true test of the acceptances would use a NLO MC.
Similar to their use with matrix element calculations,  events can be
generated once  using the central pdf and the pdf  weights stored for the error pdfs. These pdf weights then can be used to
construct  the pdf uncertainty for any  observable. Some sample code for PYTHIA is given on the benchmark website.  One
additional  complication with respect to  their use in matrix  element programs is that the  parton distributions are used to
construct the initial state parton showers through  the  backward evolution process. The space-like evolution of the initial
state partons  is guided by the ratio of parton distribution functions at different $x$ and $Q^2$ values,
c.f.~\eref{eq:sudakov}. Thus the Sudakov form
factors in parton shower Monte Carlos will be constructed using only  the central pdf and not with any of the individual error
pdfs and  this may lead  to some   errors for  the calculation of the pdf uncertainties of some observables.  However, it was 
demonstrated in Reference~\cite{Gieseke:2004tc} that the pdf uncertainty for Sudakov form factors in the  kinematic region
relevant for the LHC is minimal, and the weighting technique can be  used  just as well with parton shower Monte Carlos as with
matrix element  programs. 

\subsection{LHAPDF}
\label{sec:LHAPDF}

Libraries such as PDFLIB~\cite{Plothow-Besch:1995ci} have been established that maintain a large collection of available pdfs.
However, PDFLIB is no longer supported, making it more difficult for easy access to the most up-to-date pdfs. In addition, the
determination of the pdf uncertainty of any cross section typically involves the use of a large number of pdfs (on the order of
30-100) and PDFLIB is not set up for easy accessibility for a large number of pdfs. 

At Les Houches in 2001, representatives from a number of pdf groups were present and an interface (Les Houches Accord 2, or
LHAPDF)~\cite{Giele:2002hx} that allows the compact storage of the information needed to define a pdf was defined. Each pdf can
be determined either from a grid in $x$ and $Q^2$ or by a few lines of information (basically the starting values of the
parameters at $Q=Q_o$) and the interface carries out the evolution to any $x$ and $Q$ value, at either LO or NLO as appropriate
for each pdf. 

The interface is as easy to use as PDFLIB and consists essentially of 3 subroutine calls:
\begin{itemize}
\item	call Initpdfset({\it name}): called once at the beginning of the code; {\it name} is the file name of the
 external pdf file that defines the pdf set (for example, CTEQ, GKK~\cite{Giele:2001mr} or MRST)
\item	call Initpdf({\it mem}): {\it mem} specifies the individual member of the pdf set
\item	call evolvepdf({\it x,Q,f}): returns the pdf momentum densities for flavour {\it f} at a momentum fraction {\it x}
and scale {\it Q}
\end{itemize}

Responsibility for LHAPDF has been taken over by the Durham HEPDATA project~\cite{Whalley:2005nh} and regular updates/improvements have been
produced.  It is currently included in the matrix element program MCFM~\cite{Campbell:2000bg} and will be included in future
versions of other matrix element and parton shower Monte Carlo programs. Recent modifications make it possible to include all
error pdfs in memory at the same time. Such a possibility reduces the amount of time needed for pdf error calculations on any
observable. The matrix element result can be calculated once using the central pdf and the relative (pdf)$\times$(pdf) parton-parton
luminosity can be calculated for each of the error pdfs (or the values of $x_1$,$x_2$, the flavour of partons 1 and 2 and the
value of $Q^2$ can be stored). Such a pdf re-weighting has been shown to work both for exact matrix element calculations as well
as for matrix element+parton shower calculations. In addition, a new routine LHAGLUE~\cite{Whalley:2005nh} provides an interface
from PDFLIB to LHAPDF making it possible to use the PDFLIB subroutine calls that may be present in older programs. 
	

\section{Comparisons to Tevatron data}
\label{sec:tevatron}

\subsection{$W$/$Z$ production}
\label{sec:tevwz}

\begin{figure}[b]
\begin{center}
\includegraphics[width=9cm]{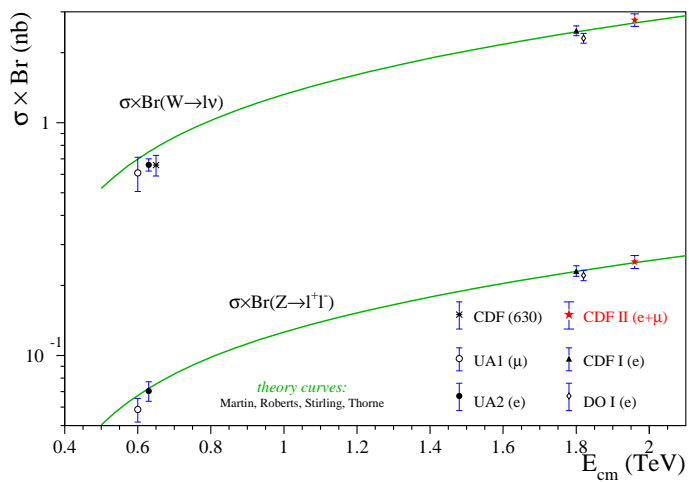}
\end{center}
\vspace*{-0.5cm}
\caption{
$W$ and $Z$ cross sections as a function of the centre-of-mass energy.
\label{fig:cs_vs_e}}
\end{figure}

As discussed earlier, $W$ production at hadron-hadron colliders serves as a precision benchmark for Standard Model physics.
Experimentally, the systematic errors are small.  The decay leptons are easy to trigger on and the backgrounds are under good
control. Theoretically, the cross section and rapidity distributions are known to NNLO. The $W$ and $Z$ cross sections measured at the
Tevatron are shown in Figure~\ref{fig:cs_vs_e} for both Run 1 and Run 2~\cite{Acosta:2004uq}.
The experimental cross sections agree well with the predictions and the Run 2 cross sections show the rise expected from the increase
in centre-of-mass energy over Run 1. In Figure~\ref{fig:wz} in Section~\ref{sec:formalism}, the Run 2 cross sections from CDF and D0 are compared to
predictions at LO, NLO and NNLO. The NNLO predictions are a few percent larger than the NLO predictions. Good agreement is observed
with both. It is noteworthy that the largest experimental uncertainty for the $W$ and $Z$ cross sections is the uncertainty in the
luminosity, typically on the order of $5\%$. The theoretical systematic errors are primarily from  the pdf uncertainty. If other
Tevatron cross sections were normalized to the $W$ cross section, then both the theoretical and experimental systematic errors for those
cross sections could be reduced.  

%
%
%

\begin{figure}[t]
\begin{center}
\includegraphics[width=9cm]{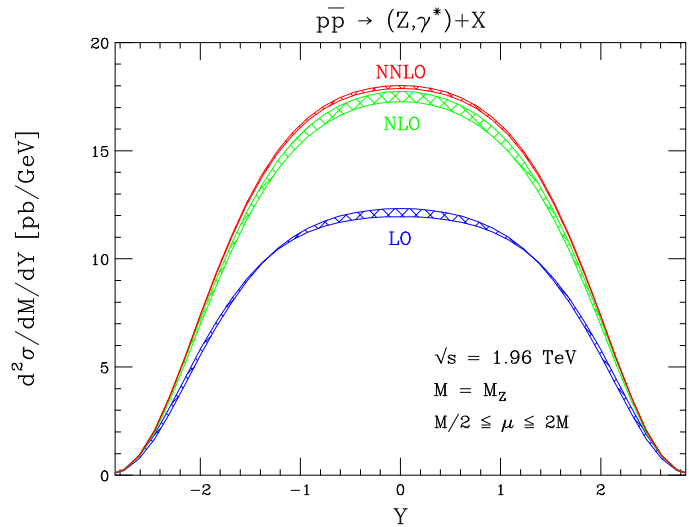}
\end{center}
\vspace*{-0.5cm}
\caption{
Predictions for the rapidity distribution of an on-shell $Z$ boson in Run 2 at the Tevatron at LO, NLO and NNLO. The bands indicate the
variation of the renormalization and factorization scales within the range $M_Z/2$ to $2M_Z$.
\label{fig:tev2z}}
\end{figure}

\begin{figure}[t]
\begin{center}
\includegraphics[width=10cm]{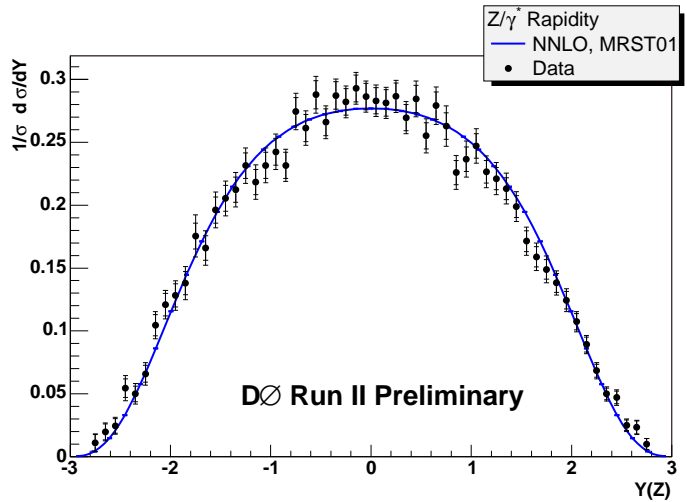}
\end{center}
\vspace*{-0.5cm}
\caption{
Z rapidity distribution from D0 in Run 2.
\label{fig:d0_z_y}}
\end{figure}

Predictions of the rapidity distribution for $Z$ ($\rightarrow e^+e^-$) production at the Tevatron are shown in 
Figure~\ref{fig:tev2z}~\cite{ Anastasiou:2003yy,Anastasiou:2003ds}.
There is a shape change in going from LO to NLO, as well as an increase in normalization, while the transition from NLO to NNLO is
essentially just a small $K$-factor with little change in shape. The $Z$ rapidity distribution measured by D0 in Run 2 is shown in
Figure~\ref{fig:d0_z_y}~\cite{D0WZ}; the measurement agrees with the NNLO prediction over the entire rapidity range.
Such cross sections may be useful as input to global pdf fits in the near future. 

\begin{figure}[t]
\begin{center}
\includegraphics[width=10cm]{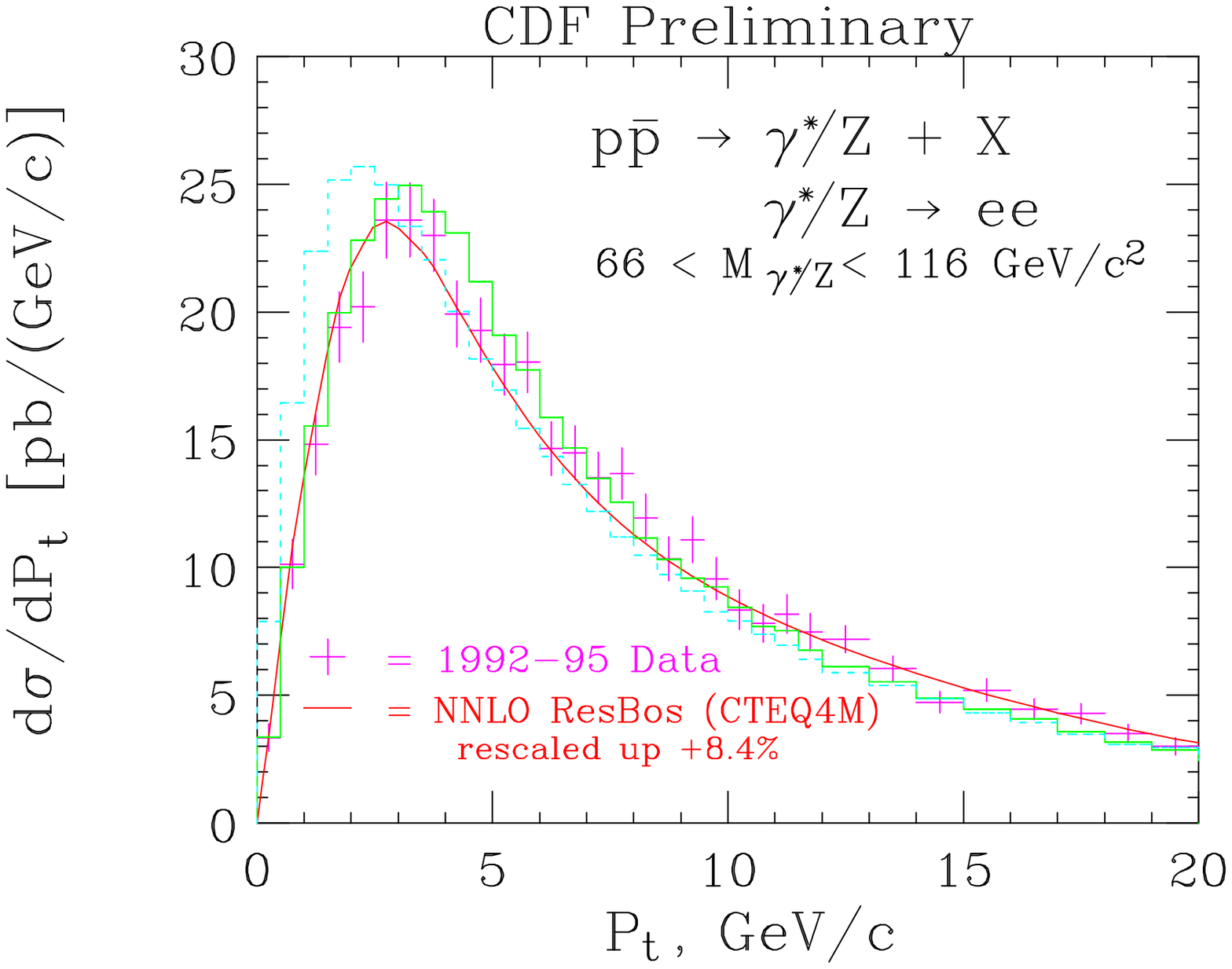}
\end{center}
\caption{
The transverse momentum distribution (low $p_T$) for $Z \rightarrow e^+e^-$ from CDF in Run 1, along with comparisons to predictions
from PYTHIA and ResBos. The dashed-blue curve is the default PYTHIA prediction. The PYTHIA solid-green curve has had an
additional $2$~GeV of $k_T$ added to the parton shower.
\label{fig:run1ee}}
\end{figure}

\begin{figure}[t]
\begin{center}
\includegraphics[width=10cm]{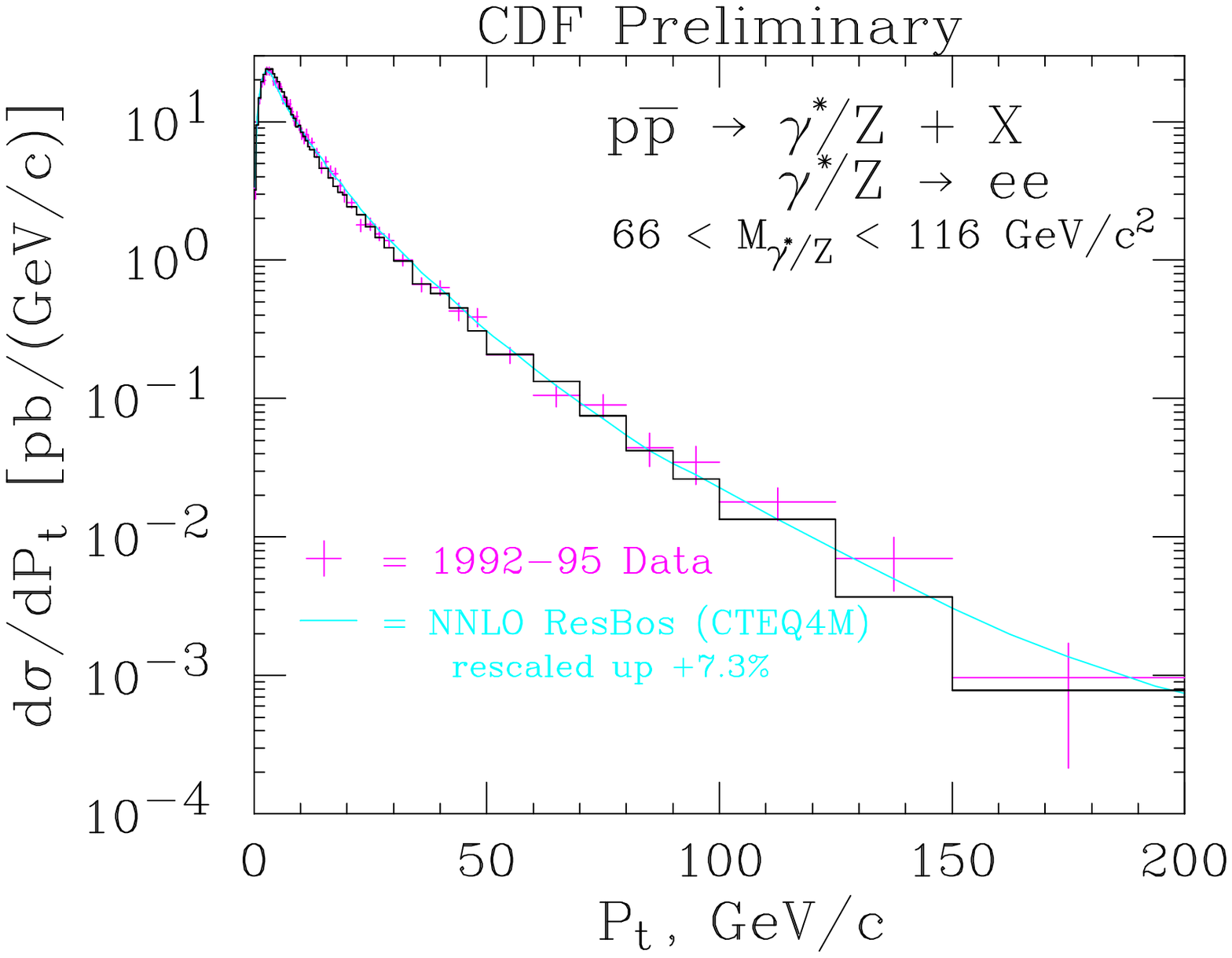}
\end{center}
\caption{
The transverse momentum distribution (full $p_T$ range) for $Z \rightarrow e^+e^-$ from CDF in Run 1, along with comparisons to
predictions from PYTHIA (solid histogram) and ResBos. 
\label{fig:zptall}}
\end{figure}

The transverse momentum distribution for $Z$ bosons at the Tevatron (CDF Run 1) is shown in Figures~\ref{fig:run1ee}
(low $p_T$) and~\ref{fig:zptall} (all $p_T$) along with comparisons to the parton showers PYTHIA and ResBos which are
discussed in Section~\ref{sec:allorders}.
The ResBos predictions agree well with the data over the entire kinematic region, while the PYTHIA predictions require the addition of
an intrinsic $k_T$ of approximately $2$~GeV for best agreement at  lower transverse momentum. This intrinsic $k_T$ is not totally due
to the finite size of the proton; it mostly results from the finite cutoff in the parton shower evolution and the need to add the
missing component back ``by-hand''. ResBos correctly describes the non-perturbative as well as the perturbative region.

\begin{figure}[t]
\begin{center}
\includegraphics[width=10cm]{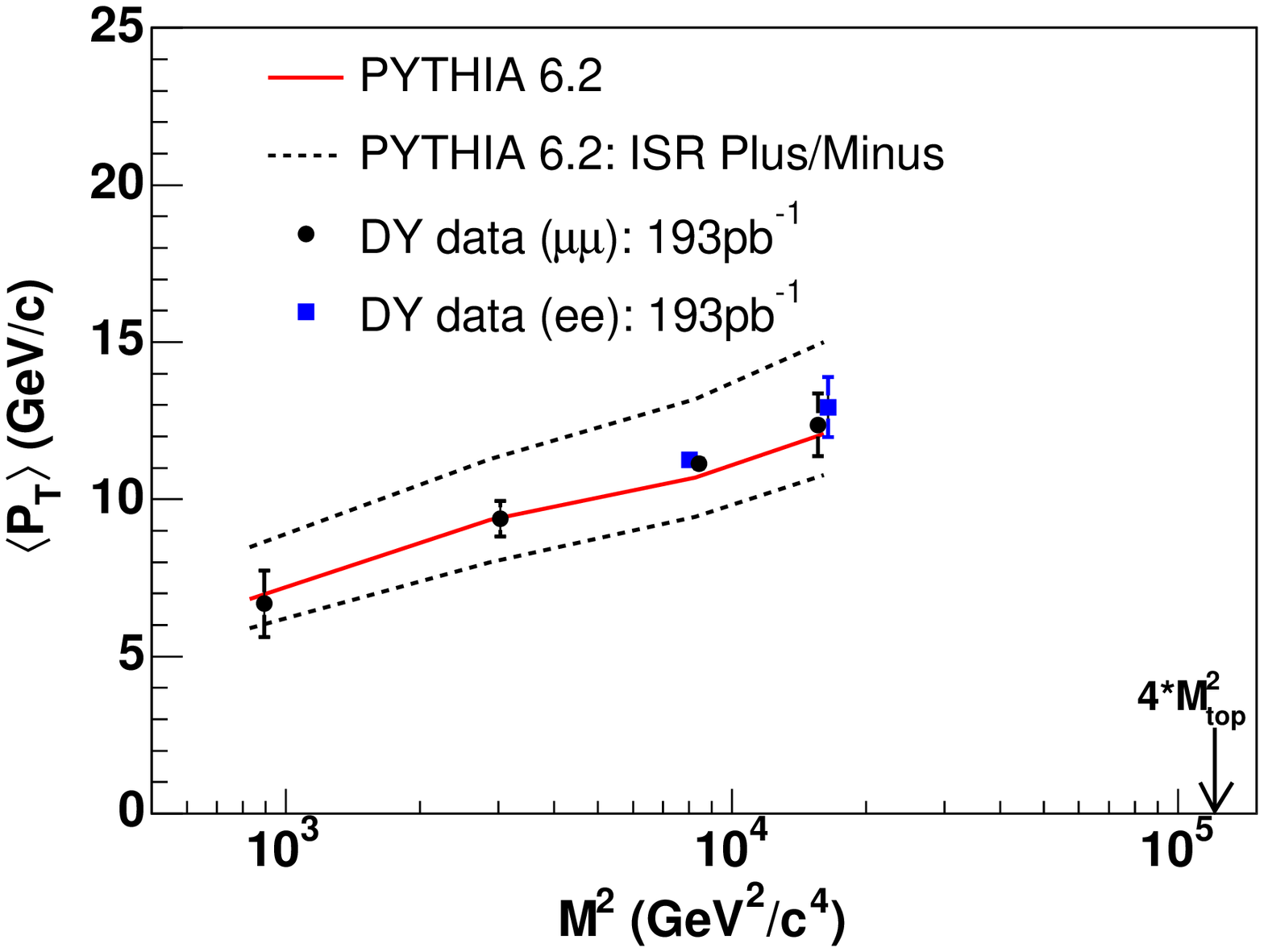}
\end{center}
\vspace*{-0.5cm}
\caption{
The average transverse momentum for Drell-Yan pairs from CDF in Run 2, along with comparisons to predictions from PYTHIA.
\label{fig:pt_evol}}
\end{figure}

The average transverse momentum for Drell-Yan production has been measured as a function of the Drell-Yan pair mass in CDF in Run 2.
This is shown in Figure~\ref{fig:pt_evol}~\cite{Abulencia:2005aj}.
The average transverse momentum (neglecting the high $p_T$ tail) increases roughly logarithmically with the square of the Drell-Yan
mass. The data agree well with the default PYTHIA~6.2 prediction using Tune A.
Also shown are two predictions involving tunes of PYTHIA that give larger/smaller values for the average Drell-Yan transverse momentum
as a function of the Drell-Yan mass. The {\it Plus/Minus} tunes were used to estimate the initial-state-radiation uncertainty for the
determination of the top mass in CDF. Most of the $t\bar{t}$ cross section at the Tevatron arises from $q\bar{q}$ initial states, so the
Drell-Yan measurements serve as a good model. (See further discussion in Section~\ref{sec:tT_tev}.) 
The information regarding these tunes is available on the benchmarks website. 

\subsection{Underlying event}
\label{sec:UE}

What is meant by a ``minimum bias event'' is somewhat murky, and the exact definition will depend on the trigger of each experiment. The
description of the underlying event energy and of minimum bias events requires a non-perturbative and/or semi-perturbative
phenomenological model. There are currently a number of models available, primarily inside parton shower Monte Carlo programs, to
predict both of these processes. We discuss several of the popular models below. An understanding of this soft physics is interesting
in its own right but is also essential for precision measurements of hard interactions where the soft physics effects need to be
subtracted. 

Perhaps the simplest model for the underlying event is the uncorrelated soft scattering model present in HERWIG. Basically, the model is
a parametrization of the minimum bias data taken by the UA5 experiment~\cite{Alner:1987wb} at the CERN $p\overline{p}$ Collider. The
model tends to predict underlying event distributions softer than measured at the Tevatron and has a questionable extrapolation to
higher centre-of-mass energies. A newer model for the underlying event in HERWIG is termed ``Jimmy''~\footnote{
http://hepforge.cedar.ac.uk/jimmy/} and describes the underlying event in terms of multiple parton interactions at a scale lower than
the hard scale and with the number of such parton scatterings depending on the impact parameter overlap of the two colliding hadrons. 

The PYTHIA model for the underlying event also utilizes a multiple parton interaction framework with the total rate for parton-parton
interactions assumed to be given by perturbative QCD. A cutoff, $p_{Tmin}$, is introduced to regularize the divergence as the
transverse momentum of the scattering goes to zero. The rate for multiple parton interactions depend strongly on the value of the gluon
distribution at low  $x$.  The cutoff, $p_{Tmin}$,  is the main free parameter of the model and basically corresponds to an inverse
colour screening distance. A tuning of the PYTHIA underlying event parameters (Tune A) basically succeeds in describing most of the
global event properties in events at the Tevatron. With the new version of
PYTHIA (version 6.4)~\cite{Sjostrand:2004ef,Sjostrand:2006za}, a new model for the underlying event is available,
similar in spirit to the old multiple parton interaction model,
but with a more sophisticated treatment of colour, flavour and momentum correlations in the remnants. 

\subsection{Inclusive jet production}
\label{sec:jet}

It is useful to consider the measurement of inclusive jet production at the Tevatron as (1) it probes the highest transverse momentum
range accessible at the Tevatron, (2) it has a  large impact on global pdf analyses, and (3) many of the subtleties regarding
measurements  with jets in the final state and the use of jet algorithms come into play.

\begin{figure}[t]
\begin{center}
\includegraphics[width=10cm]{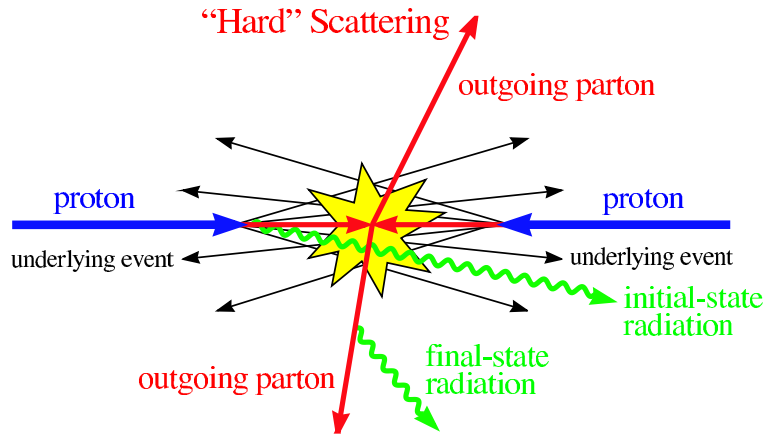}
\end{center}
\caption{
Schematic cartoon of a $2\rightarrow 2$ hard scattering event.
\label{fig:cartoon}}
\end{figure}

As shown in Figure~\ref{fig:cartoon}, a dijet event at a hadron-hadron collider consists of a hard collision of two incoming partons
(with possible gluon radiation from both  the incoming and outgoing legs) along with softer interactions from the remaining partons in
the colliding hadrons (``the underlying event energy''). 

\begin{figure}[t]
\begin{center}
\includegraphics[width=12cm]{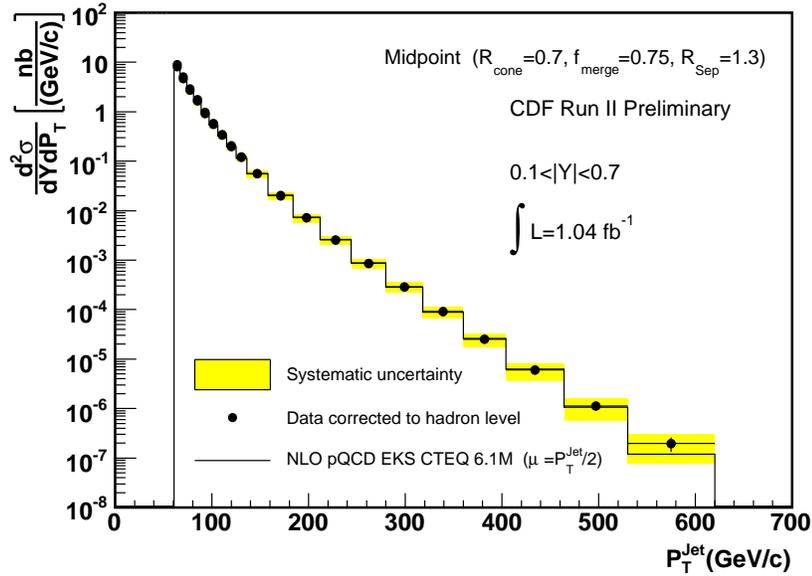}
\end{center}
\vspace*{-0.5cm}
\caption{
The inclusive jet cross section from CDF in Run 2.
\label{fig:cdf_jet_log}}
\end{figure}

The inclusive jet cross section measured by the CDF Collaboration in Run 2 is shown in  Figure~\ref{fig:cdf_jet_log}, as a function of
the jet transverse momentum~\cite{cdfqcd}.
Due to the higher statistics compared to Run 1, and the higher centre-of-mass energy,  the reach in
transverse momentum has increased by approximately $150$~GeV. The measurement uses the midpoint cone algorithm with a cone radius of
$0.7$.  As discussed in Section~\ref{sec:jetalgs}, the midpoint algorithm places additional seeds
(directions for jet cones) between stable cones 
having a separation of less than twice the size of the clustering cones.  The midpoint algorithm uses four-vector kinematics for
clustering individual partons,  particles or energies in calorimeter towers, and jets are described using rapidity ($y$) and
transverse momentum ($p_T$)~\footnote{
Self-contained versions of the CDF midpoint cone algorithm, as well as the Run 1 cone algorithm JetClu, are
available at the benchmark website, in both Fortran and C++ versions}. 

\subsubsection{Corrections}
\label{sec:corrections}

For comparison of data to theory, the calorimeter tower energies clustered into a jet  must first be corrected for the detector
response. The calorimeters in the CDF experiment  respond differently to electromagnetic  showers than to hadronic showers, and the
difference  varies as a function of the transverse momentum of the jet. The detector response corrections  are determined using a
detector simulation in which the parameters have been tuned to test-beam  and in-situ calorimeter data. PYTHIA~6.216, with Tune A,  is
used for  the production and  fragmentation of jets. 
The same clustering procedure is  applied to the final state particles in PYTHIA as is done for the data. The correction is  determined
by matching the calorimeter jet to the corresponding particle jet. An additional  correction accounts for the smearing effects due
to the finite energy resolution of the  calorimeter. At this point, the jet is said to be determined at the ``hadron level.''

For data to be compared to a parton level calculation, either the data must be corrected  from the hadron level to the parton level or
the theory must be corrected to the hadron level.  Here we describe the former; the latter just involves  the inverse corrections. It
is our ``official recommendation'' that, where possible, the data  be presented at least at the hadron level, and the corrections between hadron and parton level be clearly stated. The hadronization  corrections consist of two components:  the subtraction  from the jet of the
underlying event  energy  not associated with the hard scattering and the correction for a loss of energy outside  a jet due to the
fragmentation process. The hadronization corrections can be calculated by  comparing the results obtained from PYTHIA at the hadron
level to the results from PYTHIA when  the underlying event and the parton fragmentation into hadrons has been turned off. The
underlying  event energy is due to the  interactions of the spectator partons in the colliding hadrons and the  size of the  correction
depends on the size of the jet cone. It is approximately $0.5$~GeV for a cone of radius $0.7$ and is similar to the amount of energy
observed in minimum bias events with a  high track multiplicity. The rule-of-thumb has always been that the underlying event energy in
a jet event looks very much like that observed in minimum bias events, i.e. that there is a rough factorization of the event into a
hard scattering part and a soft physics part. 

\begin{figure}[t]
\begin{center}
\includegraphics[width=8cm]{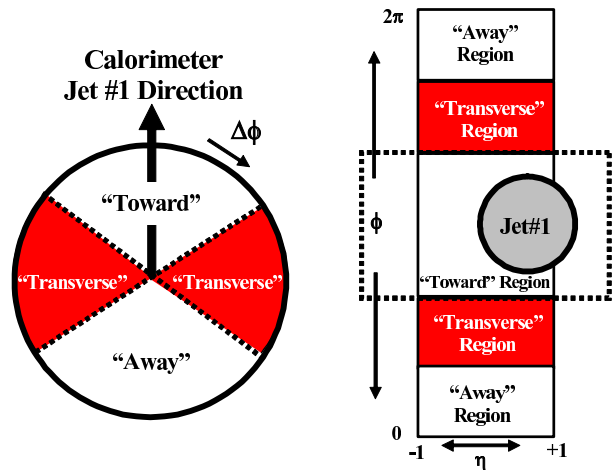}
\end{center}
\caption{
Definition of the ``toward'', ``away'' and ``transverse'' regions.
\label{fig:phase_space}}
\end{figure}

\begin{figure}[t]
\begin{center}
\includegraphics[width=12cm]{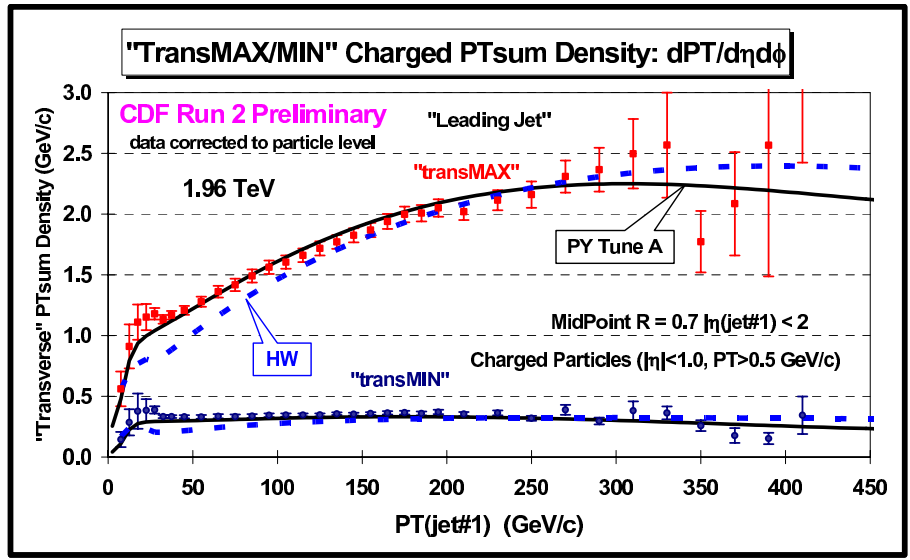}
\end{center}
\caption{
The sum of the transverse momenta of charged particles inside the TransMAX and TransMIN regions, as a function of the transverse
momentum of the leading jet.
\label{fig:pttrack}}
\end{figure}

Studies have been carried out with inclusive jet production in CDF, examining the transverse momentum  carried  by charged particles
inside and outside of jets~\cite{Acosta:2004wq,Affolder:2001xt}. For example, the geometry for one study is shown in
Figure~\ref{fig:phase_space}, where the ``towards'' and ``away'' regions have been defined with respect to the direction of the leading
jet.
The transverse momenta in the two transverse regions are shown in Figure~\ref{fig:pttrack}.
%


The region with the largest transverse momentum is designated as the max region and the one with the lowest, the  min region. As the
lead jet transverse momentum increases, the momentum in the max region increases;  the momentum in the min region does not. The amount
of transverse momentum in the min region is  consistent with that observed in minimum bias events at the Tevatron. At the partonic
level,  the max region  can receive contributions from the extra parton  present in NLO inclusive jet  calculations. The min region can
not. There is good agreement between the Tevatron data  and the PYTHIA tunes. Tune A was determined using the LO pdf CTEQ5L; an
equivalently good agreement is also observed with a tune using the NLO pdf CTEQ6.1~\footnote{
The parameters for the tune are given on the benchmarks website.}.
As discussed previously, such a tune is necessary for the use of a NLO pdf such as CTEQ6.1 with the PYTHIA Monte Carlo. 

\begin{figure}[t]
\begin{center}
\includegraphics[width=7cm]{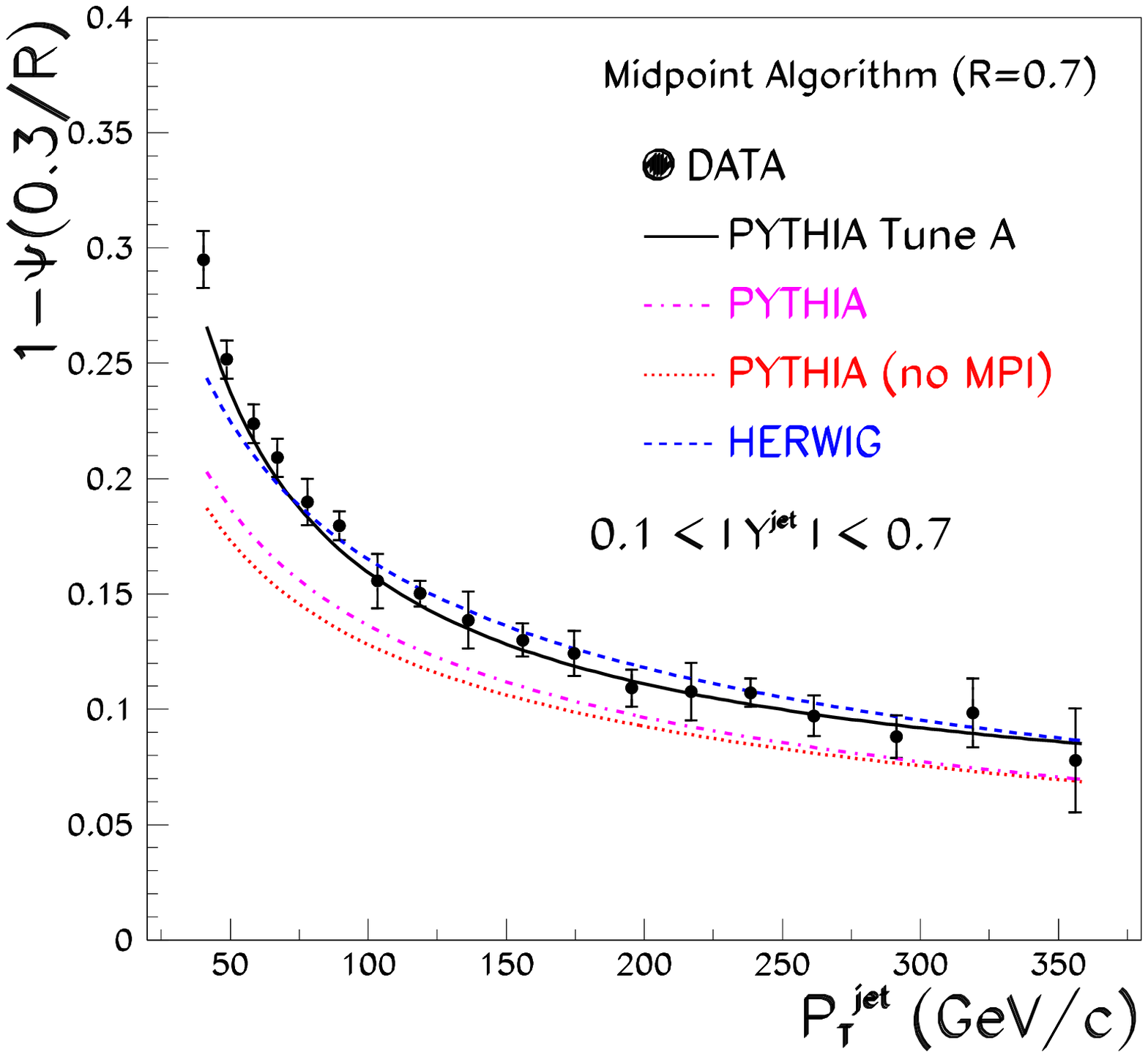}
\includegraphics[width=7cm]{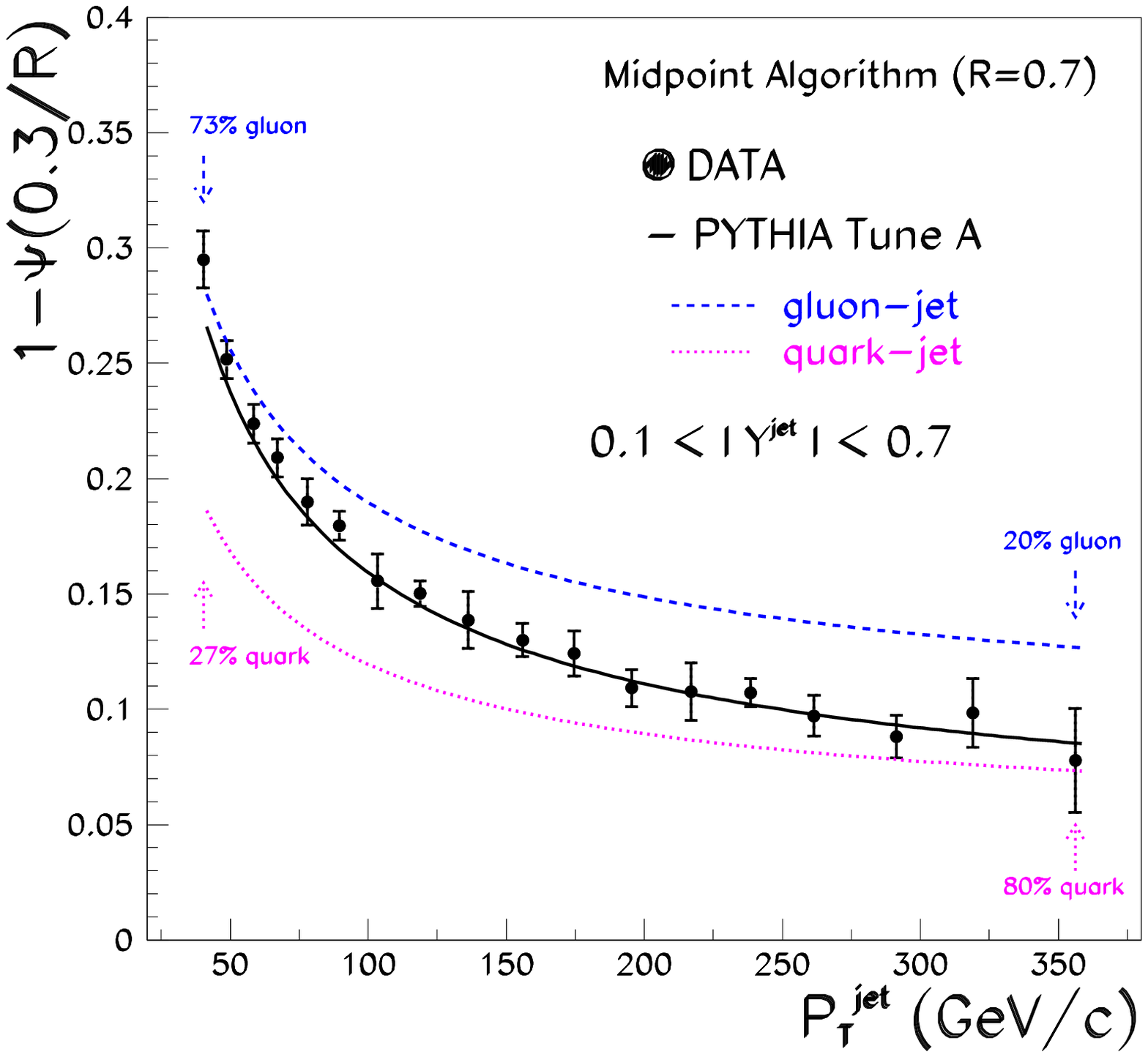}
\end{center}
\caption{The fraction of the transverse momentum in a cone jet of radius $0.7$ that lies in the annulus
from $0.3$ to $0.7$, as a function of the transverse momentum of the jet.
Comparisons are made to several tunes of PYTHIA (left) and to the separate predictions
for quark and gluon jets (right). 
\label{fig:fig6}}
\end{figure}

The jet shape is also well described by PYTHIA predictions using Tune A, as can be seen in
Figure~\ref{fig:fig6}~\cite{Acosta:2005ix}, where the jet energy away from the core of the jet (i.e. in the annulus
from 0.3 to 0.7) is plotted as a function of the transverse momentum of the jet. A better description is provided by
the use of Tune A then with the default PYTHIA prediction.  A reasonable description of the core of the jet ($<
0.3$) can also be provided by the pure NLO prediction~\cite{Ellis:1992qq}. Jets become more collimated as
the inclusive jet transverse momentum increases for three reasons: (1)) power corrections that tend to broaden the
jet decrease as $1/p_T$ or $1/p_T^2$, (2) a larger fraction of jets are quark jets rather than gluon jets and (3),
the probability of a hard gluon to be radiated (the dominant factor in the jet shape) decreases as $\as(p_T^2)$. 

The fragmentation correction accounts for the daughter hadrons ending up outside the jet cone   from mother partons whose trajectories
lie inside the cone (also known as {\it splash-out}); it  does not correct for any out-of-cone energy arising from perturbative effects
as these should  be correctly accounted for in a NLO calculation. It is purely a power correction to the cross  section. The numerical
value of the splash-out energy is roughly constant at $1$~GeV for a cone  of radius $0.7$, independent of the jet transverse momentum.
This constancy may seems surprising.  But, as the jet transverse momentum increases, the jet becomes more collimated; the result is 
that the energy in the outermost annulus (the origin of the splash-out energy) is roughly constant. The correction for splash-out derived
using parton shower Monte Carlos can be applied to a NLO  parton level calculation to the extent to which both the parton shower and
the $2$ partons in a NLO jet correctly describe the jet shape. 

\begin{figure}[t]
\begin{center}
\includegraphics[width=7cm,angle=-90]{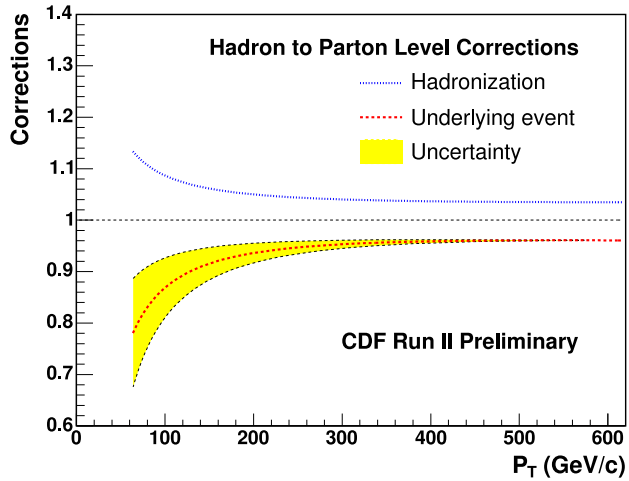}
\end{center}
\caption{
Fragmentation and underlying event corrections for the CDF inclusive jet result, for a cone size $R=0.7$.
\label{fig:cdf_jet_cor}}
\end{figure}

The two effects (underlying event and splash-out) go in opposite directions so there is a  partial cancellation in the correction to
parton level. For a jet cone of $0.7$, the underlying  event correction is larger, as seen in Figure~\ref{fig:cdf_jet_cor}, for the 
case of inclusive  jet production at CDF. 
For a jet cone radius of $0.4$, the fragmentation correction remains roughly the same size but  the underlying event corrections scales
by the ratio of the cone areas; as a result the two effects basically cancel each other out over the full transverse momentum range at
the Tevatron.  

\subsubsection{Results}
\label{sec:results}

\begin{figure}[t]
\begin{center}
\includegraphics[width=12cm]{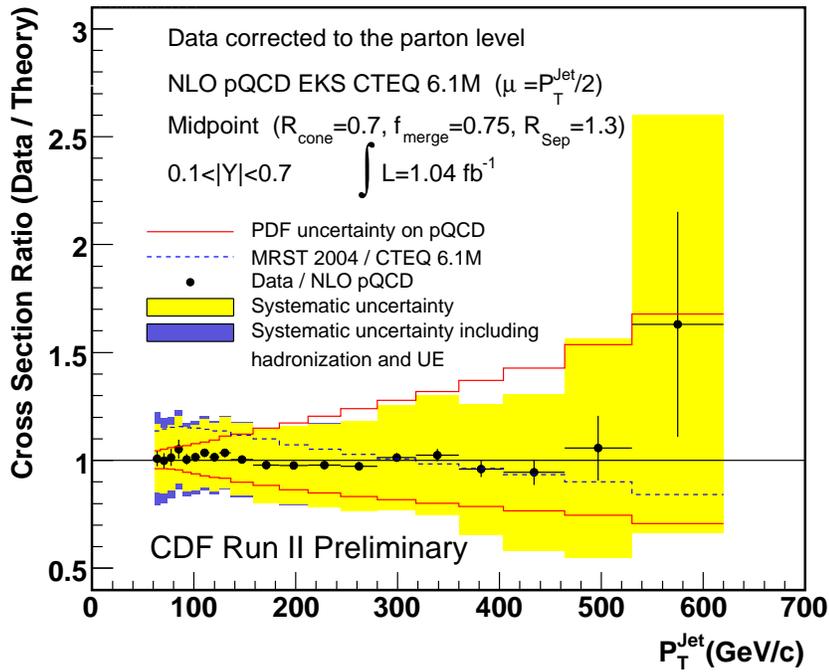}
\end{center}
\vspace*{-0.5cm}
\caption{
The inclusive jet cross section from CDF in Run 2 compared on a linear scale to NLO theoretical predictions using CTEQ6.1 and MRST2004
pdfs.
\label{fig:cdf_jet_lin}}
\end{figure}

\begin{figure}[t]
\begin{center}
\includegraphics[width=14cm]{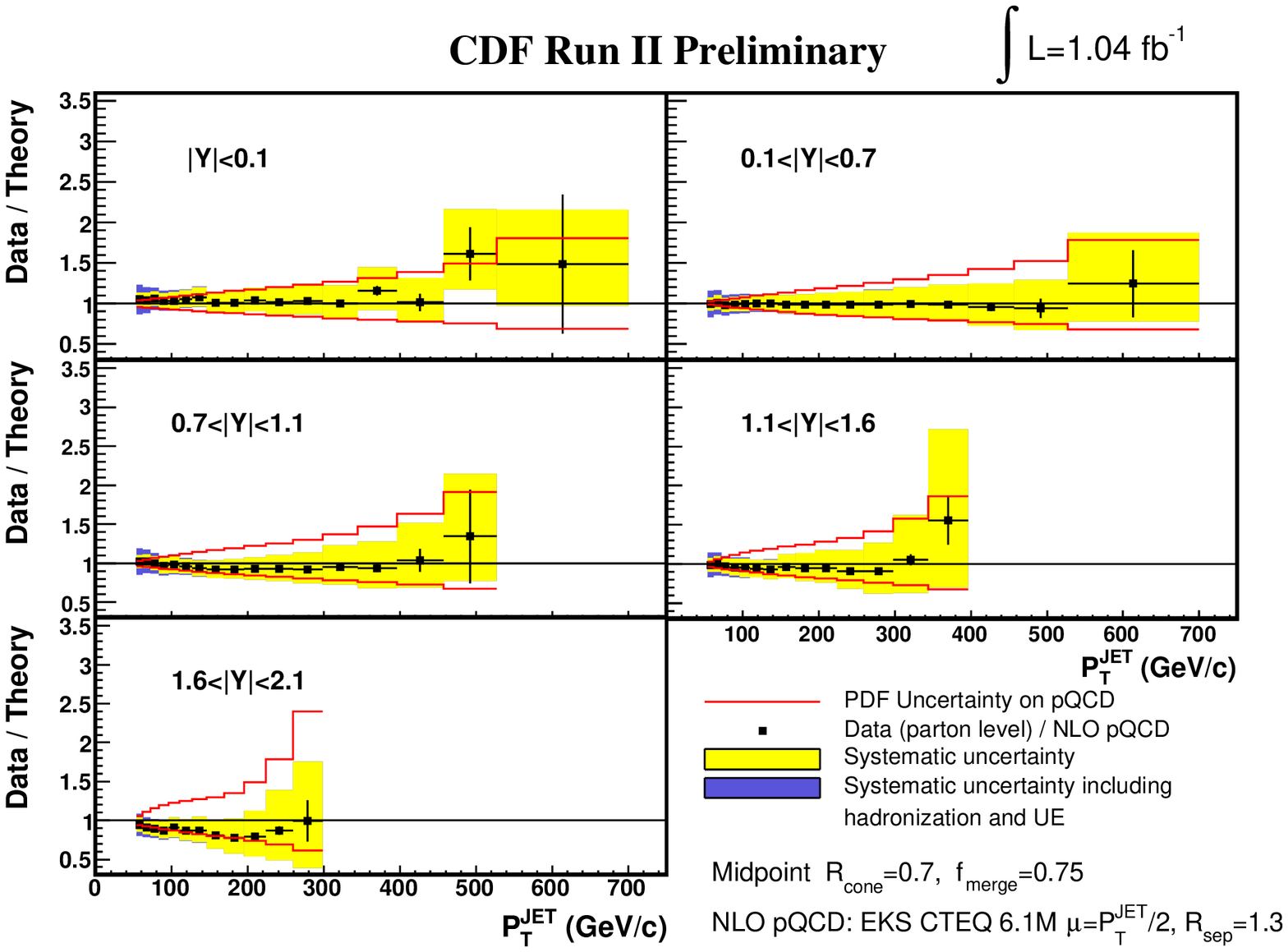}
\end{center}
\vspace*{-0.5cm}
\caption{
The inclusive jet cross section from CDF in Run 2, for several rapidity intervals using the midpoint cone algorithm,
compared on a linear scale to NLO theoretical predictions using CTEQ6.1 pdfs.
\label{fig:cdf_jet_rap}}
\end{figure}

A comparison of the inclusive jet  cross section measured by CDF in Run 2 with the midpoint cone algorithm~\cite{Abulencia:2005yg}  to
NLO pQCD predictions using the EKS~\cite{Ellis:1990ek} program  with the CTEQ6.1 and MRST2004 pdfs is shown in
Figure~\ref{fig:cdf_jet_lin}.
A renormalization/factorization scale of $(p_T^{jet}/2)$ has  been used in the calculation. Typically, this
leads to the highest predictions for inclusive jet cross sections at the Tevatron, as discussed in Section~\ref{sec:scaledep}.
There is good
agreement  with the CTEQ6.1 predictions over the transverse momentum range of the prediction. The MRST2004  predictions are slightly
higher at lower $p_T$ and slightly lower at higher $p_T$, but still in good overall agreement. As noted before, the CTEQ6.1 and
MRST2004 pdfs have a higher gluon at  large $x$ as compared to previous pdfs, due to the  influence of the Run 1 jet data from CDF
and  D0. This enhanced gluon provides a good agreement with the high $p_T$ Run 2 measurement as well which, as stated before, extends
approximately $150$~GeV higher in transverse momentum.  The red curves indicate the pdf uncertainty for the prediction using the
CTEQ6.1 pdf error set. The yellow band indicates the experimental systematic uncertainty, which is dominated by the  uncertainty in
the jet energy scale (on the order of $3\%$). The purple band shows the effect  of the uncertainty due to the hadronization and
underlying event, which is visible only for  transverse momenta below $100$~GeV. In Figure~\ref{fig:cdf_jet_rap}, the jet cross
sections measured with the midpoint cone algorithm are shown for the full rapidity coverage of the CDF experiment.
Good agreement is observed in all rapidity regions with the CTEQ6.1 predictions. It is also important to note that for much of the
kinematic range, the experimental systematic errors are less than pdf uncertainties; thus, the use of this data in future global pdf
fits should serve to further constrain the gluon pdf. 

\subsubsection{Jet algorithms and data}
\label{sec:jetalgdata}

\begin{figure}[t]
\begin{center}
\includegraphics[width=11cm]{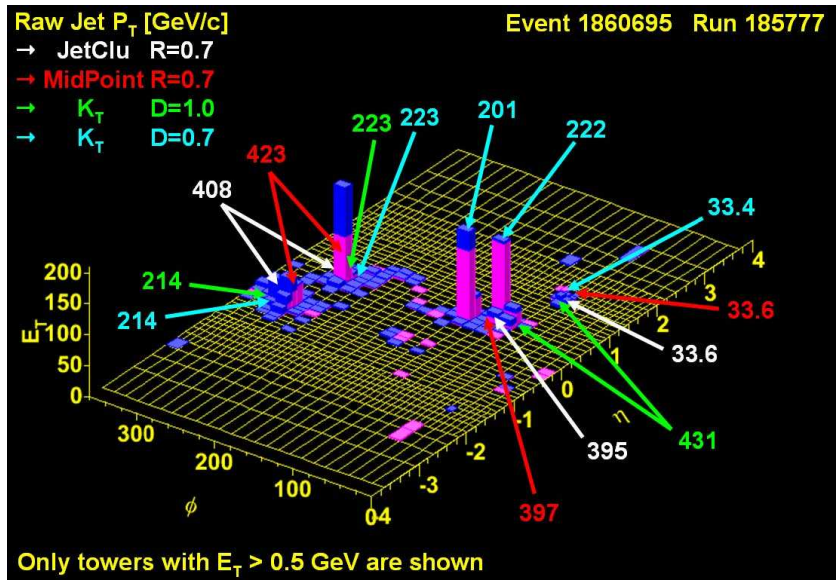}
\end{center}
\caption{
Impact of different jet clustering algorithms on an interesting event. 
\label{fig:lego}}
\end{figure}

For many events, the jet structure is clear and the jets to which the individual towers  should be assigned are fairly unambiguous.
However, in other events such as Figure~\ref{fig:lego}, the complexity of the energy  depositions means that different algorithms will
result in different assignments of towers  to the various jets.
This is no problem to the extent that a similar complexity can be matched by the theoretical calculation to which it is being
compared. This is the case, for example, for events simulated with parton shower Monte Carlos, but, as discussed in Section~\ref{sec:partonxsecs}, current NLO
calculations can place at most $2$ partons in a jet.

\begin{figure}[t]
\begin{center}
\includegraphics[width=14cm]{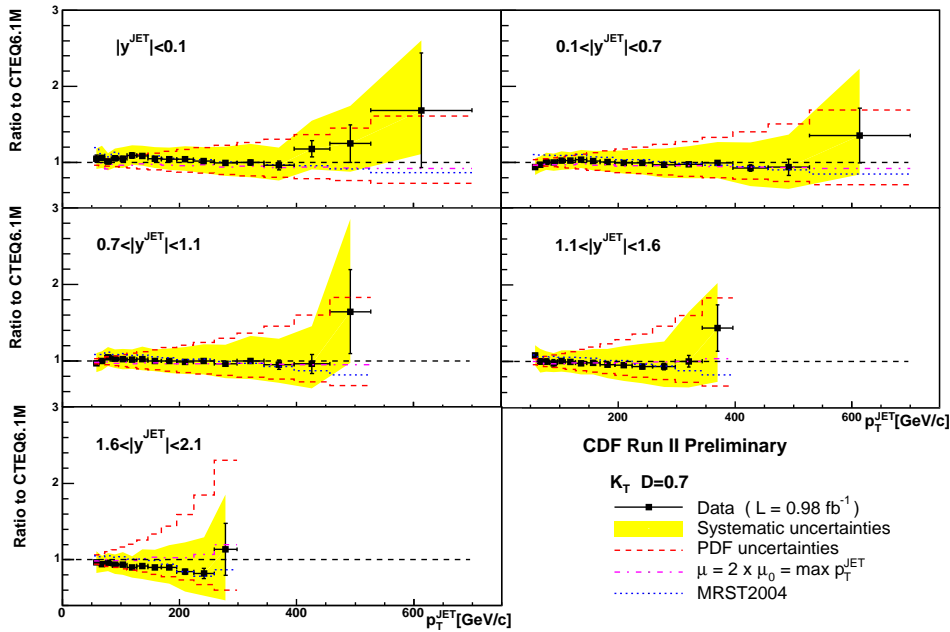}
\end{center}
\vspace*{-0.5cm}
\caption{
The inclusive jet cross section from CDF in Run 2, for several rapidity intervals using the $k_T$ jet algorithm, compared on a linear
scale to NLO theoretical predictions using CTEQ6.1 pdfs.
\label{fig:cdf_jet_kt}}
\end{figure}

In Figure~\ref{fig:cdf_jet_kt}, the experimental jet cross sections using the $k_T$ algorithm from CDF Run 2~\cite{Abulencia:2005jw}
are compared to NLO predictions using the JETRAD~\cite{Giele:1994gf} program.
Similarly good agreement to that obtained for the midpoint cone algorithm is observed. This is an important observation. The two
different jet algorithms have different strengths and weaknesses. It is our recommendation that, where possible, analyses at the
Tevatron and LHC use both algorithms in order to obtain more robust results. 

We noted in Section~\ref{sec:jetalgs} that on general principles, for NLO parton level predictions, the cone jet cross section is larger than the $k_T$ jet cross section when $R_{cone}=D$. At the hadron level, this is no longer true; the cone jet loses energy by the ``splash-out'' effect while the $k_T$ algorithm has a tendency to ``vacuum up'' contributions from the underlying event. This will be corrected at least partially by the hadron to parton level corrections for each algorithm. 

A particular complexity with the cone algorithm occurs when two jets overlap; a decision must be  made whether to merge the two jets
into one, or to separate them. This is an experimental  decision; in CDF in Run 2, the two overlapping jets are merged when more than
$75\%$ of the smaller jet energy overlaps with the larger jet. When the overlap is less, the towers are  assigned to the nearest jet. 
D0 uses a criterion of a $50\%$ fraction. NLO theory is agnostic on the subject as there is no overlap between the two partons that can
comprise a jet. Further study is needed as to which choice is best, especially for the high luminosity conditions at the LHC. 

\begin{figure}[t]
\begin{center}
\includegraphics[width=9cm]{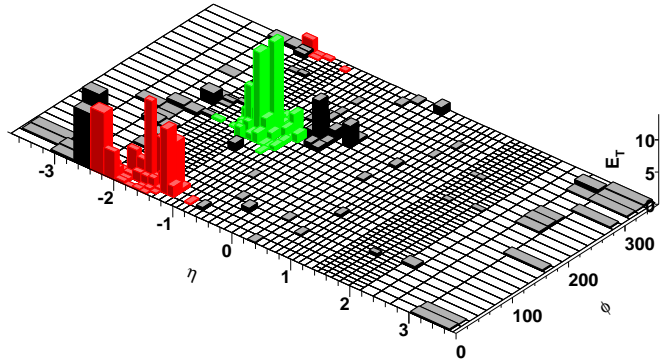}
\end{center}
\vspace*{-0.5cm}
\caption{
An example of a Monte Carlo inclusive jet event where the midpoint algorithm has left substantial energy unclustered.
\label{fig:dark_towers}}
\end{figure}

Another problem that can arise on the particle or calorimeter level, but not on the  NLO parton level, occurs when particles or
calorimeter towers remain unclustered in  any jet, due to the strong attraction of a nearby larger jet peak that will attract away  any
trial jet cone placed at the location of the original particles/calorimeter towers.  The result will be what~\cite{Ellis:2001aa}
calls ``dark towers'', i.e. clusters that have a transverse  momentum large  enough to be designated either a separate jet or to be
included in an existing  nearby jet, but which are not clustered into  either. Such a Monte Carlo event is shown in 
Figure~\ref{fig:dark_towers}, where the towers unclustered  into any jet are shaded black. 
A simple way of understanding these dark towers begins by defining
a \textquotedblleft Snowmass potential\textquotedblright\ in terms of the 2-dimensional vector
$\overrightarrow{r}=\left(  y,\phi\right)  $ via
\begin{equation}
\fl
V\left(  \overrightarrow{r}\right)  =-\frac{1}{2}\sum_{j}p_{T,j}\left(
R_{cone}^{2}-\left(  \overrightarrow{r_j}-\overrightarrow{r}\right)
^{2}\right)  \Theta\left(  R_{cone}^{2}-\left(  \overrightarrow{r_j}
-\overrightarrow{r}\right)^{2}\right) \; .
\end{equation}
The flow is then driven by the \textquotedblleft force\textquotedblright
$\overrightarrow{F}\left(  \overrightarrow{r}\right)  =-\overrightarrow
{\nabla}V\left(  \overrightarrow{r}\right)$
which is thus given by,
\begin{eqnarray}
\overrightarrow{F}\left(  \overrightarrow{r}\right) &=\sum_{j}p_{T,j} \left(\overrightarrow{r_j}-\overrightarrow{r}\right)
 \Theta\left(  R_{cone}^{2}-\left(  \overrightarrow{r_j}-\overrightarrow{r}\right)^{2}\right) \nonumber \\
  &=\left(  \overrightarrow{\overline{r}}_{C\left( \overrightarrow{r}\right) }-\overrightarrow{r}\right)
  \sum_{j\subset C\left(  r \right)  }p_{T,j},
\end{eqnarray}
where $\overrightarrow{\overline{r}}_{C(  \overrightarrow{r})
}=\left(  \overline{y}_{C(  \overrightarrow{r})  },\overline{\phi
}_{C(  \overrightarrow{r}) }\right)$ and the sum runs over
$j\subset C(\overrightarrow{r})$ such that $\sqrt{\left(  y_{j}-y\right)
^{2}+\left(  \phi_{j}-\phi\right)  ^{2}}$ $\leq R_{cone}$.
As desired, this force pushes the cone to the stable cone position.

\begin{figure}[t]
\begin{center}
\includegraphics[width=12cm]{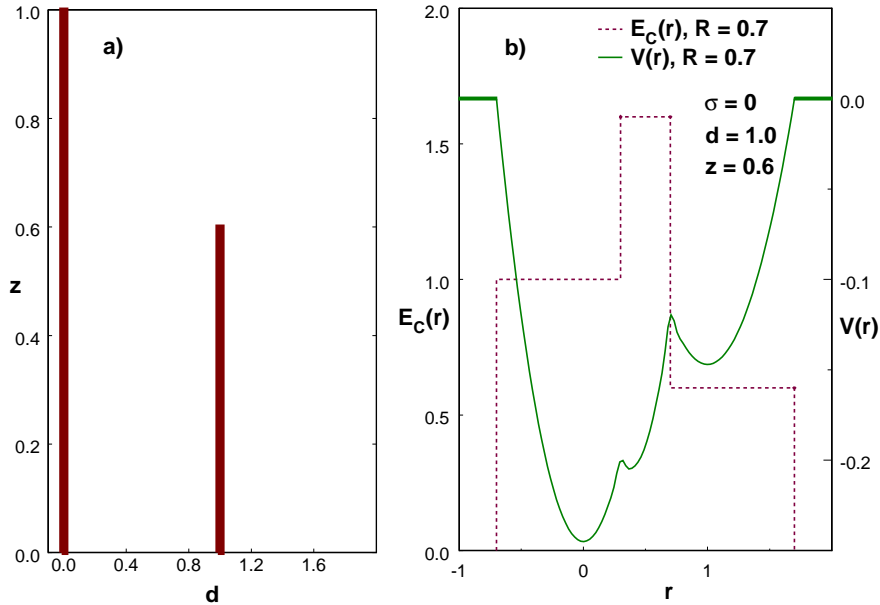}
\end{center}
\vspace*{-0.5cm}
\caption{A schematic depiction of a specific parton configuration and the results of applying the midpoint cone
jet clustering algorithm. The potential discussed in the text and the resulting energy in the jet are plotted. 
\label{fig:Edist}}
\end{figure}

\begin{figure}[t]
\begin{center}
\includegraphics[width=9cm]{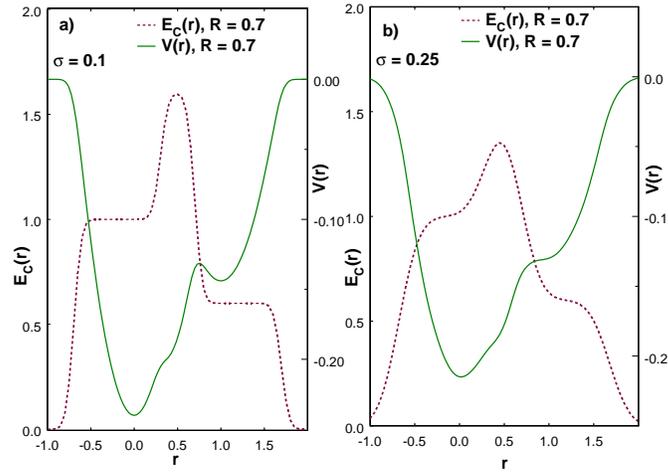}
\end{center}
\vspace*{-0.5cm}
\caption{A schematic depiction of the effects of smearing on the midpoint cone jet clustering algorithm. 
\label{fig:smearPT6}}
\end{figure}

In Figure~\ref{fig:Edist} (left), two partons are placed a distance $\Delta R=0.9$ apart; the second parton has a 
fractional energy $0.6$ that of the first parton. 
On the right, the potential $V(r)$ and the energy contained inside a cone of radius $0.7$ are plotted  for this parton
configuration. At  the parton level, there are three positions where minima are present in the potential: at the position of the left
parton, the right parton and the midpoint between the two partons. The midpoint jet algorithm applied to a NLO parton level
calculation, as discussed in Section~\ref{sec:jetalgs}, would find all three solutions.  In Figure~\ref{fig:smearPT6}(left), the spatial distributions
of the two partons' energies have been smeared with a spatial resolution  $\sigma$ of $0.1$, as would take place for example due to the effects
of parton showering and  hadronization.
The central minimum is wiped out by the smearing. On the right, the spatial distributions of the two partons'
energy  distributions have been smeared with a resolution $\sigma$ of $0.25$ and both the midpoint and right minima
have been wiped out. Any attempt to place the centroid  of a jet cone at  the position of the right parton will
result in the centroid ``sliding''  to the position of the left parton and the energy corresponding to the right
parton remaining  unclustered in any jet. This is the origin of the dark towers. The effective smearing in the data
lies between a $\sigma$ of $0.1$ and $0.25$. 

The TeV4LHC workshop~\cite{Group:2006rt} has recommended the following solution to the problem of unclustered energy with
cone jet algorithms, which we pass on as one of our recommendations as well. The standard midpoint algorithm should
be applied to the list of calorimeter towers/particle/partons, including the full split/merge procedure. The
resulting identified jets are then referred to as first pass jets and their towers/particles/partons are removed
from the list. The same algorithm is then applied to the remaining unclustered energy and any jets that result are
referred to as second pass jets. There are various possibilities for making use of the second pass jets. They can be
kept as separate jets, in addition to the first pass jets, or they can be merged with the nearest first pass jets.
The simplest solution, until further study, is to keep the second pass jets as separate jets. This coding of this
solution is available in the midpoint algorithm given on the benchmark website. 

%
%
It was originally thought that with the addition of a midpoint seed, the value of $R_{sep}$ used with the NLO theory
could be returned to its {\it natural} value of $2.0$(c.f. Section~\ref{sec:jetalgs}). Now it is realized that the effects of parton
showering/hadronization result in the midpoint solution virtually always being lost. Thus, a value of $R_{sep}$ of
$1.3$ is required  for the NLO jet algorithm to best model the experimental one. The theory cross section with
$R_{sep}=1.3$ is approximately $3-5\%$ smaller than with $R_{sep}=2.0$, decreasing slowly with the jet transverse
momentum.

\setcounter{footnote}{0}

\subsubsection{Inclusive jet production at the Tevatron and global pdf fits}
\label{sec:jetglobal}

\begin{figure}[t]
\begin{center}
\includegraphics[width=8cm,angle=-90]{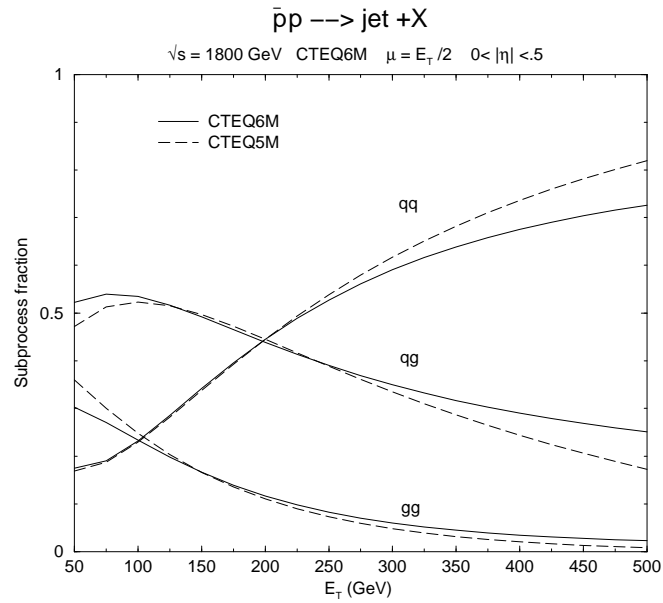}
\end{center}
\caption{
The subprocess contributions to inclusive jet production at the Tevatron for the CTEQ5M and CTEQ6M pdfs.
The impact of the larger larger gluon at high $x$ for CTEQ6 is evident.
\label{fig:JO}}
\end{figure}

\begin{figure}[t]
\begin{center}
\includegraphics[width=8cm]{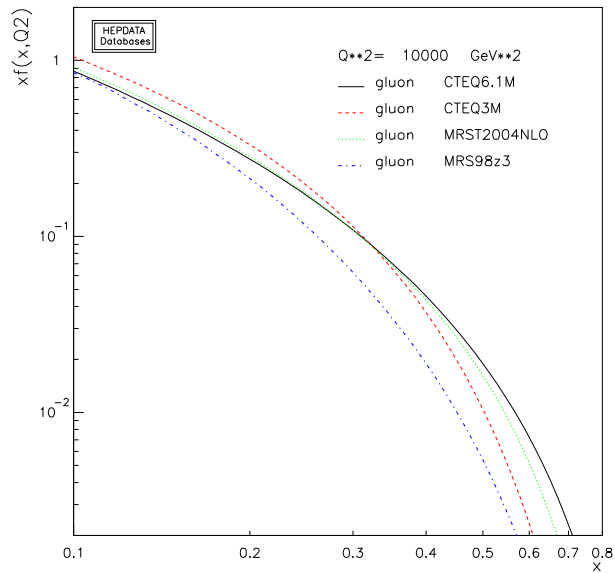}
\end{center}
\caption{
Gluon pdfs before and after the inclusion of Tevatron inclusive jet data.
\label{fig:cteq3_61}}
\end{figure}

Inclusive jet production receives contributions from $gg$, $gq$ and $qq(q\overline{q})$  initial states as shown in Figure~\ref{fig:JO}~\cite{Stump:2003yu}.
The experimental precision of the measurement, along  with the remaining theoretical uncertainties, means that the cross sections  do
not serve  as a meaningful constraint on the quark or antiquark distributions. However, they do serve  as an important constraint on
the gluon distribution, especially at high $x$. Figure~\ref{fig:cteq3_61} shows the gluon distributions for the CTEQ and MRST groups 
prior to the inclusion of any inclusive jet data and the latest sets which include the Run 1 data from both CDF and D0.
The influence of the high $E_T$ Run 1 jet cross section on the high $x$ gluon is evident.  There is always the danger of sweeping new
physics under  the rug of pdf uncertainties. Thus, it is important to measure the inclusive jet cross section over as wide a
kinematic range as possible. New physics tends to be central while a pdf explanation should be universal, i.e. fit the data in all
regions.

\subsection{$W$/$Z$ + jets}
\label{sec:wzjets}

The production of a $W$ or $Z$ boson in conjunction with jets is an interesting process in its own right as well as a background to many SM
and non-SM physics signals. Jet multiplicities of up to $7$ have been measured at the Tevatron. Production of
$W$/$Z +$~jets at the Tevatron is dominated by $gq$ initial states. The NLO cross sections have been calculated
only for $W$/$Z+2$~jets;  predictions for the higher jet multiplicity final states are accessible through matrix  element  (+ parton shower)
predictions and in fact can be considered as a prime testing  ground for the accuracy of such predictions as well as for measurements of
$\as$. Some comparisons to a recent measurement of $W \rightarrow e\nu + \ge n$ jets from CDF are shown below~\cite{cdfqcd}. In this analysis, the
data have been reconstructed using  the CDF Run 1 cone algorithm (with a midpoint cone analysis to come) with a cone  radius of $0.4$. A
smaller jet cone size is preferred  for final states that may be ``complicated'' by the presence of a large number of jets.  The data have
been compared, at the hadron level, to predictions using matrix element information from ALPGEN and parton shower and hadronization
information from PYTHIA.

\begin{figure}[t]
\begin{center}
\includegraphics[width=7cm]{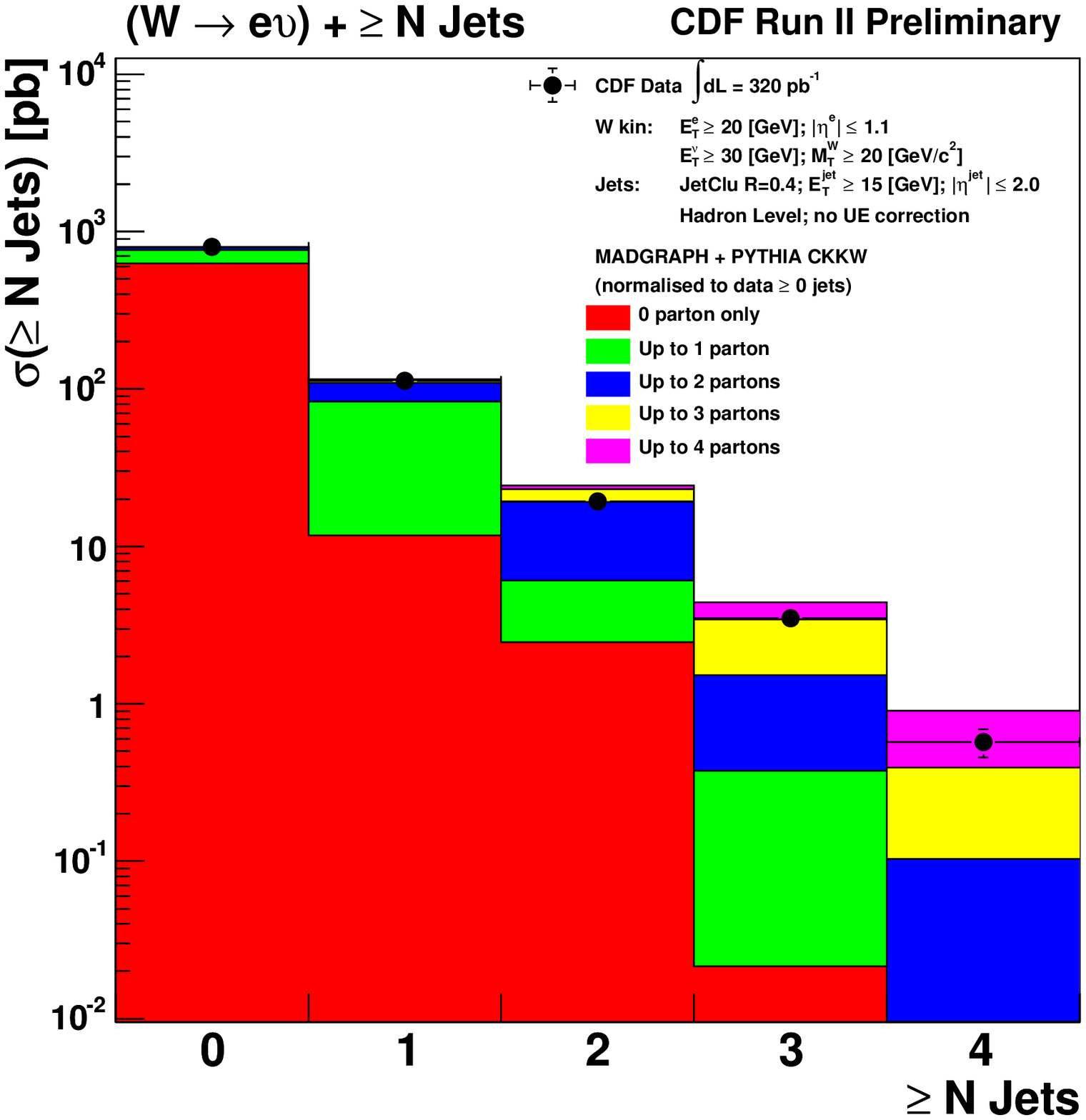}
\includegraphics[width=7cm]{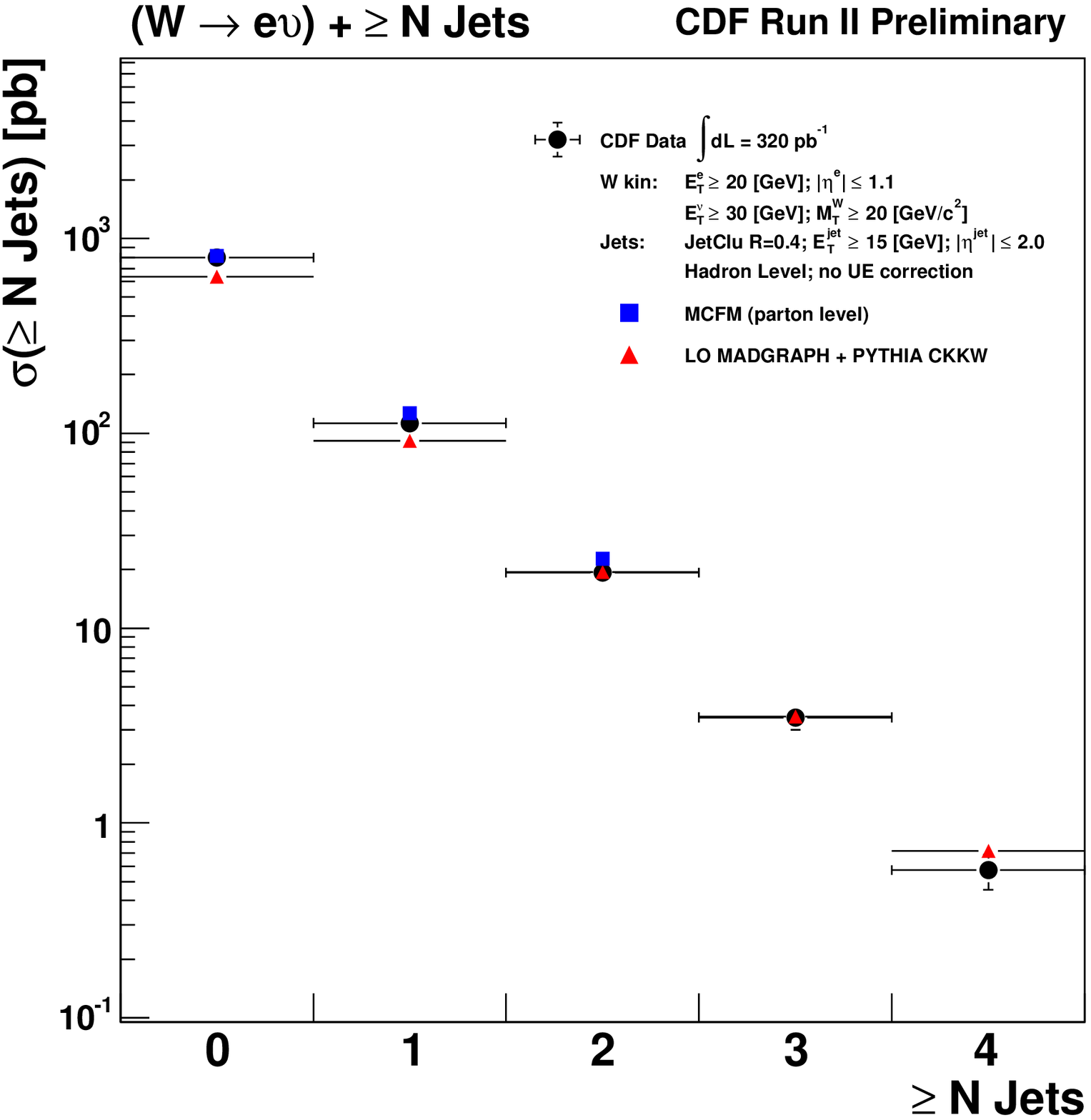}
\end{center}
\caption{
The rate of production of $W + n$~jets at CDF, compared to CKKW (left) and also NLO QCD using MCFM (right).
The measurements and predictions were performed with a cone jet of radius
$0.4$ and with a requirement of $15$~GeV or greater. The CKKW predictions are
normalized to the first bin and use a scale of $10$~GeV for the matching; the MCFM predictions are absolutely normalized.} 
\label{fig:jet_mult}
\end{figure}
The jet multiplicity distribution for $W + n$~jets measured at the Tevatron is shown in Figure~\ref{fig:jet_mult}. In the
left-hand plot, the data have been compared to the CKKW predictions from~\cite{Mrenna:2003if} (normalized to the first bin of
the data). The CKKW predictions use matrix element information for up to $4$ partons in the final state. Thus, most of the cross section in
the $\ge$ 4-jet bin has been constructed from the $4$ parton matrix element information, but a non-negligible number of events have been
generated from a $2$ or $3$ parton final state, with the extra jets coming from the parton shower. In the right-hand plot, the jet
multiplicity distribution is shown again, this time compared as well to the NLO (LO) prediction from MCFM for the $1$, $2$ ($3$) jet final
states.  The CKKW prescription agrees well with the NLO calculation for the jet multiplicities where it is available and agrees reasonably
well with the Tevatron data for the range shown. Note that the  production of each additional jet in this inclusive distribution is
suppressed by a factor of the order of $0.2$, or approximately $\as$. 

\begin{figure}
\begin{center}
\includegraphics[width=10cm]{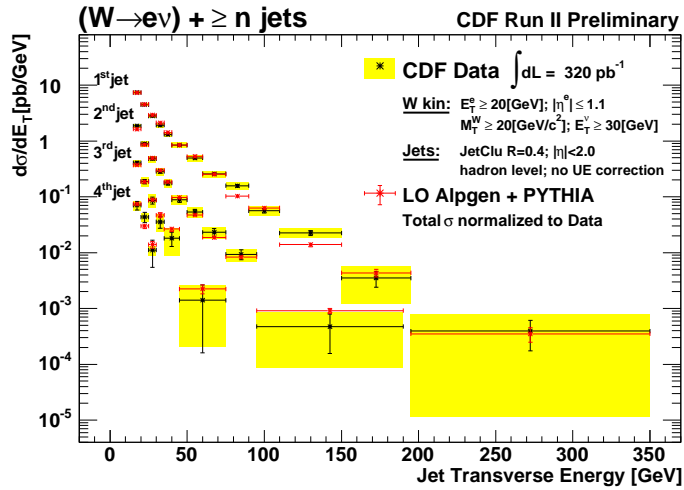}
\end{center}
\caption{
A comparison of the measured cross sections for $W + \ge n$~jets in CDF Run 2 to predictions from ALPGEN+PYTHIA. The experimental cross
sections have been corrected to the hadron level.
\label{fig:tev_w_mcfm}}
\end{figure}

A comparison of the measured cross sections for $W + \ge n$~jets in CDF Run 2 as a function of the jet transverse momentum, to predictions
from ALPGEN+PYTHIA is shown in Figure~\ref{fig:tev_w_mcfm}. The agreement is good. Note that this data is in a form (at the hadron level,
corrected for detector effects) that  makes it convenient for comparison to any hadron level Monte  Carlo
prediction~\footnote{
As mentioned before,  the corrections for underlying event and for fragmentation basically cancel each other out
for a cone of radius $0.4$, so that the hadron level predictions are essentially parton level predictions as well}.
Such a form should be the norm for measurements at both the Tevatron and LHC. 

Comparisons with the NLO predictions of MCFM will be available in the near future. There is little change in normalization in going from LO
to NLO predictions; as we saw in Section~\ref{sec:partonxsecs}, the $K$-factor for these processes is close to
unity. The major impact of the NLO corrections for the two highest
$p_T$ jets is to soften the distributions.
The NLO calculation allows some of the momentum of the hard  partons to be carried off by gluon radiation. A similar effect also occurs
with the CKKW  calculation where again there is the possibility for the parton momentum to be decreased by  additional branchings. This
is an instance of where parton showering contains some of the  physics present in NLO calculations.

\begin{figure}[t]
\begin{center}
\includegraphics[width=8cm]{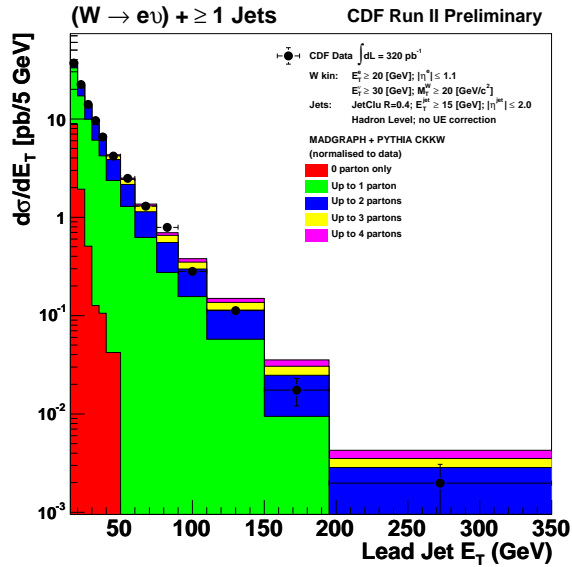}
\end{center}
\caption{
The $E_T$ distribution of the lead jet in $W + \ge 1$~jet events, along with the CKKW decomposition.
\label{fig:W_1jet}}
\end{figure}

The transverse momentum distribution for the highest $p_T$ jet in $W +$~jets events at CDF 
is shown in Figure~\ref{fig:W_1jet}, along with the CKKW predictions.
Both the total CKKW prediction and its decomposition into $W + 1$~jet, $W + 2$~jet, $\ldots$ contributions are indicated. As the
transverse momentum of the lead jet increases, more contributions are seen to come  from the higher jet multiplicity CKKW components.
The exact subdivision of contributions will  depend on where the cutoff was applied in the formation of the CKKW sample. Note  that
when the lead  jet has a transverse momentum of  the order of $200$~GeV, the events are  dominated by the higher multiplicity matrix
elements, i.e. there is a large probability for the event to contain a second (or more) jet given the large $p_T$ of the first.  
If the same transverse momentum cut were applied to all jets in the event, then the expectation would be that adding an additional jet
would lead to a penalty of $\as$ (as we have seen for the inclusive jet multiplicity distribution). In this case, since the additional
jet is soft compared to the leading jet, $\as$ is
multiplied by a log relating the energies of the leading and additional jet. The logarithm more than makes up for the factor of $\as$. 
Another way of looking at the jet multiplicity  is to say that there would be a Sudakov suppression on each of the incoming and
outgoing  parton legs  if a requirement was imposed that no extra jet be produced. An estimate of the approximate size of the Sudakov
suppression can be made using the figures given in Section~\ref{sec:partonxsecs}.
%

\begin{figure}
\begin{center}
\includegraphics[width=8cm]{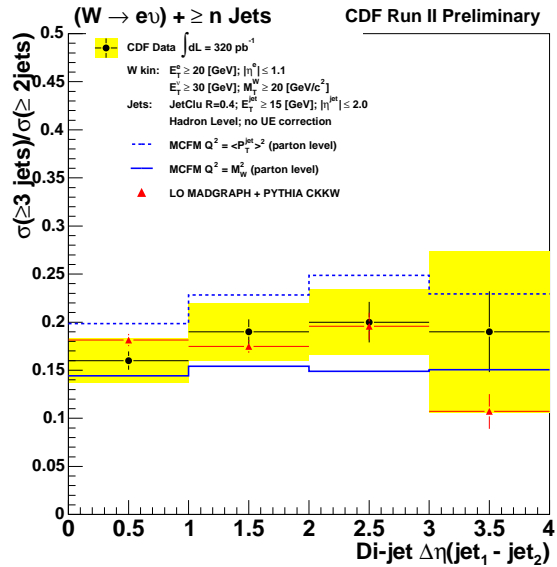}
\end{center}
\caption{Predictions and a measurement from CDF Run 2 for the rate for the production of a third jet in  $W + \ge 2$~jet events,
as a function of the rapidity separation of the two lead jets. 
\label{fig:tev_w_jets_8}}
\end{figure}

A category of $W$/$Z +$~jets event that has drawn particular interest lately is where  two jets are produced separated by a large
rapidity gap. Such a configuration (involving a $Z$) serves as a background to vector boson fusion (VBF) production of a
Higgs boson, where
the two  forward-backward jets serve to tag the event. As the Higgs boson is produced through a colour-less exchange, the rate for an
additional jet to be produced in the central region between the two  tagging jets should be suppressed with respect to the QCD
production of $Z + \ge 2$jets. The  probability for an additional jet to be emitted in QCD $W + 2$~jet events
(rather than $Z$, in order to obtain a higher rate), plus the ability of various theoretical predictions to describe  this rate, is a measurement that can be
carried out at the Tevatron prior to the turn-on of  the LHC. Such a measurement is shown in Figure~\ref{fig:tev_w_jets_8}, where the
rate for a 3rd jet to be emitted is shown versus the rapidity separation of the two tagging jets.
It is evident that (1) the rate for a 3rd jet to be produced is large and (2) that the observed rate is in agreement with the CKKW
predictions,  and is bracketed by the predictions  of MCFM for two choices of scale. Since the prediction is for $W + 3$~jets, the MCFM
calculation is at LO and retains a large scale dependence. The $W$/$Z + 3$~jets process is one to which a high priority has been given for
calculation to NLO, as will be discussed in Section~\ref{sec:wishlist}. The rate for an additional jet to be emitted is roughly
independent of the rapidity separation of the two tagging jets.
The agreement of the  data  with the CKKW predictions is heartening for two reasons: (1) it indicates that CKKW  predictions will most
likely provide accurate predictions for similar topologies at the LHC and (2) the rate for additional jet production in
$W$/$Z + 2$~widely separated jet events is high, leading to an effective veto in VBF Higgs boson searches at the LHC. 

\begin{figure}
\begin{center}
\includegraphics[width=7cm]{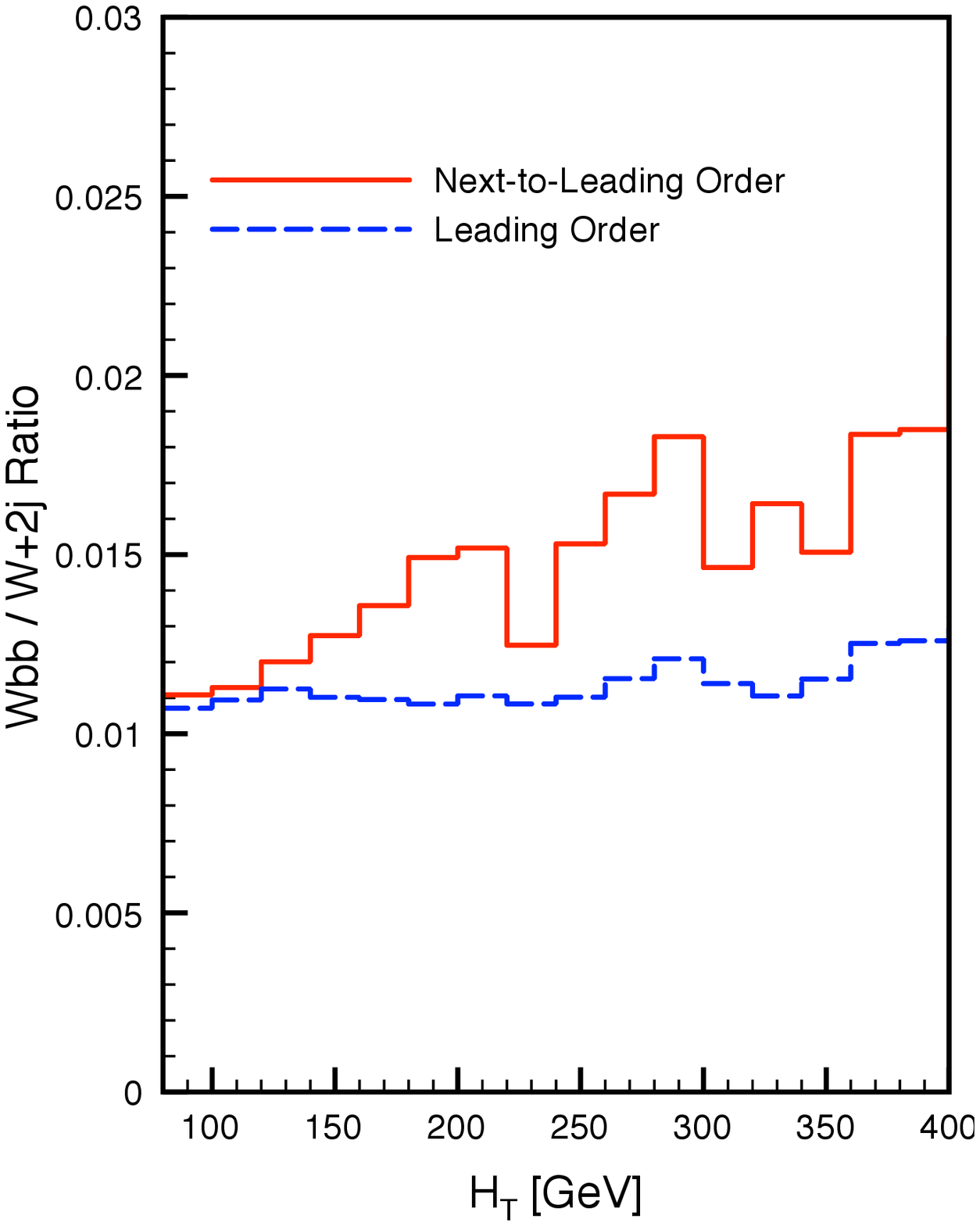}
\includegraphics[width=7cm]{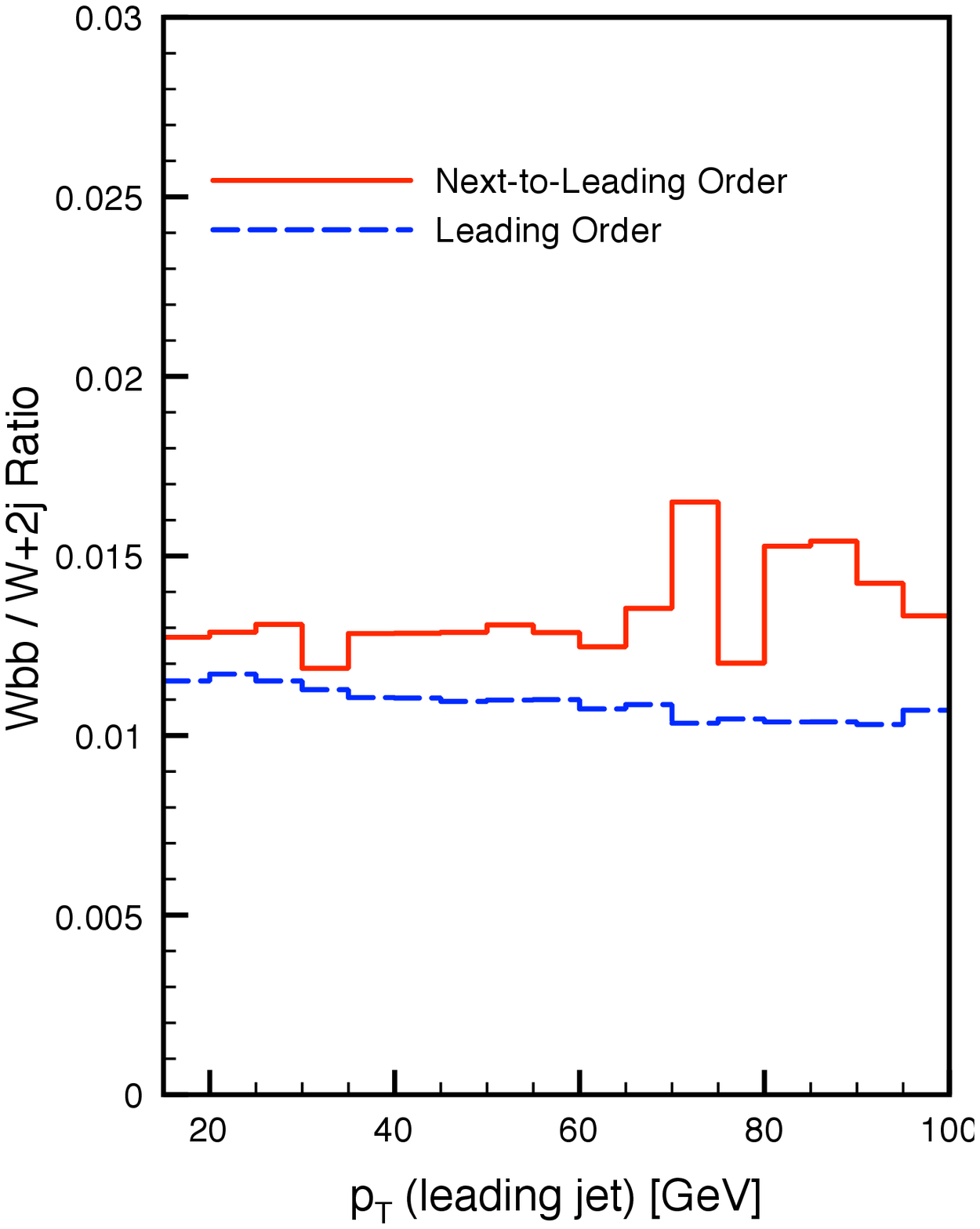}
\end{center}
\caption{
The ratio of the cross sections for $Wb\overline{b}$ and $Wjj$ is plotted at LO and NLO as a function of the variables $H_T$ (left)
and the $p_T$ of the leading jet (right). The ratio is observed to have a different slope at NLO than at LO on the left, but a similar
slope on the right.
\label{fig:htpt5ratios}}
\end{figure}

For many of the analyses at the Tevatron, it is useful to calculate the rate of leading order parton shower Monte Carlo predictions.
For example, the Method 2 technique~\cite{Acosta:2004hw} in CDF's top analysis uses the calculated ratio of
$[Wb\overline{b}+(n-2)~{\rm jets}]/[W+n~{\rm jets}]$ (for $n=3,4$)
and the measured rate for $W +  n$~jets to calculate the $Wb\overline{b}+(n-2)$ jet background to top production. The ratio of the two
processes should be more reliable than the absolute  leading order prediction for  $Wb\overline{b}+(n-2)$ jets. The above statement
should be true as long as the NLO corrections (the $K$-factors) do not have an appreciable difference in shape between the two processes.
As discussed previously, NLO corrections are available for $Wb\overline{b}$ and $Wjj$ but not for higher jet multiplicities. A recent
study has shown that the relative shape of the two processes is indeed different at NLO if highly exclusive variables (such as $H_T$,
the sum of the transverse energy of all final state objects in the event) are used for the measurement~\cite{Campbell:2004sp}.
On the other hand, inclusive variables such as the transverse momentum of the leading jet seem to be safe in
this regard. The dependence of the ratio on these two variables is contrasted in Figure~\ref{fig:htpt5ratios}.
There is some indication that the change in shape observed in going from LO to NLO can also be reproduced by including higher
jet multiplicities using the CKKW procedure.


\subsection{$t\bar{t}$ production at the Tevatron}
\label{sec:tT_tev}

Perhaps the greatest discovery at the Tevatron was that of the top quark. The production mechanism was through
$t\bar{t}$ pair production, dominated by a $q\bar{q}$ initial state. The top quark decays essentially $100$\% into a
$W$ and a $b$ quark; thus the final states being investigated depend on the decays of the two $W$'s. The most useful
(combination of rate and background) final state occurs when one of the $W$'s decays into a lepton and neutrino and
the other decays into two quarks. Thus, the final state consists of a lepton, missing transverse energy and of the
order of four jets. The number of jets may be less than $4$ due to one or more of the jets not satisfying the
kinematic cuts, or more than $4$ due to additional jets being created by gluon radiation off the initial or final
state.  Because of the relatively large number of jets, a smaller cone size ($R=0.4$) has been used for jet
reconstruction. Unfortunately, no top analysis has yet been performed using the $k_T$ jet algorithm.  There is a
sizeable background for this final state through QCD production of $W +$~jets.  Two of the jets in $t\bar{t}$ events
are created by $b$ quarks; thus there is the additional possibility of an improvement of signal purity by the
requirement of one or two $b$-tags. 

\begin{figure}
\begin{center}
\includegraphics[width=9cm]{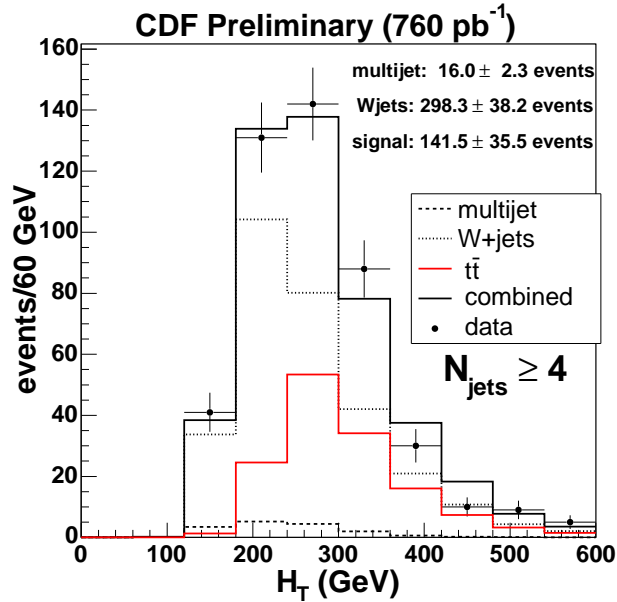}
\end{center}
\caption{The $H_T$ distribution in $W+\ge 4$ jet events at CDF, along with the fitted components for $t\overline{t}$ production and backgrounds from $W$ + jets and multijet events. 
\label{fig:fitHt_4j}}
\end{figure}

In Figure~\ref{fig:fitHt_4j} are shown the CDF measurement of the $H_T$ (sum of the transverse energies of all jets,
leptons, missing $E_T$ in the event) distribution for lepton $ + \ge 4$ jets final states, along with the
predictions for the $H_T$ distributions from $t\bar{t}$ events and from the $W$+ jets background~\cite{Acosta:2005am, cdftop}. A requirement of
high $H_T$ improves the $t\bar{t}$ signal purity, as the jets from the $t\bar{t}$ decays tend to be at higher
transverse momentum than from the QCD backgrounds. 

\begin{figure}[t]
\begin{center}
\includegraphics[width=7cm]{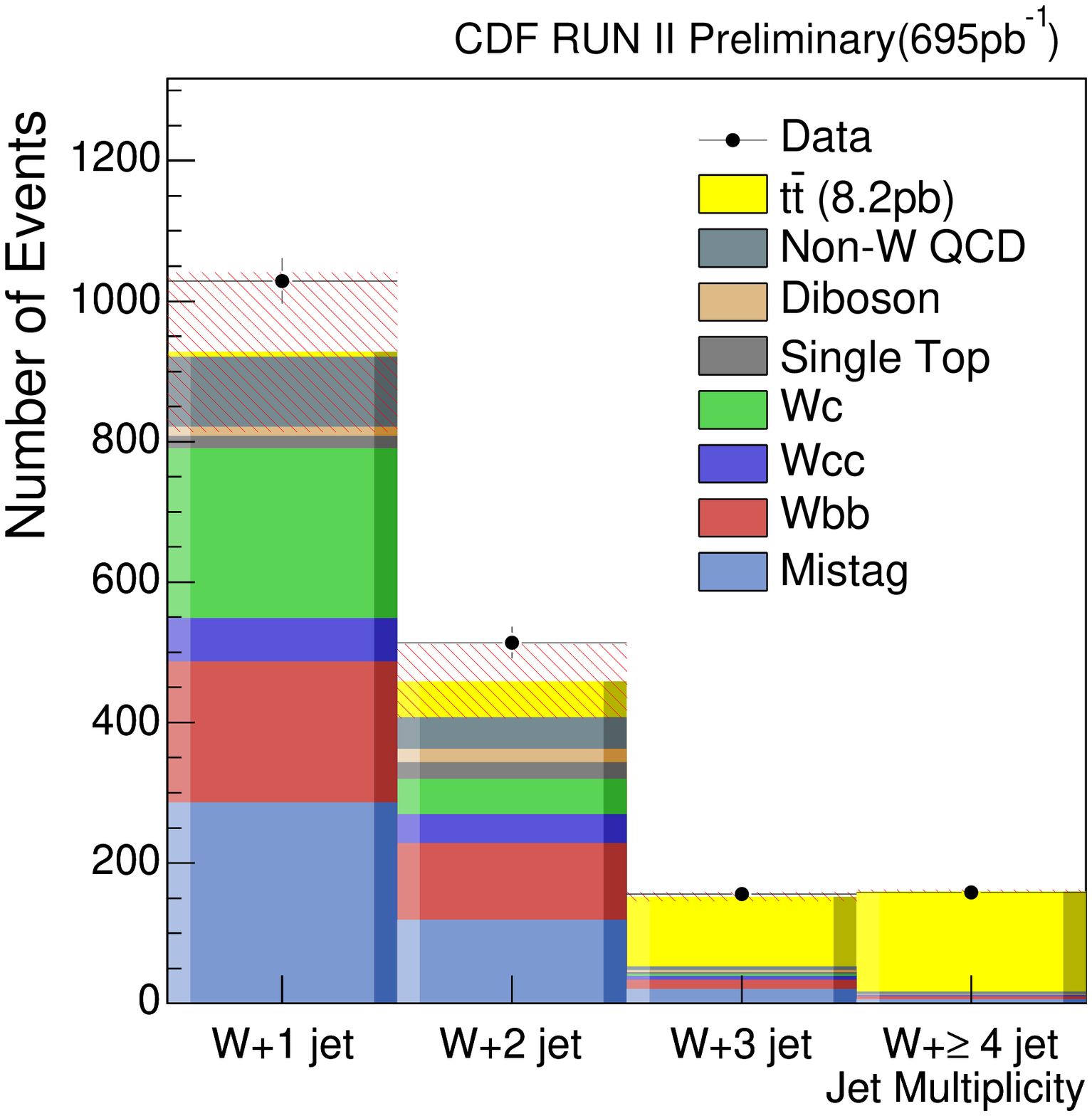}
\includegraphics[width=7cm]{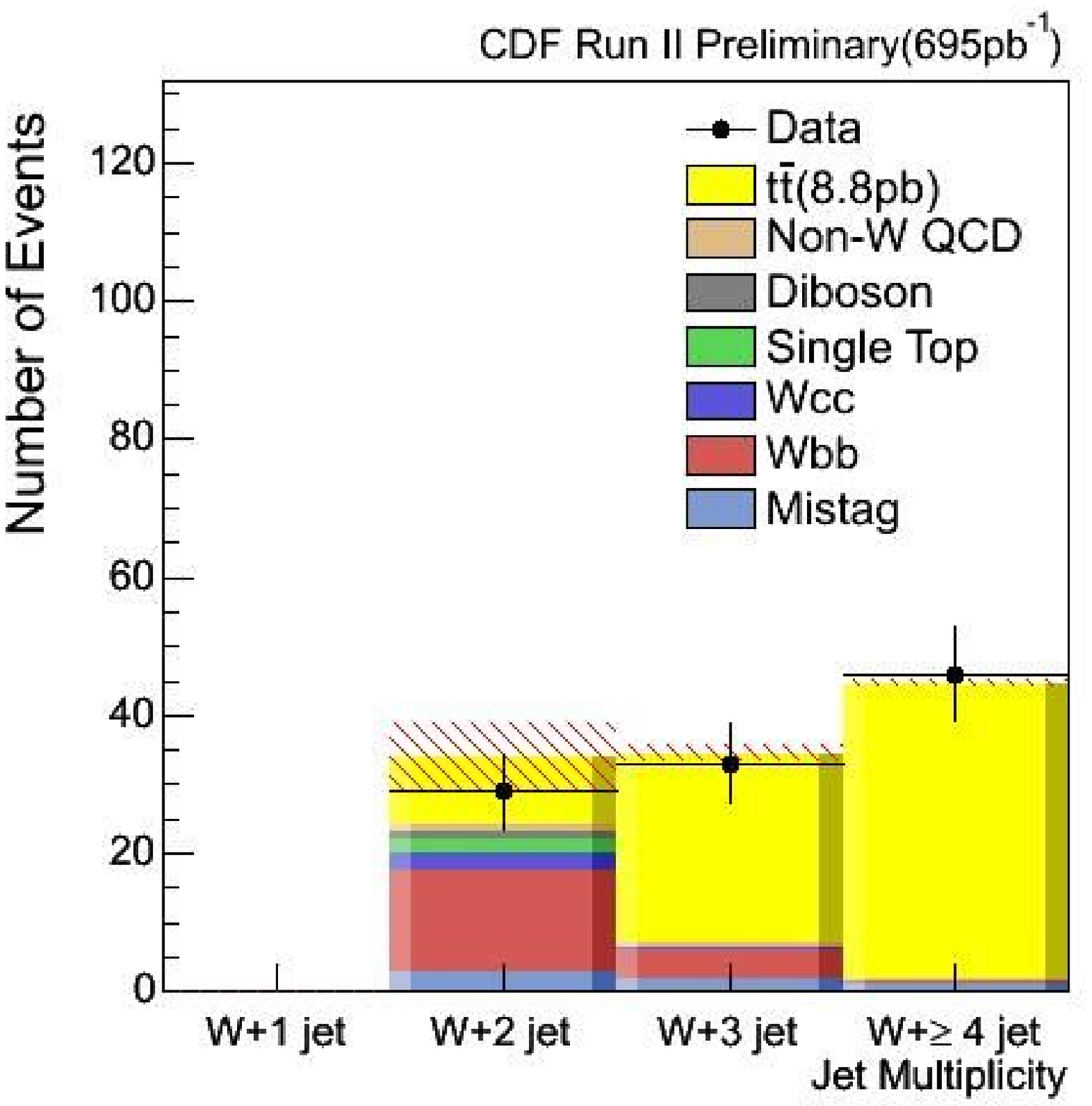}
\end{center}
\caption{The expected number of $W$+ jets tagged (left) and double-tagged (right) events indicated by source.
} 
\label{fig:b_jet_mult}
\end{figure}

The jet multiplicity distribution for the top candidate sample from CDF in Run 2 is shown in
Figure~\ref{fig:b_jet_mult} for the case of one of the jets being tagged as a $b$-jet (left)
and two of the jets being
tagged (right)~\cite{Acosta:2004hw, cdfsvx}. The requirement of one or more $b$-tags greatly reduces the $W +$~jets background in the $3$ and $4$ jet
bins, albeit with a reduction in the number of events due to the tagging efficiency. 

\begin{figure}[t]
\begin{center}
\includegraphics[width=7cm]{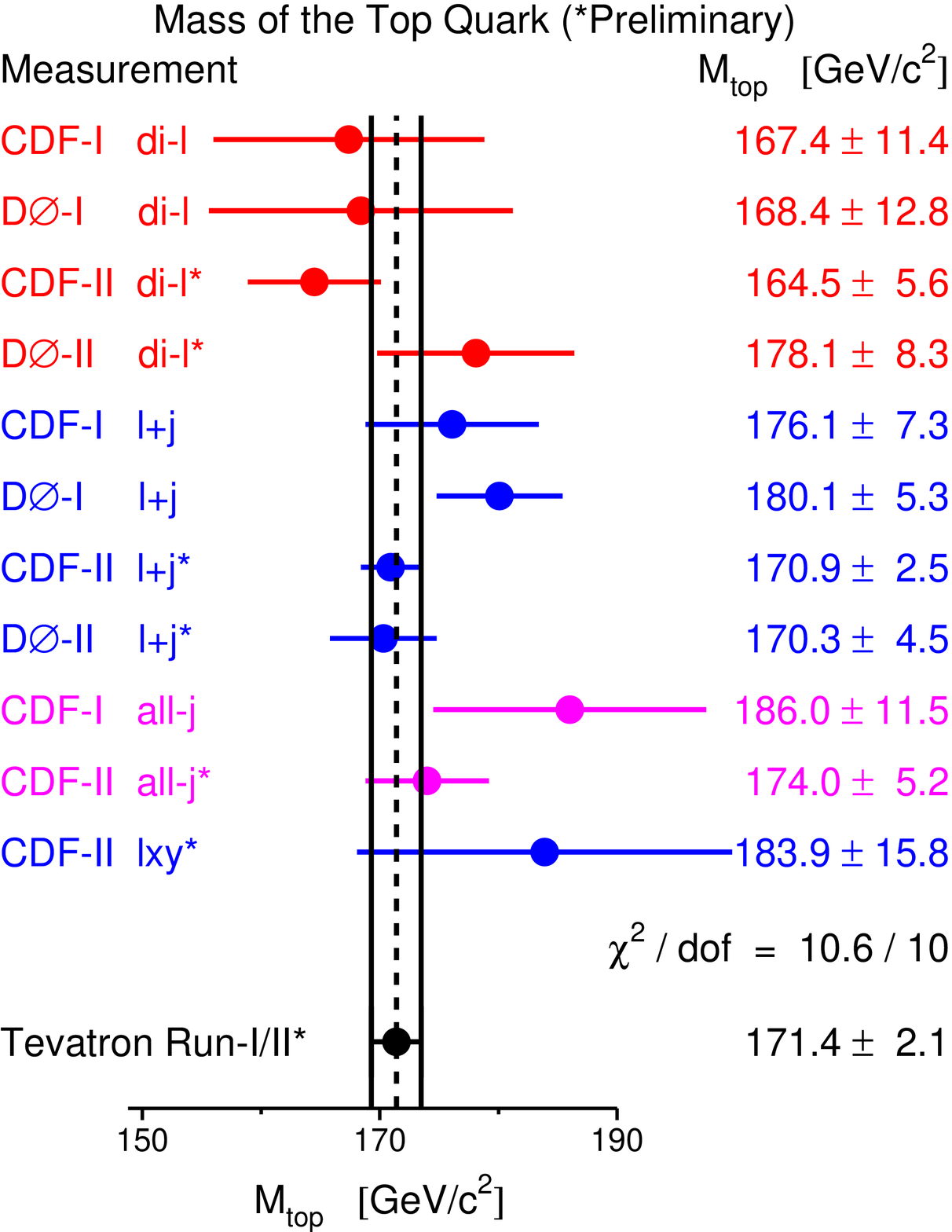}
\includegraphics[width=7cm]{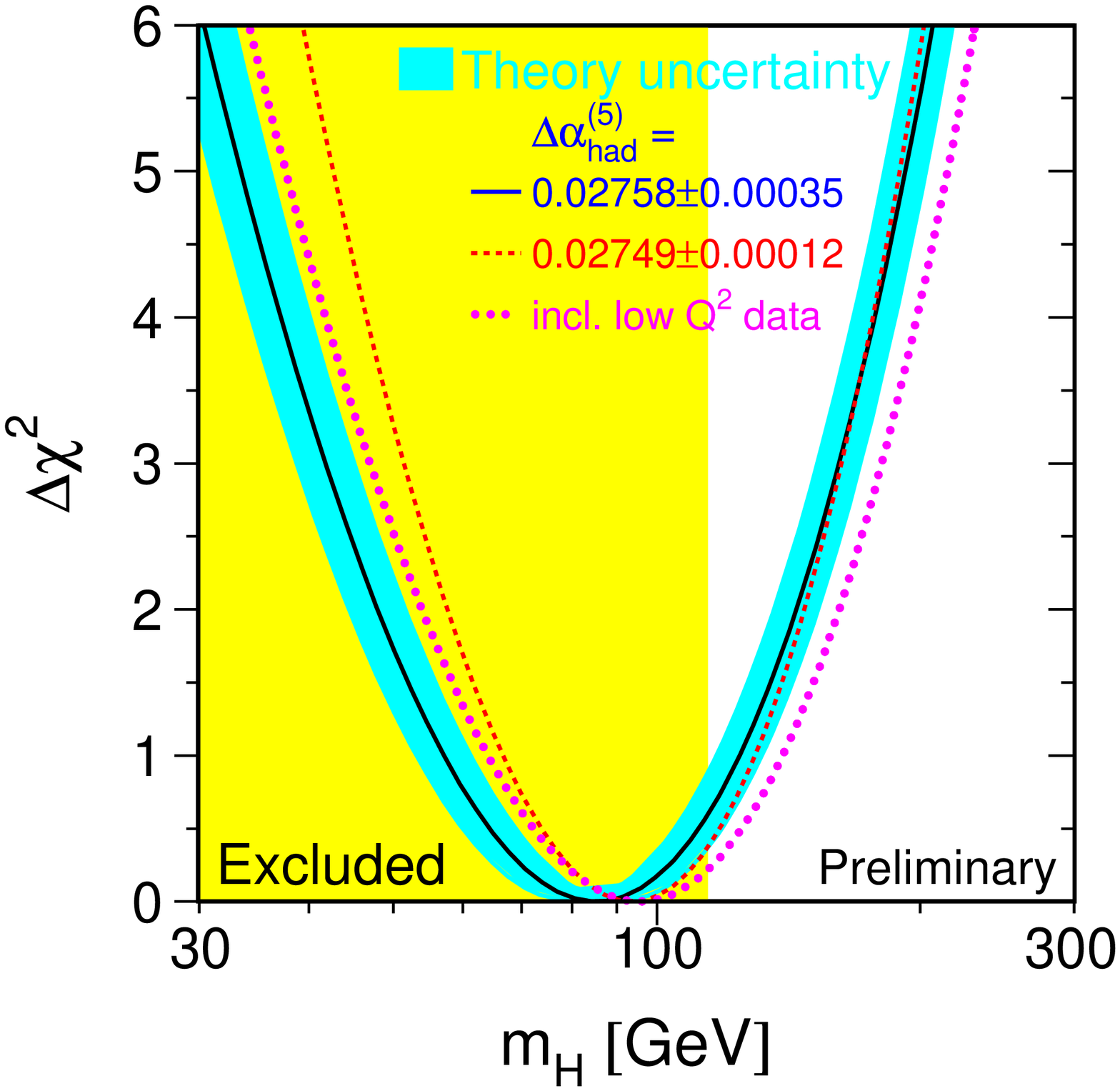}
\end{center}
\caption{A compilation of the top quark mass measurements from CDF and D0 (left) and the implications for the mass of the Higgs boson (right). } 
\label{fig:topmass}
\end{figure}

The lepton + jets final state (with one or more $b$-tags) is also the most useful for the determination of the top
mass. A compilation of top mass determinations from the Tevatron is shown in Figure~\ref{fig:topmass} (left) along
with the implications for the measured top and $W$ masses for the Higgs mass (right)~\cite{tevtopmass}. 

\begin{figure}[t]
\begin{center}
\includegraphics[width=10cm]{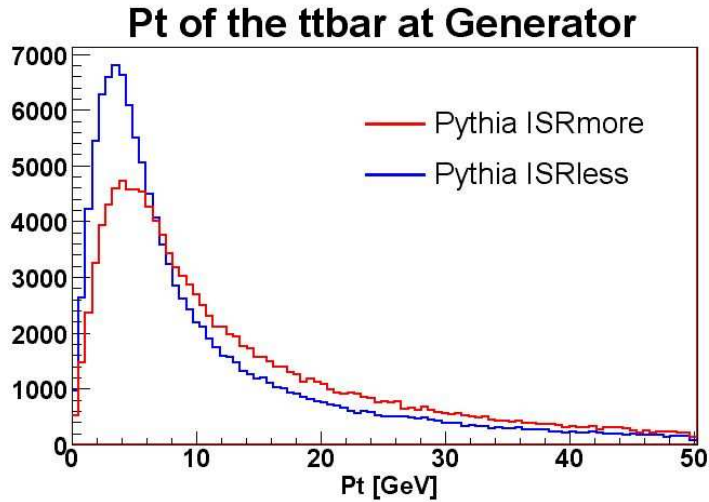}
\end{center}
\vspace*{-0.5cm}
\caption{
The  PYTHIA predictions for the $t\overline{t}$ transverse momentum using the ``Plus/Minus'' tunes.
\label{fig:pairpt}}
\end{figure}

The precision of the top mass determination has reached the point where some of the systematics due to QCD effects
must be considered with greater care. One of the larger systematics is that due to the effects of initial state
radiation. Jets created by initial state radiation may replace one or more of the jets from the top quark decays,
affecting the reconstructed top mass. In the past, the initial state radiation (ISR) systematic was determined
by turning the radiation
off/on, leading to a relatively large impact. A more sophisticated treatment was adopted in Run 2, where the tunings
for the parton shower Monte Carlos were modified leading to more/less initial state radiation, in keeping with the
uncertainties associated with Drell-Yan measurements as discussed in Section~\ref{sec:tevwz}. The resultant
$t\bar{t}$ pair transverse momentum distributions are shown above in Figure~\ref{fig:pairpt}.  The changes to the
$t\overline{t}$ transverse momentum distribution created by the tunes are relatively modest, as is the resultant
systematic error on the top mass determination. 

Note that the peak of the $t\overline{t}$ transverse momentum spectrum is slightly larger than that for $Z$
production at the Tevatron, due to the larger mass of the $t\overline{t}$ system. As both are produced primarily by
$q\overline{q}$ initial states, the differences are not large.

\begin{figure}[t]
\begin{center}
\includegraphics[width=8cm,angle=0]{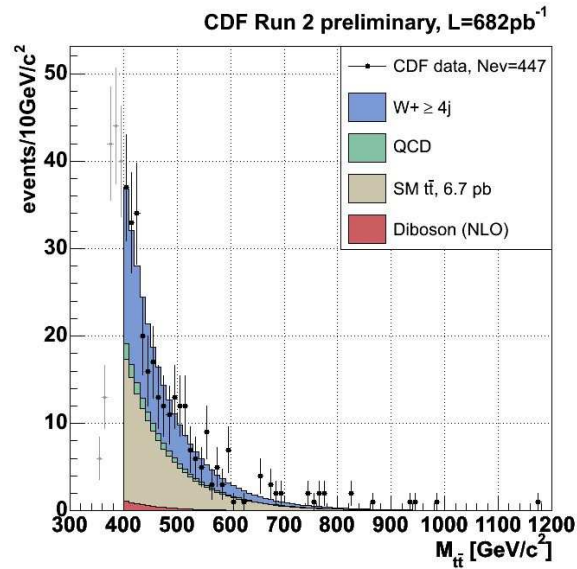}
\end{center}
\vspace*{-0.5cm}
\caption{The $t\overline{t}$ mass distribution as observed by CDF in Run 2.
\label{fig:cdftTmass}}
\end{figure}

\begin{figure}[t]
\begin{center}
\includegraphics[width=8cm,angle=90]{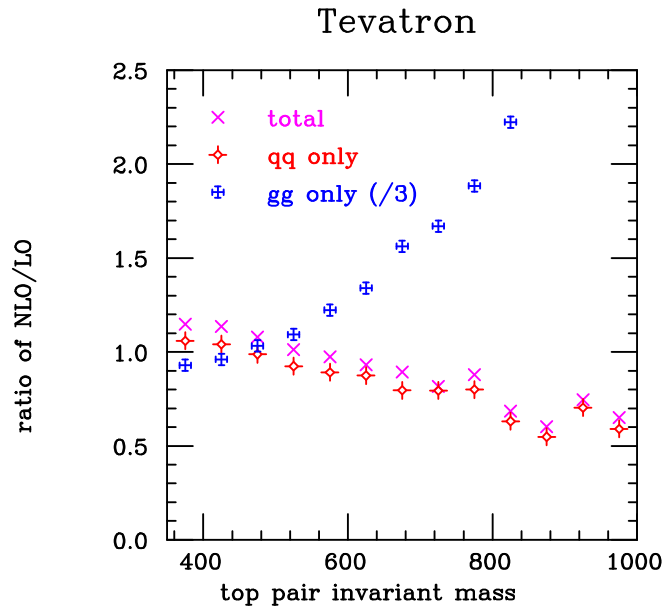}
\end{center}
\vspace*{-0.5cm}
\caption{
The ratio of the NLO to LO predictions for the $t\overline{t}$ mass at the Tevatron.
The predictions include the ratio for the total cross section and for the
specific $q\overline{q}$ and $gg$ initial states. Note that the total also includes a $gq$ contribution
(not present at LO) and that the $gg$ ratio is divided by a factor of $3$.  
\label{fig:mtt_ratio_tev}}
\end{figure}

It is also interesting to look at the mass distribution of the $t\overline{t}$ system, as new physics (such as a
$Z'$~\cite{Hill:2002ap}) might couple preferentially to top quarks. Such a comparison for CDF Run 2 is shown in
Figure~\ref{fig:cdftTmass} where no signs of a high mass resonance are evident. 
If we look at predictions for the
$t\overline{t}$ mass distribution at NLO, we see that the NLO cross section is substantially less than the LO one at
high mass (see Figure~\ref{fig:mtt_ratio_tev}). This should be taken into account in any searches for new physics at high
$t\overline{t}$ mass. 


Further investigation shows that the decrease of NLO compared to LO at high mass is found only in the
$q\overline{q}$ initial state and not in the $gg$ initial state. In fact, at the Tevatron, the ratio of NLO to LO for $gg$ initial states grows dramatically
with increasing top pair invariant mass. This effect is largely due to the increase in the gluon
distribution when going from CTEQ6L1 in the LO calculation to CTEQ6M at NLO. For instance, at $x \sim 0.4$ (and hence an invariant
mass of about $800$~GeV) the gluon distribution is about a factor two larger in CTEQ6M than in CTEQ6L1, giving a factor four
increase in the cross section. Conversely, the quark distribution is slightly decreased at such large $x$. However, the absolute contribution of the $t\bar{t}$ cross at high masses from $gg$ initial states is small, due to the rapidly falling gluon distribution at high $x$. 

\section{Benchmarks for the LHC}
\label{sec:lhc}

\subsection{Introduction}
\label{sec:lhcintro}

Scattering at the LHC is not simply {\it rescaled} scattering at the Tevatron. For many of the key processes the typical momentum fractions
$x$ are small;  thus, there is a dominance of sea quark and gluon scattering as compared to valence  quark
scattering at the Tevatron. There is a large phase space for gluon emission and thus intensive  QCD backgrounds for many of
the signatures of new physics. Many of the scales relating to  interesting processes are large compared to the $W$ mass; thus,
electroweak corrections can  become important even for nominally QCD processes. In this section, we will try to provide some
useful benchmarks for LHC predictions.  

\subsection{Parton-parton luminosities at the LHC~\protect\footnote{
Parts of this discussion also appeared in a contribution to the Les Houches 2005
proceedings~\cite{Buttar:2006zd} by A. Belyaev, J. Huston and J. Pumplin}}
\label{sec:lum}

It is useful to return to the idea of  differential parton-parton luminosities. Such luminosities, when multiplied by the dimensionless 
cross section $\hat{s}\hat{\sigma}$ for a given process, provide a useful estimate  of 
the size of an event cross section at the LHC. 
Below we define the differential parton-parton luminosity
$dL_{ij}/d\hat{s}\,dy$ and its integral $dL_{ij}/d\hat{s}$:
\begin{equation}
\frac{d L_{ij}}{d\hat{s}\,dy} = 
\frac{1}{s} \, \frac{1}{1+\delta_{ij}} \, 
[f_i(x_1,\mu) f_j(x_2,\mu) + (1\leftrightarrow 2)] \; .
\label{eq1}
\end{equation}
The prefactor with the Kronecker delta avoids double-counting in case the
partons are identical.  The generic parton-model formula 
\begin{equation}
\sigma = \sum_{i,j} \int_0^1 dx_1 \, dx_2 \, 
f_i(x_1,\mu) \, f_j(x_2,\mu) \, \hat{\sigma}_{ij}
\end{equation}
can then be written as 
\begin{equation}
\sigma = \sum_{i,j} \int \left(\frac{d\hat{s}}{\hat{s}} \, dy\right) 
\, \left(\frac{d L_{ij}}{d\hat{s}\,dy}\right) \, 
\left(\hat{s} \,\hat{\sigma}_{ij} \right) \; .
\label{eq:xseclum}
\end{equation}
(Note that this result is easily derived by defining $\tau = x_1 \, x_2 = \hat{s}/s$ 
and observing that the Jacobian $\partial(\tau,y)/\partial(x_1,x_2) = 1$.)

\begin{figure}[t]
\begin{center}
\includegraphics[width=9cm]{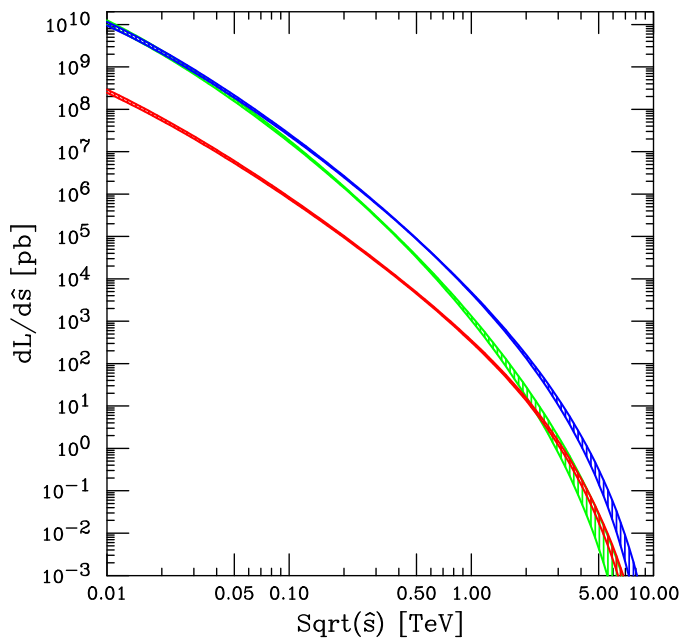}
\end{center}
\caption{
The parton-parton luminosity$\left[\frac{dL_{ij}}{d\tau}\right]$ 
in picobarns,
integrated over $y$. 
Green=$gg$, 
Blue=$\sum_i (gq_i+g{\bar q}_i+q_ig+{\bar q}_ig)$,
Red=$\sum_i (q_i{\bar q}_i+{\bar q}_iq_i)$,
where the sum runs over the five quark flavours $d$, $u$, $s$, $c$, $b$.
} 
\label{fig:figlum4}
\end{figure}

Equation \eref{eq:xseclum} can be used to estimate the production rate for a  hard scattering process at the LHC
as follows.  Figure~\ref{fig:figlum4} shows a plot of the luminosity function integrated  over rapidity,
$dL_{ij}/d\hat{s} = \int (dL_{ij}/d\hat{s}\,dy) \, dy$, at the LHC $\sqrt{s} = 14 \, \mathrm{TeV}$ for various
parton flavour combinations, calculated using the CTEQ6.1 parton distribution  functions~\cite{Stump:2003yu}. 
The widths of the curves indicate an estimate  for the pdf uncertainties.  We assume $\mu = \sqrt{\hat{s}}$ for
the scale~\footnote{
Similar plots made with earlier pdfs are shown in Ellis, Stirling, Webber~\cite{Ellis:1991qj}}.
As expected, the $gg$ luminosity is large at low $\sqrt{\hat{s}}$ but falls rapidly with respect to the other
parton luminosities. The $gq$ luminosity is large over the entire kinematic region plotted. 

\begin{figure}[t]
\begin{center}
\includegraphics[width=9cm]{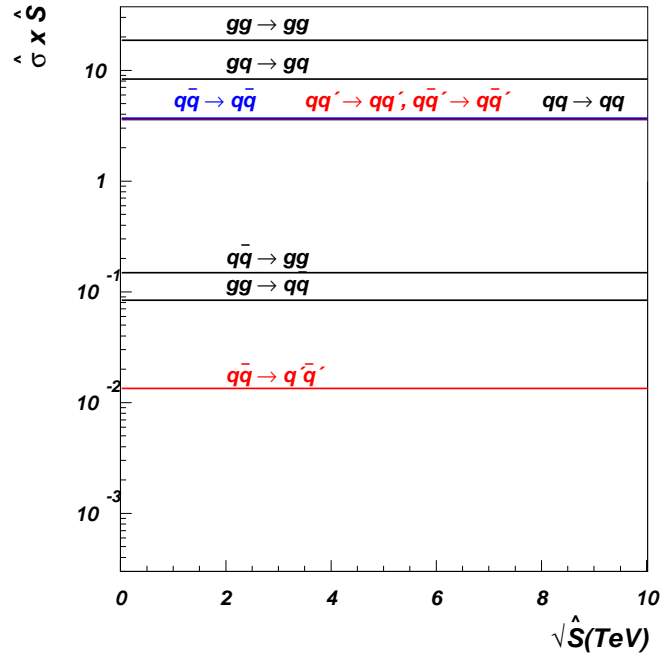}
\end{center}
\vspace*{-0.5cm}
\caption{
Parton level cross sections
$\left(\hat{s}\hat{\sigma}_{ij}\right)$
for various processes involving massless partons in the final state. 
} 
\label{fig:sigma}
\end{figure}

\begin{figure}[t]
\begin{center}
\includegraphics[width=9cm]{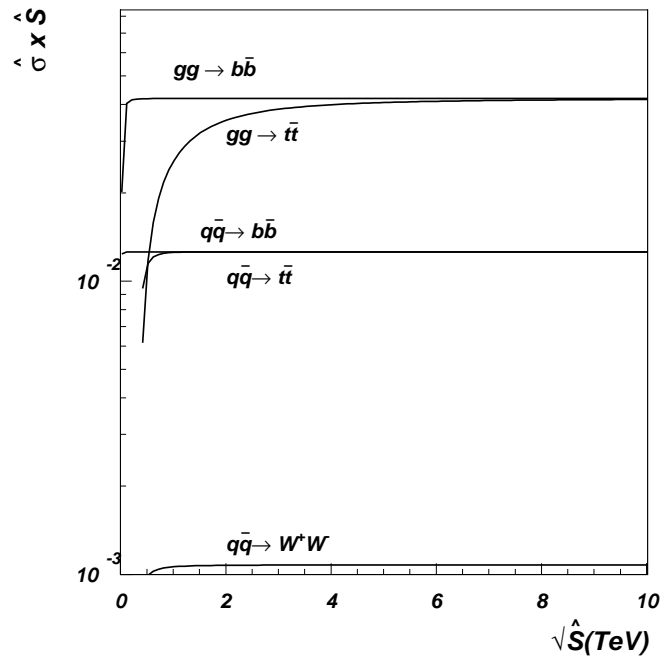}
\end{center}
\vspace*{-0.5cm}
\caption{
Parton level cross sections
$\left(\hat{s}\hat{\sigma}_{ij}\right)$
for various processes involving massive partons in the final state. 
} 
\label{fig:sigmah}
\end{figure}

Figures~\ref{fig:sigma} and~\ref{fig:sigmah} present the second product, 
$\left[\hat{s}\hat{\sigma}_{ij}\right]$, for various $2 \rightarrow 2$ partonic processes with massless and
massive partons in the final state respectively. The parton level cross sections have been calculated for a
parton $p_T> 0.1\times\sqrt{\hat{s}}$ cut and for fixed $\as=0.118$ using the CalcHEP
package~\cite{Pukhov:2004ca}. For the case of massive partons in the final state, there is a threshold behaviour
not present with massless partons. Note also that the threshold behaviour is different for $q\overline{q}$ and
$gg$ initial states. The $gg$ processes can proceed through the $t$-channel as well as the $s$-channel and this is
responsible for the extra structure. 

%
%

\begin{figure}[t]
\begin{center}
\includegraphics[width=8cm]{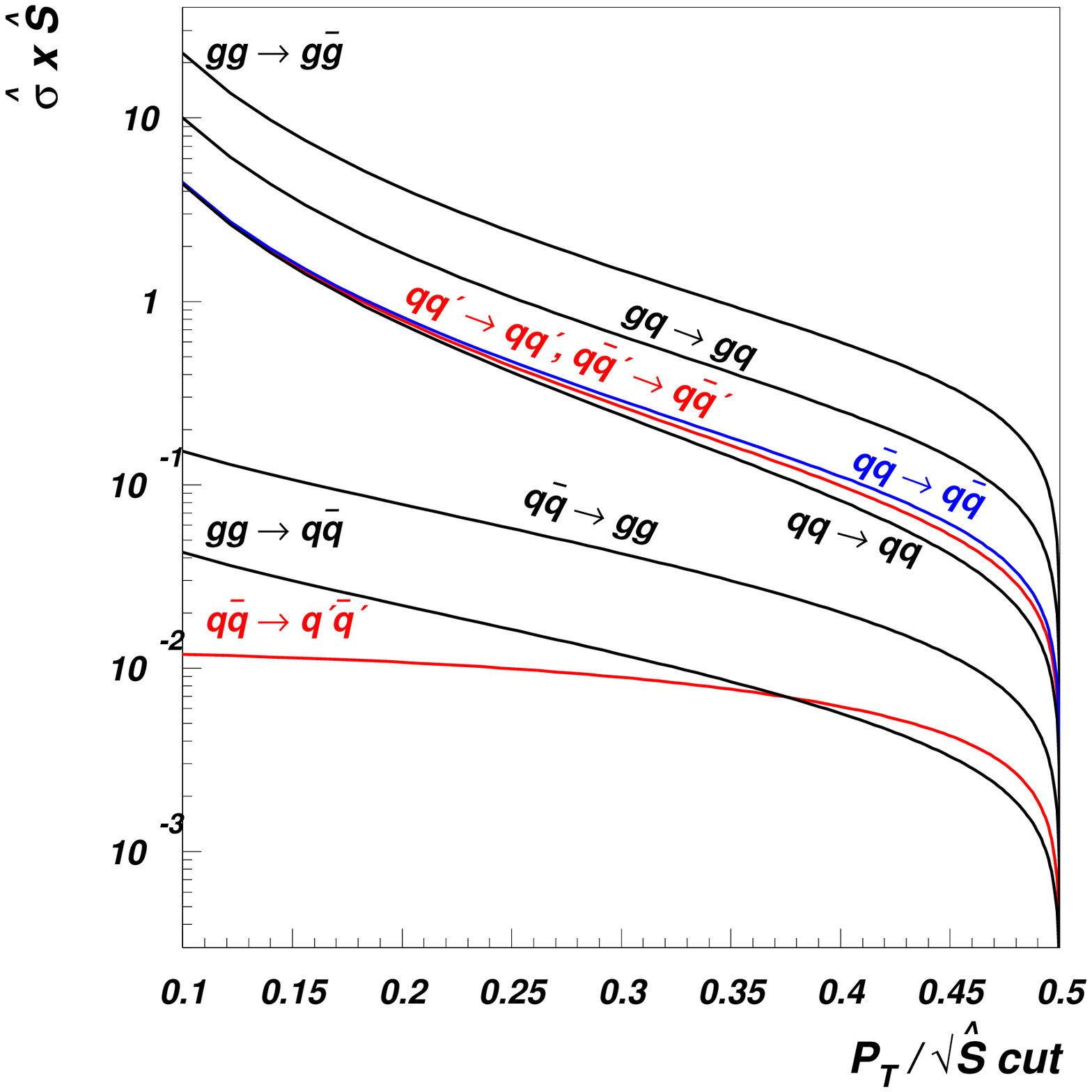}
\end{center}
\caption{
Parton level cross sections
$\left(\hat{s}\hat{\sigma}_{ij}\right)$
for various processes involving massless partons in the final state as a function of the variable $p_T/\sqrt{\hat{s}}$. 
} 
\label{fig:sigma_d}
\end{figure}

\begin{figure}[t]
\begin{center}
\includegraphics[width=8cm]{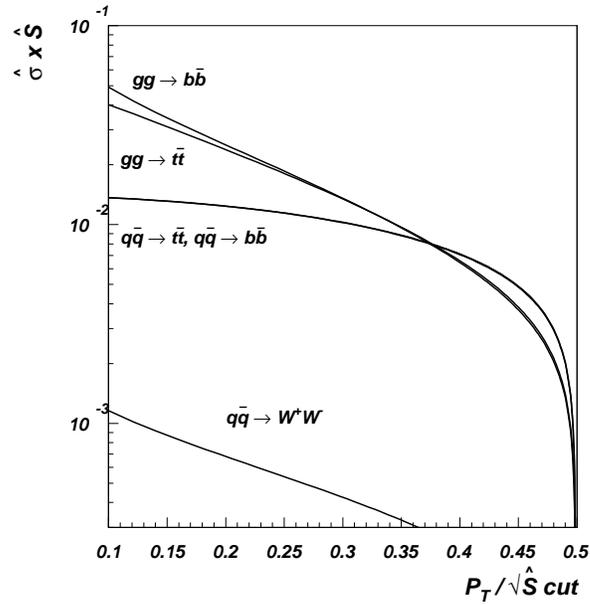}
\end{center}
\caption{
Parton level cross sections
$\left(\hat{s}\hat{\sigma}_{ij}\right)$
for various processes involving massive partons in the final state as a function of the variable
$p_T/\sqrt{\hat{s}}$. The calculations were performed  for $\sqrt{\hat{s}} = 2$~TeV, in the region
where $\hat{\sigma}\hat{s}$ has a relatively flat behaviour with $\sqrt{\hat{s}}$.
} 
\label{fig:sigmah_d}
\end{figure}

The products $\left[\hat{s}\hat{\sigma}_{ij}\right]$ are plotted for massless and massive final state partons as
a function of the ratio $p_T/\sqrt{\hat{s}}$ in Figures~\ref{fig:sigma_d} and~\ref{fig:sigmah_d}.
One can use \eref{eq:xseclum} in the form
\begin{equation}
\sigma=\frac{\Delta\hat{s}}{\hat{s}}
\left(\frac{d L_{ij}}{d \hat{s}}\right)
\left(\hat{s} \,\hat{\sigma}_{ij} \right).
\end{equation}
and Figures~\ref{fig:sigma},~\ref{fig:sigma_d},~\ref{fig:sigmah},~\ref{fig:sigmah_d}
to estimate  the QCD production cross sections 
for a given $\Delta \hat{s}$ interval and a particular cut on $p_T/\sqrt{\hat{s}}$.
For example, for the $gg \rightarrow gg$  rate
for $\hat{s}$=1~TeV  and $\Delta\hat{s}=0.01\hat{s}$,
we have  $d L_{gg}/d \hat{s}\simeq 10^3$~pb and 
$\hat{s} \,\hat{\sigma}_{gg}\simeq 20$
leading to $\sigma\simeq 200$~pb 
(for the $p_T^g> 0.1\times\sqrt{\hat{s}}$ cut we have used above).
Note that for a given small $\Delta\hat{s}/\hat{s}$ interval,
the corresponding  invariant mass 
$\Delta\sqrt{\hat{s}}/\sqrt{\hat{s}}$ interval, is 
$\Delta\sqrt{\hat{s}}/\sqrt{\hat{s}}\simeq \frac{1}{2}\Delta\hat{s}/\hat{s}$.
One should also mention  that all hard cross sections presented in Figure~\ref{fig:sigma}
are proportional  to $\as^2$ and have been calculated for
$\as=0.118$, so production rates can be easily rescaled 
for a particular $\as$ at a given scale.

\begin{figure}[t]
\begin{center}
\includegraphics[width=8cm]{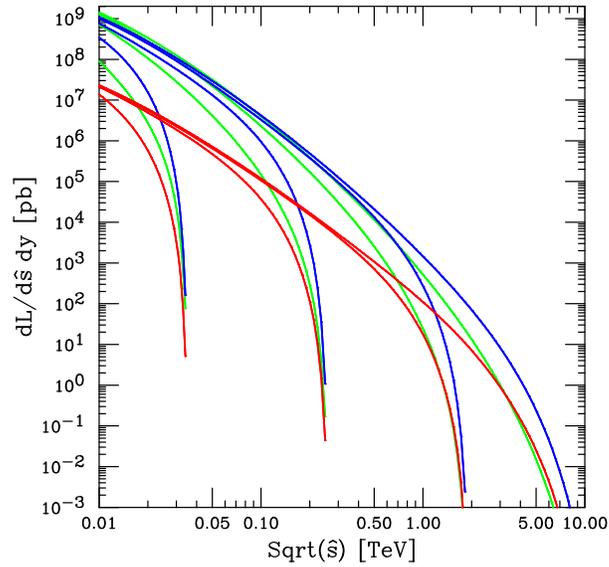}
\end{center}
\caption{
$d({\rm Luminosity})/dy$ at rapidities (right to left) $y=0, 2, 4, 6$.
Green=$gg$, 
Blue=$\sum_i (gq_i+g{\bar q}_i+q_ig+{\bar q}_ig)$,
Red=$\sum_i (q_i{\bar q}_i+{\bar q}_iq_i)$,
where the sum runs over the five quark flavours $d$, $u$, $s$, $c$, $b$.
} 
\label{fig:figlum5}
\end{figure}

One can further specify the parton-parton luminosity for a specific  rapidity $y$ and $\hat{s}$,
$dL_{ij}/d\hat{s}\, dy$.  If one is interested in a specific partonic initial state, then the resulting
differential luminosity can be displayed in families of curves as shown in Figure~\ref{fig:figlum5}, where the
differential parton-parton luminosity at the LHC is shown as a function of the subprocess centre-of-mass energy
$\sqrt{\hat{s}}$ at various values of rapidity for the produced system for several different combinations of
initial state partons.
One can read from the curves the parton-parton luminosity for a specific value of mass
fraction and  rapidity. (It is also easy to use the Durham pdf plotter to generate the pdf curve for any desired
flavour and kinematic configuration~\footnote{http://durpdg.dur.ac.uk/hepdata/pdf3.html}.) 

\begin{figure}[t]
\begin{center}
\includegraphics[width=9cm]{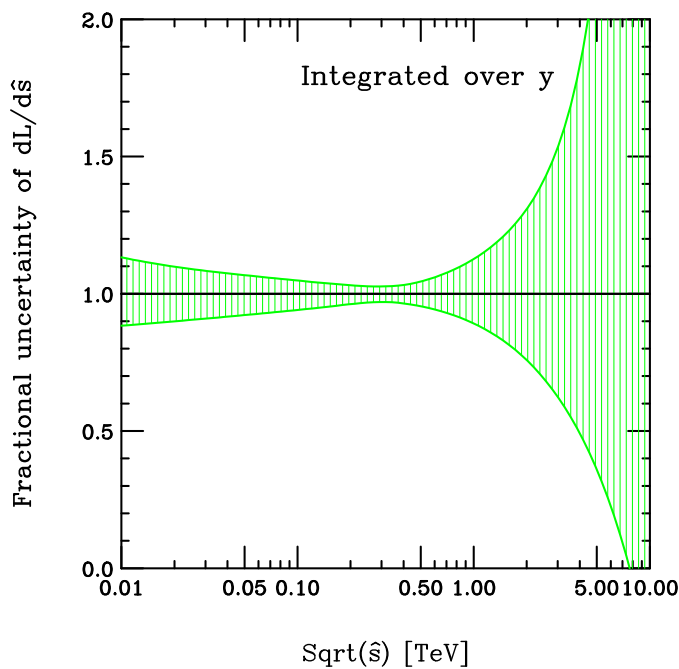}
\end{center}
\caption{
Fractional uncertainty of the $gg$ luminosity integrated over $y$. 
\label{fig:figlum6yall}
}
\end{figure}
\begin{figure}[t]
\begin{center}
\includegraphics[width=9cm]{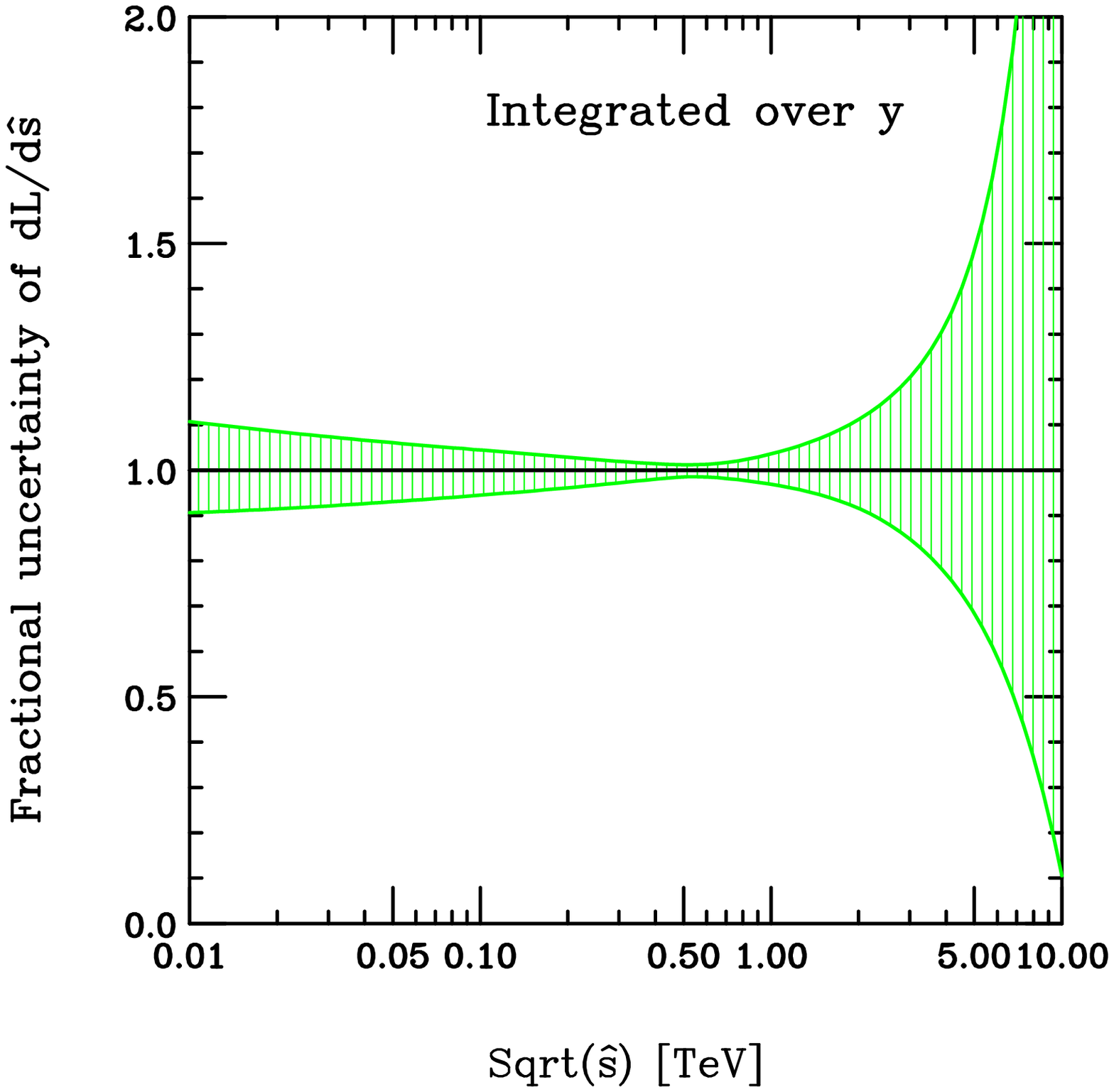}
\end{center}
\caption{
Fractional uncertainty for the parton-parton luminosity integrated over $y$
for $\sum_i (q_i{\bar q}_i+{\bar q}_iq_i)$,
where the sum runs over the five quark flavours $d$, $u$, $s$, $c$, $b$.
\label{fig:figlum7yall}
}
\end{figure}
\begin{figure}[t]
\begin{center}
\includegraphics[width=9cm]{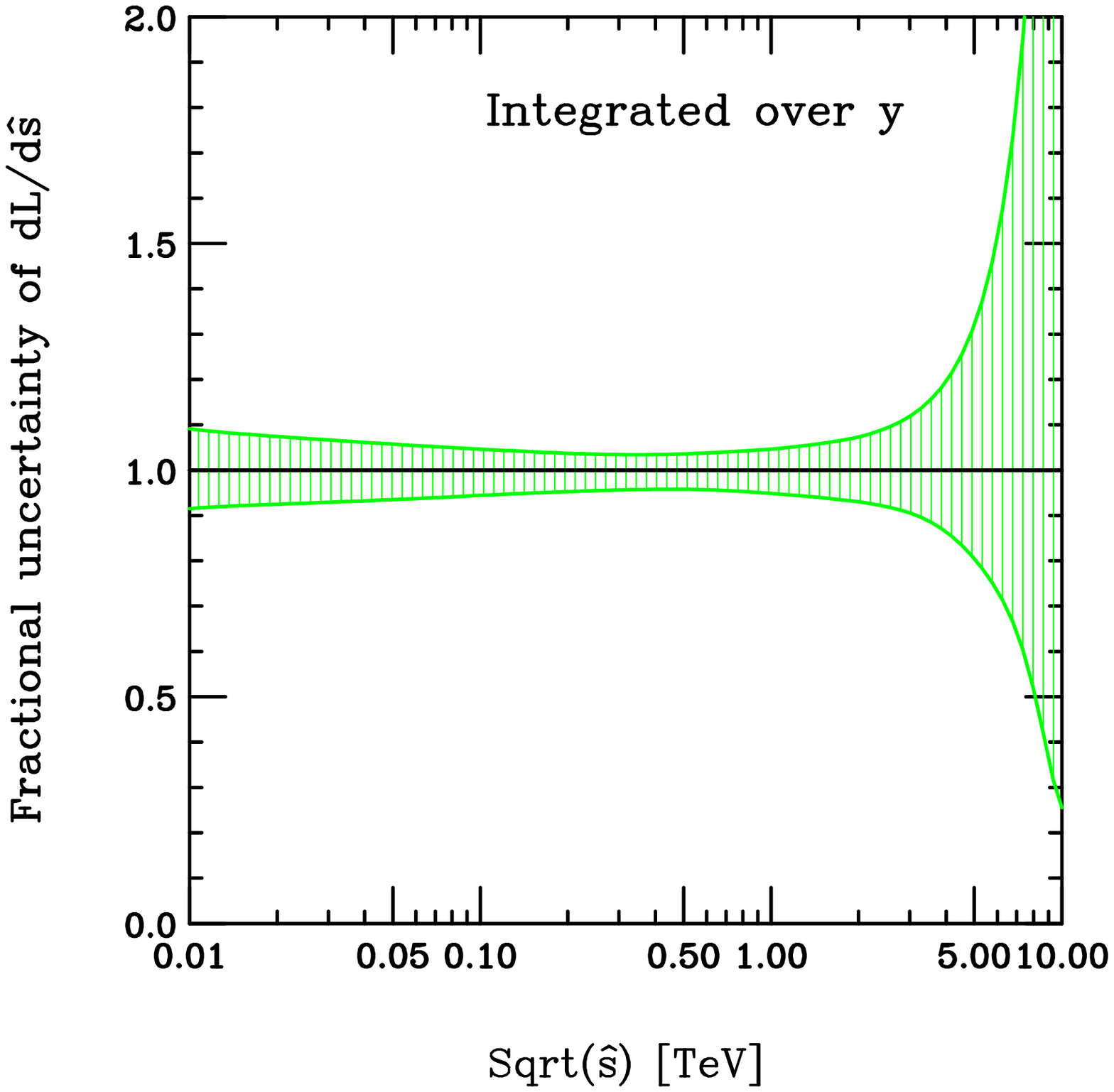}
\end{center}
\caption{
Fractional uncertainty for the luminosity integrated over $y$
for $\sum_i (q_i{\bar q}_i+{\bar q}_iq_i)$,
where the sum runs over the five quark flavours $d$, $u$, $s$, $c$, $b$.
\label{fig:figlum8yall}
}
\end{figure}

It is also of great interest to understand the uncertainty in the parton-parton luminosity for specific
kinematic configurations. Some representative parton-parton luminosity uncertainties, integrated  over
rapidity,  are shown in Figures~\ref{fig:figlum6yall}, \ref{fig:figlum7yall} and~\ref{fig:figlum8yall}. The pdf
uncertainties were generated from the CTEQ6.1 Hessian error analysis using the standard $\Delta \chi^2 = 100$
criterion. Except for kinematic  regions where one or both partons is a gluon at high $x$, the pdf uncertainties
are of  the order of $5$--$10\%$. Luminosity uncertainties for specific rapidity values are available at the SM
benchmark website. Even tighter constraints will be possible once the LHC Standard Model  data is included in
the global pdf fits. Again, the uncertainties for individual pdfs can also be calculated online using the
Durham pdf plotter. 
Often it is not the pdf uncertainty for a cross section that is required, but rather the pdf uncertainty for an acceptance for a given
final state. The acceptance for a particular process may depend on the input pdfs due to the rapidity cuts placed on the jets, leptons,
photons, etc. and the impacts of the varying longitudinal boosts of the final state caused by the different pdf pairs. An approximate
``rule-of-thumb'' is that the pdf uncertainty for the acceptance is a factor of $5$--$10$ times smaller than the uncertainty for
the cross section itself.

\begin{figure}[t]
\begin{center}
\includegraphics[width=8cm]{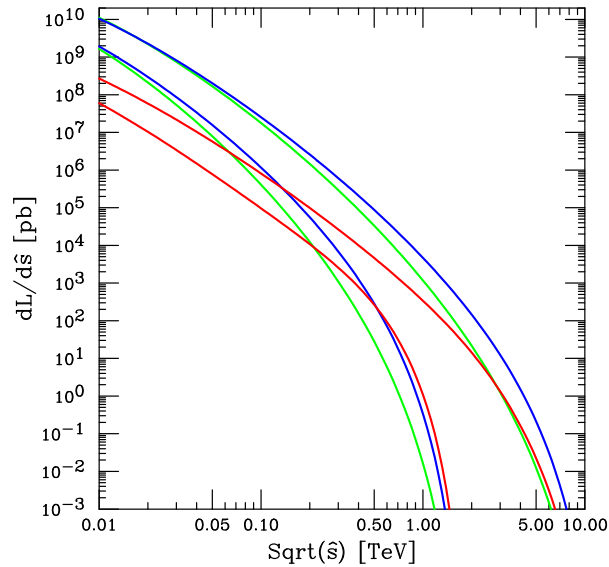}
\end{center}
\caption{
The parton-parton luminosity$\left[\frac{1}{\hat{s}}\frac{dL_{ij}}{d\tau}\right]$ in pb
integrated over $y$. 
Green=$gg$, 
Blue=$\sum_i (gq_i+g{\bar q}_i+q_ig+{\bar q}_ig)$,
Red=$\sum_i (q_i{\bar q}_i+{\bar q}_iq_i)$,
where the sum runs over the five quark flavours $d$, $u$, $s$, $c$, $b$.
The top family of curves are for the LHC and the bottom  for the Tevatron. 
} 
\label{fig:figlum12}
\end{figure}

\begin{figure}[t]
\begin{center}
\includegraphics[width=8cm]{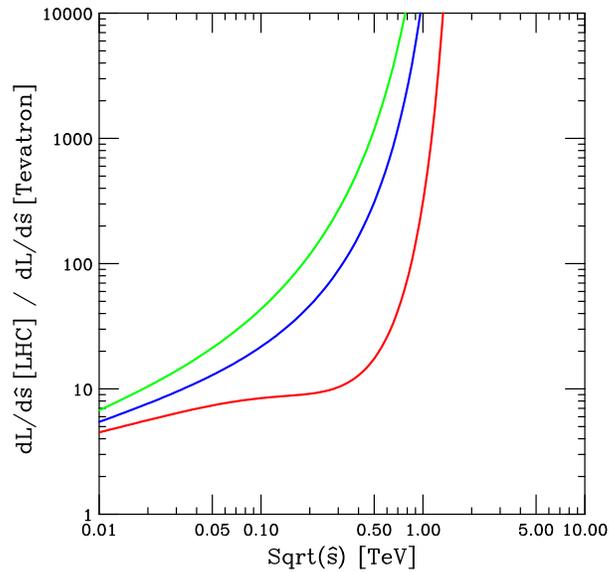}
\end{center}
\caption{
The ratio of parton-parton luminosity$\left[\frac{1}{\hat{s}}\frac{dL_{ij}}{d\tau}\right]$ in pb
integrated over $y$ at the LHC and Tevatron. 
Green=$gg$ (top), 
Blue=$\sum_i (gq_i+g{\bar q}_i+q_ig+{\bar q}_ig)$ (middle),
Red=$\sum_i (q_i{\bar q}_i+{\bar q}_iq_i)$ (bottom),
where the sum runs over the five quark flavours $d$, $u$, $s$, $c$, $b$. 
\label{fig:figlum13}}
\end{figure}

In Figure~\ref{fig:figlum12}, the pdf luminosity curves shown in Figure~\ref{fig:figlum4} are overlaid with
equivalent luminosity curves from the Tevatron.
In Figure~\ref{fig:figlum13}, the ratios of the pdf luminosities
at the LHC to those at the Tevatron are plotted.
The most dramatic increase in pdf luminosity at the LHC comes from  $gg$ initial states, followed by $gq$ initial states and
then $q\bar{q}$ initial states. The latter ratio is smallest because of the availability of valence antiquarks at the
Tevatron at moderate to large $x$. As an example, consider chargino pair production with $\sqrt{\hat{s}}=0.4$~TeV. This
process proceeds through $q\bar{q}$ annihilation; thus, there is only a factor of $10$ enhancement at the LHC compared to the
Tevatron. 

Backgrounds to interesting physics at the LHC proceed mostly through $gg$ and $gq$ initial states. Thus, there
will be a commensurate increase in the rate for background processes at the LHC.

\subsection{Stability of NLO global analyses}
\label{sec:stable}

\begin{figure}[t]
\begin{center}
\includegraphics[width=5cm,angle=90]{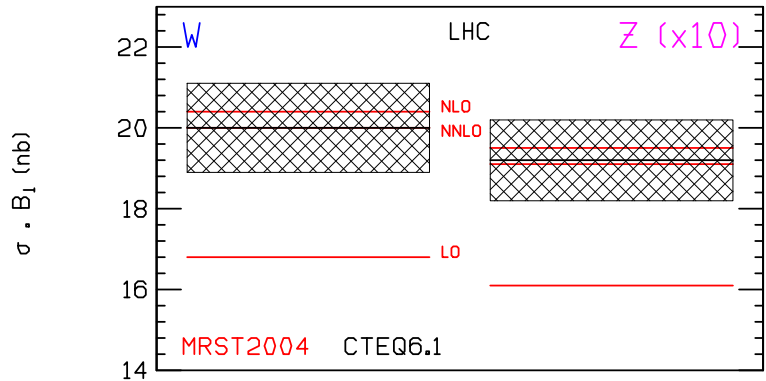}
\end{center}
\caption{Predicted cross sections for $W$ and $Z$ production at the LHC using MRST2004 and CTEQ6.1 pdfs. The
overall pdf uncertainty of the NLO CTEQ6.1 prediction is approximately $5\%$,
consistent with Figure~\ref{fig:figlum8yall}.
}
\label{fig:LHC_WZ}
\end{figure}

The $W$ and $Z$ cross sections at the LHC are shown in Figure~\ref{fig:LHC_WZ}. The MRST2004 predictions are shown
at LO, NLO and NNLO; also shown are the CTEQ6.1 predictions at NLO, along with the CTEQ6.1 pdf error band. There
is a significant increase in cross section when going from LO to NLO, and then a small decrease when going from
NLO to NNLO. The NLO MRST2004 and CTEQ6.1 predicted cross sections agree with each other, well within the
CTEQ6.1 pdf uncertainty band. 

Most of the absolute predictions for LHC observables have been carried out at  NLO, i.e. with NLO pdfs and with
NLO matrix elements. Such predictions have worked well  at the Tevatron, but the LHC explores a new region of
(small) $x$ for  a hadron-hadron collider. Thus, it is important to understand whether the NLO  formalism carries over
to  the LHC with the same degree of accuracy.  In recent years, some preliminary next-to-next-leading-order
global pdf analyses have been carried out either for DIS alone~\cite{Alekhin}, or in a global analysis
context~\cite{mrstnnlo}. The differences with respect to the corresponding NLO analyses are reasonably small.
However, most observables of interest (including  inclusive jet production) have not yet been calculated at
NNLO.  

All other considerations being equal, a global analysis at NNLO must be expected to have a higher accuracy.
However, NLO analyses can be adequate as long as their accuracy is sufficient for the task, and as long as their
predictions are stable with respect to certain choices inherent in the analysis. Examples of those choices are
the functional forms used to parametrize the initial nonperturbative parton distribution functions, and the
selection of experimental data sets included in the fit --- along with the kinematic cuts that are imposed on that
data. In a recent MRST analysis~\cite{Martin:2003sk}, a $20\%$ variation in the cross section predicted for $W$
production at the LHC -- a very important ``standard candle'' process for hadron colliders -- was observed when
data at low $x$ and $Q^2$ were removed from the global fit. This is a fundamental consideration, since the
presence of this instability would significantly impact the phenomenology of a wide range of physical processes
for the LHC. The instability was removed in the same study when the analysis was carried out at NNLO. 

\begin{figure}[t]
\begin{center}
\includegraphics[width=10cm]{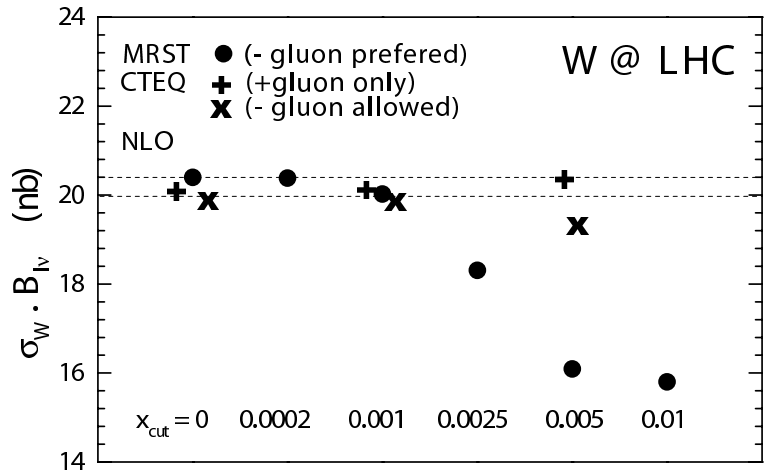}
\end{center}
\caption{
Predicted total cross section of $W^+ + W^-$ production
at the LHC for the fits obtained in the CTEQ stability study, compared
to the MRST results. The overall pdf
uncertainty of the prediction is $\sim 5\%$, as observed in Figure~\ref{fig:figlum8yall}.} 
\label{fig:WtotXs}
\end{figure}

\begin{figure}[t]
\begin{center}
\includegraphics[width=8cm]{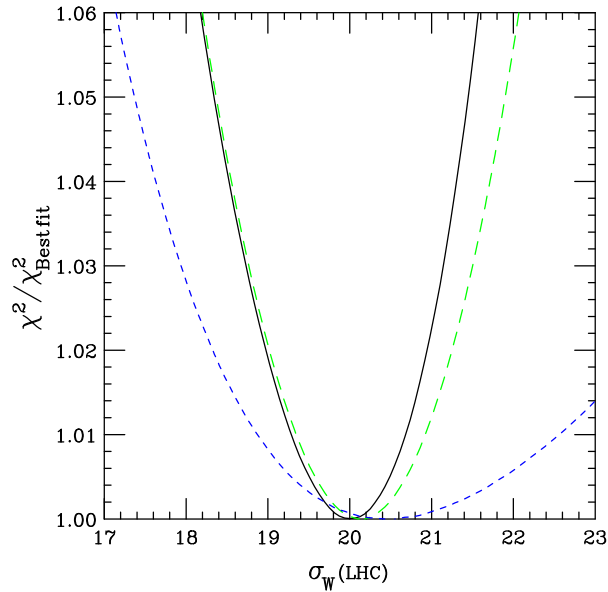}
\end{center}
\caption{
Lagrange multiplier results for the $W$ cross section 
(in $\mathrm{nb}$) at the LHC
using a positive-definite gluon. The three curves, in order of
decreasing steepness, correspond to  three sets of kinematic cuts,  standard/intermediate/strong.
\label{fig:ChiVsSigCutsPG}
}
\end{figure}

A similar study was also carried out by the CTEQ collaboration in which this  instability was not observed, and the
predictions  for the $W$ cross section at the LHC remained stable (but with increased uncertainties) when more severe low
($x,Q^2$) kinematic cuts were performed~\cite{Huston:2005jm}. The predicted cross sections, as a function of the $x$ cut on
the data removed, are shown in Figure~\ref{fig:WtotXs} for the two studies.
In Figure~\ref{fig:ChiVsSigCutsPG}, the uncertainty on the predictions for the $W$ cross section at the LHC is
shown from a study that uses the Lagrange Multiplier Method.
As more data are removed at low $x$ and $Q^2$, the resulting uncertainty on  the predicted $W$ cross section
increases greatly. If the gluon is parameterized in a form that allows it to become  negative at low $x$, such
as the parameterization that MRST used, then the uncertainty becomes even  larger, especially for the lower
range of the cross section. 

%
%

%
%

\subsection{The future for NLO calculations}
\label{sec:future}

Unfortunately, our ability to take advantage of the benefits of NLO predictions 
is severely limited by the calculations available. Currently, no complete NLO QCD
calculation exists for a process involving more than 5 particles. This means
that we are limited to the consideration of scenarios such as 4 jet production
at LEP (effectively a $1 \to 4$ process) and a number of $2 \to 3$ processes of
interest at hadron colliders.

The bottleneck for these calculations is the evaluation of the loop diagrams. The real
radiation component consists of the evaluation of known lowest order matrix elements
combined with a method for systematically removing their singularities, several of which are
well understood. In contrast, the evaluation of the virtual contribution must at present be
performed on a case-by-case basis. Even the most basic integral that can appear depends
non-trivially on the number of external legs in the loop, as well as on the masses of all the
external and internal particles. To complicate matters further, the presence of powers of loop
momenta in the integrals (from fermion propagators or multi-gluon vertices) leads to ``tensor
integrals'' which typically are expressed as the sum of many terms. Taken all together, this
means that traditional methods based on an analytical Feynman diagram approach are already
limited by the computing power currently available. It seems unlikely that further progress,
such as the NLO calculation of a $2 \to 4$ process, will be made in this fashion.

An alternative to this brute-force Feynman diagram approach is provided by
``unitarity-based'' methods, which are based on sewing together tree level amplitudes
(for a review, see~\cite{Bern:1996je}). These methods rely on the Cutkosky rules to
generate the 1-loop amplitude by summing over all possible cuts of sewn-together tree
amplitudes. Although this knowledge in itself only suffices for sufficiently
supersymmetric theories, results for 1-loop amplitudes in non-supersymmetric theories
(such as QCD) can be obtained by consideration of collinear limits of the amplitudes.
Such techniques were used to obtain relatively compact expressions for the 1-loop
amplitudes required to evaluate the process $e^+ e^- \to 4$~partons~\cite{Bern:1997sc}.

More recently the ``twistor-inspired'' methods, which have proven so efficient at
generating compact expressions for tree-level amplitudes, have been considered at the
1-loop level~\cite{Cachazo:2004zb}. The MHV rules (discussed in Section~\ref{sec:LOtools})
were shown to reproduce known
results at the 1-loop level~\cite{Brandhuber:2004yw} and their application has passed
from calculations in $N=4$ supersymmetric Yang-Mills theory to the evaluation of
amplitudes in QCD~(\cite{Brandhuber:2005jw} and references therein). Similarly, the
on-shell recursion relations have proven equally useful at one-loop~\cite{Bern:2005hs},
even allowing for the calculation of MHV amplitudes involving an arbitrary number
of gluons~\cite{Forde:2005hh}. These methods will doubtless be refined and extended
to more complicated calculations over time.

An alternative path to 1-loop amplitudes is provided by numerical approaches. Such
techniques have long been applied at lowest order, with the main attraction deriving
from the promise of a generic solution that can be applied to any process (given
sufficient computing power). The techniques are generally either completely numerical
evaluations of loop integrals~\cite{Ferroglia:2002mz,Anastasiou:2005cb}, or they
are semi-numerical approaches that perform only the tensor integral reductions
numerically~\cite{Giele:2004iy,Giele:2004ub,delAguila:2004nf,vanHameren:2005ed,Binoth:2005ff,Ellis:2005zh}.
In most cases these algorithms have only been applied to individual integrals or
diagrams, although one recent application evaluates the 1-loop matrix elements for
the Higgs$+4$~parton process (which is otherwise uncalculated)~\cite{Ellis:2005qe}~\footnote{
All partons are treated as outgoing so this process corresponds to a Higgs$+2$~parton final state}.
Once more, the study of these approaches is in its infancy and the coming years
promise significant further development.

It should also be noted that many of the recent simplifying techniques and numerical approaches
have concentrated on processes involving gluons and massless quarks. Although this is sufficient
for many hadron collider applications, there are many processes for which a calculation which
includes the quark mass is necessary. The top quark mass must clearly be taken into account and, in
addition, the experimental ability to identify $b$-quarks in events often demands that the mass of the
bottom quark is included. When including the mass, the evaluation of Feynman diagrams becomes much more
complicated and the expressions for amplitudes do not simplify as conveniently. However, the presence
of an additional hard scale renders the loop amplitudes less infrared divergent, so that some numerical
approaches may in fact benefit from the inclusion of quark masses. Due to the large role of the top quark
in much of the LHC program, advances on this front are to be expected in the near future.

\subsection{A realistic NLO wishlist for multi-parton final states at the LHC}
\label{sec:wishlist}

\setcounter{footnote}{0}

A somewhat whimsical experimenter's {\it wishlist} was first presented at the Run
2 Monte Carlo workshop at Fermilab in 2001~\footnote{
http://vmsstreamer1.fnal.gov/Lectures/MonteCarlo2001/Index.htm}. 
Since then the list has gathered a great deal of notoriety and has appeared in numerous
LHC-related theory talks. It is unlikely that $WWW+b\overline{b}+3$\,jets will be
calculated at NLO soon, no matter the level of physics motivation, but there are a
number of high priority calculations, primarily of backgrounds to new physics,
that are accessible with the present technology. However, the manpower available
before the LHC turns on is limited. Thus, it is necessary to prioritize the
calculations, both in terms of the importance of the calculation and the effort
expected to bring it to completion.

A prioritized list, determined at the Les Houches 2005 Workshop, is shown in Table~\ref{wishlist}, along with a
brief discussion of the physics motivation. Note that the list contains only
$2\rightarrow 3$ and $2\rightarrow 4$ processes, as these are feasible to be completed by the turn on of the LHC.
\begin{table}[htb]
\begin{center}
\begin{tabular}{|l|l|}
\hline
&\\
process&relevant for\\
($V\in\{Z,W,\gamma\}$)&\\
\hline
&\\
1. $pp\to V\,V+$\,jet &  $t\bar{t}H$, new physics\\
2. $pp\to H+2$~jets& $H$ production by vector boson fusion (VBF)\\
3. $pp\to t\bar{t}\,b\bar{b}$ &  $t\bar{t}H$\\
4. $pp\to t\bar{t}+2$~jets  &  $t\bar{t}H$\\
5. $pp\to V\,V\,b\bar{b}$ &  VBF$\to H\to VV$, 
$t\bar{t}H$, new physics\\
6. $pp\to V\,V+2$~jets &  VBF$\to H\to VV$\\
7. $pp\to V+3$~jets &  various new physics signatures\\
8. $pp\to V\,V\,V$ &  SUSY trilepton searches\\
&\\
\hline
\end{tabular}
\caption{The wishlist of processes for which a NLO calculation is both desired and feasible
in the near future.
\label{wishlist}
}
\end{center}
\end{table}

First, a few general statements: in general, signatures for new physics will involve high $p_T$ leptons and jets (especially
$b$ jets) and missing transverse momentum. Thus, backgrounds to new physics will tend to involve (multiple) vector boson
production (with jets) and $t\overline{t}$ pair production (with jets). The best manner in which to understand the
normalization of a cross section is to measure it; however the rates for some of the complex final states listed here may be
limited and (at least in the early days) must be calculated from NLO theory. As discussed at length previously, NLO is the first
order at which both the normalization and shape can be calculated with any degree of confidence.

We now turn to a detailed look at each of these processes:
\begin{itemize}
\item $pp \rightarrow VV$~+ jet: One of the most promising channels for Higgs boson
production in the low mass range is through the $H\rightarrow  WW^*$ channel, with
the $W$'s decaying semi-leptonically. It is useful to look both in the
$H\rightarrow  WW$ exclusive channel, along with the $H\rightarrow  WW$+jet
channel. The calculation of $pp\rightarrow  WW$+jet will be especially important in
understanding the background to the latter.

\item $pp \rightarrow  H+2$~jets: A measurement of vector boson fusion (VBF) 
production of the Higgs boson will
allow the determination of the Higgs coupling to vector bosons. One of the key
signatures for this process is the presence of forward-backward tagging jets.
Thus, QCD production of $H$ + 2 jets must be understood, especially as the rates for
the two are comparable in the kinematic regions of interest.

\item $pp \rightarrow  t\overline{t} b\overline{b}$ and $pp
\rightarrow t\overline{t}$ + 2 jets: Both of these processes serve as background to
$t\overline{t}H$, where the Higgs boson decays into a $b\overline{b}$ pair. The rate for
$t\overline{t}jj$ is much greater than that for $t\overline{t}b\overline{b}$ and
thus, even if 3 $b$-tags are required, there may be a significant chance for the
heavy flavour mistag of a $t\overline{t}jj$ event to contribute to the background.

\item $pp \rightarrow  VV b\overline{b}$: Such a signature serves as non-resonant
background to $t\overline{t}$ production as well as to possible new physics.

\item $pp \rightarrow VV + 2$~jets: The process serves as a background to VBF
production of a Higgs boson.

\item $pp \rightarrow V + 3$~jets: The process serves as background for
$t\overline{t}$ production where one of the jets may not be reconstructed, as well
as for various new physics signatures involving leptons, jets and missing
transverse momentum.

\item $pp \rightarrow VVV$: The process serves as a background for various new
physics subprocesses such as SUSY tri-lepton production.

\end{itemize}
Work on (at least) the processes 1. to 4. of Table~\ref{wishlist} is already
in progress by several groups, and clearly all of them 
aim at a setup which allows for a straightforward 
application to other processes~\footnote{Process 2 has been calculated since the first version of this list
was formulated~\cite{Campbell:2006xx}.}


From an experimentalist's point-of-view, the NLO calculations discussed thus far may be used to understand changes in
normalization and/or shape that occur  for a given process when going from LO to NLO~\cite{Campbell:2004sp}.  As discussed
previously, direct comparisons to the data require either a determination of parton-to-hadron corrections for the theory or
hadron-to-parton corrections for the data~\cite{Flanagan:2005xv}. Furthermore, for multi-parton final states it is also
necessary to model the effects of jet algorithms, when two or more partons may be combined into one jet. 

\subsection{Some Standard Model cross sections for the LHC}
\label{sec:SMLHC}

Here we discuss some of the Standard Model benchmark cross sections at the LHC. A lack of space will keep the
discussion short, but a more complete treatment can be found at the benchmark website and also in the CMS~\cite{CMS_tdr} and ATLAS~\cite{ATLAS_tdr} Physics Technical Design Reports.

\subsubsection{Underlying event at the LHC}
\label{sec:UELHC}

\begin{figure}[t]
\begin{center}
\includegraphics[width=10cm]{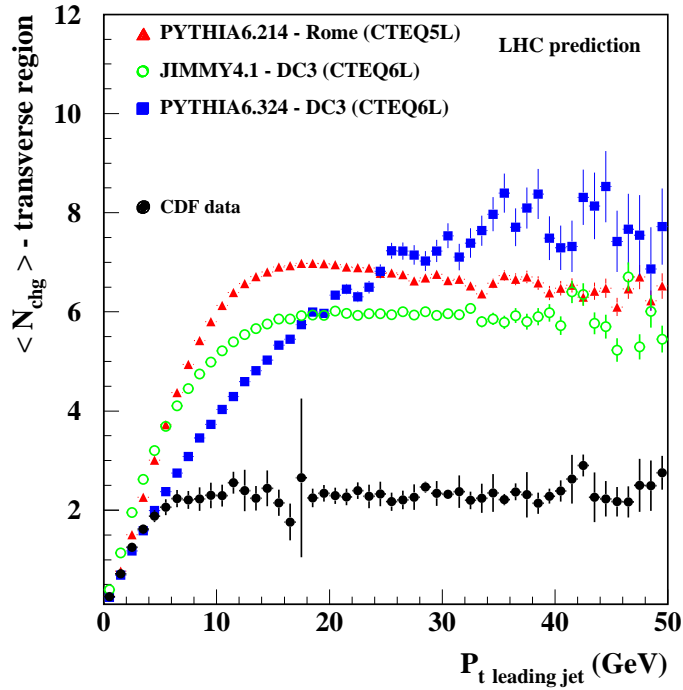}
\end{center}
\caption{PYTHIA6.2 - Tune A, Jimmy4.1 - UE and PYTHIA6.323 - UE
predictions for the average charged multiplicity in the transverse region in the underlying 
event for LHC $pp$ collisions.
\label{fig:ue_LHC}
}
\end{figure}

\begin{figure}[t]
\begin{center}
\includegraphics[width=10cm]{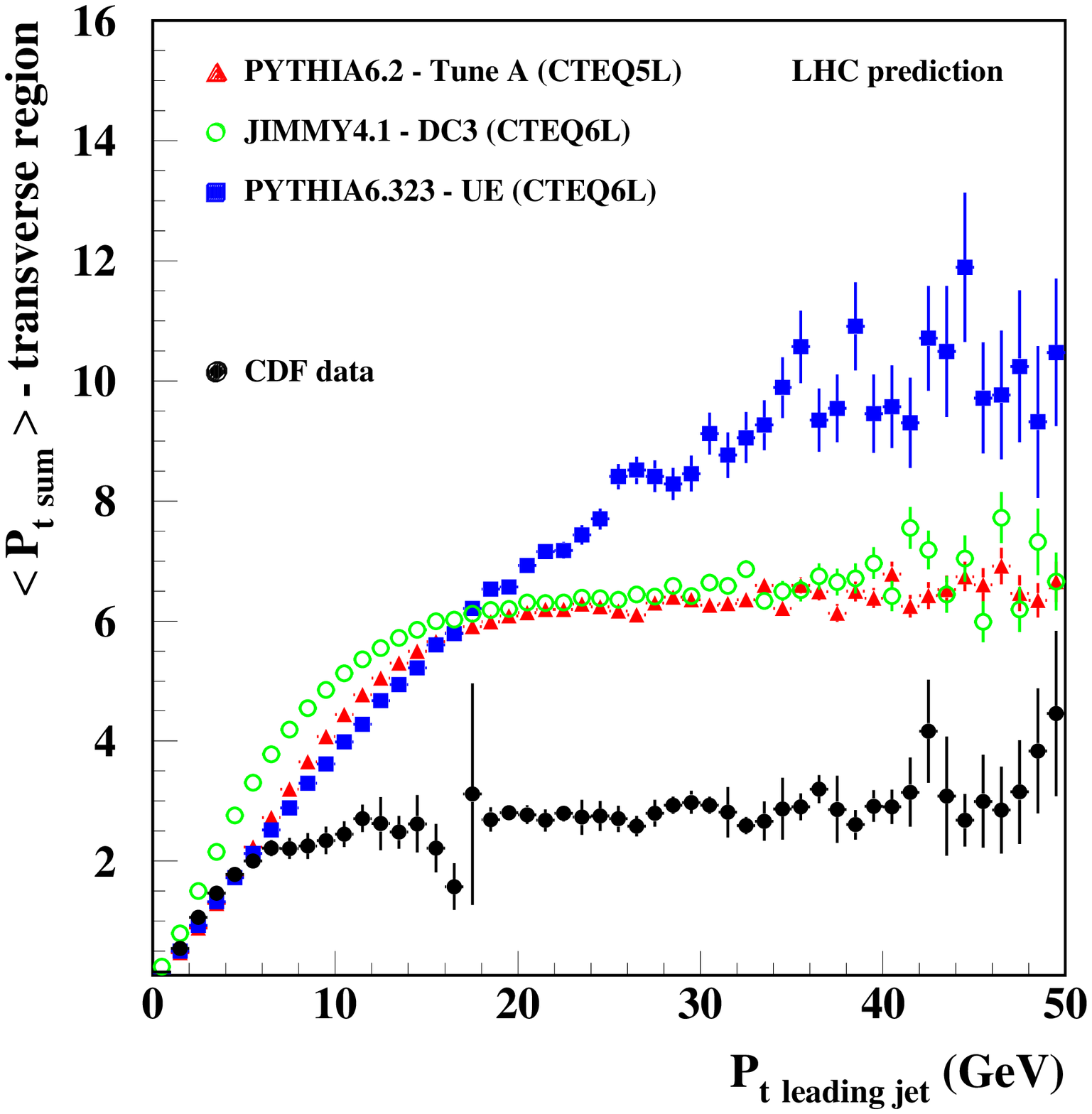}
\end{center}
\caption{PYTHIA6.2 - Tune A, Jimmy4.1 - UE and PYTHIA6.323 - UE
predictions for the average sum of the transverse momenta of charged particles in the transverse
region in the underlying event for LHC $pp$ collisions.
\label{fig:ue_LHC_pt}
}
\end{figure}

As discussed in Section~\ref{sec:formalism}, hard interactions at hadron-hadron colliders consist of a hard collision of two
incoming partons along with softer interactions from the remaining partons in the colliding hadrons (``the
underlying event energy''). The underlying event will affect almost all measurements of physics processes at the LHC. Some predictions for the underlying event charged multiplicities and charged
momentum sum (defined in Section~\ref{sec:tevatron}) at the LHC are shown in Figures~\ref{fig:ue_LHC} and~\ref{fig:ue_LHC_pt}.
It is clear that (1) all predictions lead to a substantially larger charged particle multiplicity and charged
particle momentum sum at the LHC than at the Tevatron and (2) there are large differences among the predictions
from the various models. Investigations are continuing, trying to reduce the energy  extrapolation uncertainty of
these models. This measurement will be one of the first to be performed at the LHC during the commissioning run
in 2008 and will be used for subsequent Monte Carlo tunings for the LHC.

\subsubsection{$W$/$Z$ production}
\label{sec:WZ}

The stability and pdf
uncertainties for NLO $W$ production at the LHC have been previously discussed. It is interesting to examine the
pdf uncertainties of other processes at the LHC in relation to the pdf uncertainty for $W$ production. The
understanding gained may help to reduce  the theoretical uncertainties for these processes. 

\begin{figure}[t]
\begin{center}
\includegraphics[width=9cm]{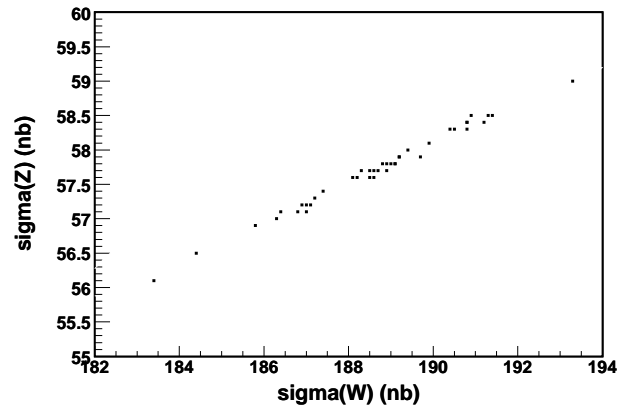}
\end{center}
\caption{
The cross section predictions for $Z$ production versus  the cross section predictions for $W$
production at the LHC plotted using the 41 CTEQ6.1 pdfs. } 
\label{fig:Z_W_LHC}
\end{figure}

In Figure~\ref{fig:Z_W_LHC},  we present cross section predictions for $Z$ production at the LHC, calculated using the 41 CTEQ6.1 pdfs. The cross section for $Z$ production  at the LHC is highly correlated with the cross
section for $W$ production. Both are sensitive to the low $x$ quark pdfs, at a similar $x$ value, which are
driven by the gluon distribution at a slightly higher $x$ value. 

\begin{figure}[t]
\begin{center}
\includegraphics[width=8cm]{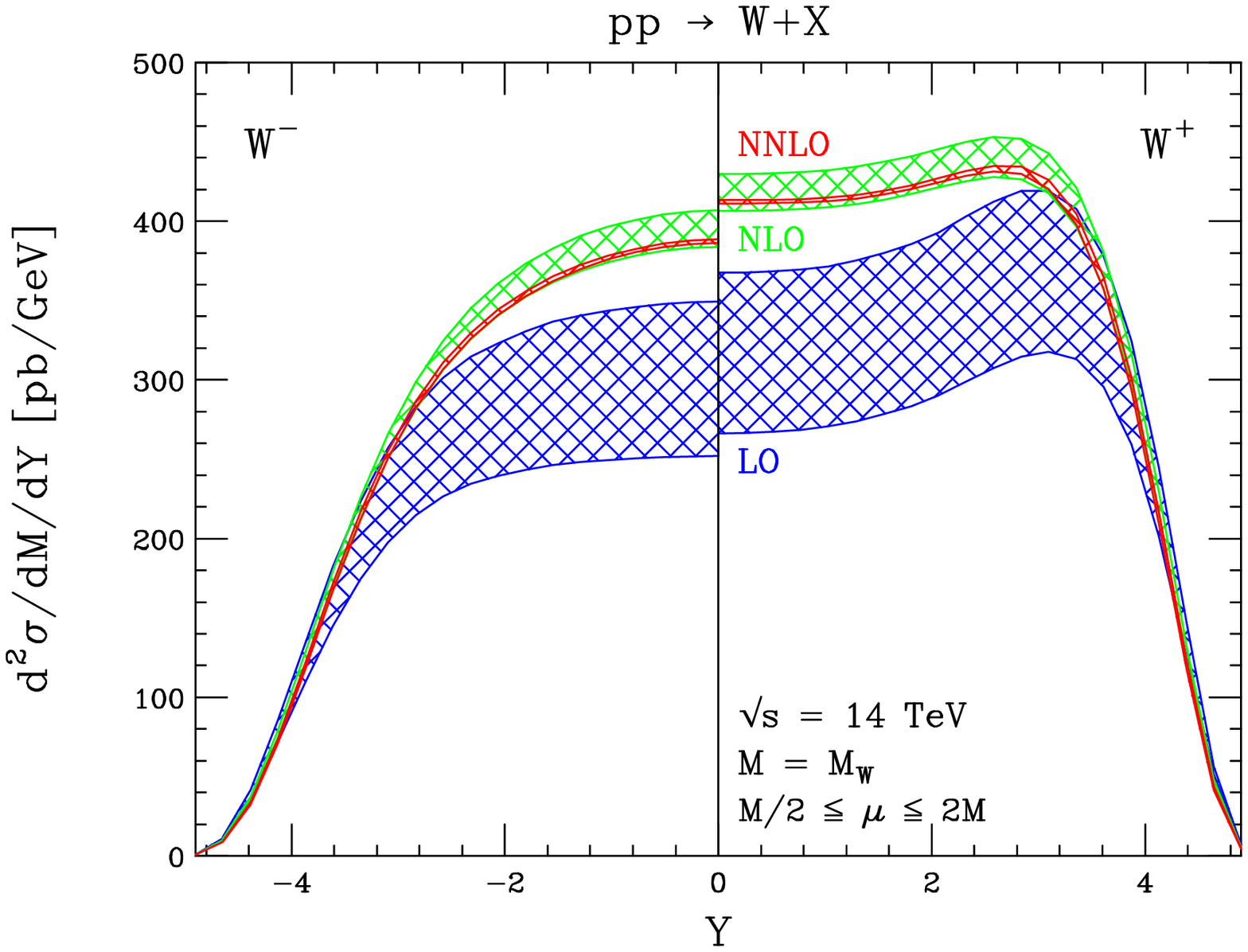}
\end{center}
\caption{
The rapidity distributions for $W^+$ and $W^-$ production at the LHC at LO, NLO and NNLO. } 
\label{fig:nnlo_w}
\end{figure}

\begin{figure}[t]
\begin{center}
\includegraphics[width=8cm]{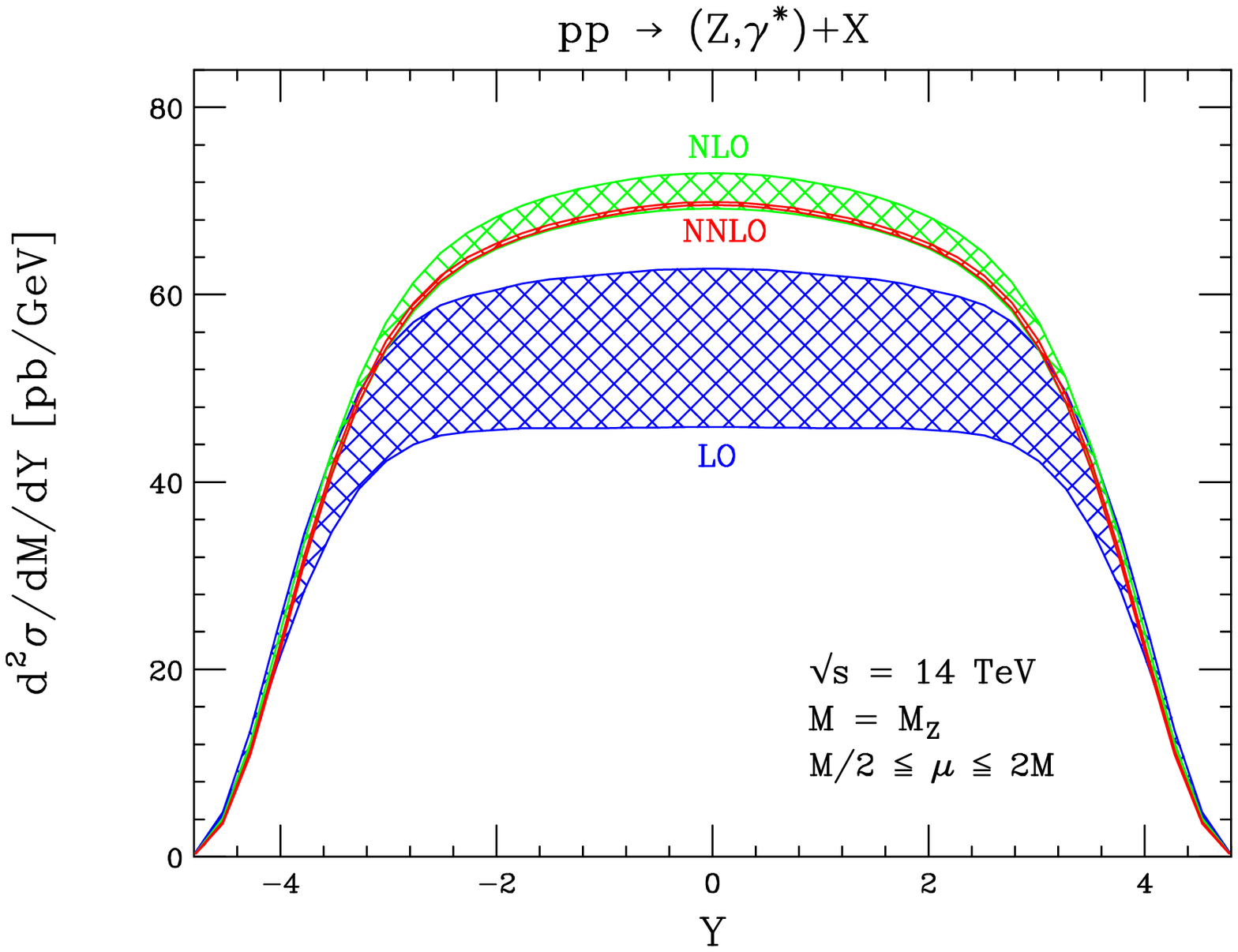}
\end{center}
\caption{
The rapidity distributions for $Z$ production at the LHC at LO, NLO and NNLO. } 
\label{fig:nnlo_z}
\end{figure}

\begin{figure}[t]
\begin{center}
\includegraphics[width=8cm]{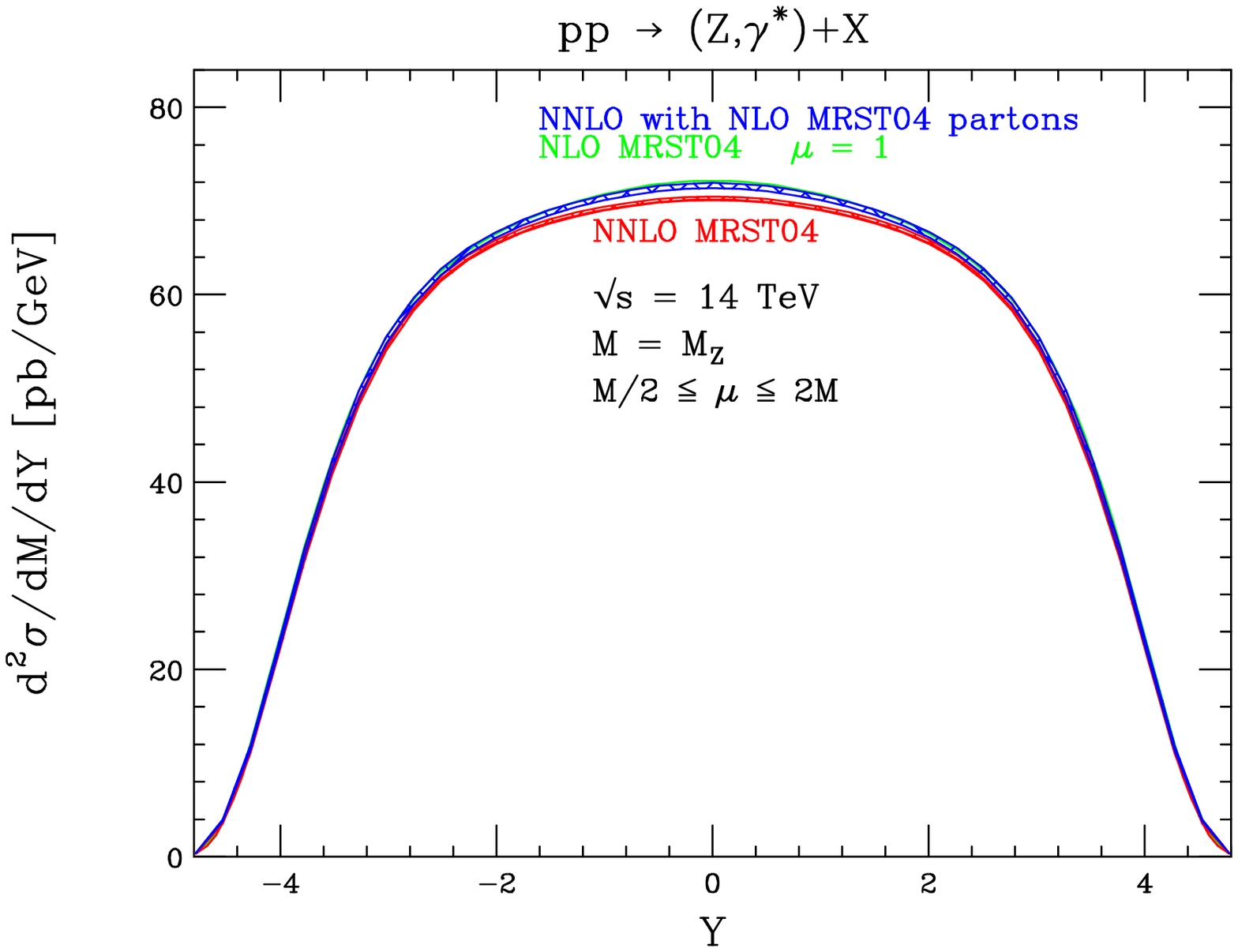}
\end{center}
\caption{
The rapidity distributions for $Z$ production at the LHC at NNLO calculated with NNLO and with NLO pdfs. } 
\label{fig:nlo_z}
\end{figure}

The rapidity distributions for $W^+$ and $W^-$ production at LO, NLO and NNLO are shown in
Figure~\ref{fig:nnlo_w}, while similar distributions for the $Z$ are shown in Figure~\ref{fig:nnlo_z}.
The widths of the curves indicate the scale uncertainty for the cross section predictions. As for the inclusive $W$ and $Z$ cross sections, the scale
dependence greatly decreases from LO to NLO to NNLO. There is a sizeable increase in the cross sections from LO
to NLO, and a slight decrease (and basically no change in shape) in the cross sections from NLO to NNLO. The
change from NLO to NNLO is within the NLO scale uncertainty band. As discussed in Section~\ref{sec:pdfs}, the NNLO pdfs are
still somewhat incomplete due to the lack of inclusion of inclusive jet production in the global fits. Although
this is formally mixing orders, the result of using the NNLO matrix element with NLO pdfs for calculation of
the $Z$ rapidity distribution is shown in Figure~\ref{fig:nlo_z}.
The predictions are similar as expected, but
the prediction using the NLO pdfs is outside the (very precise) NNLO error band~\footnote{
We would like to thank Lance Dixon for providing these curves}. 

\begin{figure}[t]
\begin{center}
\includegraphics[width=12cm]{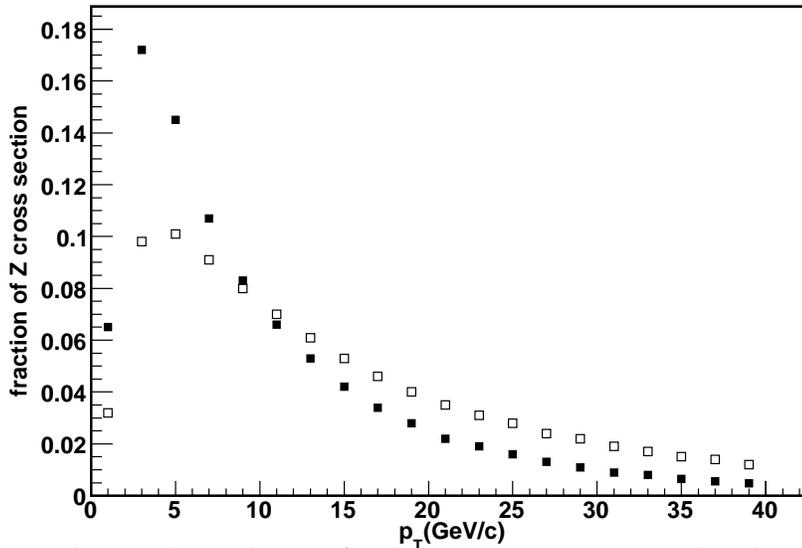}
\end{center}
\vspace*{-1cm}
\caption{
Predictions for the transverse momentum distributions for $Z$ production at the Tevatron (solid squares) and LHC (open squares).  } 
\label{fig:z_pt}
\end{figure}

The transverse momentum distributions for $W$ and $Z$ production at the LHC are also important to understand. $Z$ production will be one the Standard Model benchmark processes during the early running of the LHC. At low transverse momenta, the distributions are dominated by
the effects of multiple soft gluon emission, while at higher $p_T$, hard gluon emission is the major contribution. In
Figure~\ref{fig:z_pt}, the $Z$ $p_T$ distributions at the Tevatron and LHC are shown using predictions from ResBos.
%
The transverse momentum distribution at the LHC is similar to that at the Tevatron, although somewhat enhanced at moderate transverse momentum values. 
There is a larger  phase space for
gluon emission of incident quarks at $x=0.007$ ($Z$ production at the LHC) than for incident quarks at $x=0.05$ ($Z$
production at the Tevatron) and the enhancement at moderate transverse momentum is a result of this. There is still substantial
influence of the non-perturbative component of the parton transverse momentum near the peak region of the $Z$ transverse momentum
distribution~\cite{Balazs:2000sz}. 

\begin{figure}[t]
\begin{center}
\includegraphics[width=10cm]{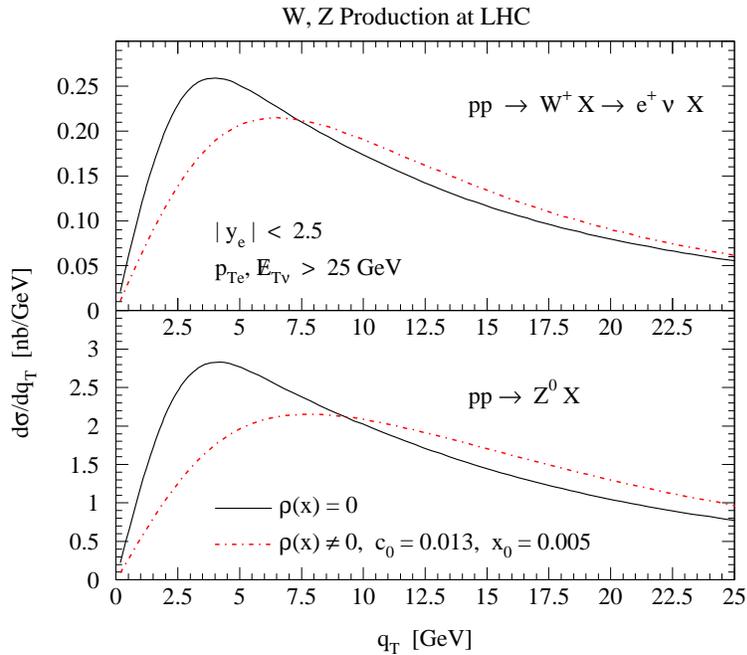}
\end{center}
\vspace*{-0.5cm}
\caption{
The predictions for the transverse momentum distributions for $W$ and $Z$ production with and
without the $p_T$-broadening effects.
\label{fig:wz_pt}
}
\end{figure}

An analysis of semi-inclusive deep-inelastic scattering hadroproduction suggests a broadening of transverse
momentum distributions for $x$ values below $10^{-3}$ to $10^{-2}$~\cite{Berge:2004nt}. (See also the discussion in Section~\ref{sec:wjets_lhc}.)  The $p_T$ broadening at
small $x$ may be due to $x$-dependent higher order contributions (like BFKL~\cite{Fadin:1975cb,Kuraev:1976ge,Kuraev:1977fs,Balitsky:1978ic}) not included in current resummation
formalisms. Such contributions are important when $\log Q^2 \ll \log(1/x)$. The BFKL formalism resums terms proportional to $\as \log(1/x)$,
retaining the full $Q^2$ dependence. The BFKL corrections would have a small impact at the Tevatron (except perhaps for $W$/$Z$ production
in the forward region) but may affect the predictions for $W$/$Z$/Higgs $p_T$ distributions for all rapidity regions at the LHC.
The $p_T$ broadening can be modeled in the Collins-Soper-Sterman formalism~\cite{Collins:1985gm} by a modification of the impact
parameter-dependent parton densities. The $p_T$ shifts for the $W$ and $Z$ transverse momentum distributions at
the LHC are shown in Figure~\ref{fig:wz_pt}~\cite{Berge:2005nm}. The observed shifts would have important
implications for the measurement of the $W$ boson mass and a measurement of the $W$/$Z$ $p_T$ distributions will be one of
the important early benchmarks to be established at the LHC. 
\subsubsection{$W/Z+$~jets}
\label{sec:wjets_lhc}

\begin{figure}
\begin{center}
\includegraphics[width=9cm]{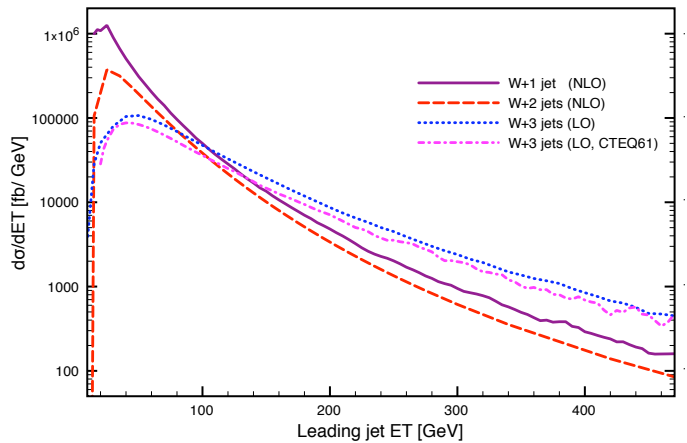}
\end{center}
\caption{
Predictions for the production of $W+\ge 1,2,3$ jets at the LHC shown as a function of the transverse energy of the lead jet. A cut of
$20$~GeV has been placed on the other jets in the prediction. 
\label{fig:lhc_Wjet}}
\end{figure}

Next we consider the production of $W/Z+$~jets at the LHC. In Figure~\ref{fig:lhc_Wjet}, the rate for production of
$W+\ge 1,2,3$ jets at the LHC is shown as a function of the transverse energy of the lead jet (calculated using the MCFM program). The rate for $3$ jet production is actually
larger than the rate for $1$ or $2$ jet production for large lead jet transverse energy due to the Sudakov effects discussed throughout this
paper. We also show the rate for $W + 3$~jet production (a LO calculation in MCFM) using both CTEQ6L1 (a leading order pdf) and CTEQ6.1
(a NLO pdf). The rate using the LO pdf is larger due to (1) the importance of the gluon distribution for production of $W + 3$~jet final
states and (2) the larger small $x$ gluon present in the LO pdf compared to the NLO pdf. 

\begin{figure}
\begin{center}
\includegraphics[width=9cm,angle=90]{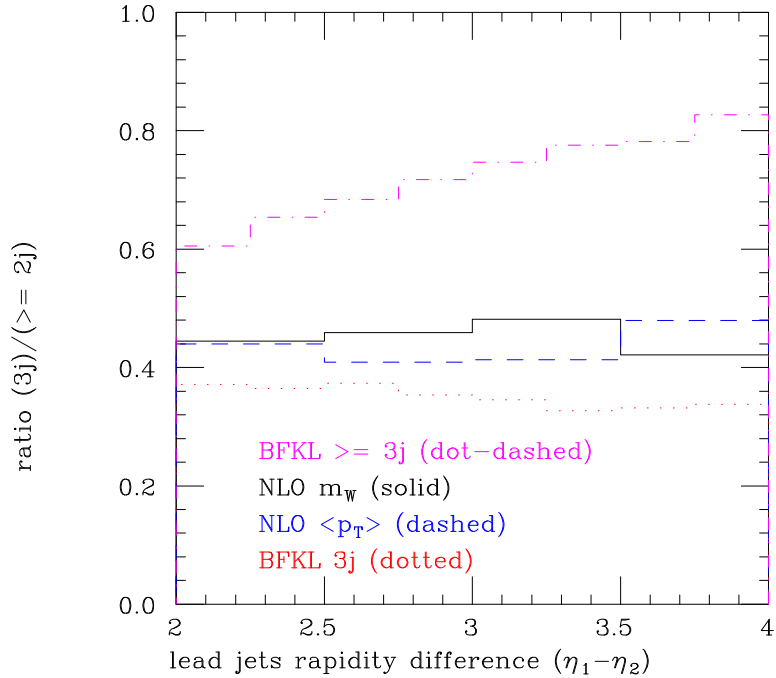}
\end{center}
\caption{The rate for production of a third (or more) jet in $W+\ge2$ jet events as a function of the rapidity separation of the two leading jets.
A cut of $20$~GeV has been placed on all jets. Predictions are shown from MCFM using two values for the renormalization
and factorization scale, and using the BFKL formalism, requiring either that there be exactly $3$ jets or $3$ or more jets. 
\label{fig:joeyrap}}
\end{figure}

In Figure~\ref{fig:joeyrap}, we show the rate for production of a third jet given a $W+\ge2$ jet event, as a function of the rapidity
separation of the two leading jets. (As a reminder, the production of $Z$ plus two jets widely separated in rapidity serves as a
background to VBF production of a Higgs.)  In Section~\ref{sec:wzjets}, the corresponding measurement at the Tevatron was compared to
predictions from MCFM and found to be bounded by the two predictions at scales of $m_W$ and the average jet $p_T$. At the LHC, we
note that the rate for emission of a third jet is considerably enhanced compared to the Tevatron
(even though the $p_T$ cut is larger, $20$~GeV compared to $15$~GeV at the Tevatron) and is reasonably flat as a function
of the two lead jets rapidity separation.  The scale dependence is also observed to be smaller than at the Tevatron. Predictions for
this rate are also available using the BFKL formalism. The BFKL formalism has potentially large logarithms proportional to the
rapidity separation of the two lead jets.
The resultant ratio in Figure~\ref{fig:joeyrap} for exactly 3 jets is also flat, slightly below the prediction
from MCFM but the ratio for greater than or equal to 3 jets in the BFKL formalism is appreciably larger and rises with the
rapidity separation of the lead jets~\cite{Andersen:2001ja,Andersen:2006sp}~\footnote{We thank Jeppe Andersen for providing us with the BFKL
predictions}.
There is a sizeable production of 4, 5 and larger number of jets in the final state. It is expected
that next-to-leading logarithmic (NLL) corrections may somewhat dampen the growth in the jet
multiplicity predicted by the BFKL formalism as the tagging jet separation increases.  As discussed in the previous section, early
measurements such as these will establish in what kinematic regions, BFKL effects are important at the LHC, and what kinematic
regions can be well-described by the DGLAP formalism alone. 

\subsubsection{Top quark production}
\label{sec:top_lhc}

\begin{figure}[t]
\begin{center}
\includegraphics[width=9cm]{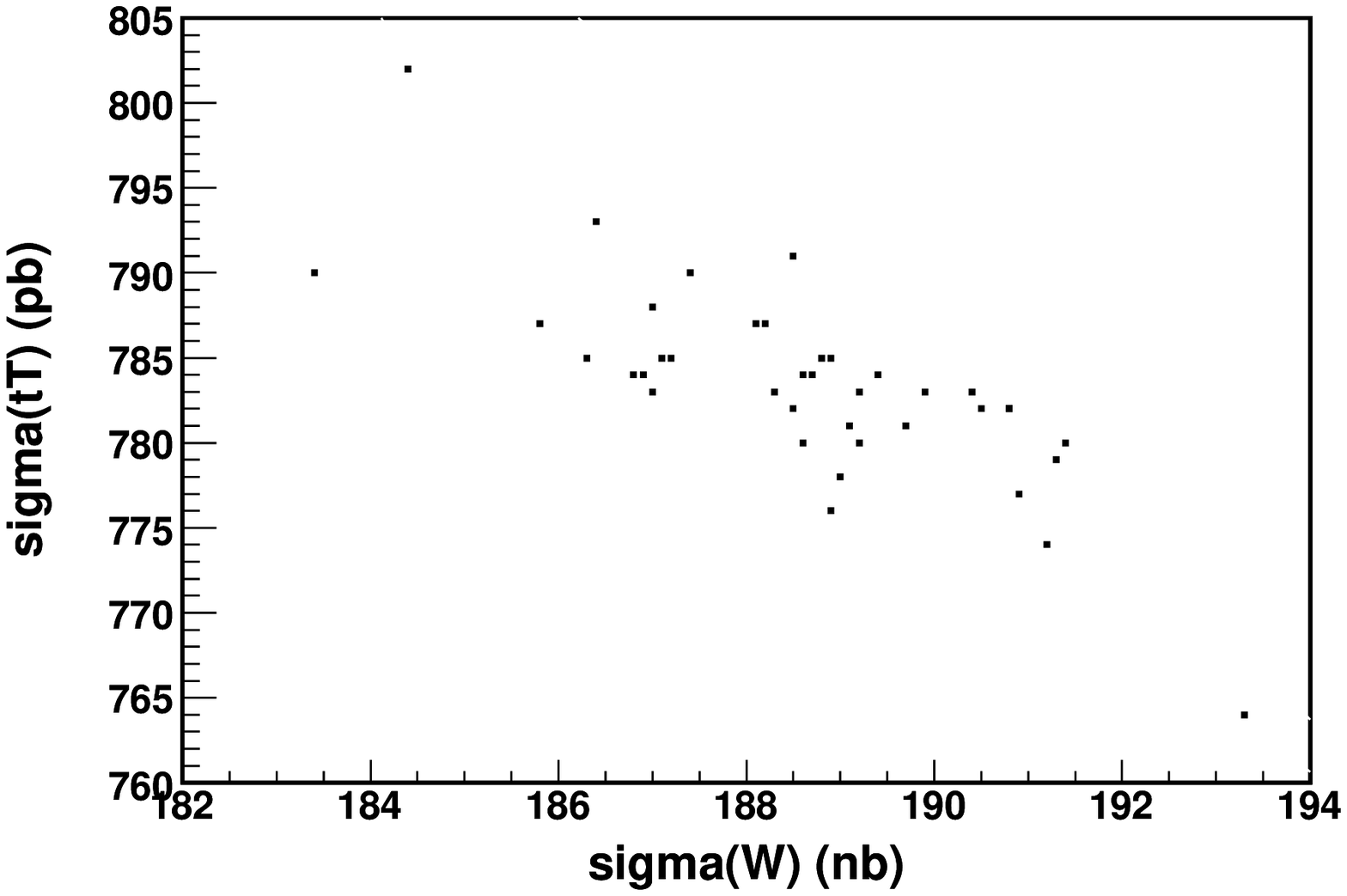}
\end{center}
\caption{
The cross section predictions for $t\overline{t}$ production versus  the cross section predictions for
$W$ production at the LHC plotted using the 41 CTEQ6.1 pdfs. } 
\label{fig:tT_W_LHC}
\end{figure}

As at the Tevatron, $t\bar{t}$ production at the LHC proceeds through both $q\bar{q}$ and $gg$ initial states. 
Consider a specific value of $\sqrt{\hat{s}}$ of $0.4$~TeV (near $t\bar{t}$ threshold); from Figure~\ref{fig:figlum13}, the $q\bar{q}$ annihilation component
is only a factor of $10$ larger at the LHC than at the Tevatron.  The $gg$ component, on the other hand, is over a
factor of $500$ larger, leading to (1) the large dominance of $gg$ scattering for top pair production at the LHC, in contrast
to the situation at the Tevatron and (2) a total $t\bar{t}$ cross section a factor of $100$ larger than at the Tevatron. 
Interestingly, as shown in
Figure~\ref{fig:tT_W_LHC}, the cross section for $t\overline{t}$ production is anti-correlated with the $W$
cross section. An increase in the $W$ cross section is correlated with a decrease in the $t\overline{t}$ cross section and vice versa. This
is due to the dominance of the  $gg$ fusion subprocess for $t\bar{t}$ production,  while $W$ production is
still predominantly quark-antiquark. An increase in the gluon distribution in the $x$ range relevant for $t\overline{t}$
production leads to a decrease in  the quark  distributions in the (lower) $x$ range relevant  for $W$ production. In fact, 
the extremes for both cross sections are produced by CTEQ6.1 eigenvector $5$ (pdfs $9$ and $10$) which is most sensitive to the low
$x$ behaviour of the gluon distribution

\begin{figure}[t]
\begin{center}
\includegraphics[width=8cm,angle=90]{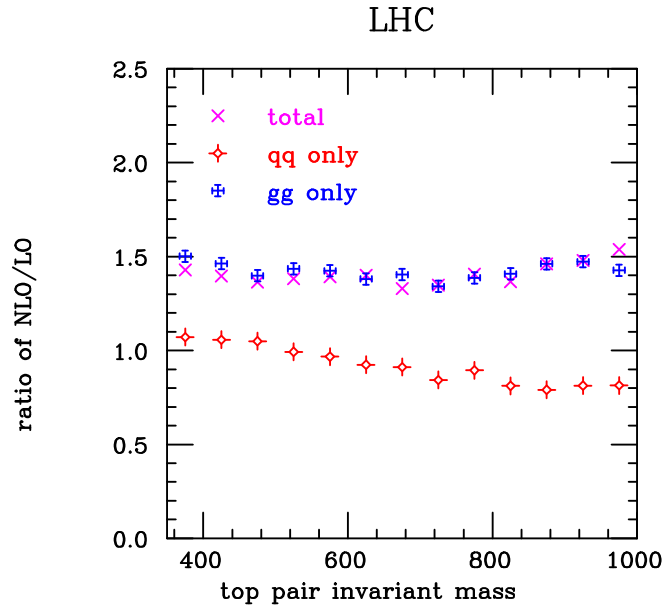}
\end{center}
\vspace*{-0.5cm}
\caption{
The ratio of the NLO to LO predictions for the $t\overline{t}$ mass at the LHC.
The predictions include the ratio for the total cross section and for the
specific $q\overline{q}$ and $gg$ initial states. Note that the total also includes a $gq$ contribution
(not present at LO).  
\label{fig:mtt_ratio_lhc}}
\end{figure}

The ratio of the NLO to LO predictions for $t\overline{t}$ production at the LHC is shown in Figure~\ref{fig:mtt_ratio_lhc}. In contrast to the
Tevatron, the NLO cross section is larger than the LO one by an approximately constant factor of 1.5. We also plot the NLO/LO ratio specifically for
the $q\overline{q}$ initial state where we observe a behaviour similar to that observed at the Tevatron, but not as pronounced.

\begin{figure}[t]
\begin{center}    
\includegraphics[width=7cm,angle=90]{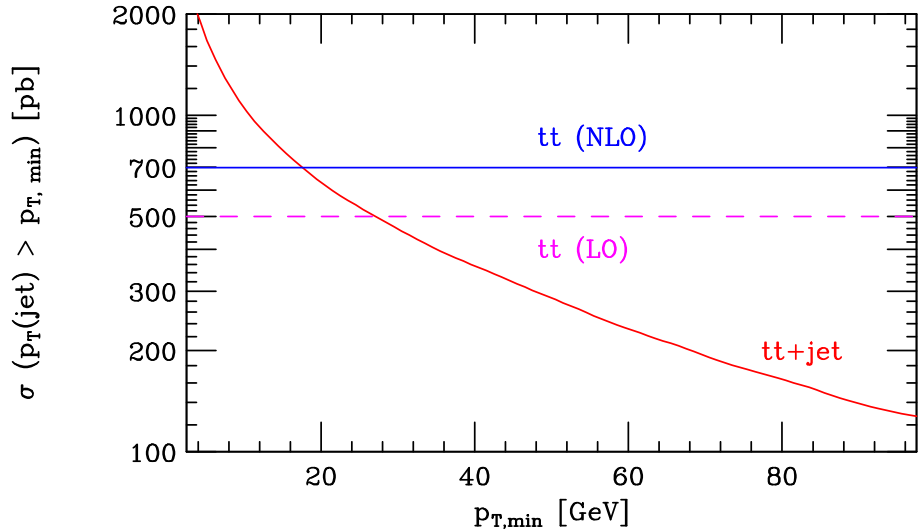}    
\end{center}    
\caption{The dependence of the LO $t{\bar t}+$jet cross section on the jet-defining parameter $p_{T, {\rm min}}$,
together with the top pair production cross sections at LO and NLO.}
\label{fig:ttjet}
\end{figure}

It is also evident that because  of the higher percentage of $gg$ production and  the lower average $x$ of the incident partons,
the jet multiplicity will be significantly higher for  $t\overline{t}$ production at the LHC than at the Tevatron. Consider the production of
a pair of top quarks in association with an additional jet at the LHC. Defining the cross section for this process to
only include events with a jet of transverse momentum greater than some minimum value, $p_{T, {\rm min}}$, yields
the dependence on $p_{T, {\rm min}}$ shown in Figure~\ref{fig:ttjet}.
The reason for this behaviour is discussed at
length in Section~\ref{sec:localcs}. Overlaid on this figure is the cross section for top pair production at LO and NLO,
which clearly has no dependence on the parameter $p_{T, {\rm min}}$. As the minimum jet transverse momentum is decreased
the cross section for $t{\bar t}+$jet production increases rapidly and in fact saturates the total LO $t{\bar t}$
cross section at around $28$~GeV. On the one hand, this appears to be a failing of the leading order predictions. When the
$t{\bar t}$ rate is calculated at NLO the cross section increases and the saturation does not occur until around $18$~GeV
(and presumably higher orders still would relax it further). On the other hand transverse momenta of this size, around $20$~GeV,
are typical values used to define jets in the LHC experiments. Based on these results, one might certainly expect that jets
of these energies might often be found in events containing top quark pairs at the LHC. 

\begin{figure}[t]
\begin{center}
\includegraphics[width=10cm]{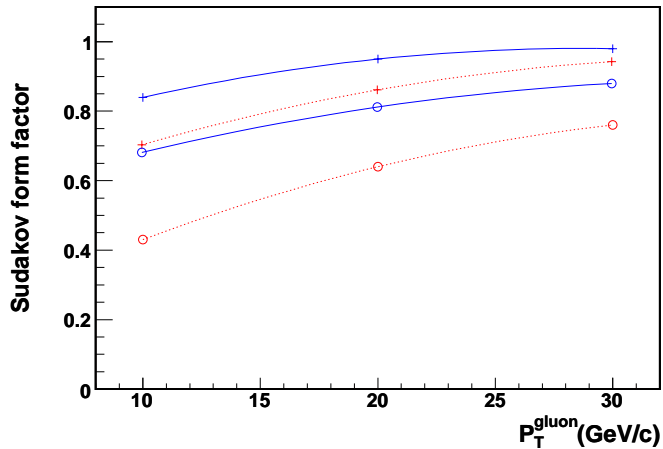}
\end{center}
\vspace*{-0.5cm}
\caption{
The Sudakov form factors for initial state quarks and gluons at a hard scale of $200$~GeV as a function of the
transverse momentum of the emitted gluon. The form factors are for quarks (blue-solid) and gluons (red-dashed)
at parton $x$ values of $0.3$ (crosses) and $0.03$ (open circles).
} 
\label{fig:sudakov_tT}
\end{figure}

Another way of estimating the probability for extra jets is to look at the Sudakov form factors for the initial state partons. 
It is interesting to 
compare the Sudakov form factors for $t\overline{t}$ production at the Tevatron and  LHC. At the Tevatron,
$t\overline{t}$ production proceeds primarily ($85\%$) through $q\overline{q}$ with  $gg$ being responsible for
$15\%$, with the partons evaluated near an average $x$ value of $0.3$. At the LHC, the percentages are roughly reversed
(or more precisely $90\%$ for $gg$) and the scattering takes place at an average $x$ value of a factor of $7$
lower (which we approximate here  as $x=0.03$). The relevant Sudakov form factors are shown in
Figure~\ref{fig:sudakov_tT}, as a function of the minimum transverse momentum of the emitted gluon, at a hard scale of
$200$~GeV (roughly appropriate for $t\overline{t}$ production).
We
can make some rough estimates from these plots. The probability for no gluon of $10$~GeV or greater to be radiated from an
initial quark leg with $x=0.3$ is $0.85$. The probability for no such gluon to be radiated from either quark leg at the
Tevatron is $0.85 \times 0.85=0.72$, i.e. a $0.28$ chance of radiating such a gluon. A similar exercise for two incident
gluons of $x=0.3$ gives a chance of radiating a $10$~GeV gluon of $0.51$. As the $q\overline{q}$ initial state makes up
$85\%$ of the Tevatron cross section, with $gg$ only $15\%$, the total probability of emitting at least one $10$~GeV gluon
is $0.3$. Using $90\%$ for $gg$ at the LHC and $10\%$ for $q\overline{q}$, gives a $0.8$ probability of radiating such a hard
gluon.

The $W$+jets background to $t\overline{t}$ production proceeds primarily through the $gq$ channel and so receives a factor of 500 enhancement. Thus, the signal to background ratio for $t\overline{t}$ production in a lepton + jets final state is significantly worse at the LHC than at the Tevatron, if the same cuts on the jet transverse momenta as at the Tevatron are used. 
Thus, the jet cuts
applied to $t\bar{t}$ analyses at the LHC need to be set larger than at the Tevatron in order, (1) to reduce
the backgrounds from $W + 4$~jet production relative to the lepton + $4$~jets final state from $t\bar{t}$ decay, (2) to reduce the number of jets produced by ISR in $t\bar{t}$ events, and (3) to reduce the likelihood of additional jets produced by fluctuations in the underlying event.
The signal to
background for $t\overline{t}$ is substantially improved at the LHC by increasing the minimum transverse momentum
cut for each jet from $15$~GeV (Tevatron) to $30$~GeV (CMS) or $40$~GeV (ATLAS). The cross section for the
production of the lowest $p_T$ jet in $W + 4$~jet events falls roughly as $1/p_T^n$ (where $n$ is in the range
$2.5-3$)
while the distribution for the 4th jet transverse momentum is essentially flat from $15$--$40$~GeV.
The background is
reduced by a factor of $15$ while the signal is reduced by a factor of $5$. This reduction in signal is acceptable because of the
large $t\overline{t}$ cross section available at the LHC. There are $2.5$~million $t\overline{t}$ pairs produced with
a lepton + jets final state for a $10$fb$^{-1}$ data sample. The requirement for two of the jets to be tagged as
$b$-jets (and the kinematic cuts on the jets ($40$~GeV) and on the lepton and missing transverse momentum) reduces the
event sample to $87,000$, but with a signal to background ratio of $78$. A requirement of only one b-tag reduces the
signal/background ratio to $28$ but with a data sample a factor of $3$ larger.

\subsubsection{Higgs boson production}
\label{ref:higgs}

\begin{figure}[t]
\begin{center}
\includegraphics[width=7cm,angle=0]{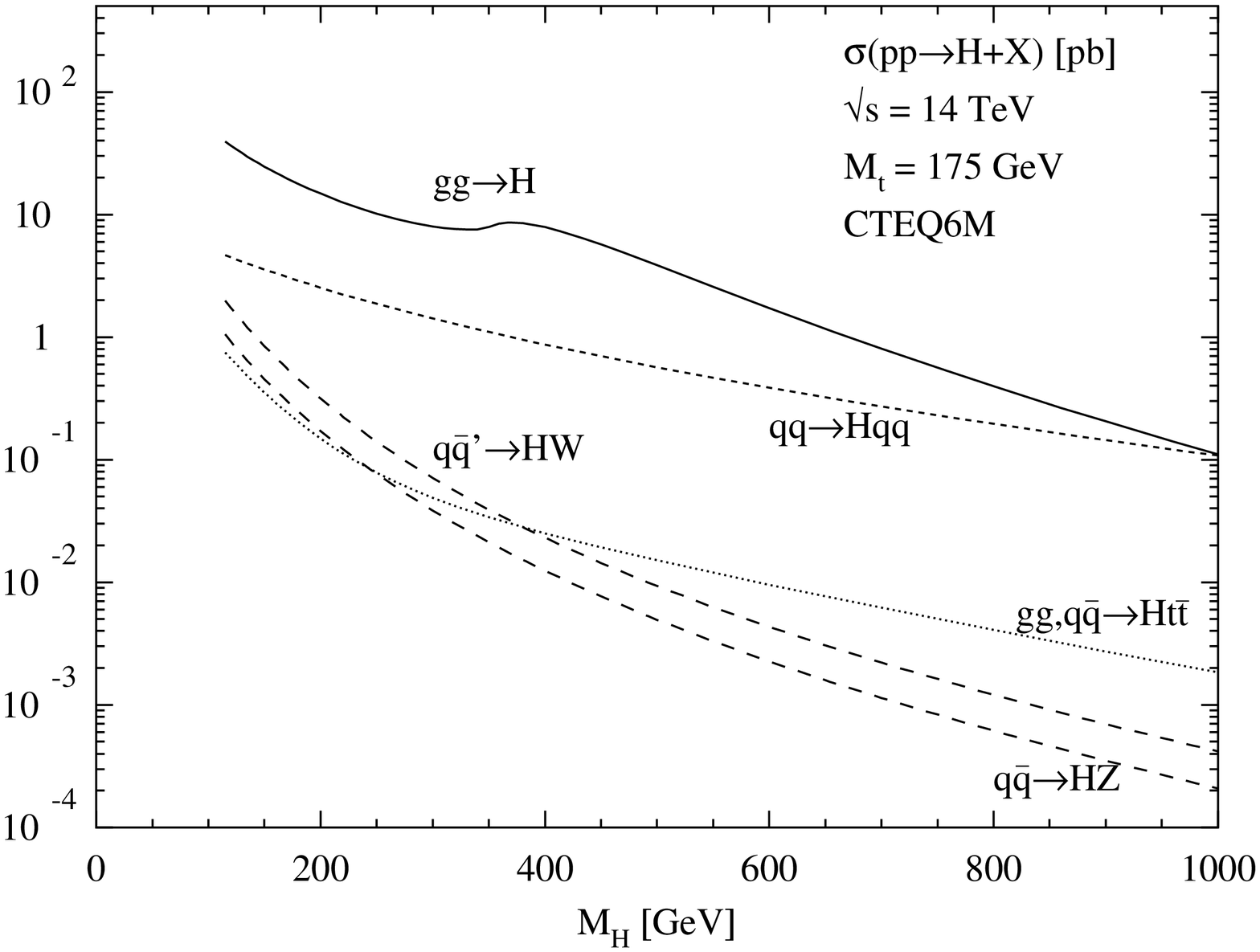}
\includegraphics[width=7cm,angle=0]{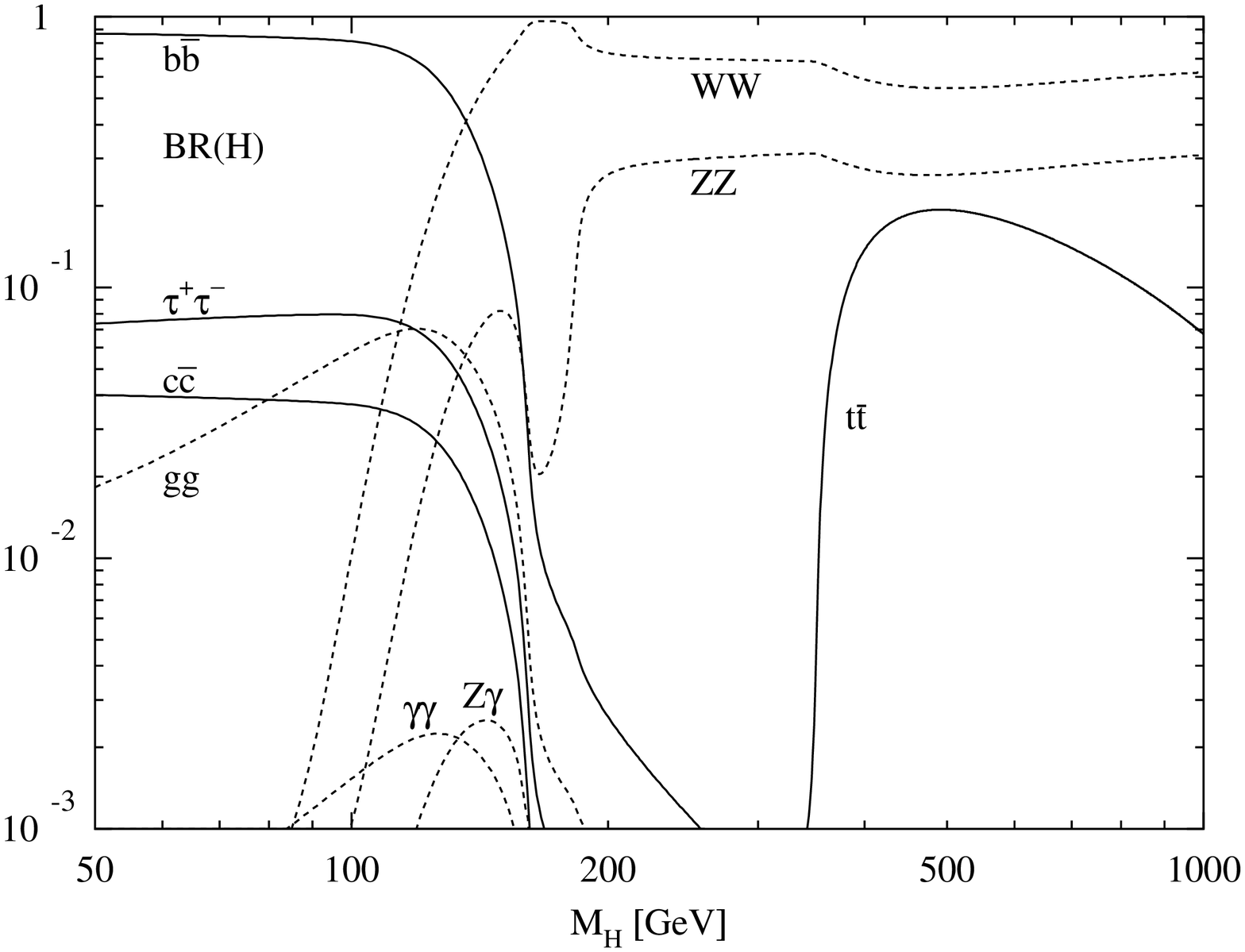}
\end{center}
\vspace*{-0.5cm}
\caption{The production subprocess cross sections for the Standard Model Higgs at the LHC, as a function of the Higgs boson mass(left); the branching ratios for the Standard Model Higgs as a function of the Higgs boson mass(right).  
\label{fig:higgs}}
\end{figure}

As discussed in Section~\ref{sec:formalism}, the Higgs boson is the cornerstone of the Standard Model and its supersymmetric extensions, and its discovery will be one of the primary physics goals at the LHC. The Standard Model Higgs cross sections for various production mechanisms are shown in Figure~\ref{fig:higgs} (left) as a function of the Higgs mass~\cite{CMS_tdr}. The branching ratios of the dominant decay modes for the Standard Model Higgs boson are shown on the right side of the figure. The primary production mechanism is through $gg$ fusion and the primary decay mode is $b\overline{b}$ (low mass) and $WW$ (high mass).  

\begin{figure}[t]
\begin{center}    
\includegraphics[width=7cm,angle=90]{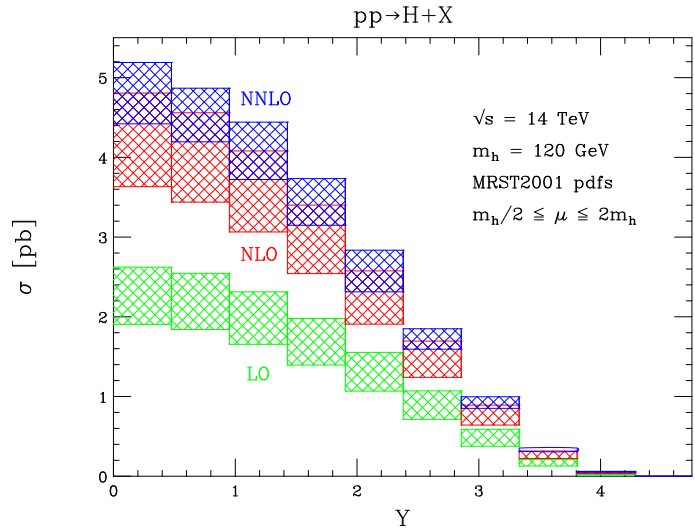}    
\end{center}    
\caption{The rapidity distributions, at LO, NLO and NNLO,  for the production of a 120 GeV mass Higgs at the LHC.}
\label{fig:higgs_y}
\end{figure}

Inclusive Higgs production through $gg$ scattering is one of the few processes known to NNLO. The rapidity distribution for Higgs boson production at the LHC is shown in Figure~\ref{fig:higgs_y} at LO, NLO and NNLO~\cite{Anastasiou:2004xq}. The NLO cross section is a factor of two larger than the leading order cross section and the NNLO cross section is a factor of 20\% larger than the NLO one. The scale dependence at NNLO is of the order of 15\%, so the cross section uncertainties are under reasonable control. As observed in Figure~\ref{fig:figlum6yall}, the $gg$ parton luminosity uncertainty is less than 10\% over the expected Higgs mass range. 

\begin{figure}[t]
\begin{center}
\includegraphics[width=9cm]{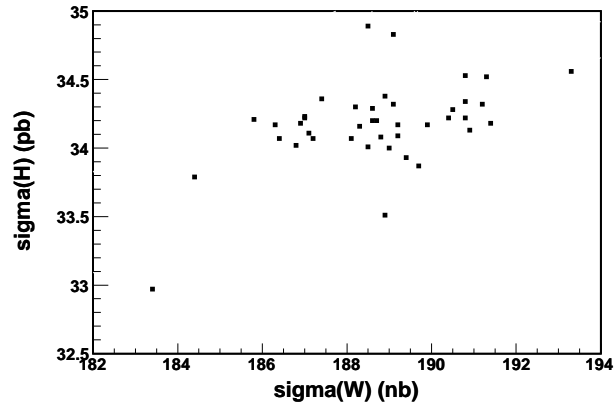}
\end{center}
\caption{
The cross section predictions for Higgs production versus the cross section predictions for $W$ production
at the LHC plotted using the 41 CTEQ6.1 pdfs. } 
\label{fig:Higgs_W_LHC}
\end{figure}

The NLO predictions from  the 41 CTEQ6.1 pdfs for the production of a $125$~GeV Higgs boson through gluon fusion at the LHC are
plotted versus similar predictions for the $W$ cross section in Figure~\ref{fig:Higgs_W_LHC}. 
The low cross section point
comes from error pdf 30, which is responsible for the high jet cross section predictions at high $p_T$; the flow
of momentum to high $x$ removes it from the kinematic region responsible for producing a Higgs boson of this mass. 

\begin{figure}[t]
\begin{center}    
\includegraphics[width=7cm,angle=90]{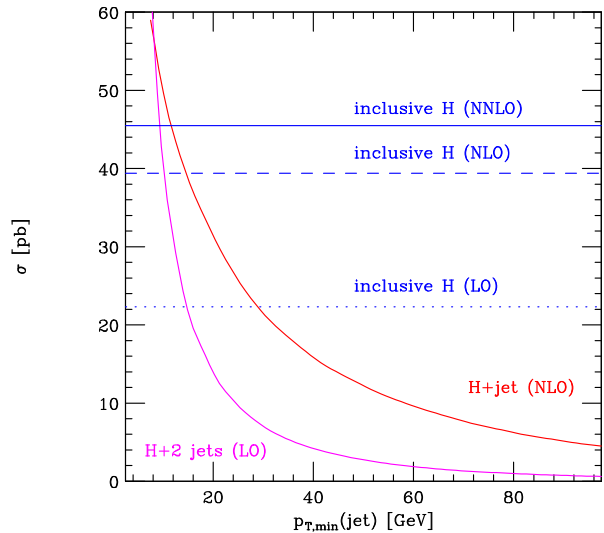}    
\end{center}    
\caption{The dependence of the LO $t{\bar t}+$jet cross section on the jet-defining parameter $p_{T, {\rm min}}$,
together with the top pair production cross sections at LO and NLO.}
\label{fig:higgs_jet}
\end{figure}

Consider the rate for production (through $gg$ fusion) of a Higgs boson in association with an additional jet at the LHC. Defining the cross section for this process to
only include events with a jet of transverse momentum greater than some minimum value, $p_{T, {\rm min}}$, yields
the dependence on $p_{T, {\rm min}}$ shown in Figure~\ref{fig:higgs_jet}. Overlaid on this figure is the cross section for Higgs boson production at LO, NLO and NNLO. Similar to the case for $t\bar{t}$ production at the LHC, as the minimum jet transverse momentum is decreased,
the cross section for Higgs+jet production increases rapidly and in fact saturates the total LO Higgs boson
cross section at around $28$~GeV. When the
Higgs boson rate is calculated at NLO, the cross section increases and the saturation does not occur until around $15$~GeV and at NNLO, the saturation occurs at around 12 GeV. 

\begin{figure}[t]
\begin{center}
\includegraphics[width=12cm]{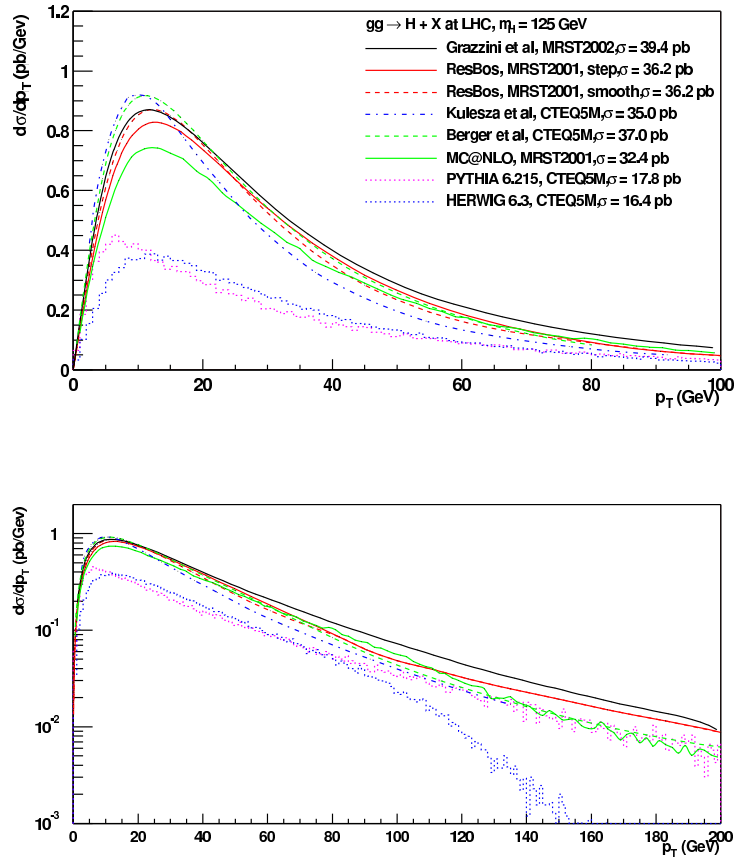}
\end{center}
\vspace*{-1cm}
\caption{
The predictions for the transverse momentum distribution for a $125$~GeV mass Higgs boson at the LHC
from a number of theoretical predictions. The predictions are shown with an absolute normalization.
This figure can also be viewed in colour on the benchmark website. 
\label{fig:higgs_no_norm}
}
\end{figure}

\begin{figure}[t]
\begin{center}
\includegraphics[width=12cm]{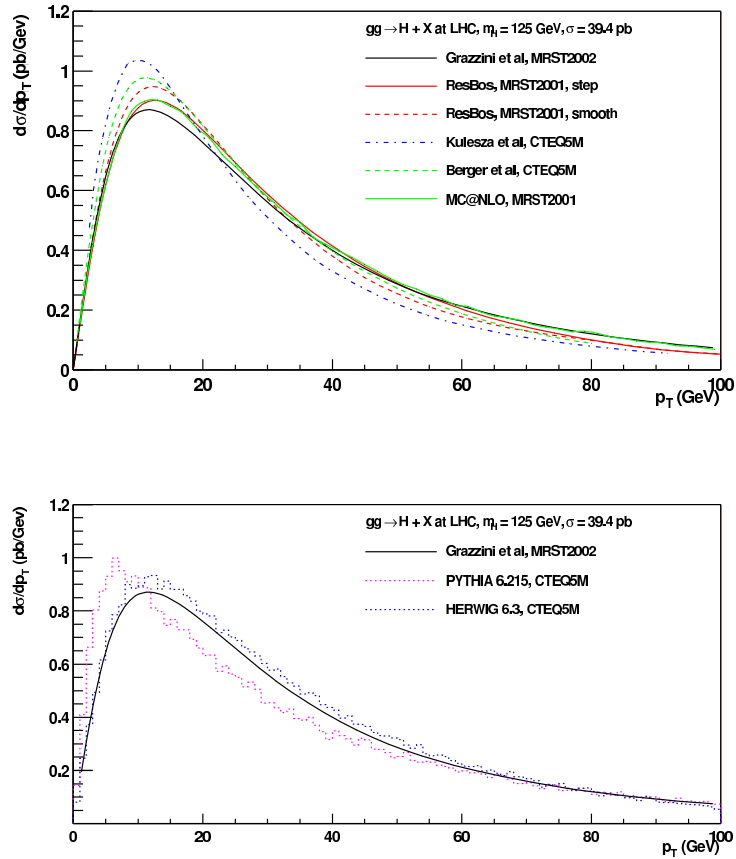}
\end{center}
\vspace*{-1cm}
\caption{
The predictions for the transverse momentum distribution for a $125$~GeV mass Higgs boson at the LHC
from a number of theoretical predictions. The predictions have all been normalized to the same cross
section for shape comparisons. This figure can also be viewed in colour on the benchmark website. 
\label{fig:higgs_norm}
}
\end{figure}

As for the case of $t\bar{t}$ production at the LHC, the combination of a $gg$ initial state, the high mass of the final state, and the relatively low $x$ values of the initial state parton results in (1) a large probability for the Higgs to be produced with extra jets and (2) for the Higgs to be produced with relatively high transverse momentum. 
In Figure~\ref{fig:higgs_no_norm} (un-normalized) and Figure~\ref{fig:higgs_norm} (normalized) are shown a large
number of predictions for the transverse momentum distributions for the production of a $125$~GeV mass Higgs boson
at the LHC, through the $gg$ fusion process.
The impact of higher order corrections (both NLO and NNLO) on the
normalization of the Higgs cross section is evident in Figure~\ref{fig:higgs_no_norm}. But in
Figure~\ref{fig:higgs_norm}, it can be observed that the shapes for the $p_T$ distributions are basically the
same. In particular, parton shower Monte Carlos provide a reasonable approximation to the more rigorous
predictions shown. Two PYTHIA predictions are indicated: PYTHIA~6.215 is a virtuality-ordered parton shower
while PYTHIA~6.3 is a $p_T$-ordered shower (and as different from PYTHIA~6.215 as HERWIG is from PYTHIA).  The
peak of the PYTHIA~6.3 prediction is more in line with the other theoretical predictions. 

Note that the $p_T$ distribution of the Higgs boson peaks at approximately $12$~GeV while the $p_T$ distribution for the $Z$ peaks
at $3$~GeV. The peak is slightly higher because of the higher mass ($125$~GeV compared to $90$~GeV) but the major shift is
due to the different production mechanisms. The colour factor for the gluon is greater than that for the quark and the
phase space available for gluon radiation for $x=0.01$ gluons is much larger due to the $z \rightarrow 0$ pole in
the splitting function. 

\begin{figure}[t]
\begin{center}
\includegraphics[width=6cm,angle=-90]{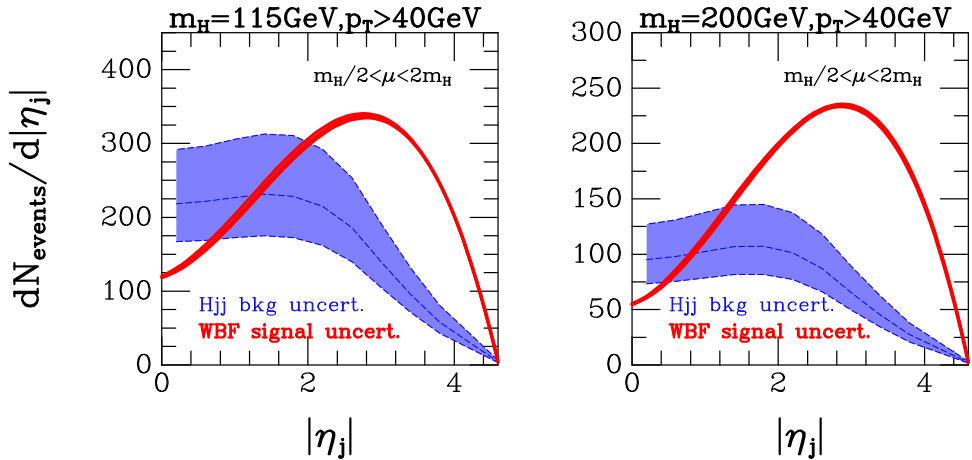}
\end{center}
\vspace*{-0.5cm}
\caption{The tagging jet $\eta$ distributions in QCD $Hjj$ and VBF events. The band indicates the range of
predictions corresponding to variation of renormalization and factorization scales by a factor of two about
the central value of $m_H$.
 
\label{fig:Hjj}}
\end{figure}

The production of a Higgs boson through the vector boson fusion (VBF) process is important, first of all, as a discovery
channel and second, as a way of measuring the Higgs couplings to $W$ and $Z$ bosons. There are backgrounds to this process
from (1) Higgs$ + 2$~jet production through $gg$ fusion and (2) $Z + 2$~jet production, where the $Z$ decays into $\tau$
pairs (one of the primary  search modes for Higgs production from VBF)~\cite{Berger:2004pc}. VBF production of a Higgs
boson proceeds through the diagrams shown in Figure~\ref{fig:wbf}. The emission of vector bosons from the initial state
quark lines results in the quarks acquiring a transverse momentum of the order of half of the $W$ mass (as the weak boson
mass provides a natural cutoff for the weak boson propagator) with a rapidity distribution peaked in the forward direction
(the quarks still retain an appreciable fraction of their original longitudinal momentum, corresponding to $x$ values of
$0.1--0.2$). The two leading jets in Higgs$+ 2$~jets (from $gg$ fusion) 
are peaked at central rapidities as
shown in Figure~\ref{fig:Hjj} and have a softer jet transverse momentum.  Thus, a relatively hard cut on the tagging jet
transverse momenta of $40$~GeV and a requirement that the tagging jets have large rapidity is effective at reducing the
background without a large impact on the VBF signal. The recent calculation of the Higgs$ + 2$~jet rate
at NLO~\cite{Campbell:2006xx} (one of the calculations on the wishlist) has somewhat~\footnote{
The reduction of the scale dependence is not dramatic since there
is a large contribution from the (LO)  Higgs$+ 3$~jet final state.} reduced the scale dependence, and thus the uncertainty,
on the prediction~\footnote{It is interesting that the NLO corrections for Higgs$+ 2$~jets are relatively small, on the
order of $20$\%. This follows the pattern of the NLO corrections decreasing in size as the jet multiplicity increases; the
corrections are $80$\% for inclusive Higgs boson production (through $gg$ fusion) and 50\% for Higgs$ + 1$~jet.}. The rate for
additional jets to be produced between the two tagging jets is high for the background processes~ \cite{DelDuca:2004wt}.
The colourless exchange in the hard scattering in VBF leads to a suppression of extra jets in the central rapidity region.
Thus, as discussed earlier, a veto on extra jets in the central region is effective at decreasing the background to VBF
production of the Higgs boson. 

\subsubsection{Inclusive jet production}
\label{sec:jetlhc}

\begin{figure}[t]
\begin{center}
\includegraphics[width=9cm]{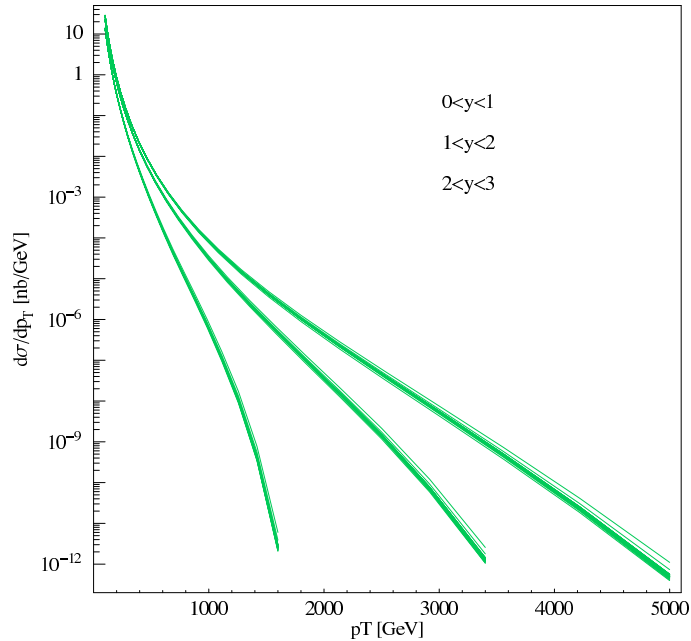}
\end{center}
\caption{
Inclusive jet cross section predictions for the LHC using the 
CTEQ6.1 central pdf and the $40$ error pdfs.} 
\label{fig:lhc_jet}
\end{figure}

\begin{figure}[t]
\begin{center}
\includegraphics[width=10cm,angle=0]{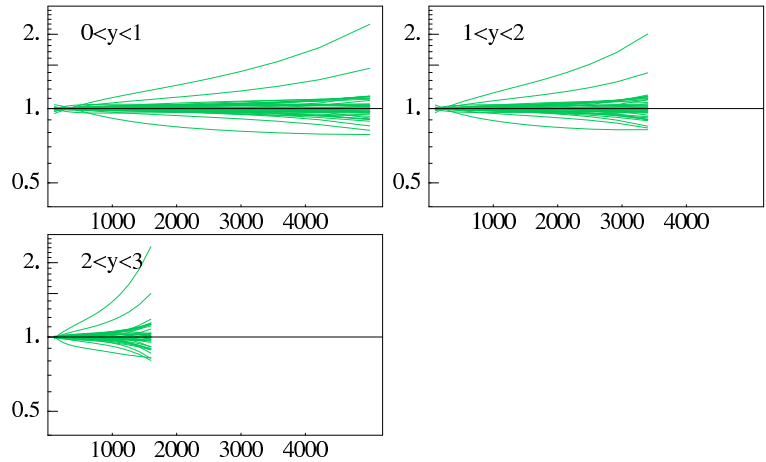}
\end{center}
\caption{
The ratios of the jet cross section predictions for the LHC using the 
CTEQ6.1 error pdfs to the prediction using the central pdf. The extremes are produced by eigenvector 15.} 
\label{fig:lhc_ratios}
\end{figure}

The increase of the centre-of-mass energy to $14$~TeV at the LHC will result in a  dramatically larger accessible
kinematic range. Inclusive jet cross sections can be measured out to transverse momentum values of order
$4$~TeV in the central region and $1.5$~TeV in  the forward region. The predictions with the CTEQ6.1 central
pdfs and the $40$ error pdfs are shown in Figure~\ref{fig:lhc_jet} and~\ref{fig:lhc_ratios} for three
different rapidity regions~\cite{Stump:2003yu}.
The cross sections were generated with a renormalization and
factorization scale equal to $p_T^{jet}/2$. The cross section predictions have a similar sensitivity  to the
error pdfs as do the jet cross sections at the Tevatron for similar $x_T$ values, and  the uncertainties on the
predicted cross sections remain up to a factor of $2$ at the highest $p_T$ values. Measurements of the jet cross
section over the full rapidity range at the LHC  will serve to further constrain the high $x$ gluon pdf and
distinguish between possible new  physics and uncertainties in pdfs. 

\begin{figure}[t]
\begin{center}
\includegraphics[width=9cm]{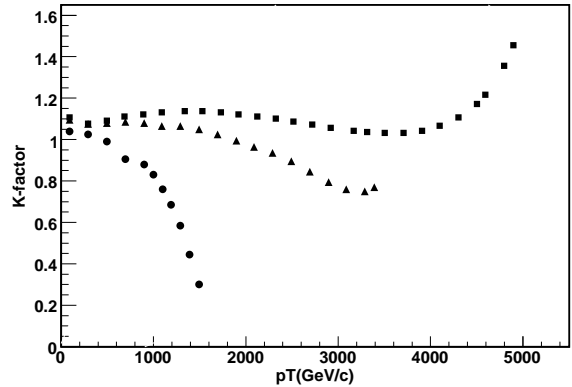}
\end{center}
\vspace*{-0.5cm}
\caption{
The ratios of the NLO to LO jet cross section predictions for the LHC using the 
CTEQ6.1 pdfs for the three different rapidity regions ($0$--$1$ (squares), $1$--$2$ (triangles),
$2$--$3$ (circles)). 
\label{fig:Kfactors}
}
\end{figure}

It is useful to plot the $K$-factors (the ratio of the NLO to LO cross sections) for  the three  different
rapidity intervals shown above. As discussed previously in Section~\ref{sec:kfac}, the value of the $K$-factor is a scale-dependent
quantity; the $K$-factors shown in Figure~\ref{fig:Kfactors} are calculated with the nominal scale of $p_T^{jet}/2$.
The $K$-factors have a somewhat complicated shape due to the interplay between the different subprocesses comprising
inclusive jet production and the behaviours of the relevant pdfs in the different regions of parton momentum
fraction $x$. In the central region, the $K$-factor is within $10\%$ of unity for the observable range. There are
no new parton-parton subprocesses that contribute at NLO but not at LO. Thus a LO prediction, using the NLO
CTEQ6.1 pdfs, will reproduce fairly closely the NLO calculation. For rapidities between $1$ and $2$, the
$K$-factor is within $20\%$ of unity, dropping below one at higher transverse momentum. For forward rapidities,
the $K$-factor drops almost immediately below one, due to the behaviour of the high-$x$ pdfs that contribute  to
the cross  section in this region. There is nothing wrong with the NLO prediction in this region; its
relationship to the LO cross section has just changed due to the kinematics. LO predictions in this region will
provide an overestimate of the NLO cross section.  

We saw in Section~\ref{sec:corrections} that jets in the upper range of transverse momentum values at the Tevatron were very collimated. This will be even more the case at the LHC, where in the multi-TeV range, a large fraction of the jet's momentum will be contained in a single calorimeter tower. 
Jet events at the LHC will be much more active than events in a similar $p_T$ range at the Tevatron. The
majority of the dijet production for the transverse momentum range less than $1$~TeV will be with a $gg$ 
initial state. As discussed previously, the larger colour factor associated with the gluon and the greater
phase space available at the LHC for gluon emission  will result in an increased production of additional soft
jets. In addition, there is an increased probability for the production of ``mini-jets'' from multiple-parton
scattering among the spectator partons. At full design luminosity, on the order of $25$ additional minimum bias interactions
will be present at each crossing. Such events, either singly or in combination, may create additional jets. As a result,
the minimum jet transverse momentum requirement may need to be increased for most analyses; in addition, it may be
advantageous to use smaller cone sizes than used for similar analyses at the Tevatron. 

\begin{figure}[t]
\begin{center}
\includegraphics[width=7cm,angle=90]{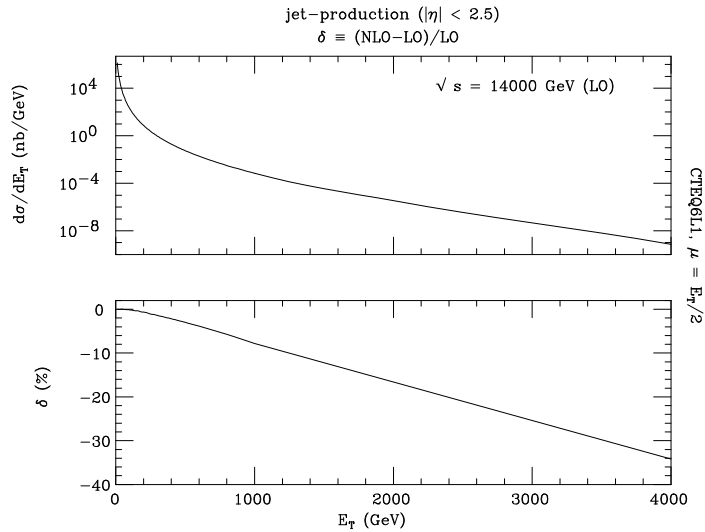}
\end{center}
\caption{
The effect of electroweak logarithms on jet cross sections at the LHC.} 
\label{fig:lhc_logs}
\end{figure}

As discussed above, the LHC will produce jets at transverse momenta far larger than the 
$W$ mass. Thus, virtual electroweak corrections of the form $\alpha_W \log^2(p_T^2/M_W^2)$, 
where $\alpha_W=\alpha_{EM}\sin^2\theta_W$, may become important. For example, at $p_T=3$~TeV,
$\log^2(p_T^2/m_W^2)=10$. These double logs are a result of the lack of cancellation 
between real and virtual $W$ emission in higher order calculations and may become competitive 
in size with NLO QCD corrections. The impact of a subset of the full EW corrections (i.e. those
involving virtual electroweak boson emission only) reaches of the order of 
$40\%$ at $4$~TeV, as shown in Figure~\ref{fig:lhc_logs} below~\cite{Moretti:2005ut}.
This is an indication that  all Standard Model corrections, and not just those from higher order QCD, need to
be understood in order to correctly interpret any new physics that  may be present at the LHC.

\section{Summary}
\label{sec:summary}

The LHC will present an unprecedented potential for the discovery of new physics. Thus, it is even more important to understand both
the possible production mechanisms for the new physics and the Standard Model backgrounds for their signals. The experiences gained
at the Tevatron, and the phenomenological tools developed in the last few years, will serve as a basis towards the establishment of
the relevant Standard Model benchmarks at the LHC. In this paper, we have put together a primer to aid this establishment. 

For convenience, we summarize below the rules-of-thumb and official recommendations discussed in our paper.
First, the rules-of-thumb:

\begin{itemize}
\item In Section~\ref{sec:kfac}, we present $K$-factors for some important processes at the Tevatron and the LHC. 

\item In Section~\ref{sec:sudakov}, we present some plots of Sudakov form factors useful for calculating the probability for producing jets from initial state radiation. 

\item In Section~\ref{sec:lum}, we provide pdf luminosities and their uncertainties for the LHC. 
\end{itemize}

Next, our official recommendations:

\begin{itemize}
\item In Section~\ref{sec:uncertainties}, we give our official recommendation for the form of the Master Equation to be used to calculate pdf uncertainties. 

\item In Section~\ref{sec:nlolopdfs}, we recommend that NLO pdfs be used for Monte Carlo event generations. 

\item In Section~\ref{sec:corrections}, we recommend that experimental cross section measurements be presented at the hadron level, and that corrections between the parton and hadron levels be clearly stated. 

\item In Section~\ref{sec:jetalgdata}, we present a revised version of the midpoint jet cone algorithm and recommend its use at the Tevatron and LHC. We also recommend the use of both cone and $k_T$ jet algorithms for all analyses.  


\end{itemize}

\section*{Acknowledgements}

We would like to thank J.~Andersen, A.~Belyaev, B.~Cooper, A.~de Roeck, L.~Dixon, S.~Ellis, R.~Field, S.~Frixione, S.~Gieseke,
J.~Heyninck, M.~Mangano, A.~Messina, S.~Mrenna, S.~Moretti, I.~Puljak, J.~Pumplin, P.~Skands, R.~Thorne, M.~Toennesmann and S.~Tsuno.


\section*{References}

\begin{thebibliography}{99}

\bibitem{DRELL}
S.D.~Drell and T.M.~Yan, Ann. Phys. {\bf 66} (1971) 578.

\bibitem{FACT}
J.C.~Collins and D.E.~Soper, Ann. Rev. Nucl. Part. Sci. {\bf 37} (1987) 33.


\bibitem{DelDuca:2001fn}
  V.~Del Duca, W.~Kilgore, C.~Oleari, C.~Schmidt and D.~Zeppenfeld,
  Nucl.\ Phys.\ B {\bf 616} (2001) 367 
  [arXiv:hep-ph/0108030].

\bibitem{Harlander:2002wh}
  R.~V.~Harlander and W.~B.~Kilgore,
  Phys.\ Rev.\ Lett.\  {\bf 88}, 201801 (2002)
  [arXiv:hep-ph/0201206].
  
\bibitem{Anastasiou:2002yz}
  C.~Anastasiou and K.~Melnikov,
  Nucl.\ Phys.\ B {\bf 646}, 220 (2002)
  [arXiv:hep-ph/0207004].

\bibitem{Zeppenfeld:2000td}
 D.~Zeppenfeld, R.~Kinnunen, A.~Nikitenko and E.~Richter-Was,
 Phys.\ Rev.\ D {\bf 62} (2000) 013009
 [arXiv:hep-ph/0002036].

\bibitem{Djouadi:2005gi}
  A.~Djouadi,
  arXiv:hep-ph/0503172.

\bibitem{Ellis:1991qj}
  R.~K.~Ellis, W.~J.~Stirling and B.~R.~Webber,
  Camb.\ Monogr.\ Part.\ Phys.\ Nucl.\ Phys.\ Cosmol.\  {\bf 8}, 1 (1996).


\bibitem{DGLAP}
L.N.~Lipatov, Sov. J. Nucl. Phys. {\bf 20} (1975) 95; \\
V.N.~Gribov and L.N.~Lipatov, Sov. J. Nucl. Phys. {\bf 15} (1972) 438; \\
G.~Altarelli and G.~Parisi, Nucl. Phys. {\bf B126} (1977) 298; \\
Yu.L.~Dokshitzer, Sov. Phys. JETP {\bf 46} (1977) 641.

\bibitem{Martin:2004ir}
  A.~D.~Martin, R.~G.~Roberts, W.~J.~Stirling and R.~S.~Thorne,
  Phys.\ Lett.\ B {\bf 604}, 61 (2004)
  [arXiv:hep-ph/0410230].



\bibitem{Stump:2003yu}
  D.~Stump, J.~Huston, J.~Pumplin, W.~K.~Tung, H.~L.~Lai, S.~Kuhlmann and J.~F.~Owens,
  JHEP {\bf 0310}, 046 (2003)
  [arXiv:hep-ph/0303013].

\bibitem{Acosta:2004uq}
  D.~Acosta {\it et al.}  [CDF II Collaboration],
  Phys.\ Rev.\ Lett.\  {\bf 94}, 091803 (2005)
  [arXiv:hep-ex/0406078].

\bibitem{D0WZ} http://www-d0.fnal.gov/Run2Physics/WWW/results/prelim/EW/


\bibitem{Ellis:1992qq}
  S.~D.~Ellis, Z.~Kunszt and D.~E.~Soper,
  Phys.\ Rev.\ Lett.\  {\bf 69}, 3615 (1992)
  [arXiv:hep-ph/9208249].

\bibitem{Corcella:2000bw}
  G.~Corcella {\it et al.},
  JHEP {\bf 0101}, 010 (2001)
  [arXiv:hep-ph/0011363].

\bibitem{Sjostrand:2006za}
  T.~Sjostrand, S.~Mrenna and P.~Skands,
  JHEP {\bf 0605}, 026 (2006)
  [arXiv:hep-ph/0603175].

\bibitem{Mangano:2001xp}
  M.~L.~Mangano, M.~Moretti and R.~Pittau,
  Nucl.\ Phys.\ B {\bf 632}, 343 (2002)
  [arXiv:hep-ph/0108069].

\bibitem{Mangano:2002ea}
  M.~L.~Mangano, M.~Moretti, F.~Piccinini, R.~Pittau and A.~D.~Polosa,
  JHEP {\bf 0307}, 001 (2003)
  [arXiv:hep-ph/0206293].

\bibitem{Pukhov:1999gg}
  A.~Pukhov {\it et al.},
  arXiv:hep-ph/9908288.

\bibitem{Boos:2004kh}
  E.~Boos {\it et al.}  [CompHEP Collaboration],
  Nucl.\ Instrum.\ Meth.\ A {\bf 534}, 250 (2004)
  [arXiv:hep-ph/0403113].
  
\bibitem{Stelzer:1994ta}
  T.~Stelzer and W.~F.~Long,
  Comput.\ Phys.\ Commun.\  {\bf 81}, 357 (1994)
  [arXiv:hep-ph/9401258].

\bibitem{Maltoni:2002qb}
  F.~Maltoni and T.~Stelzer,
  JHEP {\bf 0302}, 027 (2003)
  [arXiv:hep-ph/0208156].

\bibitem{Caravaglios:1998yr}
  F.~Caravaglios, M.~L.~Mangano, M.~Moretti and R.~Pittau,
  Nucl.\ Phys.\ B {\bf 539}, 215 (1999)
  [arXiv:hep-ph/9807570].

\bibitem{Murayama:1992gi}
  H.~Murayama, I.~Watanabe and K.~Hagiwara,
KEK-91-11

\bibitem{Maltoni:2002mq}
  F.~Maltoni, K.~Paul, T.~Stelzer and S.~Willenbrock,
  Phys.\ Rev.\ D {\bf 67}, 014026 (2003)
  [arXiv:hep-ph/0209271].

\bibitem{Witten:2003nn}
  E.~Witten,
  Commun.\ Math.\ Phys.\  {\bf 252}, 189 (2004)
  [arXiv:hep-th/0312171].

\bibitem{Cachazo:2004kj}
  F.~Cachazo, P.~Svrcek and E.~Witten,
  JHEP {\bf 0409}, 006 (2004)
  [arXiv:hep-th/0403047].

\bibitem{Georgiou:2004wu}
  G.~Georgiou and V.~V.~Khoze,
  JHEP {\bf 0405}, 070 (2004)
  [arXiv:hep-th/0404072].
 
\bibitem{Wu:2004jx}
  J.~B.~Wu and C.~J.~Zhu,
  JHEP {\bf 0409}, 063 (2004)
  [arXiv:hep-th/0406146].
  
\bibitem{Georgiou:2004by}
  G.~Georgiou, E.~W.~N.~Glover and V.~V.~Khoze,
  JHEP {\bf 0407}, 048 (2004)
  [arXiv:hep-th/0407027].
  
\bibitem{Dixon:2004za}
  L.~J.~Dixon, E.~W.~N.~Glover and V.~V.~Khoze,
  JHEP {\bf 0412}, 015 (2004)
  [arXiv:hep-th/0411092].
    
\bibitem{Badger:2004ty}
  S.~D.~Badger, E.~W.~N.~Glover and V.~V.~Khoze,
  JHEP {\bf 0503}, 023 (2005)
  [arXiv:hep-th/0412275].

\bibitem{Bern:2004ba}
  Z.~Bern, D.~Forde, D.~A.~Kosower and P.~Mastrolia,
  Phys.\ Rev.\ D {\bf 72}, 025006 (2005)
  [arXiv:hep-ph/0412167].
    
\bibitem{Britto:2004ap}
  R.~Britto, F.~Cachazo and B.~Feng,
  Nucl.\ Phys.\ B {\bf 715}, 499 (2005)
  [arXiv:hep-th/0412308].

\bibitem{Britto:2005fq}
  R.~Britto, F.~Cachazo, B.~Feng and E.~Witten,
  Phys.\ Rev.\ Lett.\  {\bf 94}, 181602 (2005)
  [arXiv:hep-th/0501052].

\bibitem{Bloch:1937pw}
  F.~Bloch and A.~Nordsieck,
  Phys.\ Rev.\  {\bf 52}, 54 (1937).
  
\bibitem{Kinoshita:1962ur}
  T.~Kinoshita,
  J.\ Math.\ Phys.\  {\bf 3}, 650 (1962).
  
\bibitem{Lee:1964is}
  T.~D.~Lee and M.~Nauenberg,
  Phys.\ Rev.\  {\bf 133}, B1549 (1964).

\bibitem{Kajantie:1978qv}
  K.~Kajantie, J.~Lindfors and R.~Raitio,
  Nucl.\ Phys.\ B {\bf 144}, 422 (1978).

\bibitem{Kunszt:1992tn}
  Z.~Kunszt and D.~E.~Soper,
  Phys.\ Rev.\ D {\bf 46}, 192 (1992).

\bibitem{Ellis:1980wv}
  R.~K.~Ellis, D.~A.~Ross and A.~E.~Terrano,
  Nucl.\ Phys.\ B {\bf 178}, 421 (1981).
  
\bibitem{Frixione:1995ms}
  S.~Frixione, Z.~Kunszt and A.~Signer,
  Nucl.\ Phys.\ B {\bf 467}, 399 (1996)
  [arXiv:hep-ph/9512328].
  
\bibitem{Catani:1996vz}
  S.~Catani and M.~H.~Seymour,
  Nucl.\ Phys.\ B {\bf 485}, 291 (1997)
  [Erratum-ibid.\ B {\bf 510}, 503 (1997)]
  [arXiv:hep-ph/9605323].

\bibitem{Nagy:1996bz}
  Z.~Nagy and Z.~Trocsanyi,
  Nucl.\ Phys.\ B {\bf 486}, 189 (1997)
  [arXiv:hep-ph/9610498].
  
\bibitem{Fabricius:1981sx}
  K.~Fabricius, I.~Schmitt, G.~Kramer and G.~Schierholz,
  Z.\ Phys.\ C {\bf 11}, 315 (1981).
  
\bibitem{Giele:1991vf}
  W.~T.~Giele and E.~W.~N.~Glover,
  Phys.\ Rev.\ D {\bf 46}, 1980 (1992).


\bibitem{Gehrmann-DeRidder:2005cm}
  A.~Gehrmann-De Ridder, T.~Gehrmann and E.~W.~N.~Glover,
  JHEP {\bf 0509}, 056 (2005)
  [arXiv:hep-ph/0505111].

\bibitem{Glover:1996eh}
  E.~W.~N.~Glover and D.~J.~Miller,
  Phys.\ Lett.\ B {\bf 396}, 257 (1997)
  [arXiv:hep-ph/9609474].
  
\bibitem{Bern:1996ka}
  Z.~Bern, L.~J.~Dixon, D.~A.~Kosower and S.~Weinzierl,
  Nucl.\ Phys.\ B {\bf 489}, 3 (1997)
  [arXiv:hep-ph/9610370].
  
\bibitem{Campbell:1997tv}
  J.~M.~Campbell, E.~W.~N.~Glover and D.~J.~Miller,
  Phys.\ Lett.\ B {\bf 409}, 503 (1997)
  [arXiv:hep-ph/9706297].
  
\bibitem{Glover:2002gz}
  E.~W.~N.~Glover,
  Nucl.\ Phys.\ Proc.\ Suppl.\  {\bf 116}, 3 (2003)
  [arXiv:hep-ph/0211412].
  
\bibitem{Bern:1997sc}
  Z.~Bern, L.~J.~Dixon and D.~A.~Kosower,
  Nucl.\ Phys.\ B {\bf 513}, 3 (1998)
  [arXiv:hep-ph/9708239].

\bibitem{Badger:2004uk}
  S.~D.~Badger and E.~W.~N.~Glover,
  JHEP {\bf 0407}, 040 (2004)
  [arXiv:hep-ph/0405236].

\bibitem{H1}
H1 Collaboration: C.\ Adloff {\it et al.},
Eur.\ Phys.\ J.\ {\bf C 13} (2000) 609 [hep-ex/9908059]; %
Eur.\ Phys.\ J.\ {\bf C 19} (2001) 269 [hep-ex/0012052]; %
Eur.\ Phys.\ J.\ {\bf C 21} (2001) 33 [hep-ex/0012053].

\bibitem{ZEUS}
ZEUS Collaboration: S. Chekanov {\it et al.},
Eur.\ Phys.\  J.\ {\bf C 21} (2001) 443 [hep-ex/0105090];
A.M. Cooper-Sarkar,
Proceedings of International Europhysics Conference
on HEP 2001, Budapest [hep-ph/0110386].


\bibitem{E866}
E866 Collaboration: R.\ S.\ Towell {\it et al.},
Phys.\ Rev.\ {\bf D 64} (2001) 052002 [hep-ex/0103030].

\bibitem{CCFR2}
CCFR Collaboration: U.\ K.\ Yang {\it et al.},
Phys.\ Rev.\ Lett.\ {\bf 86} (2001) 2742 [hep-ex/0009041].

\bibitem{BCDMSp}
BCDMS Collaboration: A.\ C.\ Benvenuti {\it et al.},
Phys.\ Lett.\ {\bf B 223} (1989) 485.

\bibitem{BCDMSd}
BCDMS Collaboration: A.\ C.\ Benvenuti {\it et al.},
Phys.\ Lett.\ {\bf B 236} (1989) 592.

\bibitem{NMC} 
New Muon Collaboration: M.\ Arneodo {\it et al.},
Nucl.\ Phys.\ {\bf B 483} (1997) 3 [hep-ph/9610231]; and
M.~Arneodo {\it et al.},
Nucl.\ Phys.\ B {\bf 487} (1997) 3
[hep-ex/9611022].

\bibitem{CCFR3}
CCFR Collaboration: W.\ G.\ Seligman {\it et al.},
Phys.\ Rev.\ Lett.\ {\bf 79} (1997) 1213 [hep-ex/970107].

\bibitem{E605}
E605 Collaboration: G.\ Moreno {\it et al.},
Phys.\ Rev.\ {\bf D 43} (1991) 2815.


\bibitem{CDFjet}
CDF Collaboration: T. Affolder {\it et al.},
Phys.\ Rev.\ {\bf D 64} (2001) 032001 [hep-ph/0102074].


\bibitem{D0jet} D\O\ Collaboration: B. Abbott {\it et al.},
Phys.\ Rev.\ Lett.\ {\bf 86} (2001) 1707 [hep-ex/0011036];
and Phys.\ Rev.\ {\bf D 64} (2001) 032003 [hep-ex/0012046].


\bibitem{Moch:2004pa}
  S.~Moch, J.~A.~M.~Vermaseren and A.~Vogt,
  Nucl.\ Phys.\ B {\bf 688}, 101 (2004)
  [arXiv:hep-ph/0403192].
  
\bibitem{Vogt:2004mw}
  A.~Vogt, S.~Moch and J.~A.~M.~Vermaseren,
  Nucl.\ Phys.\ B {\bf 691}, 129 (2004)
  [arXiv:hep-ph/0404111].

\bibitem{Hamberg:1990np}
  R.~Hamberg, W.~L.~van Neerven and T.~Matsuura,
  Nucl.\ Phys.\ B {\bf 359}, 343 (1991)
  [Erratum-ibid.\ B {\bf 644}, 403 (2002)].
  
\bibitem{Anastasiou:2003yy}
  C.~Anastasiou, L.~J.~Dixon, K.~Melnikov and F.~Petriello,
  Phys.\ Rev.\ Lett.\  {\bf 91}, 182002 (2003)
  [arXiv:hep-ph/0306192].
  
\bibitem{Anastasiou:2004xq}
  C.~Anastasiou, K.~Melnikov and F.~Petriello,
  Phys.\ Rev.\ Lett.\  {\bf 93}, 262002 (2004)
  [arXiv:hep-ph/0409088].

\bibitem{Sterman:1986aj}
  G.~Sterman,
  Nucl.\ Phys.\ B {\bf 281}, 310 (1987).

\bibitem{Catani:1989ne}
  S.~Catani and L.~Trentadue,
  Nucl.\ Phys.\ B {\bf 327}, 323 (1989).

\bibitem{Catani:1990rp}
  S.~Catani and L.~Trentadue,
  Nucl.\ Phys.\ B {\bf 353}, 183 (1991).

\bibitem{Parisi:1979se}
  G.~Parisi and R.~Petronzio,
  Nucl.\ Phys.\ B {\bf 154}, 427 (1979).

\bibitem{Altarelli:1984pt}
  G.~Altarelli, R.~K.~Ellis, M.~Greco and G.~Martinelli,
  Nucl.\ Phys.\ B {\bf 246}, 12 (1984).

\bibitem{Collins:1981uk}
  J.~C.~Collins and D.~E.~Soper,
  Nucl.\ Phys.\ B {\bf 193}, 381 (1981)
  [Erratum-ibid.\ B {\bf 213}, 545 (1983)].

\bibitem{Collins:1981va}
  J.~C.~Collins and D.~E.~Soper,
  Nucl.\ Phys.\ B {\bf 197}, 446 (1982).

\bibitem{Collins:1984kg}
  J.~C.~Collins, D.~E.~Soper and G.~Sterman,
  Nucl.\ Phys.\ B {\bf 250}, 199 (1985).

\bibitem{Sterman:2004yk}
  G.~Sterman and W.~Vogelsang,
  Phys.\ Rev.\ D {\bf 71}, 014013 (2005)
  [arXiv:hep-ph/0409234].

\bibitem{Kulesza:2003wn}
  A.~Kulesza, G.~Sterman and W.~Vogelsang,
  Phys.\ Rev.\ D {\bf 69}, 014012 (2004)
  [arXiv:hep-ph/0309264].

\bibitem{Kulesza:2002vk}
  A.~Kulesza, G.~Sterman and W.~Vogelsang,
  arXiv:hep-ph/0207148.

\bibitem{Balazs:1997xd}
  C.~Balazs and C.~P.~Yuan,
  Phys.\ Rev.\ D {\bf 56}, 5558 (1997)
  [arXiv:hep-ph/9704258].
  
\bibitem{Banfi:2005mt}
  A.~Banfi, G.~Corcella, M.~Dasgupta, Y.~Delenda, G.~P.~Salam and G.~Zanderighi,
  arXiv:hep-ph/0508096.
 
\bibitem{Gieseke:2006ga}
  S.~Gieseke {\it et al.},
  arXiv:hep-ph/0609306.

\bibitem{Gleisberg:2003xi}
  T.~Gleisberg, S.~Hoche, F.~Krauss, A.~Schalicke, S.~Schumann and J.~C.~Winter,
  JHEP {\bf 0402}, 056 (2004)
  [arXiv:hep-ph/0311263].

\bibitem{Sjostrand:2004ef}
  T.~Sjostrand and P.~Z.~Skands,
  Eur.\ Phys.\ J.\ C {\bf 39}, 129 (2005)
  [arXiv:hep-ph/0408302].

\bibitem{Gieseke:2004tc}
  S.~Gieseke,
  JHEP {\bf 0501}, 058 (2005)
  [arXiv:hep-ph/0412342].

\bibitem{Blazey:2000qt}
G.~C.~Blazey {\it et al.},
arXiv:hep-ex/0005012.

\bibitem{Giele:1994gf}
  W.~T.~Giele, E.~W.~N.~Glover and D.~A.~Kosower,
  Phys.\ Rev.\ Lett.\  {\bf 73}, 2019 (1994)
  [arXiv:hep-ph/9403347].

\bibitem{Ellis:2001aa}
  S.~D.~Ellis, J.~Huston and M.~Tonnesmann,
in {\it Proc. of the APS/DPF/DPB Summer Study on the Future of Particle Physics (Snowmass 2001) } ed. N.~Graf,
  eConf {\bf C010630}, P513 (2001)
  [arXiv:hep-ph/0111434].

\bibitem{Boos:2001cv}
  E.~Boos {\it et al.},
  arXiv:hep-ph/0109068.

\bibitem{Catani:2001cc}
  S.~Catani, F.~Krauss, R.~Kuhn and B.~R.~Webber,
  JHEP {\bf 0111}, 063 (2001)
  [arXiv:hep-ph/0109231].

\bibitem{Gleisberg:2004hm}
  T.~Gleisberg, S.~Hoeche, F.~Krauss, A.~Schaelicke, S.~Schumann, G.~Soff and J.~Winter,
  Czech.\ J.\ Phys.\  {\bf 55}, B529 (2005)
  [arXiv:hep-ph/0409122].

\bibitem{Group:2006rt}
  T.~Q.~W.~Group {\it et al.},
  arXiv:hep-ph/0610012.
  
\bibitem{Mrenna:2003if}
  S.~Mrenna and P.~Richardson,
  JHEP {\bf 0405}, 040 (2004)
  [arXiv:hep-ph/0312274].

\bibitem{Chen:2001nf}
  Y.~J.~Chen, J.~Collins and X.~M.~Zu,
  JHEP {\bf 0204}, 041 (2002)
  [arXiv:hep-ph/0110257].

\bibitem{Collins:2004vq}
  J.~C.~Collins and X.~Zu,
  JHEP {\bf 0503}, 059 (2005)
  [arXiv:hep-ph/0411332].

\bibitem{Frixione:2002ik}
  S.~Frixione and B.~R.~Webber,
  JHEP {\bf 0206}, 029 (2002)
  [arXiv:hep-ph/0204244].


\bibitem{Frixione:2003ei}
  S.~Frixione, P.~Nason and B.~R.~Webber,
  JHEP {\bf 0308}, 007 (2003)
  [arXiv:hep-ph/0305252].


\bibitem{Frixione:2005gz}
  S.~Frixione and B.~R.~Webber,
  arXiv:hep-ph/0506182.


\bibitem{Kurihara:2002ne}
  Y.~Kurihara, J.~Fujimoto, T.~Ishikawa, K.~Kato, S.~Kawabata, T.~Munehisa and H.~Tanaka,
  Nucl.\ Phys.\ B {\bf 654}, 301 (2003)
  [arXiv:hep-ph/0212216].

\bibitem{Kramer:2003jk}
  M.~Kramer and D.~E.~Soper,
  Phys.\ Rev.\ D {\bf 69}, 054019 (2004)
  [arXiv:hep-ph/0306222].


\bibitem{Soper:2003ya}
  D.~E.~Soper,
  Phys.\ Rev.\ D {\bf 69}, 054020 (2004)
  [arXiv:hep-ph/0306268].


\bibitem{Kramer:2005hw}
  M.~Kramer, S.~Mrenna and D.~E.~Soper,
  Phys.\ Rev.\ D {\bf 73}, 014022 (2006)
  [arXiv:hep-ph/0509127].

\bibitem{Nagy:2005aa}
  Z.~Nagy and D.~E.~Soper,
  JHEP {\bf 0510}, 024 (2005)
  [arXiv:hep-ph/0503053].

\bibitem{Nagy:2006kb}
  Z.~Nagy and D.~E.~Soper,
  arXiv:hep-ph/0601021.

\cite{Frixione:2005vw}
\bibitem{Frixione:2005vw}
  S.~Frixione, E.~Laenen, P.~Motylinski and B.~R.~Webber,
  arXiv:hep-ph/0512250.





\bibitem{Alekhin:2005gq}
  S.~Alekhin,
  JETP Lett.\  {\bf 82}, 628 (2005)
  [Pisma Zh.\ Eksp.\ Teor.\ Fiz.\  {\bf 82}, 710 (2005)]
  [arXiv:hep-ph/0508248].

\bibitem{Adloff:2000qk}
  C.~Adloff {\it et al.}  [H1 Collaboration],
  Eur.\ Phys.\ J.\ C {\bf 21}, 33 (2001)
  [arXiv:hep-ex/0012053].

\bibitem{Adloff:2003uh}
  C.~Adloff {\it et al.}  [H1 Collaboration],
  Eur.\ Phys.\ J.\ C {\bf 30}, 1 (2003)
  [arXiv:hep-ex/0304003].


\bibitem{Chekanov:2002pv}
  S.~Chekanov {\it et al.}  [ZEUS Collaboration],
  Phys.\ Rev.\ D {\bf 67}, 012007 (2003)
  [arXiv:hep-ex/0208023].

\bibitem{Chekanov:2005nn}
  S.~Chekanov {\it et al.}  [ZEUS Collaboration],
  Eur.\ Phys.\ J.\ C {\bf 42}, 1 (2005)
  [arXiv:hep-ph/0503274].

\bibitem{Moch:2004sf}
  S.~Moch, J.~A.~M.~Vermaseren and A.~Vogt,
  Nucl.\ Phys.\ Proc.\ Suppl.\  {\bf 135}, 137 (2004)
  [arXiv:hep-ph/0408075].

\bibitem{Martin:2002dr}
  A.~D.~Martin, R.~G.~Roberts, W.~J.~Stirling and R.~S.~Thorne,
  Phys.\ Lett.\ B {\bf 531}, 216 (2002)
  [arXiv:hep-ph/0201127].

\bibitem{Gluck:1998xa}
  M.~Gluck, E.~Reya and A.~Vogt,
  Eur.\ Phys.\ J.\ C {\bf 5}, 461 (1998)
  [arXiv:hep-ph/9806404].

\bibitem{Abe:1998rv}
  F.~Abe {\it et al.}  [CDF Collaboration],
  Phys.\ Rev.\ Lett.\  {\bf 81}, 5754 (1998)
  [arXiv:hep-ex/9809001].



\bibitem{Stump:2001gu}
  D.~Stump {\it et al.},
  Phys.\ Rev.\ D {\bf 65}, 014012 (2002)
  [arXiv:hep-ph/0101051].

\bibitem{Pumplin:2001ct}
  J.~Pumplin {\it et al.},
  Phys.\ Rev.\ D {\bf 65}, 014013 (2002)
  [arXiv:hep-ph/0101032].

\bibitem{Martin:2002aw}
  A.~D.~Martin, R.~G.~Roberts, W.~J.~Stirling and R.~S.~Thorne,
  Eur.\ Phys.\ J.\ C {\bf 28}, 455 (2003)
  [arXiv:hep-ph/0211080].

\bibitem{Eidelman:2004wy}
  S.~Eidelman {\it et al.}  [Particle Data Group],
  Phys.\ Lett.\ B {\bf 592}, 1 (2004).

\bibitem{Pumplin:2005rh}
  J.~Pumplin, A.~Belyaev, J.~Huston, D.~Stump and W.~K.~Tung,
  JHEP {\bf 0602}, 032 (2006)
  [arXiv:hep-ph/0512167].


\bibitem{Collins:2002ey}
  J.~C.~Collins and X.~m.~Zu,
  JHEP {\bf 0206}, 018 (2002)
  [arXiv:hep-ph/0204127].


\bibitem{Plothow-Besch:1995ci}
  H.~Plothow-Besch,
  Int.\ J.\ Mod.\ Phys.\ A {\bf 10}, 2901 (1995).

\bibitem{Giele:2001mr}
  W.~T.~Giele, S.~A.~Keller and D.~A.~Kosower,
  arXiv:hep-ph/0104052.

\bibitem{Giele:2002hx}
  W.~Giele {\it et al.},
  arXiv:hep-ph/0204316.

\bibitem{Whalley:2005nh}
  M.~R.~Whalley, D.~Bourilkov and R.~C.~Group,
  arXiv:hep-ph/0508110.

\bibitem{Campbell:2000bg}
  J.~M.~Campbell and R.~K.~Ellis,
  Phys.\ Rev.\ D {\bf 62}, 114012 (2000)
  [arXiv:hep-ph/0006304].






\bibitem{Abulencia:2005aj}
  A.~Abulencia {\it et al.}  [CDF Collaboration],
  Phys.\ Rev.\ D {\bf 73}, 032003 (2006)
  [arXiv:hep-ex/0510048].

\bibitem{Alner:1987wb}
  G.~J.~Alner {\it et al.}  [UA5 Collaboration],
  Phys.\ Rept.\  {\bf 154}, 247 (1987).



\bibitem{Anastasiou:2003ds}
  C.~Anastasiou, L.~J.~Dixon, K.~Melnikov and F.~Petriello,
  Phys.\ Rev.\ D {\bf 69}, 094008 (2004)
  [arXiv:hep-ph/0312266].


\bibitem{cdfqcd} See http://www-cdf.fnal.gov/physics/new/qcd/QCD.html. 

\bibitem{Acosta:2004wq}
  D.~Acosta {\it et al.}  [CDF Collaboration],
  Phys.\ Rev.\ D {\bf 70}, 072002 (2004)
  [arXiv:hep-ex/0404004].

\cite{Affolder:2001xt}
\bibitem{Affolder:2001xt}
  A.~A.~Affolder {\it et al.}  [CDF Collaboration],
  Phys.\ Rev.\ D {\bf 65}, 092002 (2002).


\bibitem{Ellis:1990ek}
  S.~D.~Ellis, Z.~Kunszt and D.~E.~Soper,
  Phys.\ Rev.\ Lett.\  {\bf 64}, 2121 (1990).


\bibitem{Acosta:2005ix}
  D.~Acosta {\it et al.}  [CDF Collaboration],
  arXiv:hep-ex/0505013.


\bibitem{Abulencia:2005yg}
  A.~Abulencia {\it et al.}  [CDF Run II Collaboration],
  arXiv:hep-ex/0512020.

\bibitem{Abulencia:2005jw}
  A.~Abulencia {\it et al.}  [CDF II Collaboration],
  Phys.\ Rev.\ Lett.\  {\bf 96}, 122001 (2006)
  [arXiv:hep-ex/0512062].



\bibitem{Acosta:2004hw}
  D.~Acosta {\it et al.}  [CDF Collaboration],
  Phys.\ Rev.\ D {\bf 71}, 052003 (2005)
  [arXiv:hep-ex/0410041].

\bibitem{Campbell:2006xx}
  J.~M.~Campbell, R.~Keith Ellis and G.~Zanderighi,
  arXiv:hep-ph/0608194.

\bibitem{Campbell:2004sp}
  J.~Campbell and J.~Huston,
  Phys.\ Rev.\ D {\bf 70}, 094021 (2004).







\bibitem{Acosta:2005am}
  D.~Acosta {\it et al.}  [CDF Collaboration],
  Phys.\ Rev.\ D {\bf 72}, 052003 (2005)
  [arXiv:hep-ex/0504053].

\bibitem{cdftop}	http://www-cdf.fnal.gov/physics/new/top/2006/xs\_ann/public.html


\bibitem{cdfsvx} $http://www-cdf.fnal.gov/physics/new/top/2006/xs\_ljetsvx/public.html$

\bibitem{tevtopmass} http://tevewwg.fnal.gov/top/

\bibitem{Hill:2002ap}
  C.~T.~Hill and E.~H.~Simmons,
  Phys.\ Rept.\  {\bf 381}, 235 (2003)
  [Erratum-ibid.\  {\bf 390}, 553 (2004)]
  [arXiv:hep-ph/0203079].

\bibitem{Andersen:2001ja}
  J.~R.~Andersen, V.~Del Duca, F.~Maltoni and W.~J.~Stirling,
  JHEP {\bf 0105}, 048 (2001)
  [arXiv:hep-ph/0105146].

\bibitem{Andersen:2006sp}
  J.~R.~Andersen,
  Phys.\ Lett.\ B {\bf 639}, 290 (2006)
  [arXiv:hep-ph/0602182].

\bibitem{CMS_tdr} http://cms.cern.ch/iCMS/

\bibitem{ATLAS_tdr} http://atlas.web.cern.ch/Atlas/GROUPS/PHYSICS/TDR/access.html

\bibitem{Buttar:2006zd}
  C.~Buttar {\it et al.},
  arXiv:hep-ph/0604120.


\bibitem{Pukhov:2004ca}
  A.~Pukhov,
  arXiv:hep-ph/0412191.


\bibitem{Alekhin} 
S.~Alekhin,
[arXiv:hep-ph/0311184];
Phys.\ Rev.\ D {\bf 68}, 014002 (2003)
[arXiv:hep-ph/0211096].

\bibitem{Martin:2003sk}
A.~D.~Martin, R.~G.~Roberts, W.~J.~Stirling and R.~S.~Thorne,
Eur.\ Phys.\ J.\ C {\bf 35}, 325 (2004)
[arXiv:hep-ph/0308087].

\bibitem{mrstnnlo} 
A.~D.~Martin, R.~G.~Roberts, W.~J.~Stirling and R.~S.~Thorne,
Phys.\ Lett.\ B {\bf 531}, 216 (2002)
[arXiv:hep-ph/0201127].

\bibitem{Huston:2005jm}
  J.~Huston, J.~Pumplin, D.~Stump and W.~K.~Tung,
  JHEP {\bf 0506}, 080 (2005)
  [arXiv:hep-ph/0502080].

\bibitem{Bern:1996je}
  Z.~Bern, L.~J.~Dixon and D.~A.~Kosower,
  Ann.\ Rev.\ Nucl.\ Part.\ Sci.\  {\bf 46}, 109 (1996)
  [arXiv:hep-ph/9602280].


\bibitem{Cachazo:2004zb}
  F.~Cachazo, P.~Svrcek and E.~Witten,
  JHEP {\bf 0410}, 074 (2004)
  [arXiv:hep-th/0406177].

\bibitem{Brandhuber:2004yw}
  A.~Brandhuber, B.~J.~Spence and G.~Travaglini,
  Nucl.\ Phys.\ B {\bf 706}, 150 (2005)
  [arXiv:hep-th/0407214].

\bibitem{Brandhuber:2005jw}
  A.~Brandhuber, S.~McNamara, B.~J.~Spence and G.~Travaglini,
  JHEP {\bf 0510}, 011 (2005)
  [arXiv:hep-th/0506068].

\bibitem{Bern:2005hs}
  Z.~Bern, L.~J.~Dixon and D.~A.~Kosower,
  Phys.\ Rev.\ D {\bf 71}, 105013 (2005)
  [arXiv:hep-th/0501240].

\bibitem{Forde:2005hh}
  D.~Forde and D.~A.~Kosower,
  arXiv:hep-ph/0509358.

\bibitem{Ferroglia:2002mz}
  A.~Ferroglia, M.~Passera, G.~Passarino and S.~Uccirati,
  Nucl.\ Phys.\ B {\bf 650}, 162 (2003)
  [arXiv:hep-ph/0209219].
  
\bibitem{Anastasiou:2005cb}
  C.~Anastasiou and A.~Daleo,
  arXiv:hep-ph/0511176.
  
\bibitem{Giele:2004iy}
  W.~T.~Giele and E.~W.~N.~Glover,
  JHEP {\bf 0404}, 029 (2004)
  [arXiv:hep-ph/0402152].
  
\bibitem{Giele:2004ub}
  W.~Giele, E.~W.~N.~Glover and G.~Zanderighi,
  Nucl.\ Phys.\ Proc.\ Suppl.\  {\bf 135}, 275 (2004)
  [arXiv:hep-ph/0407016].

\bibitem{delAguila:2004nf}
  F.~del Aguila and R.~Pittau,
  JHEP {\bf 0407}, 017 (2004)
  [arXiv:hep-ph/0404120].
  
\bibitem{vanHameren:2005ed}
  A.~van Hameren, J.~Vollinga and S.~Weinzierl,
  Eur.\ Phys.\ J.\ C {\bf 41}, 361 (2005)
  [arXiv:hep-ph/0502165].
  
\bibitem{Binoth:2005ff}
  T.~Binoth, J.~P.~Guillet, G.~Heinrich, E.~Pilon and C.~Schubert,
  JHEP {\bf 0510}, 015 (2005)
  [arXiv:hep-ph/0504267].
  
\bibitem{Ellis:2005zh}
  R.~K.~Ellis, W.~T.~Giele and G.~Zanderighi,
  arXiv:hep-ph/0508308.
  
\bibitem{Ellis:2005qe}
  R.~K.~Ellis, W.~T.~Giele and G.~Zanderighi,
  Phys.\ Rev.\ D {\bf 72}, 054018 (2005)
  [arXiv:hep-ph/0506196].


\bibitem{Flanagan:2005xv}
  G.~U.~Flanagan,
FERMILAB-THESIS-2005-58

\bibitem{Balazs:2000sz}
  C.~Balazs, J.~Huston and I.~Puljak,
  Phys.\ Rev.\ D {\bf 63}, 014021 (2001)
  [arXiv:hep-ph/0002032].

\bibitem{Collins:1985gm}
  J.~C.~Collins, D.~E.~Soper and G.~Sterman,
  Nucl.\ Phys.\ B {\bf 263}, 37 (1986).

\bibitem{Berge:2004nt}
  S.~Berge, P.~Nadolsky, F.~Olness and C.~P.~Yuan,
  Phys.\ Rev.\ D {\bf 72}, 033015 (2005)
  [arXiv:hep-ph/0410375].

\bibitem{Fadin:1975cb}
  V.~S.~Fadin, E.~A.~Kuraev and L.~N.~Lipatov,
  Phys.\ Lett.\ B {\bf 60}, 50 (1975).

\bibitem{Kuraev:1976ge}
  E.~A.~Kuraev, L.~N.~Lipatov and V.~S.~Fadin,
  Sov.\ Phys.\ JETP {\bf 44}, 443 (1976)
  [Zh.\ Eksp.\ Teor.\ Fiz.\  {\bf 71}, 840 (1976)].

\bibitem{Kuraev:1977fs}
  E.~A.~Kuraev, L.~N.~Lipatov and V.~S.~Fadin,
  Sov.\ Phys.\ JETP {\bf 45}, 199 (1977)
  [Zh.\ Eksp.\ Teor.\ Fiz.\  {\bf 72}, 377 (1977)].

\bibitem{Balitsky:1978ic}
  I.~I.~Balitsky and L.~N.~Lipatov,
  Sov.\ J.\ Nucl.\ Phys.\  {\bf 28}, 822 (1978)
  [Yad.\ Fiz.\  {\bf 28}, 1597 (1978)].


\bibitem{Berge:2005nm}
  S.~Berge, P.~M.~Nadolsky, F.~I.~Olness and C.~P.~Yuan,
  AIP Conf.\ Proc.\  {\bf 792}, 722 (2005)
  [arXiv:hep-ph/0508215].


\bibitem{Berger:2004pc}
  E.~L.~Berger and J.~Campbell,
  Phys.\ Rev.\ D {\bf 70}, 073011 (2004)
  [arXiv:hep-ph/0403194].

\cite{DelDuca:2004wt}
\bibitem{DelDuca:2004wt}
  V.~Del Duca, A.~Frizzo and F.~Maltoni,
  JHEP {\bf 0405}, 064 (2004)
  [arXiv:hep-ph/0404013].

\bibitem{Moretti:2005ut}
  S.~Moretti, M.~R.~Nolten and D.~A.~Ross,
  arXiv:hep-ph/0503152.


\end{thebibliography}
\end{document}